%% file: tools_for_tree_amplitudes_penult.tex
\documentclass[12pt,letterpaper,showkeys]{article}
\usepackage{jheplikemod,tocloft,ifpdf,tocloft,rotating,amsmath,mathrsfs,multirow,mathtools,xcolor,cancel,shuffle,tikz,bbm}
\usetikzlibrary{calc,decorations.markings,shapes.geometric}
\usepackage[backgroundcolor=green!5, linecolor=blue!60]{todonotes}
\hypersetup{pdftitle={Tools for Tree Amplitudes},pdfcreator={},linkcolor=[rgb]{0.15,0.35,0.75},colorlinks=true,citecolor=[rgb]{0.675,0,0.2},urlcolor=[rgb]{0.15,0.35,0.65}}
\usepackage[T1]{fontenc}
\usepackage{lmodern}
\usepackage{makeidx}
\usepackage[totoc,itemlayout=singlepar,column sep=50pt]{idxlayout}

\input{header_data.tex}

\makeindex

\title{\texorpdfstring{~\\[-20pt]{\huge Computational Tools for Trees}\\[-6pt]in Gauge Theory and Gravity}{Tools for Trees in Gauge Theory and Gravity}}
\author[]{\vspace{-12pt}Jacob~L.~Bourjaily\!}\emailAdd{bourjaily@psu.edu}
\affiliation[]{Institute for Gravitation and the Cosmos, Department of Physics,\\Pennsylvania State University, University Park, PA 16802, USA}
\abstract{%
We describe a new set of public, self-contained, and versatile computational tools for the investigation, manipulation, and evaluation of tree-level amplitudes in pure (super)Yang-Mills and (super)Gravity, $\phi^p$-scalar field theory, and various other theories related to these through the double-copy.

The package brings together a diverse set of frameworks for representing amplitudes, from twistor string theory and scattering equations, to KLT and the double-copy, to on-shell recursion and the (oriented) positroid geometry of the amplituhedron. In addition to checking agreement across frameworks, we have made it easy to test many of the non-trivial relations satisfied by amplitudes, their components and building blocks, including: Ward identities, KK and BCJ relations, soft theorems, and the $E_{7(7)}$ structure of maximal supergravity. 

Beyond providing a coherent and consistent implementation of many well known (if not publicly available) results, our package includes a number of results well beyond what has existed in the present literature---from local, covariant, manifestly color-kinematic-dual representations of amplitudes for gluons/gravitons at \emph{arbitrary multiplicity}, to a complete classification of Yangian invariants via \emph{oriented} homology in the amplituhedron.

The \textsc{Mathematica} package \package\, and a notebook illustrating its functionality are available as ancillary files attached to this work's page on the \texttt{arXiv}.
}
\preprint{}

\begin{document}
\maketitle
\addtocontents{toc}{\protect\thispagestyle{empty}}
%
\thispagestyle{empty}
\newpage\setcounter{page}{0}\pagenumbering{arabic}
\vspace{0pt}\section{\texorpdfstring{Introduction and Overview}{Introduction and Overview}}\label{sec:introduction}\vspace{-0pt}

\subsection{\emph{Spiritus Movens}}\label{subsec:spiritus_movens}

Recent years have been witness to enormous advances in both our \emph{understanding of} and our \emph{ability to use} quantum field theory to make predictions for experiments (both real and imagined). This has been especially true for quantum field theories of \emph{massless} particles in \emph{four} spacetime dimensions. Arguably, the present revolution owes its beginning to the discovery by Parke and Taylor in \cite{Parke:1986gb} that `color-stripped'(/`partial') tree-amplitudes involving particular sets of of external gluon helicities (`MHV' type) displayed a universal, shocking simplicity for \emph{arbitrary multiplicity}. For example, they had conjectured that:\vspace{5pt}
\mathematicaBox{
\mathematicaSequence{\funL[]{nice}@\funL[]{componentAmp}\brace{\stateYM{m},\stateYM{p},\stateYM{p},\stateYM{p},\stateYM{m},\stateYM{p},\stateYM{p},\stateYM{p},\stateYM{p}}}{
$\displaystyle\frac{\ab{1\,5}^4}{\ab{1\,2}\ab{2\,3}\ab{3\,4}\ab{4\,5}\ab{5\,6}\ab{6\,7}\ab{7\,8}\ab{8\,9}\ab{9\,1}}$}
}\\[-4pt]
Such shocking simplicity would seem impossibly obscured by the Feynman rules, which suggest that complexity amplitudes should (and rapidly!) with multiplicity---regardless of which external polarizations were chosen. (In the 9-particle example above, this amplitude would involve (\funL{numberOfYMfeynmanDiagrams}\brace{9}\,=\!)559\,405 Feynman diagrams!)

It is worth noting that Parke and Taylor were only able to make their discovery because they both \emph{knew the correct result} for several non-trivial examples of such amplitudes from their previous work \cite{Parke:1985pn,Parke:1985ax} \emph{and they had them on hand, in machine-usable form}, making it easy to numerically check their inspired guess. The value of having the `right answer' readily available to scrutinize should not be underestimated.\\[-5pt]

It would take many years for Parke and Taylor's discovered simplicity to be explained and ultimately generalized to all helicity configurations of gluons. One such explanation would come from the discovery of on-shell recursion relations by Britto, Cachazo, and Feng in \cite{BCF} (see also \cite{BCFW}). Not only were these new, recursive formulae \emph{dramatically} more efficient than previous methods (\emph{immediately} giving the Parke-Taylor `guess' for so-called MHV amplitudes, for example), but they exposed many previously-unimagined symmetries and structures underlying these amplitudes. Among other things, the particular representations of tree-level scattering amplitudes resulting from on-shell recursion played a key role in: the discovery of dual-conformal (and ultimately, Yangian-)invariance \mbox{\cite{Drummond:2006rz,Alday:2007hr,Drummond:2008vq,Drummond:2009fd}} of amplitudes in planar, maximally supersymmetric Yang-Mills theory; the connection between on-shell diagrams and subspaces of Grassmannian manifolds \cite{ArkaniHamed:2009dn,Kaplan:2009mh,ArkaniHamed:2012nw,Bourjaily:2012gy,Arkani-Hamed:2014bca} and the amplituhedron \cite{ArkaniHamed:2010gg,Arkani-Hamed:2013jha,Arkani-Hamed:2013kca,Bourjaily:2018bbb}; and remarkable new formulae for perturbative scattering amplitudes at the integrand especially in the planar limit (see e.g.~\cite{ArkaniHamed:2010gh,Bourjaily:2013mma,Bourjaily:2015jna,Bourjaily:2023apy}) and beyond (see e.g.~\cite{Bourjaily:2016mnp,Bourjaily:2018omh,Bourjaily:2019gqu,Bourjaily:2019iqr,Bourjaily:2021hcp,Bourjaily:2021iyq}).

\newpage

This story and many others like it are often presented as inspiration and evidence for the existence of a possible revolutionary reformulation of the basic rules of quantum field theory and our understanding thereof. Although this revolution is arguably still underway, many of the key results have matured to the point that they can be found in a variety of introductions to quantum field theory---from mainstream textbooks (e.g.~\cite{Schwartz:2014sze}) to the more casual introductions (e.g.~\cite{Zee:2010qce}). Today, there are a wide variety of pedagogical introductions, lecture series, and reviews to help new students learn about the many developments, ideas, and remarkable forms that amplitudes takes (see e.g.~\cite{Dixon:1996wi,Bedford:2005yy,Alday:2008zza,Dixon:2013uaa,Cheung:2017pzi,HennPlefka,Strominger:2017zoo,Travaglini:2022uwo,Bern:2022wqg,Badger:2023eqz})---and even an excellent textbook \emph{Scattering Amplitudes in Gauge Theory and Gravity} by Elvang and Huang \cite{Elvang:2013cua} for a more thorough treatment.

One goal of this work is to provide a supplemental resource for any student first encountering these ideas through a text such as \cite{Elvang:2013cua}. We hope the \package\, package will serve as a practical playground in which a student can easily test and develop her understanding, and with which she may verify for herself many of the most remarkable results discovered in recent years. As a source of reliable reference material, sewn together using a consistent set of conventions and the tools to scrutinize and manipulate results, we hope it may someday help her check her own inspired guess---one just as surprising as impactful as that made by Parke and Taylor in \cite{Parke:1986gb}.\\[-5pt]

Another goal of this work is to provide a concrete consolidation of a wide variety of well-established facts, formulae, and techniques used in scattering amplitudes research. Even the most beautiful, compact formulae can involve considerable subtlety in notation, often requiring involved translations from more familiar ideas, notation, conventions, variables, etc. And even the most reliable researchers can make mistakes when typesetting results for printed publication. Indeed, it would not be hard to point to examples where subtle typos---most often signs, factors of 2, or ranges of indices for summation---have propagated verbatim through the literature for years without correction. (To be clear, the present author is not free of such blunders.) 

Because of this, it has become common to expect truly reliable formulae to be found in code attached to work as ancillary files. And it is not hard to find examples where an ancillary file represents the near-entirety of a critical result (see e.g.~\cite{Carrasco:2021otn,Bourjaily:2016evz} among many). But such attachments are rarely more than snippets of code, with occasional hints toward a broader set of (private, rarely documented) computational tools; and such code as may be provided, frequently relies upon idiosyncratic conventions or quixotic choices that must be unpacked to use.

Of course, there are a number of excellent examples of public, versatile code bases being made available to great effect both targeted towards the amplitudes community (see e.g.~\mbox{\cite{Maitre:2007jq,Dixon:2010ik,Bourjaily:2010wh,Bern:2013pya,Cullen:2013cka,Cullen:2013mza,vanDeurzen:2014uaa,Kuczmarski:2014ara,Panzer:2014caa,AccettulliHuber:2023ldr}}) and tools more motivated by phenomenological applications (e.g., to name a small number of the many good examples,~\mbox{\cite{Mertig:1990an,Shtabovenko:2020gxv,Gleisberg:2003xi,Alwall:2014hca}}). But a general tool box versatile enough to span both amplitudes in both gauge theory and gravity (even at tree-level) has been lacking. Building tools into single framework with a coherence of conventions makes their correctness easy to verify.

It would be worthwhile to extend this project to include also the many results known beyond leading order of perturbation theory, and to a broader class of quantum field theories. A consolidation of results in amplitudes similar to \cite{Bogner:2018bvz} for period integrals would be an incredibly worthwhile endeavor. But the line must be drawn somewhere, and so we must leave such ambitions to future work (and to others).\\[-5pt]

\vspace{-5pt}\subsection{Organization and Outline}

The goal of this work is to document and explain the conventions used in the \package\, package. We start in \mbox{section~\ref{section:novel_features}} with a \emph{brief} summary of the new features and functionality provided by the \package\, package (especially in the context of existing public codes). 

In \mbox{section~\ref{sec:basic_aspects_of_amplitudes}} we review the key variables, notation, conventions, and related concepts used by the \package. In addition to providing a coherent (if concise) summary of how the key ingredients are defined, we hope \mbox{section~\ref{sec:basic_aspects_of_amplitudes}} will serve as a clear motivation for and illustration of the notation used by the package to denote kinematic invariants, \fun{superFunction}s, and the external particle states. 

Even for those readers familiar with the kinematic language of spinor variables, momentum twistors, etc., we suspect that the unified language for labelling states of $\mathcal{N}$-supersymmetric sYM and sGR require some clarity and explanation; these are clarified in \mbox{section~\ref{subsec:labeling_states}}. Also, our use of \funL{superFunction} objects is likely to be idiosyncratic, and a helpful summary can be found there.

In \mbox{section~\ref{sec:using_the_package}} we describe how the \package\, package can be downloaded and installed, and illustrate a few examples of its typical use. More thorough explanations can be found within the demonstration notebook provided along with the primary source-code of \package\!\textbf{\t{.m}}. 

We conclude in \mbox{section~\ref{sec:conclusions}} with a short discussion of some of the most important additional functionality that remains to be added to the package. 

The primary documentation of the package is included in this work's appendices. Importantly, the outline of these appendices is identical to that of the source-code in \package\!\textbf{\t{.m}}. In particular, the functions provided are organized exactly according to those here---which can be accessed using \funL{definedFunctionCategories}, as illustrated in the \mbox{\hyperlink{context_organization_of_appendix}{appendix}}. 

Finally, for the sake of easy cross-referencing, we have provided an alphabetical index of all functions and symbols used by the \package\, package in \mbox{the \hyperlink{index}{index}}. 

\newpage
\section[\texorpdfstring{Novel Features and Functionality of the \package\, Package}{Novel Features and Functionality of the tree\_amplitudes Package}]{Novel Features \& Functionality of \package\, Package}\label{section:novel_features}

The \package\, package aims to cover a wide range of results at tree-level related to amplitudes in pure (or supersymmetric) gauge and gravity theories, and with a handful of other theories included as well---all results implemented with a great deal of redundancy. Because the `best' formulae often depends on context or intended use, we have implemented many distinct algorithms, strategies, and implementations thereof---with many potential use cases in mind.

For example, the \emph{fastest} way to \emph{abstractly represent} the results of on-shell recursion in sYM must certainly be its combinatorial form defined directly in terms of the labels of positroids (implemented in \mbox{\funL{ymPositroidAmp}\brace{\var{$n$},\var{$k$}}\,);} and such efficiency allows us to literally explore \emph{all} possible recursive choices \mbox{(\funL{allRecursedYMpositroidTrees}\brace{\var{$n$},\var{$k$}}\,)}---and show, for example that the famous parity/cyclic `$T{+}U{+}V$' formula given in the original paper of \cite{BCF} is \emph{not} among them. (See the walkthrough file for a demonstration of this arguably `fun' fact.) 

But positroid labels are many steps removed from analytic expressions, let alone numerics. And it is hard to imagine a use case where someone interested in the 10-particle N$^3$MHV amplitude, say, would begrudge the one second required to obtain it from from `scratch' as an analytic expression of momenta (using \funL{ymAmp}\brace{\var{$n$},\var{$k$}}\,)---even if one second is 100 times slower than positroid-label recursion:  
\mathematicaBox{
\mathematicaSequence{\funL[]{timed}\brace{\funL[]{ymPositroidAmp}\brace{10,3}};\\
\funL[]{timed}\brace{\funL[]{ymAmp}\brace{10,3}};}{}
&\scalebox{0.6}{\hspace{-30pt}Evaluation of the function \texttt{\textbf{ymPositroidAmp[]}} for $\mathcal{A}^{(3)}_{10}$ required \textbf{9 ms, 671 $\mu$s} to complete.}\\[-5pt]
&\scalebox{0.6}{\hspace{-30pt}Evaluation of the function \texttt{\textbf{ymAmp[]}} for $\mathcal{A}^{(3)}_{10}$ required \textbf{931 ms, 412 $\mu$s} to complete.}
}\vspace{-5pt}

Similarly, if one were primarily interested in amplitudes involving external \emph{gluons}, it is easy to imagine a user willing to invest the time required to have a local, covariant (and color-kinematic-dual) representation of the result (as obtained using the function \mbox{\funL{ymLocalCovariantAmp}\brace{\var{$n$}}\,)}---even if this exceeds (by a factor of $>\!25$) the time required to construct analytic expressions for literally \emph{every} helicity amplitude separately:
\mathematicaBox{
\mathematicaSequence{\funL[]{timed}\brace{\funL[]{ymLocalCovariantAmp}\brace{7}};}{}\\[-22pt]
&\scalebox{0.6}{\hspace{-30pt}Evaluation of the function \texttt{\textbf{ymLocalCovariantAmp[]}} for $\mathcal{A}_{7}$ required \textbf{8 seconds, 351 ms} to complete.}\\[-5pt]
\mathematicaSequence[1]{helicities=ReplacePart[\stateYM{p}\&/@Range[7],\#$\to$\stateYM{m}\&/@\#]\&/@Subsets[Range[7]\!]\!\!;\\
\funL[]{timed}\brace{\funL[]{componentAmp}/@helicities};}{}\\[-18pt]
&\scalebox{0.6}{\hspace{-30pt}Evaluation of the function \texttt{\textbf{Map}} required \textbf{364 ms, 878 $\mu$s} to complete.}\\[-5pt]
}
As I hope is becoming increasingly clear, the \package\, package has implemented a number of distinct, independent strategies for the representation, determination, and evaluation of tree amplitudes in gauge and gravitational theories, among many others. 
 
\newpage
To better appreciate the variety of implementations made available by the package---and to confirm the consistency between them---consider some of the ways in which a particular helicity amplitude of gluons could be determined by the package:
\mathematicaBox{
\mathematicaSequence[2]{\{egN,egK\}=\{7,2\};\\
\funL[]{useRandomKinematics}\brace{egN};\\
egCmpt=\funL[]{randomHelicityComponent}\brace{egN,egK};}{\{\stateYM{m},\stateYM{p},\stateYM{m},\stateYM{p},\stateYM{p},\stateYM{m},\stateYM{m}\}}
\mathematicaSequence{distinctEvals=List[\\[-4pt]
\scalebox{0.8}{{\color{hteal}(*Numerically-implemented helicity-amplitude recursion*)}}\\[-2pt]
\funL[]{timed}\brace{\funL[]{componentAmpN}\brace{egCmpt}},\\[-4pt]
\scalebox{0.8}{{\color{hteal}(*Analytic helicity-amplitude recursion, followed by evaluation*)}}\\[-2pt]
\funL[]{timed}\brace{\funL[]{evaluate}\brace{\funL[]{componentAmp}\brace{egCmpt}}},\\[-4pt]
\scalebox{0.8}{{\color{hteal}(*Numerically optimized superAmplitude recursion, with component extraction*)}}\\[-2pt]
\funL[]{timed}\brace{Total\brace{\funL[]{component}\brace{egCmpt}\,\brace{\funL[]{ymAmpN}\brace{egN,egK}}}},\\[-4pt]
\scalebox{0.8}{{\color{hteal}(*Numeric \textbf{randomized} superAmplitude recursion, with component extraction*)}}\\[-2pt]
\funL[]{timed}\brace{Total\brace{\funL[]{component}\brace{egCmpt}\,\brace{\funL[]{ymAmpRandomN}\brace{egN,egK}}}},\\[-4pt]
\scalebox{0.8}{{\color{hteal}(*Analytic superAmplitude recursion, with evaluation and component extraction*)}}\\[-2pt]
\funL[]{timed}\brace{Total\brace{\funL[]{component}\brace{egCmpt}\,\brace{\funL[]{evaluate}\brace{\funL[]{ymAmp}\brace{egN,egK}}}}},\\[-4pt]
\scalebox{0.8}{{\color{hteal}(*Analytic \textbf{random} superAmplitude recursion, evaluation and component extraction*)}}\\[-2pt]
\funL[]{timed}\brace{Total\brace{\funL[]{component}\brace{egCmpt}\,\brace{\funL[]{evaluate}\brace{\funL[]{ymAmpRandom}\brace{egN,egK}}}}},\\[-4pt]
\scalebox{0.8}{{\color{hteal}(*Positroid recursion, uplifted to superFunctions, with component extraction*)}}\\[-2pt]
\funL[]{timed}\brace{Total\brace{\funL[]{component}\brace{egCmpt}\,\brace{\funL[]{fromPositroidsToSuperFunctionsN}\brace{\\\phantom{\{}\hspace{220pt}\funL[]{ymPositroidAmp}\brace{egN,egK}}}}},\\[-4pt]
\scalebox{0.8}{{\color{hteal}(*Positroid randomized recursion, to superFunctions, component extraction*)}}\\[-2pt]
\funL[]{timed}\brace{Total\brace{\funL[]{component}\brace{egCmpt}\,\brace{\funL[]{fromPositroidsToSuperFunctionsN}\brace{\\\phantom{\{}\hspace{190pt}\funL[]{ymPositroidAmpRandom}\brace{egN,egK}}}}},\\[-4pt]
\scalebox{0.8}{{\color{hteal}(*Scattering equations rep, with direct (numeric) integration of CHY integral*)}}\\[-2pt]
\funL[]{timed}\brace{\funL[]{directCHYintegration}\brace{\funL[]{evaluate}\brace{\\\phantom{\{}\hspace{50pt}\funL[]{ymCHYintegrand}\brace{egN}/\!\!.\funL[]{helicityPolarizationRules}\brace{egCmpt}}}},\\[-4pt]
\scalebox{0.8}{{\color{hteal}(*Local, covariant, color-kinematic-dual representation*)}}\\[-2pt]
\funL[]{timed}\brace{\funL[]{evaluate}\brace{\\\phantom{\{}\hspace{15pt}\funL[]{ymLocalCovariantAmp}\brace{egN}/\!\!.\funL[]{helicityPolarizationRules}\brace{egCmpt}}}]\!;
}{}
\\[-16pt]
&\scalebox{0.6}{\hspace{-30pt}Evaluation of the function \texttt{\textbf{componentAmpN[]}} for $\mathcal{A}^{(2)}_{7}(-,+,-,+,+,-,-)$ required \textbf{9 ms, 574 $\mu$s} to complete.}\\[-5pt]
&\scalebox{0.6}{\hspace{-30pt}Evaluation of the function \texttt{\textbf{componentAmp[]}} for $\mathcal{A}^{(2)}_{7}(-,+,-,+,+,-,-)$ required \textbf{18 ms, 242 $\mu$s} to complete.}\\[-5pt]
&\scalebox{0.6}{\hspace{-30pt}Evaluation of the function \texttt{\textbf{ymAmpN[]}} for $\mathcal{A}^{(2)}_{7}$ required \textbf{7 ms, 736 $\mu$s} to complete.}\\[-5pt]
&\scalebox{0.6}{\hspace{-30pt}Evaluation of the function \texttt{\textbf{ymAmpRandomN[]}} for $\mathcal{A}^{(2)}_{7}$ required \textbf{8 ms, 938 $\mu$s} to complete.}\\[-5pt]
&\scalebox{0.6}{\hspace{-30pt}Evaluation of the function \texttt{\textbf{ymAmp[]}} for $\mathcal{A}^{(2)}_{7}$ required \textbf{14 ms, 704 $\mu$s} to complete.}\\[-5pt]
&\scalebox{0.6}{\hspace{-30pt}Evaluation of the function \texttt{\textbf{ymAmpRandom[]}} for $\mathcal{A}^{(2)}_{7}$ required \textbf{16 ms, 307 $\mu$s} to complete.}\\[-5pt]
&\scalebox{0.6}{\hspace{-30pt}Evaluation of the function \texttt{\textbf{ymPositroidAmp[]}} for $\mathcal{A}^{(2)}_{7}$ required \textbf{378 $\mu$s} to complete.}\\[-5pt]
&\scalebox{0.6}{\hspace{-30pt}Evaluation of the function \texttt{\textbf{ymPositroidAmpRandom[]}} for $\mathcal{A}^{(2)}_{7}$ required \textbf{247 $\mu$s} to complete.}\\[-5pt]
&\scalebox{0.6}{\hspace{-30pt}Evaluation of the function \texttt{\textbf{directCHYintegration}} required \textbf{3 seconds, 547 ms} to complete.}\\[-5pt]
&\scalebox{0.6}{\hspace{-30pt}Evaluation of the function \texttt{\textbf{ymLocalCovariantAmp[]}} for $\mathcal{A}_7$ required \textbf{1 minutes, 26 seconds} to complete.}\\[-5pt]
\mathematicaSequence{SameQ@@distinctEvals}{True}\\[-20pt]
}
Notice that the example above makes use of \emph{seven} distinct versions of on-shell recursion: numeric and analytic algorithms encoded separately for both helicity amplitudes of gluons and entire superAmplitudes, and an entirely combinatorial form of recursion via positroid labels---with several allowing for randomized recursive choices to be made at every iterated stage. In addition, we have implemented the CHY representation of gluon amplitudes (see e.g.~\cite{Cachazo:2013gna,Cachazo:2013hca,Cachazo:2013iaa,Cachazo:2013iea,Dolan:2013isa,Dolan:2014ega,Cachazo:2014xea}) and used the strategy described in \cite{Bjerrum-Bohr:2016axv} (see also e.g.~\cite{Baadsgaard:2015hia,Baadsgaard:2015ifa,Baadsgaard:2015voa,Bjerrum-Bohr:2016juj}) to extract from this a local, covariant (color-kinematic-dual) representation of tree amplitudes for any multiplicity. Although the evaluation times vary considerably---and therefore will become ineffective at varying multiplicities---it is clear from above that all algorithms are sufficiently fast through at least $n{=}7$.

For gravitational scattering amplitudes, we have similarly implemented a variety of algorithms for multiplicity $n$ and N${}^k$MHV degrees. Consider for example:
\mathematicaBox{
\mathematicaSequence{
egCmptGR=\funL[]{randomHelicityComponentGR}\brace{egN,egK};}{\{\stateGR{pp},\stateGR{pp},\stateGR{mm},\stateGR{pp},\stateGR{mm},\stateGR{mm},\stateGR{mm}\}}\\[-20pt]
\mathematicaSequence{distinctEvalsGR=List[\\[-4pt]
\scalebox{0.8}{{\color{hteal}(*Numerically-implemented, helicity-amplitude recursion*)}}\\[-2pt]
\funL[]{timed}\brace{\funL[]{componentAmpN}\brace{egCmptGR}},\\[-4pt]
\scalebox{0.8}{{\color{hteal}(*Analytic helicity-amplitude recursion, followed by evaluation*)}}\\[-2pt]
\funL[]{timed}\brace{\funL[]{evaluate}\brace{\funL[]{componentAmp}\brace{egCmptGR}}},\\[-4pt]
\scalebox{0.8}{{\color{hteal}(*Numerically optimized superAmplitude recursion, with component extraction*)}}\\[-2pt]
\funL[]{timed}\brace{Total\brace{\funL[]{component}\brace{egCmptGR}\,\brace{\funL[]{grAmpN}\brace{egN,egK}}}},\\[-4pt]
\scalebox{0.8}{{\color{hteal}(*Analytic superAmplitude recursion, with evaluation and component extraction*)}}\\[-2pt]
\funL[]{timed}\brace{Total\brace{\funL[]{component}\brace{egCmptGR}\,\brace{\funL[]{evaluate}\brace{\funL[]{grAmp}\brace{egN,egK}}}}},\\[-4pt]
\scalebox{0.8}{{\color{hteal}(*Analytic KLT representation, with evaluation and component extraction*)}}\\[-2pt]
\funL[]{timed}\brace{Total\brace{\funL[]{component}\brace{egCmptGR}\,\brace{\funL[]{evaluate}\brace{\funL[]{grAmpKLT}\brace{egN,egK}}}}},\\[-4pt]
\scalebox{0.8}{{\color{hteal}(*Scattering equation repn, with numeric, direct integration of CHY integral*)}}\\[-2pt]
\funL[]{timed}\brace{\funL[]{directCHYintegration}\brace{\funL[]{evaluate}\brace{\\\phantom{\{}\hspace{35pt}\funL[]{grCHYintegrand}\brace{egN}/\!\!.\funL[]{helicityPolarizationRules}\brace{egCmptGR}}}},\\[-4pt]
\scalebox{0.8}{{\color{hteal}(*Local, covariant, color-kinematic-dual representation*)}}\\[-2pt]
\funL[]{timed}\brace{\funL[]{evaluate}\brace{\\\phantom{\{}\hspace{6pt}\funL[]{grLocalCovariantAmp}\brace{egN}/\!\!.\funL[]{helicityPolarizationRules}\brace{egCmptGR}}}]\!;
}{}
\\[-16pt]
&\scalebox{0.6}{\hspace{-30pt}Evaluation of the function \texttt{\textbf{componentAmpN[]}} for $\mathcal{A}^{(2)}_{7}(++,++,--,++,--,--,--)$ required \textbf{122 ms, 18 $\mu$s} to complete.}\\[-5pt]
&\scalebox{0.6}{\hspace{-30pt}Evaluation of the function \texttt{\textbf{componentAmp[]}} for $\mathcal{A}^{(2)}_{7}(++,++,--,++,--,--,--)$ required \textbf{204 ms, 826 $\mu$s} to complete.}\\[-5pt]
&\scalebox{0.6}{\hspace{-30pt}Evaluation of the function \texttt{\textbf{grAmpN[]}} for $\mathcal{A}^{(2)}_{7}$ required \textbf{142 ms, 431 $\mu$s} to complete.}\\[-5pt]
&\scalebox{0.6}{\hspace{-30pt}Evaluation of the function \texttt{\textbf{grAmp[]}} for $\mathcal{A}^{(2)}_{7}$ required \textbf{138 ms, 768 $\mu$s} to complete.}\\[-5pt]
&\scalebox{0.6}{\hspace{-30pt}Evaluation of the function \texttt{\textbf{grAmpKLT[]}} for $\mathcal{A}^{(2)}_{7}$ required \textbf{1 seconds, 531 ms} to complete.}\\[-5pt]
&\scalebox{0.6}{\hspace{-30pt}Evaluation of the function \texttt{\textbf{directCHYintegration}} required \textbf{2 seconds, 405 ms} to complete.}\\[-5pt]
&\scalebox{0.6}{\hspace{-30pt}Evaluation of the function \texttt{\textbf{grLocalCovariantAmp[]}} for $\mathcal{A}_7$ required \textbf{1 minutes, 50 seconds} to complete.}\\[-8pt]
\mathematicaSequence{SameQ@@distinctEvalsGR}{True}\\[-20pt]
}
In addition to these, we have implemented the formulae for MHV amplitudes of both Hodges \cite{Hodges:2011wm} (\funL{grAmpHodges}\brace{\var{$n$}}\,) and Berends, Giele, Kuijf `BGK' (\funL{grAmpBGK}\brace{\var{$n$}}\,) \cite{Berends:1988zp}.

(A close inspection of the time-data in the above examples also exposes a degree of `hysteresis': many of the algorithms of \package\, build upon each-other. Thus, for example, much of the work required for the evaluation of the CHY integral for graviton amplitude was already done when computing the YM integral earlier; and similarly, the reason the entire local, covariant graviton amplitude could be computed in less time than a single partial/ordered amplitude of YM is precisely because the formula is a manifest double-copy.)

Such a variety of algorithms for determining the same amplitude may seem excessive, but it is in line with our goal of providing concrete context for the variety of powerful results to be found in a textbook on modern methods in scattering amplitudes such as \cite{Elvang:2013cua}. 

It would be worthwhile to briefly highlight some of the key novelties in scope and functionality provided by the \package\, package, with more thorough documentation provided in the \hyperlink{context_organization_of_appendix}{appendices}.

\subsection{Scattering Amplitudes of ($\mathcal{N}$-)superYang-Mills Theory (`sYM')}

\vspace{0pt}\subsubsection{Explicit and Abstract Representations of Color-Dressed Amplitudes}\vspace{-5pt}

Color-dressed tree-amplitudes in sYM are represented \emph{abstractly} in terms of (`color-stripped') \emph{partial} amplitudes \funL{amp}\brace{\var{legOrdering}\patternTwo}\, and color-dependent tensors in two different ways:\vspace{-8pt}
\begin{itemize}
\item\funL{colorDressedAmp}\brace{\var{$n$}}\,: returns the complete \var{$n$}-particle tree as a sum over color-stripped, partial amplitudes \funL{amp}\brace{\var{legOrdering}\patternTwo}\, and a minimal basis (independent under Jacobi relations) of \funL{colorFactor} objects. This is the form that results directly from on-shell recursion, and was originally presented by Del Duca, Dixon, and Maltoni in \cite{DelDuca:1999rs}.
\item\funL{colorTraceDressedAmp}\brace{\var{$n$}}\,: returns the complete \var{$n$}-particle tree as a sum over color-stripped, partial amplitudes \funL{amp}\brace{\var{legOrdering}\patternTwo}\, and `color-traces' \funL{tr} (see e.g.~\cite{Cvitanovic:1980bu,Berends:1987cv,Mangano:1987xk,Kosower:1987ic,Zeppenfeld:1988bz,Mangano:1988kk,Mangano:1990by}).
\end{itemize}\vspace{-8pt}
We have automated the expansion of \funL{colorFactor} objects into \funL{tr} objects, and the projection of any \funL{colorFactor} tensor into our chosen basis via \builtL{colorFactorReductionRule}.

Concretely---as a list of minimal list of \funL{superFunction} contributions---the color-dressed \var{$n$}-particle N${}^\var{k}$MHV amplitude is given by \funL{colorDressedYMamp}\brace{\var{$n$},\var{$k$}}\, (with a numerically optimized version thereof given by \funL{colorDressedYMampN}\brace{\var{$n$},\var{$k$}}\,), from which component amplitudes can be easily extracted. In addition, for helicity amplitudes involving gluons, a local, covariant, color-kinematic dual representation is given by \funL{ymLocalCovariantColorDressedAmp}\brace{\var{$n$}}\,.

The partial/ordered (color-stripped) amplitudes of sYM are not independent, and their relations are enumerated as follows. 

\vspace{5pt}\subsubsection{Sets of and Relations Among Partial/Ordered Amplitudes of sYM}\vspace{-5pt}

For the sake of making conventions explicit, our preferred subsets/`bases' of partial amplitudes are enumerated in:\vspace{-8pt}
\begin{itemize}
\item\funL{cyclicAmps}\brace{\var{$n$}}\,: a choice of $(\var{n}{-}1)!$ cyclically-distinct partial amplitudes. 
\item\funL{dihedralAmps}\brace{\var{$n$}}\,: a choice of $(\var{n}{-}1)!/2$ dihedrally-distinct partial amplitudes. 
\item\funL{kkBasisAmps}\brace{\var{$n$}}\,: a choice of $(\var{n}{-}2)!$ partial amplitudes independent under all Kleiss-Kuijf (`KK') relations \cite{KK} (including $U(1)$ decoupling identities \cite{Mangano:1990by}). 
\item\funL{bcjBasisAmps}\brace{\var{$n$}}\,: a choice of $(\var{n}{-}3)!$ partial amplitudes independent under all BCJ relations \cite{Bern:2008qj}. 
\end{itemize}\vspace{-8pt}
Relative to these subsets, we have enumerated complete lists of independent (KK and BCJ) identities they satisfy:\vspace{-8pt}
\begin{itemize}
\item\funL{kkIdentities}\brace{\var{$n$}\pattern}\,: a list of $(\var{n}{-}3)(\var{n}{-}2)!/2$ independent relations satisfied by the partial amplitudes of \funL{dihedralAmps}\brace{$\var{n}$}\,.
\item\funL{bcjIdentities}\brace{\var{$n$}\pattern}\,: a list of $(\var{n}{-}3)(\var{n}{-}3)!$ independent relations satisfied by the partial amplitudes of \funL{kkBasisAmps}\brace{$\var{n}$}\,.
\end{itemize}\vspace{-8pt}
And any of the $\var{n}$! partial amplitudes may be readily projected into these bases via:\vspace{-8pt}
\begin{itemize}
\item\builtL{cyclicAmpReductionRule}\,: a \built{Rule} which automatically replaces any partial amplitude by an element of \funL{cyclicAmps}\brace{\var{$n$}}\,.
\item\builtL{dihedralAmpReductionRule}\,: a \built{Rule} which automatically replaces any partial amplitude by an element of \funL{dihedralAmps}\brace{$\var{n}$}\, (possibly with a sign). 
\item\builtL{kkAmpReductionRule}\,: a \built{Rule} while automatically replaces any partial amplitude by a sum of elements of \funL{kkBasisAmps}\brace{\var{$n$}}\, with coefficients $\pm1$ as dictated by the KK relations. 
\item\builtL{bcjAmpReductionRule}\,: a \built{Rule} which automatically replaces any partial amplitude by a sum of elements of \funL{bcjBasisAmps}\brace{\var{$n$}}\, with coefficients dictated by the BCJ relations. 
\end{itemize}\vspace{-8pt}
To see how these reductions work, consider the following example:\vspace{-0pt}
\mathematicaBox{
\mathematicaSequence{egPartialAmp=\funL{amp}@@\funL[]{randomPerm}\brace{8}}{\funL{amp}\brace{4,2,6,8,3,7,1,5}}\\[-24pt]
\mathematicaSequence[1]{reductions=egPartialAmp/\!\!.\{\{\},\\[-2pt]
\builtL{cyclicAmpReductionRule},\\[-2pt]
\builtL{dihedralAmpReductionRule},\\[-2pt]
\builtL{kkAmpReductionRule},\\[-2pt]
\built{bcjAmpReductionRule}\};\\[5pt]
Count[\#,\funL{amp}\brace{\var{$x$}\patternTwo},\{0,$\infty$\}]\&/@\textnormal{\%}}{\{1,1,1,15,120\}}
\mathematicaSequence[1]{\mbox{\funL[]{evaluate}\brace{reductions/\!\!.\{\funL{amp}\brace{\var{$x$}\patternTwo}$\,\mapsto$\funL[]{componentAmp}\brace{\stateYM{m},\stateYM{p},\stateYM{m},\stateYM{p},\stateYM{m},\stateYM{p},\stateYM{m},\stateYM{p},\{\var{$x$}\}\!}\}\!};}\\
SameQ@@\textnormal{\%}}{True}~\\[-20pt]
}\vspace{-40pt}

Relative to other, existing software (e.g. \textbf{\t{bcfw.m}} \cite{Bourjaily:2010wh}), the \package\, package is primarily designed to directly generate expressions in terms of spinor variables (see \mbox{section~\ref{subsec:kinematics}}), as these make no reference (implicit or otherwise) to the ordering of the external particles' labels, allowing for us to more directly determine amplitudes without any planar ordering such as those of color-dressed sYM or sGR.

\vspace{-0pt}\subsubsection[Ordered superAmplitudes Computed via On-Shell Recursion]{Partial/Ordered superAmplitudes Computed via On-Shell Recursion}\vspace{-0pt}
As all component amplitudes of $\mathcal{N}$-supersymmetric Yang-Mills theory (sYM) can be obtained from those of maximally supersymmetric ($\mathcal{N}\!=\!4$) sYM, these superAmplitudes encode all particular cases of $\var{n}$-particle scattering at tree-level. These are primarily obtained via on-shell recursion (see e.g.~\cite{Cachazo:2005ca,Benincasa:2007qj,Bjerrum-Bohr:2005xoa,Spradlin:2008bu,Bianchi:2008pu,Elvang:2007sg,Drummond:2009ge,Mason:2009afn}), and expressed in terms of \funL{superFunction} objects depending on spinor/momentum variables:\vspace{-6pt}
\begin{itemize}
\item\funL{ymAmp}\brace{\var{$n$}\pattern,\var{$k$}\pattern}\,: gives a list of \funL{superFunction} contributions to the partial/ordered (color-stripped) \var{$n$}-point N${}^{\var{k}}$MHV tree amplitude in sYM, as obtained via on-shell recursion using a default scheme. 
\item\funL{ymAmpRandom}\brace{\var{$n$}\pattern,\var{$k$}\pattern}\,: gives a list of \funL{superFunction} contributions to the partial/ordered (color-stripped) \var{$n$}-point N${}^{\var{k}}$MHV tree amplitude in sYM, as obtained by using a \textbf{\emph{randomized}} choice of bridge-parity and adjacent legs to deform at each stage of on-shell recursion. 
\end{itemize}\vspace{-6pt}
And numerically-optimized, evaluated forms of these \funL{superFunction} contributions using the globally-defined kinematic data  (set by e.g.~\funL{useReferenceKinematics}\brace{\var{$n$}}\,---see \mbox{appendix \ref{appendix:evaluation}}) can be obtained using:\vspace{-6pt}
\begin{itemize}
\item\funL{ymAmpN}\brace{\var{$n$}\pattern,\var{$k$}\pattern}\,: gives a list of \funL{superFunction} contributions to the partial/ordered (color-stripped) \var{$n$}-point N${}^{\var{k}}$MHV tree amplitude in sYM, as obtained via on-shell recursion using a default scheme. 
\item\funL{ymAmpRandomN}\brace{\var{$n$}\pattern,\var{$k$}\pattern}\,: gives a list of \funL{superFunction} contributions to the partial/ordered (color-stripped) \var{$n$}-point N${}^{\var{k}}$MHV tree amplitude in sYM, as obtained by using a \textbf{\emph{randomized}} choice of bridge-parity and adjacent legs to deform at each stage of on-shell recursion. 
\end{itemize}\vspace{-0pt}

Like the algorithms used by the \textbf{\t{bcfw.m}} package \cite{Bourjaily:2010wh} (see also \cite{Drummond:2008cr}), these amplitudes are ultimately found starting with expressions involving momentum-twistor \cite{Hodges:2009hk} \funL{R}-invariants, uplifted to spinor-variable expressions using e.g.\mbox{~\funL{fromTwistorsToSpinors}.} These expressions may be useful to some readers, and so are accessible via:\vspace{-6pt}
\begin{itemize}
\item\funL{ymTwistorAmp}\brace{\var{$n$}\pattern,\var{$k$}\pattern}\,: returns a list of contributions to the partial/ordered (color-stripped) \var{$n$}-point N${}^{\var{k}}$MHV tree amplitude in sYM as expressed in terms of momentum-twistor \funL{R}-invariants. 
\item\funL{ymTwistorAmpRandom}\brace{\var{$n$}\pattern,\var{$k$}\pattern}\,: returns a list of contributions to the partial/ordered (color-stripped) \var{$n$}-point N${}^{\var{k}}$MHV tree amplitude in sYM as expressed in terms of momentum-twistor \funL{R}-invariants, as obtained using a \textbf{\emph{randomized}} choice of bridge-parity and adjacent legs to deform at each stage of on-shell recursion. 
\item\funL{ymSuperTwistorAmp}\brace{\var{$n$}\pattern,\var{$k$}\pattern}\,: returns a list of contributions to the partial/ordered (color-stripped) \var{$n$}-point N${}^{\var{k}}$MHV tree amplitude in sYM as expressed in terms of  \funL{superFunction}${}^{\star}$ objects of momentum-superTwistor variables. 
\item\funL{ymSuperTwistorAmpRandom}\brace{\var{$n$}\pattern,\var{$k$}\pattern}\,: returns a list of contributions to the partial/ordered (color-stripped) \var{$n$}-point N${}^{\var{k}}$MHV tree amplitude in sYM as expressed in terms of  \funL{superFunction}${}^{\star}$ objects of momentum-superTwistor variables, as obtained using a \textbf{\emph{randomized}} choice of bridge-parity and adjacent legs to deform at each stage of on-shell recursion. 
\end{itemize}\vspace{-6pt}
${}^{\star}$\textbf{Note}: the momentum-twistor \funL{superFunction} objects appearing in these expressions (e.g.~in~\funL{ymSuperTwistorAmp}\brace{\var{$n$},\var{$k$}}\,) are fundamentally different than those of spinor-variables: they make reference of coherent states labelled by momentum-twistor $\eta$'s, which should not be confused with spinor-expressions, which label states by spinor $\tilde\eta$'s. These expressions can be `uplifted' to spinor-variable expressions using the mapping functions \funL{fromTwistorsToSpinors} or \funL{toSuperSpinorFunctions}.

Finally, we have implemented on-shell recursion directly in terms of the \emph{positroid labels} \funL{$\sigma$}\brace{\var{perm}\patternTwo}\, of on-shell functions:
\eq{\fig{-40pt}{0.2}{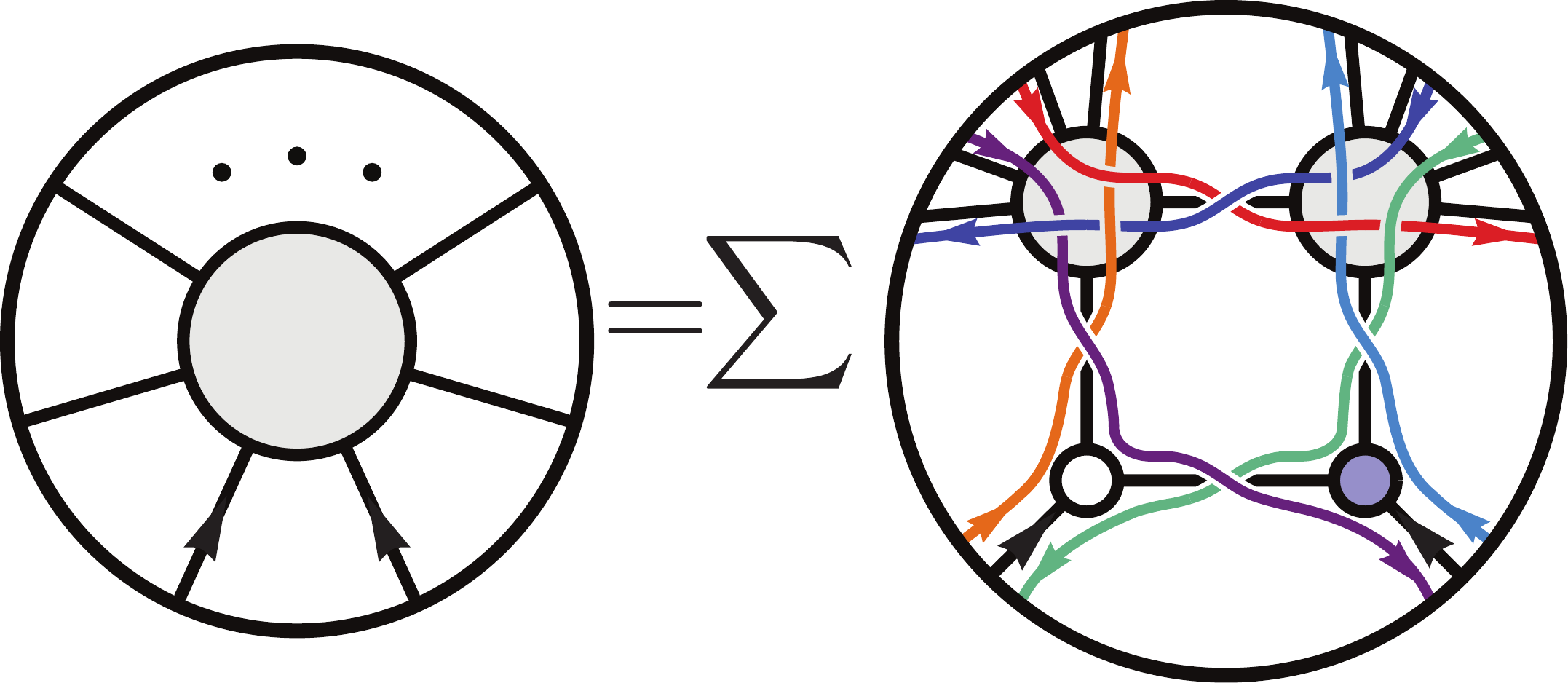}}
Although several steps removed from analytic expressions, the combinatorial recursion relations are unquestionably the most efficiently determined, and can be readily converted into analytic expressions via \funL{fromPositroidsToSuperFunctions}. These forms can be obtained via:\vspace{-6pt}
\begin{itemize}
\item\funL{ymPositroidAmp}\brace{\var{$n$}\pattern,\var{$k$}\pattern}\,: returns a list of contributions to the partial/ordered (color-stripped) \var{$n$}-point N${}^{\var{k}}$MHV tree amplitude in sYM as expressed in terms of positroid labels \funL{$\sigma$}\brace{\var{permutation}\patternTwo} following the default recursion scheme. 
\item\funL{ymPositroidAmpRandom}\brace{\var{$n$}\pattern,\var{$k$}\pattern}\,: returns a list of contributions to the partial/ordered (color-stripped) \var{$n$}-point N${}^{\var{k}}$MHV tree amplitude in sYM as expressed in terms of positroid labels \funL{$\sigma$}\brace{\var{permutation}\patternTwo} using a \textbf{\emph{randomized}} choice of bridge-parity and adjacent legs to deform at each stage of on-shell recursion. 
\item\funL{allRecursedYMpositroidTrees}\brace{\var{$n$}\pattern,\var{$k$}\pattern}\,: returns a complete list of \emph{all} distinct on-shell formulae for the partial/ordered (color-stripped) \var{$n$}-point N$^{\var{k}}$MHV tree amplitude in sYM in terms of positroid labels. This is done by literally exploring all possible recursive choices at every iterated stage of recursion. 
\end{itemize}\vspace{-6pt}
The positroid labels of any \funL{superFunction} can be obtained using \funL{positroidLabel}.

\vspace{-0pt}\subsubsection{Helicity Amplitudes and Amplitudes for Component States}\vspace{-5pt}
The \emph{component} functions of any \funL{superFunction}---those involving specified choices of external states---can be extracted using \mbox{\funL{component}\brace{\var{stateList}}\,\brace{\var{expression}}\,.} In particular, when applied to any superAmplitude, this will give the particular (Bosonic) amplitude involving those particular external states. However, it turns out that---especially for gluonic, helicity amplitudes---more compact, elegant expressions can be obtained using non-supersymmetric, recursion relations directly in terms of helicity amplitudes directly (see e.g.~\cite{ArkaniHamed:2008yf,Cohen:2010mi}). 

With this in mind, we have implemented a number of similar functions:\vspace{-6pt}
\begin{itemize}
\item\funL{componentAmpTerms}\brace{\var{componentList}\patternTwo}\,: returns a list of terms contributing to the particular component amplitude sYM.\\
\textbf{Note}: when \var{componentList} consists of only gluon \stateYM{m},\stateYM{p}, it calls \funL{ymHelicityAmp} \emph{but for a possibly rotated set of external helicities} in order to optimize the resulting expression. For all other component states, this function is equivalent to \funL{component}\brace{\var{componentList}}\,\brace{\funL{ymAmp}\brace{\var{$n$},\var{$k$}}}\,.
\item\funL{componentAmp}\brace{\var{componentList}\patternTwo}\,: returns the summed contributions from the output of \mbox{\funL{componentAmpTerms}\brace{\var{componentList}}\,.}
\item\funL{ymHelicityAmp}\brace{\var{gluonHelicities}\patternTwo}\,: uses a specific scheme for on-shell recursion \emph{directly} in terms of helicity amplitudes.
\end{itemize}\vspace{-6pt}
The reason for the separate implementation of \funL{ymHelicityAmp} is that our default recursion scheme \emph{guarantees} that every term is a ratio of products of monomials of Lorentz-invariant spinor brackets, thus guaranteeing relatively compact expressions. 

To appreciate the distinction in output of these functions, consider the following:
\mathematicaBox{
\mathematicaSequence[1]{formula1=\funL[]{componentAmpTerms}\brace{\stateYM{p},\stateYM{p},\stateYM{m},\stateYM{m},\stateYM{m},\stateYM{p},\stateYM{p}};\\
\{Length[\textnormal{\%}],LeafCount[\textnormal{\%}]\}}{\{3,158\}}
\mathematicaSequence[1]{formula2=\funL[]{ymHelicityAmp}\brace{\stateYM{p},\stateYM{p},\stateYM{m},\stateYM{m},\stateYM{m},\stateYM{p},\stateYM{p}};\\
\{Length[\textnormal{\%}],LeafCount[\textnormal{\%}]\}}{\{4,238\}}
\mathematicaSequence[1]{formula3=\funL[]{component}\brace{\stateYM{p},\stateYM{p},\stateYM{m},\stateYM{m},\stateYM{m},\stateYM{p},\stateYM{p}}\,\brace{\funL[]{ymAmp}\brace{7,1}};\\
\{Length[\textnormal{\%}],LeafCount[\textnormal{\%}]\}}{\{6,369\}}
\mathematicaSequence{Total/@\funL[]{evaluate}\brace{\{formula1,formula2,formula3\}}}{\rule{0pt}{22pt}\!\!$\displaystyle\left\{\rule{0pt}{20pt}\right.\text{-}\frac{2419788926288}{6978203418375},\text{-}\frac{2419788926288}{6978203418375},\text{-}\frac{2419788926288}{6978203418375}\left.\rule{0pt}{20pt}\right\}$}
}
To be clear, in the example above, \t{formula3} above is essentially identical to \t{formula2}. The sources of their apparent discrepancy in complexity are two-fold: first, \t{formula3} involves two terms that vanish \emph{only one of which vanishes manifestly} (i.e.~prior to evaluation); and secondly, \t{formula3} involves the product of minors of the coefficients of $\tilde\eta$ for its extraction, and these do not manifestly simplify into products of Lorentz-invariant spinor brackets (or cancel poles in the expression). 

Finally, we have \emph{separately} implemented a purely-numeric recursion scheme for gluon helicity amplitudes in \funL{ymHelicityAmpN}. This is called by e.g.~\funL{componentAmpN} when appropriate. For non-gluonic component amplitudes, \funL{componentAmpTermsN} and \funL{componentAmpN} makes use of the numerically-optimized \funL{ymAmpN}. 

For the evaluation of a single helicity amplitude for a single set of specified kinematics, \funL{ymHelicityAmpN} is vastly faster than the evaluation of \funL{ymHelicityAmp}:
\mathematicaBox{
\mathematicaSequence[2]{\funL[]{useRandomKinematics}\brace{15};\\
egCmpt=\{\stateYM{m},\stateYM{p},\stateYM{m},\stateYM{p},\stateYM{m},\stateYM{p},\stateYM{m},\stateYM{p},\stateYM{m},\stateYM{p},\stateYM{m},\stateYM{p},\stateYM{m},\stateYM{p},\stateYM{m}\};\\
\funL[]{timed}@\funL[]{ymHelicityAmpN}\brace{egCmpt};\\
\funL[]{timed}@\funL[]{ymHelicityAmp}\brace{egCmpt};\\
\funL[]{timed}@\funL[]{evaluate}\brace{\textnormal{\%}};}{}
&\scalebox{0.58}{\hspace{-45pt}Evaluation of the function \texttt{\textbf{ymHelicityAmpN[]}} for $\mathcal{A}^{(6)}_{15}(-,+,-,+,-,+,-,+,-,+,-,+,-,+,-)$ required \textbf{27 seconds, 357 ms} to complete.}\\[-5pt]
&\scalebox{0.58}{\hspace{-45pt}Evaluation of the function \texttt{\textbf{ymHelicityAmp[]}} for $\mathcal{A}^{(6)}_{15}(-,+,-,+,-,+,-,+,-,+,-,+,-,+,-)$ required \textbf{5 minutes, 45 seconds} to complete.}\\[-5pt]
&\scalebox{0.58}{\hspace{-45pt}Evaluation of the function \texttt{\textbf{evaluate}} required \textbf{4 seconds, 855 ms} to complete.}\\[-5pt]
}
However, the analytic expression of \funL{ymHelicityAmp} is \emph{evaluated} much more efficiently. Thus, if more than a single evaluation is desired, it is more efficient to invest in the analytic expression first.

\vspace{-5pt}\paragraph{Local, Covariant (Color-Kinematic-Dual) Helicity Amplitudes}~\\[-32pt]

Finally, we have automated the algorithm described in \cite{Bjerrum-Bohr:2016axv} (see also \cite{Baadsgaard:2015hia,Baadsgaard:2015ifa,Baadsgaard:2015voa,Bjerrum-Bohr:2016juj}) to generate local, covariant, color-kinematic-dual helicity amplitudes for arbitrary multiplicity starting from the scattering equation description of Yang-Mills amplitudes of Cachazo He and Yuan in \cite{Cachazo:2013gna,Cachazo:2013hca,Cachazo:2013iaa,Cachazo:2013iea,Dolan:2013isa,Dolan:2014ega,Cachazo:2014xea}. These formulae can be obtained by calling the function \funL{ymLocalCovariantAmp}\brace{\var{$n$}\pattern}\,. To evaluate this representation, polarizations with definite helicities must be specified via \mbox{\funL{helicityPolarizationRules}\brace{\var{gluonHelicities}\patternTwo}\,.}

\vspace{0pt}\subsection{Scattering Amplitudes of $(\mathcal{N}$-)superGravity Theory (`sGR')}\vspace{-0pt}

\vspace{-0pt}\subsubsection{Abstract/Explicit Representations of Amplitudes in sGR}\vspace{-5pt}

Tree amplitudes in superGravity theory can be computed in many ways, including on-shell recursion (see e.g.~\cite{Heslop:2016plj,Herrmann:2016qea,Paranjape:2022ymg,Trnka:2020dxl,Brown:2022wqr,Paranjape:2023qsq,Bourjaily:2023ycy,structureOfOnShellGR}), as a double-copy of Yang-Mills theory (see e.g.~\cite{Bern:2010ue,Bern:2019prr,Adamo:2022dcm,Bern:2023zkg}), and via the `stringy' relations described by Kawai, Lewellen, and Tye in \cite{KLT} (see also e.g.~\cite{Mizera:2016jhj,Mizera:2019blq,Chi:2021mio}).

\newpage
\paragraph{Representations of Amplitudes via Kawai-Lewellen-Tai Relations}~\\[-32pt]

Each of these formalisms can lead to a variety of distinct formulae for specific amplitudes: by varying the choices made during iterated, on-shell recursion, by the choices of particular color-kinematic satisfying numerators, or by choosing different bases for partial amplitudes appearing in across the `KLT kernel' 
\vspace{-2pt}\eq{\mathcal{A}^{\text{L}\raisebox{0.75pt}{\scalebox{0.8}{$\otimes$}}\text{R}}=\mathcal{A}^{\text{L}}(\r{\vec{\r{\alpha}}}).m(\r{\vec{\r{\alpha}}}|\b{\vec{\b{\beta}}}).\mathcal{A}^{\text{R}}(\b{\vec{\b{\beta}}})\,.\label{general_klt_relation}\vspace{-2pt}}
where the matrix $(n{-}3)!\!\times\!(n{-}3)!$ matrix $m(\r{\vec{\alpha}}|\b{\vec{\beta}})$ can be computed via the function \mbox{\funL{kltKernel}\brace{\{\r{$\alpha\rule[-1.05pt]{14.5pt}{.65pt}$}\funL{amp}\},\{\b{$\beta\rule[-1.05pt]{14.5pt}{.65pt}$}\funL{amp}\}}\,.} Among the most famous examples is where $\mathcal{A}^{\text{L,R}}$ are tree-level partial/ordered amplitudes of sYM; in which case the (\ref{general_klt_relation}) results in a representation of the amplitude for sGR. (As reviewed below in \mbox{section \ref{}}, \emph{tree-}amplitudes in sGR with any degree of supersymmetry $\mathcal{N}\!<\!8$ are identical to those of $\mathcal{N}\!{=}8$ sGR, restricted to the appropriate external states of interest/relevance for the less supersymmetric theory.)

If the sets of partial amplitudes are chosen to be the same, then $m(\r{\vec{\alpha}}|\r{\vec{\alpha}})$ will be \emph{dense}---and the formula (\ref{general_klt_relation}) will involve $((n{-}3)!)^2$ terms added together. \vspace{-0pt}
\mathematicaBox{
\mathematicaSequence[1]{egKernel=\funL[]{kltKernel}\brace{\#,\#}\,\&@\funL[]{bcjBasisAmps}\brace{6};\\[-2pt]
\funL[]{fuzzify}\brace{\textnormal{\%}}}{$\displaystyle\left(\begin{array}{@{}cccccc@{}}\b{\rstar}&\b{\rstar}&\b{\rstar}&\b{\rstar}&\b{\rstar}&\b{\rstar}\\[-6pt]
\b{\rstar}&\b{\rstar}&\b{\rstar}&\b{\rstar}&\b{\rstar}&\b{\rstar}\\[-6pt]
\b{\rstar}&\b{\rstar}&\b{\rstar}&\b{\rstar}&\b{\rstar}&\b{\rstar}\\[-6pt]
\b{\rstar}&\b{\rstar}&\b{\rstar}&\b{\rstar}&\b{\rstar}&\b{\rstar}\\[-6pt]
\b{\rstar}&\b{\rstar}&\b{\rstar}&\b{\rstar}&\b{\rstar}&\b{\rstar}\\[-6pt]
\b{\rstar}&\b{\rstar}&\b{\rstar}&\b{\rstar}&\b{\rstar}&\b{\rstar}
\end{array}\right)$}\\[-20pt]
\mathematicaSequence{egGRstates=\funL[]{randomComponentStatesGR}\brace{6,1}}{\{\stateGR{psiBar}\![2\!,\!4\!,\!7]\!,\!\stateGR{phiBar}\![1\!,\!4\!,\!5\!,\!8]\!,\!\stateGR{psi}\![1\!,\!5\!,\!7]\!,\!\stateGR{psiBar}\![1\!,\!3\!,\!4\!,\!8]\!,\!\stateGR{phi}\![1\!,\!2\!,\!4\!,\!8]\}}\\[-20pt]
\mathematicaSequence{\{leftYMstates,rightYMstates\}=\funL[]{randomStateSplitting}\brace{egGRstates}}{\{\{\stateYM{psi}\![4]\!,\!\stateYM{phi}\![1\!,\!3]\!,\!\stateYM{phiBar}\![1\!,\!3]\!,\!\stateYM{psi}\![3]\!,\!\stateYM{psiBar}\![4]\!,\!\stateYM{psiBar}\![3]\},\\[-2pt]
\phantom{\{}\{\stateYM{m},\stateYM{phiBar}\![1\!,\!3]\!,\!\stateYM{psi}\![2]\!,\!\stateYM{phi}\![1\!,\!2]\!,\!\stateYM{phiBar}\![1\!,\!2]\!,\!\stateYM{psi}\![1]\}}\\[-20pt]
\mathematicaSequence{left=\funL[]{bcjBasisAmps}\brace{6}/\!\!.\{\funL{amp}\brace{\var{$x$}\patternTwo}$\,\mapsto\!$\funL[]{componentAmpN}\brace{leftYMstates\!,\!\{\var{$x$}\}\!}\}}{\scalebox{0.9}{$\displaystyle\left\{\rule{0pt}{20pt}\right.\!\!\!\frac{237355}{3779622},\frac{1137558068750}{33204415816341},\text{-}\frac{6479509}{31384584},\frac{33685063}{82384533},\text{-}\frac{640654950455}{272977557090777},\text{-}\frac{7772885}{46609101}\left.\rule{0pt}{20pt}\!\!\!\right\}$}}\\[-20pt]
\mathematicaSequence{\mbox{right=\funL[]{bcjBasisAmps}\brace{6}/\!\!.\{\funL{amp}\brace{\var{$x$}\patternTwo}$\,\mapsto\!$\funL[]{componentAmpN}\brace{rightYMstates\!,\!\{\var{$x$}\}\!}\}\hspace{-10pt}}}{\scalebox{0.9}{$\displaystyle\left\{\rule{0pt}{20pt}\right.\!\!\!\frac{256436810}{1269661203},\text{-}\frac{75514280}{40956813},\frac{134885}{364374},\text{-}\frac{120026980}{282642561},\text{-}\frac{1280}{5877},\frac{138550720}{282642561}\left.\rule{0pt}{20pt}\!\!\!\right\}$}}
\mathematicaSequence{left\,.\funL[]{evaluate}\brace{egKernel}\,\,.right}{$\displaystyle\text{-}\frac{66141815771}{21610062414}$}
\mathematicaSequence{\textnormal{\%}===\funL[]{componentAmpN}\brace{egGRstates}}{True}\vspace{-20pt}
}
But as we can see, this does give the correct result for component amplitudes of sGR. Notice that we have used \funL{randomStateSplitting} to express the sGR component state as a random division of left/right component amplitudes.

A better choice of bases are the ones defined by the package as \mbox{\funL{kltBasisAmpsAlpha}\brace{\var{$n$}}} and \funL{kltBasisAmpsBeta}\brace{\var{$n$}}\,, for these give rise to a KLT kernel that is maximally sparse and block-diagonal---the default for \funL{kltKernel}\brace{\var{$n$}}\,:
\mathematicaBox{
\mathematicaSequence[1]{egKernelDefault=\funL[]{kltKernel}\brace{6};\\
\funL[]{fuzzify}\brace{\textnormal{\%}}}{$\displaystyle\left(\begin{array}{@{}cccccc@{}}\b{\rstar}&\b{\rstar}&\deemph{0}&\deemph{0}&\deemph{0}&\deemph{0}\\[-6pt]
\b{\rstar}&\b{\rstar}&\deemph{0}&\deemph{0}&\deemph{0}&\deemph{0}\\[-6pt]
\deemph{0}&\deemph{0}&\b{\rstar}&\b{\rstar}&\deemph{0}&\deemph{0}\\[-6pt]
\deemph{0}&\deemph{0}&\b{\rstar}&\b{\rstar}&\deemph{0}&\deemph{0}\\[-6pt]
\deemph{0}&\deemph{0}&\deemph{0}&\deemph{0}&\b{\rstar}&\b{\rstar}\\[-6pt]
\deemph{0}&\deemph{0}&\deemph{0}&\deemph{0}&\b{\rstar}&\b{\rstar}
\end{array}\right)$}
\mathematicaSequence[1]{\mbox{grCmpt=\{\stateGR{psiBar}\![2\!,\!4\!,\!7]\!\!,\!\stateGR{phiBar}\![1\!,\!4\!,\!5\!,\!8]\!\!,\!\stateGR{psi}\![1\!,\!5\!,\!7]\!\!,\!\stateGR{psiBar}\![1\!,\!3\!,\!4\!,\!8]\!,\!\stateGR{phi}\![1\!,\!2\!,\!4\!,\!8]\hspace{-2pt}\}\!;}\\
\{leftCmpt,rightCmpt\}=\funL[]{randomStateSplitting}\brace{grCmpt}}{\{\{\stateYM{psiBar}\![2]\!,\!\stateYM{phiBar}\![1\!,\!4]\!,\!\stateYM{psi}\![4]\!,\!\stateYM{psi}\![1]\!,\!\stateYM{phiBar}\![1\!,\!4]\!,\!\stateYM{psiBar}\![3]\},\\
\phantom{\{}\{\stateYM{phi}\![1\!,\!3]\!,\!\stateYM{phi}\![1\!,\!4]\!,\!\stateYM{phi}\![1\!,\!2]\!,\!\stateYM{phiBar}\![1\!,\!2]\!,\!\stateYM{psiBar}\![1]\!,\!\stateYM{psi}\![2]\}}
\mathematicaSequence{left=\funL[]{kltBasisAmpsAlpha}\brace{6}/\!\!.\{\funL{amp}\brace{\var{$x$}\patternTwo}$\,\mapsto\!$\funL[]{componentAmpN}\brace{leftCmpt\!,\!\{\var{$x$}\}\!}\}}{\scalebox{0.9}{$\displaystyle\left\{\rule{0pt}{20pt}\right.\!\!\!\text{-}\frac{72376149425}{136019356704},\frac{296215765}{448908768},\frac{663548308825}{1385971428996},\frac{599597300}{1143540783},\text{-}\frac{1844065}{5430348},\text{-}\frac{371600}{1357587}\left.\rule{0pt}{20pt}\!\!\!\right\}$}}
\mathematicaSequence{\mbox{right=\funL[]{kltBasisAmpsBeta}\brace{6}/\!\!.\{\funL{amp}\brace{\var{$x$}\patternTwo}$\,\mapsto\!$\funL[]{componentAmpN}\brace{rightCmpt\!,\!\{\var{$x$}\}\!}\}\hspace{-10pt}}}{\scalebox{0.9}{$\displaystyle\left\{\rule{0pt}{20pt}\right.\!\!\!\text{-}\frac{19376}{1675707},\text{-}\frac{7100}{3081743},\text{-}\frac{3594699584}{204010065853},\frac{321360}{180061841},\text{-}\frac{34297324702304}{543277807762305},\frac{3047116144}{86910543555}\left.\rule{0pt}{20pt}\!\!\!\right\}$}}
\mathematicaSequence{left\,.\funL[]{evaluate}\brace{egKernelDefault}\,\,.right}{$\displaystyle\text{-}\frac{66141815771}{21610062414}$}
\mathematicaSequence{\textnormal{\%}===\funL[]{componentAmpN}\brace{egGRstates}}{True}
}
(Note that we have used a \emph{different}, randomized choice of state-splitting above relative to the previous illustration.)

When expressing amplitudes as in (\ref{general_klt_relation}), it is typical to simply write the first row---taking the first element of the basis $\{\r{\alpha}\}$ to be \funL{amp}\brace{1,\ldots,$n$}\,. This is encoded in the function \funL{kltRepresentationSeeds}\brace{\var{$n$}}\,; summing over the $(\var{n}{-}3)!$ permutations of the set $[2,\var{n}{-}2]$ for the legs will give the sum over terms in the `KLT' representation of $\mathcal{A}^{\text{L}\raisebox{0.75pt}{\scalebox{0.8}{$\otimes$}}\text{R}}$. 

The \var{$n$}-particle N${}^{\var{k}}$MHV superAmplitude computed in this way via KLT is given by \funL{grAmpKLT}\brace{\var{$n$}\pattern,\var{$k$}\pattern}\,.

\newpage
\vspace{-0pt}\subsubsection{Amplitudes in sGR via On-Shell Recursion Relations}\vspace{-5pt}

On-shell recursion for gravity has been described in many places (see e.g.~\mbox{\cite{Heslop:2016plj,Herrmann:2016qea,Paranjape:2022ymg,Trnka:2020dxl,Brown:2022wqr,Paranjape:2023qsq,Bourjaily:2023ycy,structureOfOnShellGR}}), and as described in \cite{Bourjaily:2023ycy} leads \emph{directly} to a gauge-invariant double-copy (and color-kinematic-duality satisfying (if trivially)) representation of amplitudes via the relationship between terms in recursion for sYM and those of sGR---although these representations do \emph{not} involve local denominators. 

In particular, if we represent the $i$th term in the on-shell recursed formula for the partial amplitude of sYM by $\mathfrak{f}_i$, then the corresponding term in the on-shell recursion for sGR will be $\mathfrak{f}_i^2\,D_i$ for some Bosonic `denominator' $D_i$ determined recursively. Thus, if we say that $\mathfrak{n}_i\equivR\mathfrak{f}_i\,D_i$, then the term in sYM is $\mathfrak{n}_i/D_i$ and that for sGR will be $\mathfrak{n}_i^2/D_i$---a double-copy. 

For the default recursion scheme (corresponding to the output of \funL{ymAmp}\brace{\var{$n$},\var{$k$}}\,), these denominators are computed by \funL{grAmpDenominators}\brace{\var{$n$},\var{$k$}}\,. Thus, amplitudes in sGR can be computed as follows:
\mathematicaBox{
\mathematicaSequence[4]{ymTerms=\funL[]{ymAmp}\brace{6,1};\\
grDens=\funL[]{grAmpDenominators}\brace{6,1};\\
grPartialAmp=Power[ymTerms,2].grDens/\!\!.\builtL{superFunctionProductRule};\\
\funL[]{evaluate}\brace{\funL[]{permuteLegs}\brace{\#}\,\brace{\textnormal{\%}}\,\&/@Permutations[Range[2,5]]};\\
Total[\funL[]{component}\brace{\stateGR{mm},\stateGR{pp},\stateGR{mm},\stateGR{pp},\stateGR{mm},\stateGR{pp}}\,\brace{\textnormal{\%}}]}{$\rule{0pt}{20pt}\displaystyle\text{-}\frac{66141815771}{21610062414}$}
\mathematicaSequence{\textnormal{\%}===\funL[]{componentAmpN}\brace{\stateGR{mm},\stateGR{pp},\stateGR{mm},\stateGR{pp},\stateGR{mm},\stateGR{pp}}}{True}
}
This is not quite the same as what is generated by the function \mbox{\funL{grAmpBCFW}\brace{\var{$n$},\var{$k$}}\,,} because we have grouped together terms which have \emph{equivalent} $\tilde\eta$-coefficients as \funL{superFunction} objects. This fact is more transparent for the numerically-optimized \mbox{\funL{grAmpBCFWN}\brace{\var{$n$},\var{$k$}}}---for which the Bosonic terms are naturally combined into a single expression. This will be explained in greater detail in the forthcoming \cite{structureOfOnShellGR}.

We expect the most useful reference will be \funL{grAmp}\brace{\var{$n$},\var{$k$}}\,, which provides an analytic list of \funL{superFunction} contributions for the \var{$n$}-particle N${}^{\var{k}}$MHV tree-amplitude of sGR. This function calls \funL{grAmpHodges}\brace{\var{$n$}} for MHV (and $\overline{\text{MHV}}$) amplitudes, and returns \funL{grAmpBCFW}\brace{\var{$n$},\var{$k$}} for all other N${}^{\var{k}}$MHV-degrees.

\vspace{5pt}\paragraph{Arbitrary-Multiplicity MHV Amplitudes}\vspace{-0pt}

For the case of MHV amplitudes of sGR, we have also implemented the two famous formulae from Berends, Giele, and Kuijf (`BGK') \cite{Berends:1988zp} in \funL{grAmpBGK}\brace{\var{$n$}}\, and that of Hodges \cite{Hodges:2011wm} in \funL{grAmpHodges}\brace{\var{$n$}}\,.

\newpage\vspace{-0pt}\subsubsection{Helicity Amplitudes and Amplitudes for Component States}\vspace{-5pt}

The \emph{component} functions of any \funL{superFunction}---those involving specified choices of external states---can be extracted using \mbox{\funL{component}\brace{\var{stateList}}\,\brace{\var{expression}}\,.} In particular, when applied to any superAmplitude, this will give the particular (Bosonic) amplitude involving those particular external states. However, it turns out that---especially for graviton helicity amplitudes---more compact, elegant expressions can be obtained using non-supersymmetric, recursion relations directly in terms of helicity amplitudes directly (see e.g.~\cite{ArkaniHamed:2008yf,Cohen:2010mi}). 

With this in mind, we have implemented a number of similar functions:\vspace{-6pt}
\begin{itemize}
\item\funL{componentAmpTerms}\brace{\var{componentList}\patternTwo}\,: returns a list of terms contributing to the particular component amplitude sGR.\\
\textbf{Note}: when \var{componentList} consists of only gravitons \stateGR{mm},\stateGR{pp}, it calls \funL{grHelicityAmp} \emph{but for a possibly rotated set of external helicities} in order to optimize the resulting expression. For all other component states, this function is equivalent to \funL{component}\brace{\var{componentList}}\,\brace{\funL{grAmp}\brace{\var{$n$},\var{$k$}}}\,.
\item\funL{componentAmp}\brace{\var{componentList}\patternTwo}\,: returns the summed contributions from the output of \mbox{\funL{componentAmpTerms}\brace{\var{componentList}}\,.}
\item\funL{grHelicityAmp}\brace{\var{gravitonHelicities}\patternTwo}\,: uses a specific scheme for on-shell recursion \emph{directly} in terms of helicity amplitudes.
\end{itemize}\vspace{-6pt}

Finally, we have \emph{separately} implemented a purely-numeric recursion scheme for graviton helicity amplitudes in \funL{grHelicityAmpN}. This is called by e.g.~\funL{componentAmpN} when appropriate. For non-graviton component amplitudes of sGR, \funL{componentAmpTermsN} and \funL{componentAmpN} makes use of the numerically-optimized \funL{grAmpN}.

\vspace{5pt}\paragraph{Local, Covariant (Color-Kinematic-Dual) Helicity Amplitudes}~\\[-32pt]

Finally, we have automated the algorithm described in \cite{Bjerrum-Bohr:2016axv} (see also \cite{Baadsgaard:2015hia,Baadsgaard:2015ifa,Baadsgaard:2015voa,Bjerrum-Bohr:2016juj}) to generate local, covariant, color-kinematic-dual helicity amplitudes for arbitrary multiplicity starting from the scattering equation description of Yang-Mills amplitudes of Cachazo He and Yuan in \cite{Cachazo:2013gna,Cachazo:2013hca,Cachazo:2013iaa,Cachazo:2013iea,Dolan:2013isa,Dolan:2014ega,Cachazo:2014xea}. These formulae can be obtained by calling the function \funL{grLocalCovariantAmp}\brace{\var{$n$}\pattern}\,. To evaluate this representation, polarizations with definite helicities must be specified via \mbox{\funL{helicityPolarizationRules}\brace{\var{gravitonHelicities}\patternTwo}\,.}

\newpage
\vspace{-0pt}\subsection{Scattering Amplitudes in Other Field Theories}\vspace{-0pt}

\vspace{-0pt}\subsubsection{Amplitudes of Scalar \texorpdfstring{$\phi^p$}{phi-p} Field Theory}\vspace{-5pt}

Being combinatorial in nature, we have implemented the following forms of amplitudes in scalar field theories:\vspace{-6pt}
\begin{itemize}
\item\funL{scalarAmp}\brace{\var{$p$}\pattern}\,\brace{\var{$n$}\pattern}\,: returns the sum of terms contributing to the \var{$n$}-particle ordered/partial amplitude of scalar $\phi^{\var{p}}$ field theory.
\item\funL{phi3Amp}\brace{\var{$n$}\pattern}\,: returns the \var{$n$}-particle ordered/partial amplitude of scalar $\phi^{3}$ field theory. 
\item\funL{phi3AmpCyclicSeeds}\brace{\var{$n$}\pattern}\,: returns the cyclic seeds of the ordered/partial amplitude in $\phi^3$ theory. These terms, when summed over non-cyclic permutations, give the ordered/partial amplitude of scalar $\phi^3$ field theory. 
\item\funL{nlsmAmp}\brace{\var{$n$}\pattern}\,: returns the ordered/partial \var{$n$}-particle tree amplitude of the non-linear sigma model as obtained from scalar $\phi^3$ tree amplitudes via the results of \cite{hiddenZeroes,hiddenZeroes2,hiddenZeroes3}\,.
\item\funL{nlsmAmpCyclicSeeds}\brace{\var{$n$}\pattern}\,: returns the cyclic seeds of the ordered/partial amplitude in the non-linear sigma model. These terms, when summed over non-cyclic permutations, give the ordered/partial amplitude of the NLSM.
\end{itemize}\vspace{-6pt}

\vspace{-0pt}\subsubsection{Amplitudes of in Field Theories Generated as Double-Copies}\vspace{-0pt}

Finally, for the sake of comparison against representations in the scattering equation formalism of CHY \cite{Cachazo:2014xea}, we have provided KLT-representations of\vspace{-2pt}
\begin{itemize}
\item\funL{specialGalileonAmp}\brace{\var{$n$}\pattern}\,: returns the sum of terms contributing to the \var{$n$}-particle amplitude of the special Galileon field theory as represented by (\ref{general_klt_relation}) where both left and right amplitudes are those of the non-linear sigma model \funL{nlsmAmp}\brace{\var{$n$}}\,.
\item\funL{diracBornInfeldAmp}\brace{\var{$n$}\pattern}\,: sum of terms contributing to the \var{$n$}-particle amplitude of the (super)DBI field theory as represented by (\ref{general_klt_relation}) where the left amplitudes are taken to be the parity-even superAmplitudes \funL{ymAmp}\brace{\var{$n$},\var{$(n{-}4)/2$}} and the right amplitudes are taken to be those of the NLSM \funL{nlsmAmp}\brace{\var{$n$}}\,.
\end{itemize}

\vspace{-0pt}\subsection{Scattering Equations and Twistor String Theory}\vspace{-0pt}

\vspace{-0pt}\subsubsection{Representing Amplitudes in the Scattering Equation Formalism}\vspace{-0pt}

When external kinematics are taken to be 4 dimensional, the scattering equation formalism of Cachazo, He, and Yuan (see e.g.~\cite{Cachazo:2013gna,Cachazo:2013hca,Cachazo:2013iaa,Cachazo:2013iea,Dolan:2013isa,Dolan:2014ega,Cachazo:2014xea}) is very closely related to that of the twistor string theory (see e.g.~\cite{Witten:2003nn,RSV,Roiban:2004yf,Dixon:2005cf}). It is also closely related to a manifestation of KLT (\ref{general_klt_relation}) and the double-copy. However, being dimensionally agnostic, it is more typically viewed as a formalism for scattering amplitudes involving kinematics in general $d$ dimensions, and therefore speaks directly to amplitudes to pure (non-supersymmetric) versions of gauge and gravitational theories. 

Key ingredients in the representation of amplitudes in the CHY formalism include\\[-12pt]\vspace{-4pt}
\begin{itemize}
\item\funL{scatteringEquations}\brace{\var{$n$}\pattern}\,: the list $\{S_{\var{a}}\}$ of scattering equations relating the Mandelstams \funL{s}\brace{\var{$a$},\var{$b$}} to the auxiliary variables \fun{$\sigma$}\brace{\var{$a$}}\,.
\item\funL{scatteringEquationsJacobian}\brace{\var{$n$}\pattern}\,: the (reduced) Jacobian matrix of the scattering equations when viewed as constraints. Their `reudction' is related to the $SL(2,\mathbb{C})$ redundancy that exists for the auxiliary parameters \fun{$\sigma$}\brace{\var{$a$}}\,.
\item\funL{chyCycle}\brace{\{\var{$a$},\var{$b$},\ldots,\var{$c$}\}}\,: gives 1/(\fun{$\sigma$}\brace{\var{$a$},\var{$b$}}$\,\cdots$\fun{$\sigma$}\brace{\var{$c$},\var{$a$}}) where $\fun{$\sigma$}\brace{\var{a},\var{b}}\equivR\fun{$\sigma$}\brace{\var{a}}\,{-}\fun{$\sigma$}\brace{b}$\,.
\item\funL{chyAmatrix}\brace{\var{$n$}\pattern}\,: gives the $\var{n}\!\times\!\var{n}$ CHY matrix $\mathbf{A}_\var{n}$.
\item\funL{chyPsiMatrix}\brace{\var{$n$}\pattern}\,: gives the $(2\var{n})\!\times\!(2\var{n})$ CHY matrix ${\Psi_{\var{n}}}$.
\end{itemize}\vspace{-4pt}
These appear in the representation of amplitudes via the reduced Pfaffians \mbox{(see \funL{pfaffian})} of the $\mathbf{A}$ and $\Psi$ matrices,\vspace{-4pt}
\begin{itemize}
\item\funL{chyPsiPfaffianPrime}\brace{\var{$n$}\pattern}\,: gives $\mathrm{Pf'}(\mathbf{A})$ for $\var{n}$-particles.
\item\funL{chyPsiPfaffianPrimeInCycles}\brace{\var{$n$}\pattern}\,: returns $\mathrm{Pf'}({\Psi})$ for $\var{n}$-particles but decomposed (using the algorithm described in \cite{Bjerrum-Bohr:2016axv}) into coefficients times the `cycle' objects \mbox{\funL{chyCycle}\brace{\var{permutation}}\,.}
\item\funL{chyAPfaffianPrime}\brace{\var{$n$}\pattern}\,: gives $\mathrm{Pf'}({\Psi})$ for $\var{n}$-particles.
\item\funL{chyAPfaffianPrime2InCycles}\brace{\var{$n$}\pattern}\,: returns $\mathrm{Pf'}(\mathbf{A})$ for $\var{n}$-particles but decomposed (using the algorithm described in \cite{Bjerrum-Bohr:2016axv}) into coefficients times the `cycle' objects \mbox{\funL{chyCycle}\brace{\var{permutation}}\,.}
\end{itemize}\vspace{-4pt}

In terms of these building blocks, various tree-amplitudes may be represented as an integral
\eq{\mathcal{A}_\var{n}=\int\!\!\!\Omega[\vec{\sigma}]\,\,\,\delta^{\var{n}{-}3}\!(\vec{S})\,\,\mathcal{I}_\var{n}\,.\label{general_CHY_integrand}}
which is fully localized by the $\delta$ functions involving the scattering equations. The integrands of specific field theories in this formalism are given by\vspace{-6pt}
\begin{itemize}
\item\funL{ymCHYintegrand}\brace{\var{$n$}\pattern}\,: gives the CHY integrand for the ordered/partial tree-amplitudes involving \var{$n$} gluons in YM theory.
\item\funL{grCHYintegrand}\brace{\var{$n$}\pattern}\,: gives the CHY integrand for the tree-level amplitude involving $\var{n}$ gravitons in pure GR theory.
\item\funL{specialGalileonCHYintegrand}\brace{\var{$n$}\pattern}\,: gives the CHY integrand for tree-level scattering of \var{$n$} particles in the special Galileon theory.
\item\funL{diracBornInfeldCHYintegrand}\brace{\var{$n$}\pattern}\,: gives the CHY integrand for tree-level scattering in the non-supersymmetric Dirac-Born-Infeld theory \cite{Cachazo:2014xea}.
\end{itemize}

To evaluate these integral representations of amplitudes, we have included the function \funL{directCHYintegration} which will solve the scattering equations (via the function \funL{solveScatteringEquations}) and sum over the isolated contributions of (\ref{general_CHY_integrand}). 

\newpage
\vspace{-0pt}\subsubsection{Connected Prescription of the Twistor String \& the Grassmannian}\vspace{-0pt}

The connected prescription of Roiban, Spradlin, and Volovich \cite{RSV} (see also \mbox{\cite{Roiban:2004yf,Dixon:2005cf,Dolan:2007vv,Dolan:2009wf,Dolan:2010xv}}) of the twistor string theory \cite{Witten:2003nn} is closely related to the scattering equation formalism of CHY, when the external kinematics are four-dimensional. Indeed, for four-dimensional kinematics the solutions are grouped into `sectors' indexed by $k\!\in\![0,n{-}4]$, for which amplitude integrands in (\ref{general_CHY_integrand}) involving particles with N${}^{k}$MHV-degree helicity assignments will only have non-vanishing support on the corresponding sector of solutions to the scattering equations. Moreover, it is possible to `uplift' the integral expression into a manifestly (maximally) supersymmetric \funL{superFunction}---from which partial amplitudes for any external states of sYM can be extracted. 

We do not have room to sufficiently review this story here, but we have implemented the main ingredients in this representation:\vspace{-4pt}
\begin{itemize}
\item\funL{rsvMeasure}\brace{\var{$n$}\pattern,\var{$k$}\pattern}\,: gives the measure $\Omega[\vec{\sigma},\vec{\xi}]$ (as a \funL{superFunction} object) over the auxiliary variables \funL{$\xi$}\brace{\var{$a$}} and \fun{$\sigma$}\brace{\var{$a$}} which parameterize the \funL{rsvMatrix}\brace{\var{$n$},\var{$k$}} in the RSV prescription for the partial \var{$n$}-particle N${}^{\var{k}}$MHV amplitude in sYM.
\item\funL{rsvMatrix}\brace{\var{$n$}\pattern,\var{$k$}\pattern}\,: gives the matrix $C(\vec{\sigma},\vec{\xi})$ appearing in the \funL{rsvMeasure} and constrained by the \funL{grassmannianKinematicEquations}\brace{$C$}\,.
\item\funL{solveRSVequations}\brace{\var{$n$}\pattern,\var{$k$}\pattern}\,: solves the constraints appearing in the integration measure over $\Omega[\vec{\sigma},\vec{\xi}]$  \mbox{\funL{grassmannianKinematicEquations}\brace{\funL{rsvMatrix}\brace{\var{$n$},\var{$k$}}}}\, for the current, globally-specified kinematic data. 
\item\funL{fromRSVtoScatteringEquationSolutions}\brace{\var{$n$}\pattern}\,: gives the map from the particular solutions to the RSV constraints to particular solutions to the scattering equations. 
\end{itemize}

This story is smoothly connected to description of tree amplitudes in sYM as contour integrals over the (`positive') Grassmannian of \cite{ArkaniHamed:2009dn}. As described in \cite{ArkaniHamed:2009dg} (see also \cite{Nandan:2009cc,Bourjaily:2010kw}), the connected prescription of twistor string theory can be interpreted as an integral constrained by `Veronese' contour constraints, and these constraints may be smoothly deformed to result in a contour integral over the top-dimensional form of on the (what is now known as the positive) Grassmannian. The key ingredients in this story are made available via\vspace{-4pt}
\begin{itemize}
\item\funL{gaugeFixedGrassmannianRepresentative}\brace{\var{$n$}\pattern,\var{$k$}\pattern}\,: gives a particular representative \mbox{$(\var{k}{+}2)\!\times\!\var{n}$} matrix $C$ parameterized by an independent set of Pl\"ucker coordinates expressed as `minors' \funL{m}\brace{\var{$a$},\ldots,\var{$b$}}\,. 
\item\funL{grassmannianTwistorStringMeasure}\brace{\var{$n$}\pattern,\var{$k$}\pattern}\,: gives the measure $\Omega[C]$ as a particular \funL{superFunction} object over the Grassmannian representative above for the `Grassmannian' form of the representation of the partial/ordered, \var{$n$}-particle N${}^{\var{k}}$MHV tree amplitude in sYM. 
\item\funL{veroneseContourConditions}\brace{\var{$n$}\pattern,\var{$k$}\pattern}\,\: gives the `Veronese' contour conditions which---together with \funL{grassmannianKinematicEquations}\brace{$C$}---localize the integral over the measure above. 
\item\funL{solveTwistorStringEquations}\brace{\var{$n$},\var{$k$}}\,: solves the constraints of corresponding to the vanishing of the \funL{veroneseContourConditions} combined with those of \funL{grassmannianKinematicEquations}\brace{$C$} which isolates the integral appearing in this representation.
\item\funL{fromGrassmannianToScatteringEqnSolns}\brace{\var{$n$}\pattern,\var{$k$}\pattern}\,: returns the general map from the particular solutions to the constraints above to those of the scattering equations.
\end{itemize}\vspace{-4pt}
In practice, it turns out to be numerically more efficient to solve the scattering equations in the formalism of the Grassmannian form of twistor string theory and then to subsequently map these solutions to those of the scattering equation formalism as above. This is of course only valid when the massless particles' momenta are four-dimensional---but as this is the only situation for which we have built our tools for evaluation, it suffices for the purposes of the \package\, package.

\vspace{-0pt}\subsection{Positroids: the Grassmannian Geometry of On-Shell Functions}\vspace{-0pt}

The final theme in our rapid discussion of the new functionality provided by the \package\, package relates to connection between \funL{superFunction} objects appearing in the representation of ordered/partial tree (and loop) amplitudes of sYM and specific volume integrals over certain subspaces of the `positive' Grassmannian. We refer the reader to \cite{ArkaniHamed:2009dg} \emph{and also} \cite{ArkaniHamed:book} for a thorough introduction and discussion of these ideas.

Of course, there already exists an effective and thorough \textsc{Mathematica} package, \textbf{\t{positroids}} \cite{Bourjaily:2012gy} which implements many of the key functions and algorithms relevant to this story---and its connection to scattering amplitudes in sYM more specifically. Since the release of the \textbf{\t{positroids}} package, there have been a number of key improvements---both algorithmic and conceptual---that have great value for the study of scattering amplitudes. For our purposes, it was clear that additional, very useful functionality was warranted for the \package\, package; and rather than building this functionality upon the existing code-base of the \textbf{\t{positroids}} package, we have decided instead to \emph{completely rebuild} all the algorithms most relevant for our present purposes---in part, to avoid any issues regarding namespace, and to allow for a large number of ancillary improvements to core algorithms.

There are two main sources of improvement offered here relative to the \textbf{\t{positroids}} package, which we outline below. 

\vspace{-0pt}\subsubsection{\emph{Direct} Analytic \fun{superFunctions} for `Tree-Like' Positroids}\vspace{-0pt}

Many of the most relevant on-shell functions are those which are `tree-like' (as tested by \funL{treePositroidQ}\brace{\funL{$\sigma$}\brace{\var{perm}\patternTwo}}\,): these are on-shell functions that have iterated boundaries which are `factorizations' (\funL{positroidFactorizations}\brace{\funL{$\sigma$}\brace{\var{perm}\patternTwo}}\,) recursively connecting it to three-particle amplitudes.\footnote{These are vastly more common at low multiplicity: of the 19\,954 distinct (non-vanishing) on-shell functions that exist for $n\!\leq10$ particles (at any N${}^k$MHV-degree), 19\,240 are tree-like.} For tree-like positroids, we can directly construct its corresponding \emph{sign-unambiguous} \funL{superFunction} via on-shell recursion. 

While this is not a conceptually new idea, such a construction was never implemented in the \textbf{\t{positroids}} package. This new functionality is associated with the following new functions of the \package\, package\vspace{-4pt}
\begin{itemize}
\item\funL{fromPositroidsToSuperFunctions}\brace{\var{expression}\pattern}\,: converts all positroid labels \funL{$\sigma$}\brace{\var{permutation}\patternTwo} to spinor-dependent \funL{superFunction} objects using the function \funL{spinorFormOfPositroid}.
\item\funL{spinorFormOfPostiroid}\brace{\funL{$\sigma$}\brace{\var{permutation}\patternTwo}}\,: for any tree-like positroid labelled by \mbox{\funL{$\sigma$}\brace{\var{permutation}}}, returns its corresponding analytic \fun{superFunction} depending on spinor variables.
\item\funL{twistorFormOfPositroid}\brace{\funL{$\sigma$}\brace{\var{permutation}\patternTwo}}\,: for any tree-like positroid labelled by \mbox{\funL{$\sigma$}\brace{\var{permutation}}}, returns a product of momentum-twistor \funL{R}-invariants for the tree for its corresponding momentum-twistor space \funL{superFunction}.
\end{itemize}
These functions allow the user to rapidly convert from positroid labels \mbox{\funL{$\sigma$}\brace{\var{permutation}}} and analytic \funL{superFunction} objects for the rapid confirmation of identities. As the overall \emph{signs}/orientation of these expressions is determined \emph{recursively} in terms of the three-particle amplitudes, it should be the case, for example that any list of positroid labels generated by \funL{ymPositroidAmpRandom}, when converted to analytic \funL{superFunction} objects via \funL{fromPositroidsToSuperFunctions}, should \emph{sum} to the correct result (without any sign ambiguities):
\mathematicaBox{
\mathematicaSequence[1]{treeList=\funL[]{allRecursedYMpositroidTrees}\brace{8,2};\\
Length[\textnormal{\%}]}{2624}
\mathematicaSequence{\funL[]{fromPositroidsToSuperFunctions}\brace{treeList};\\
\funL[]{evaluate}\brace{\textnormal{\%}};\\
Total[\funL[]{component}\brace{\stateYM{m},\stateYM{p},\stateYM{m},\stateYM{p},\stateYM{m},\stateYM{p},\stateYM{m},\stateYM{p}}\,\brace{\#}]\,\&/@\textnormal{\%};\\
DeleteDuplicates[\textnormal{\%}]}{$\displaystyle\left\{\rule{0pt}{20pt}\right.\!\!\!\text{-}\frac{6787653230559005806372923435157}{28036956340009088805349336320000}\!\!\!\left.\rule{0pt}{20pt}\right\}$}\\[-20pt]
}

\vspace{-0pt}\subsubsection{Positroid \emph{Oriented} Homology and the Amplituhedron}\vspace{-5pt}

The biggest \emph{conceptual} advance relative to the posted work of \cite{ArkaniHamed:2012nw} \emph{but documented in an additional chapter in the published book} \cite{ArkaniHamed:book}, deals with how canonical coordinate charts are \emph{relatively oriented}. We refer the reader to \cite{ArkaniHamed:book} for a more thorough discussion (see also \cite{Olson:2014pfa}); but the ability to compare canonical coordinate charts' orientations allows for the construction of \emph{signed} covering relations that satisfy $\partial^2(\funL{$\sigma$}\brace{\var{permutation}}){=}0$. These are implemented in
\begin{itemize}
\item\builtL{positroidOrientedBoundaryRule}\,: is a rule that returns the boundary of any positroid label as a sum of lower-dimensional positroid labels with definite signs such that applying \builtL{positroidOrientedBoundaryRule} to its image will be 0.
\item\builtL{positroidOrientedInverseBoundaryRule}\,: is a rule that returns the boundary of any positroid label as a sum of higherI-dimensional positroid labels with definite signs such that applying \builtL{positroidOrientedInverseBoundaryRule} to its image will be 0.
\end{itemize}\vspace{-4pt}
We can verify this easily enough by considering all (12-dimensional) positroid labels relevant to 8-particle N${}^2$MHV amplitudes as follows:
\mathematicaBox{
\mathematicaSequence[1]{allCells=\funL[]{allPositroids}\brace{8,2};\\
Length[\textnormal{\%}]}{330}
\mathematicaSequence[1]{allCells/\!/\!\!.\builtL[]{positroidOrientedBoundaryRule};\\
DeleteDuplicates[\textnormal{\%}]}{\{0\}}
\mathematicaSequence[1]{allCells/\!/\!\!.\builtL[]{positroidOrientedInverseBoundaryRule};\\
DeleteDuplicates[\textnormal{\%}]}{\{0\}}
}

These \emph{signed} covering relations require that some choice of reference orientation be chosen for every positroid (equivalently, a canonically oriented default coordinate chart). In principle, these choices are completely arbitrary. However, for \emph{tree-like} positroids, it is natural to define their reference orientation recursively in terms of oriented charts on  \funL{$\sigma$}\brace{2,3,4} and \funL{$\sigma$}\brace{3,4,5}\,. This will have the effect that the signed boundary of any higher-dimensional positroid will correspond to a \emph{functional identity} among the \funL{superFunction}s associated with its boundary:
\mathematicaBox{
\mathematicaSequence{egSeed=\funL[]{randomPositroid}\brace{8,2,13}}{\funL{$\sigma$}\brace{5,6,7,9,8,12,10,11}}
\mathematicaSequence{egID=egSeed/\!\!.\builtL{positroidOrientedBoundary}}{\phantom{\text{+}}\mbox{\funL{$\sigma$}\brace{3,6,7,9,8,12,10,13}}\,\text{-}\mbox{\funL{$\sigma$}\brace{4,6,7,9,8,13,10,11}}\,\text{-}\mbox{\funL{$\sigma$}\brace{5,6,7,9,8,12,11,10}}\\
\text{+}\mbox{\funL{$\sigma$}\brace{5,6,7,9,10,12,8,11}}\,\text{+}\mbox{\funL{$\sigma$}\brace{5,6,7,9,12,8,10,11}}\,\text{+}\mbox{\funL{$\sigma$}\brace{5,6,7,10,8,12,9,11}}\\
\text{-}\mbox{\funL{$\sigma$}\brace{5,6,7,12,8,9,10,11}}\,\text{+}\mbox{\funL{$\sigma$}\brace{5,6,8,9,7,12,10,11}}\,\text{-}\mbox{\funL{$\sigma$}\brace{5,6,9,7,8,12,10,11}}\\
\text{+}\mbox{\funL{$\sigma$}\brace{5,7,6,9,8,12,10,11}}\,\text{-}\mbox{\funL{$\sigma$}\brace{6,5,7,9,8,12,10,11}}\,}
\mathematicaSequence{\funL[]{component}\brace{\stateYM{m},\stateYM{p},\stateYM{m},\stateYM{p},\stateYM{m},\stateYM{p},\stateYM{m},\stateYM{p}}\,\brace{\funL{fromPositroidsToSuperFunctionsN}\brace{egID}}}{0}
}
This allows us to systematically identify all functional relations, and even to compute the \built{MatrixRank} of the identities they span (something that would not be possible with sign-ambiguous boundary relations). In particular, it allows us to compute, for example, the \funL{numberOfIndependentSuperFunctions}\brace{\var{$n$},\var{$k$}} via \emph{oriented} homology: the total number of non-vanishing \fun{superFunction} objects\footnote{A more concise definition of this number would be: the sum over all positroids of dimension $2{n}{-}4$ weighted by their \funL{positroidKinematicSupport}.} minus the \built{Rank} of the relations implied by the boundaries of higher-dimensional cells.

\newpage
\vspace{-0pt}\section{Review: Essential Ingredients of On-Shell Scattering}\label{sec:basic_aspects_of_amplitudes}\vspace{-0pt}
%

The observable (single-particle) states of any quantum field theory are labelled by a representation of Poincar\`e group, which encodes their (mass,) momenta and spin, and also any number of discrete labels (`species', `color', `$R$-charge', etc.) that distinguishes between the states which have the same mass and spin. 

For a process involving $n$ external states, we may arbitrarily assign each particle an index $a\!\in\![n]\!\equivR\{1,\ldots,n\}$ and speak of \emph{the} momentum, spin, color, species, etc.\ of the \emph{$a$th} {particle}; and we may denote these quantities by $p_a$, $\sigma_a$, $c_a$, and the multi-index `$\{I_a\}$', respectively. To be clear, the index $a$ assigned to each particle has nothing intrinsic to do with color ordering, planarity, or related concepts; it is merely a requirement to make sense of any \emph{particular} state among the $[n]$ involved.

The momentum of any of external state must satisfy Einstein's relation $p_a^\mu\,p_a^\nu\,g_{\mu\nu}{=}m_a^2$, which defines a degree-two hypersurface in energy-momentum-coordinates referred to as the `mass-shell'. Thus, external states are said to be \emph{on-(the-mass-)shell}. Moreover, any process involving $n$ external states must satisfy overall momentum conservation; by \emph{convention}, we may without any loss of generality suppose that all the particles' momenta are taken to be `incoming' so that we may write this constraint as
\eq{\sum_{a\in[n]}p_a=0\in\mathbb{R}^{d}\,.\label{momentum_conserve_v0}}
Note that this represents $d$ constraints, often imposed upon the phase space of external momenta as $d$ (in number) $\delta$ functions; to emphasize the number of constraints, we write this as
\eq{\delta^{d}\!\big(\sum_ap_a^\mu\big)\,.\label{four_component_p_conserve}}

For massive particles, spin can point in any spatial direction, and thus furnishes a representation of the little group $SO(d{-}1)$; while for massless particles, spin must point along the direction of motion, corresponding to a little group of $SO(d{-}2)$ (see, \mbox{e.g.~\cite{Weinberg:2010fx,Weinberg:2020nsn}}). 

In four dimensions, representations of the little group for massless particles are two dimensional, labelled simply by the total spin $s_a\!\in\!(\mathbb{Z}/2)$ and helicity $h_a\!=\!\pm s_a$, indicating the component of spin $\vec{\sigma}_a$ along the direction of the particle's spatial momentum. (This is always well defined for massless particles, as all inertial frames will agree on a particle's helicity.) We may refer to the wave function of the $a$th external particle by `$|a\rangle^{h_a}_{\{I_a\}}$', with any color labels left implicit.

\newpage
\vspace{-0pt}\subsection{Kinematics of On-Shell, Massless Particles in Four Dimensions}\label{subsec:kinematics}\vspace{-0pt}
%

\vspace{-0pt}\subsubsection[\textbf{Spinor Variables} for On-Shell Four-Momenta]{Spinor Variables for On-Shell, Massless Particles in $4d$}\label{subsubsec:spinors}\vspace{-0pt}

Any four-dimensional momentum $p_\mu\equivL(p_0,p_1,p_2,p_3)\equivL(E,\vec{p})$ can always, unambiguously and uniquely be represented by the $2\!\times\!2$ matrix
\eq{p^\mu\mapsto p^{\alpha\,\dot{\alpha}}\equivR p^\mu\sigma_\mu^{\alpha\,\dot\alpha}=\left(\begin{array}{cc}p_0\text{+}p_3&p_1{-}i\,p_2\\p_1{+}i\,p_2&p_0{-}p_3\end{array}\right)\,.}
If the components $p_\mu$ are \emph{real}, then $p^{\alpha\,\dot\alpha}$ is a Hermitian matrix. Notice that 
\eq{\det(p^{\alpha\,\dot{\alpha}})=p_0^2{-}p_1^2{-}p_2^2{-}p_3^2\,.}
Thus, the four-momentum of a massless particle corresponds to a $2\!\times\!2$ matrix $p^{\alpha\,\dot\alpha}$ with vanishing determinant. The requirement that $\det(p^{\alpha\,\dot\alpha}){=}0$ is still non-trivial: it is a degree-two constraint in terms of the individual components $p^{\alpha\,\dot\alpha}$; but any such matrix with vanishing determinant can be written (even if non-uniquely) in the form 
\eq{p^{\alpha\,\dot\alpha}\equivL\lambda^{\alpha}\tilde\lambda^{\dot{\alpha}}\equivR\left(\begin{array}{@{}cc@{}}\lambda^1\,\tilde\lambda^{\dot1}&\lambda^1\,\tilde\lambda^{\dot2}\\\lambda^2\,\tilde\lambda^{\dot1}&\lambda^2\,\tilde\lambda^{\dot2}\end{array}\right)=\text{\built{TensorProduct}\brace{\{\funL{$\lambda$}\brace{1}\,,\funL{$\lambda$}\brace{2}\,\},\{\funL{$\lambda$b}\brace{1}\,,\funL{$\lambda$b}\brace{2}\,\}}}\,.\label{p_as_lambdas}}
(Here, we have merely given the \textsc{Mathematica} syntax for this `outer-product' as an illustration of some of the functions that the reader may find useful.)

The variables $\lambda$ and $\tilde\lambda$ each carry two components, and are called left and right-handed (Weyl) \emph{spinors}. Under local (complexified) Lorentz transformations, $\lambda$ and $\tilde\lambda$ transform as $\mathbf{2}$-dimensional (`spin-$\tfrac{1}{2}$' or fundamental) representations of each $SU(2)_L$ and $SU(2)_R$, separately. 

One may wonder about the counting of degrees of freedom here, as it would seem that \emph{arbitrary} pairs of 2-vectors $\{\lambda,\tilde\lambda\}$ define a $2\!\times\!2$ matrix $p^{\alpha\,\dot\alpha}$ via (\ref{p_as_lambdas}) describing a massless four-momentum. while the four components of any on-shell momentum are \emph{constrained} by the on-shell condition $p^2{=}0$, leaving only three independent degrees of freedom. The mismatch between the \emph{four} numbers in $\{\lambda^\alpha,\tilde\lambda^{\dot\alpha}\}$ and the three degrees of freedom available to any on-shell $p^{\alpha\,\dot\alpha}$ is accounted for by a simple $GL(1)$ \emph{redundancy}:
\eq{t\!:\!(\lambda,\tilde\lambda)\!\mapsto\!(t\,\lambda\,,t^{\text{-}1}\tilde\lambda\,)\,,\label{little_group_action}}
leaves the momentum $p^{\alpha\,\dot\alpha}$ defined by (\ref{p_as_lambdas}) invariant. The transformation (\ref{little_group_action}) thus represents the action of the \textbf{little group}\,, and can be used to eliminate any one degree of freedom from the apparent four in $\{\lambda^\alpha,\tilde\lambda^{\dot\alpha}\}$ (by setting one component to 1, say).

When $p^{\alpha\,\dot\alpha}$ is Hermitian, the {spinors} $\lambda,\tilde\lambda$ are not independent: they must be related by $\tilde\lambda{=}\pm\lambda^{*}$, and the little group (\ref{little_group_action}) rescaling becomes restricted to the case where $t$ is a a simple \emph{phase}: $t\equivR e^{i\,\theta}$ for $\theta\!\in\!\mathbb{R}/(2\pi)$; choosing to use the $GL(1)$ redundancy to make $\lambda^1$ \emph{real}, the `$\pm$' sign relating $\tilde\lambda$ and the conjugate of $\lambda$ will be the sign of the particle's energy---whether it is incoming or outgoing.

In terms of spinors, momentum conservation for $n$ massless particles (\ref{four_component_p_conserve}) becomes a $2\!\times\!2$ system of constraints, which we choose to write as:
\eq{\delta^4\!\big(\sum_{a\in[n]}p_a^\mu\big)=\delta^{2\!\times\!2}\!\big(\sum_{a\in[n]}p_a^{\alpha\,\dot\alpha}\big)=\delta^{2\!\times\!2}\!\big(\sum_{a\in[n]}\lambda_a^{\alpha}\tilde\lambda_a^{\dot\alpha}\big)\equivL\delta^{2\!\times\!2}\!\big(\lambda\!\cdot\!\tilde\lambda\big)\label{spinor_p_conserve}}
where we have labeled the spinors associated with the $a$th particle in the obvious way, introduced the notation `$\cdot$' to denote \emph{the sum over the particle index $a$}, and introduced a slight abuse of notation by writing $\lambda,\tilde\lambda$ to collectively denote the complete set of spinor components:
\eq{\lambda\equivR(\lambda_1,\ldots,\lambda_n)\equivR\left(\begin{array}{@{}cccc@{}}\lambda^1_1&\lambda_2^1&\cdots&\lambda_n^1\\
\lambda_1^2&\lambda_2^2&\cdots&\lambda_n^2\end{array}\right)\equivL\left(\begin{array}{@{}c@{}}\lambda^1\\\lambda^2\end{array}\right)\quad\text{with}\quad \lambda_a\in\mathbb{C}^2\quad\text{and}\quad\lambda^{\alpha}\in\mathbb{C}^n}
and similarly for $\tilde\lambda$. 

Although spinor variables trivialize the on-shell condition, momentum conservation (\ref{spinor_p_conserve}) is still non-trivial. Moreover, for \emph{real} momenta---for which $\tilde\lambda_a=\pm(\lambda_a)^*$---momentum conservation is a degree-two constraint on the components of the spinors. This makes it difficult to find `nice' solutions (involving rational components, say) to momentum conservation in terms of spinor variables for real momenta. This turns out to have fairly significant consequences for our preferences \emph{not} to default to real four-momenta in the \package\, package.

\paragraph{\textbf{Aside:} the \emph{Computational} Disadvantages of Real Four-Momenta}

When evaluating any expression---an amplitude, say---there is a dramatic difference between rational operations on rational numbers and rational operations on any field extension over $\mathbb{Q}$. While rational operations on rational numbers are extremely efficient and maintain infinite precision throughout intermediate steps, even simple operations on field extensions over $\mathbb{Q}$ suffer from either non-unique representations as expressions or require time-consuming (and complexity-inducing) canonicalization---e.g.
\eq{\frac{a{+}b\sqrt{\Delta}}{c{+}d\sqrt{\Delta}}=\frac{1}{c^2{-}d^2\Delta}\left((a\,c{-}b\,d\,\Delta){+}(b\,c{-}a\,d)\sqrt{\Delta}\right)\,.}
This makes it either difficult or time-consuming to compare results, check identities, etc. This is in stark contrast to the case of rational evaluations---for which two expressions may be compared and identities may be checked efficiently, and to infinite precision. 

To overcome the complexity of using field extensions over $\mathbb{Q}$, one may be tempted to simply consider functions as acting on \built{Real}-valued (or floating-point) inputs. This has the advantage of leading to equally efficient evaluations as for rational input, but comes at the not-inconsiderable cost of fixed precision. Moreover, as the kinds of functions typically encountered in quantum field theory tend to evaluate to extremely small or extremely large numbers, it can require significant delicacy to ensure that sufficient internal precision is maintained. Also, it is worth bearing in mind that considerably more than $n$-digit precision can be required to identify a rational number whose denominator is $n$ digits.

Stepping back, it is easy to see that the appearance of square-root-containing components of the spinors arises due to the fact that $\lambda,\tilde\lambda$ are not independent variables in (\ref{spinor_p_conserve}). If they \emph{were} independent, then obviously (\ref{spinor_p_conserve}) would be a linear constraint, and we could consider any rational (or even \built{Integer}) matrix of spinor components $\lambda\equivR\{\lambda_a^\alpha\}$ and easily find rational $\tilde\lambda$ which satisfy (\ref{spinor_p_conserve}): let $\lambda^\perp$ be the $(n{-}2)\!\times\!n$ matrix defined by
\eq{\lambda^{\perp}\equivR\text{\built{NullSpace}\brace{\funL{$\lambda$}}}\,\,;\label{naive_generator_of_momentum_conserving_spinors}}
then for \emph{any} random-chosen $(2\!\times\!(n{-}2))$ matrix of numbers $\rho$, $\tilde\lambda\equivR\rho\!\cdot\!\lambda^\perp$ will satisfy momentum conservation (\ref{spinor_p_conserve}). As such, it is clear that there will be no obstruction to having entirely rational kinematic data for use in evaluating expressions. 

What would it mean for us to ignore the requirement that $\lambda_a$ and $\tilde\lambda_a$ be related by complex conjugation? Clearly, without this requirement, the matrix $p_a^{\alpha\,\dot\alpha}$ would not be Hermitian, and $p^\mu$ would not correspond to a momentum in $\mathbb{R}^{1,3}$ with \emph{real} components.

What if we chose $\lambda,\tilde\lambda$ to be \emph{real}-valued (or even \built{Integer}-valued)? This would give rise to a \emph{real} $2\!\times\!2$ matrix $p^{\alpha\,\cdot\alpha}$ which only modify the identification of the component $p_2$---rendering it pure-imaginary. Thus, this is the case of (2,2) spacetime signature---to have the external kinematics live in $p_a^\mu\!\in\!\mathbb{R}^{2,2}$.

This case is preferential from a computational complexity and efficiency point of view (if not for other reasons as well). This is the primary reason why many computations in amplitudes are done using ``complex'' kinematics; but rest assured, functions of spinor variables may always be evaluated for kinematics in real-spacetime $\mathbb{R}^{1,3}$ (or any other signature the user prefers, for whatever reason). 

In practice, it is much easier to evaluate expressions for rational-valued spinors, and more easier to compare expressions. Moreover, the identities discovered at rational-input with infinite precision can be extremely convincing. (It is worth pausing to note that any rational function in any number of variables can be uniquely determined by a single evaluation. But this can require \emph{absurdly} large values of the input parameters. In practice, it is always possible to convince oneself in the equality between two expressions beyond any Boltzmann brain's Bayesian threshold by considering a few randomly chosen evaluations.)

Consider the following comparison between two formulae for a particular helicity amplitude in YM, and checking an identity among differently ordered partial amplitudes using randomly generated $\mathbb{Q}$-valued spinors in the package
\mathematicaBox{
\mathematicaSequence[1]{\funL[]{useRandomKinematics}\brace{6};\\
\builtL[]{showSpinors}}{\rule{0pt}{30pt}}~\\[-50pt]&\scalebox{0.825}{$\displaystyle\begin{array}{@{}r|cccccc|}\multicolumn{1}{c}{~}
&\fwbox{20pt}{\lambda_1}&\fwbox{20pt}{\lambda_2}&\fwbox{20pt}{\lambda_3}&\fwbox{20pt}{\lambda_4}&\fwbox{20pt}{\phantom{\tilde{\lambda}_5}\lambda_5\phantom{\tilde{\lambda}_5}}&\multicolumn{1}{c}{\fwbox{20pt}{\lambda_6}}\\\cline{2-7}\lambda^1&1944&4452&11916&3488&1960&0\\
\lambda^2&427&1101&3078&1584&1340&200\\\cline{2-7}
\end{array}\hspace{20pt}\begin{array}{@{}r|cccccc|}\multicolumn{1}{c}{~}
&\fwbox{20pt}{\tilde\lambda_1}&\fwbox{20pt}{\tilde\lambda_2}&\fwbox{20pt}{\tilde\lambda_3}&\fwbox{20pt}{\tilde\lambda_4}&\fwbox{20pt}{\phantom{\tilde{\lambda}_5}\tilde\lambda_5\phantom{\tilde{\lambda}_5}}&\multicolumn{1}{c}{\fwbox{20pt}{\tilde\lambda_6}}\\\cline{2-7}\tilde{\lambda}^{\dot{1}}&\text{-}\frac{5197}{96932700}&\frac{341}{86242180}&\frac{8}{22911795}&\frac{463}{46195680}&\frac{13}{536375}&\text{-}\frac{2627}{17010000}\\
\tilde\lambda^{\dot{2}}&\text{-}\frac{2839}{387730800}&\frac{19}{51745308}&\frac{7}{183294360}&\frac{29}{23097840}&\frac{17}{4291000}&\text{-}\frac{3193}{136080000}\\\cline{2-7}
\end{array}$}\\[10pt]
\mathematicaSequence{\funL[]{timed}\brace{\funL[]{componentAmpN}\brace{\stateYM{m},\stateYM{p},\stateYM{m},\stateYM{p},\stateYM{m},\stateYM{p}}}}{\rule{0pt}{28pt}$\displaystyle{-}\frac{303905478694851114256700324213831390323258883}{7984839395151622586850827651481600}\rule{0pt}{30pt}$}
\\[-60pt]&\scalebox{0.6}{\hspace{-30pt}Evaluation of the function \texttt{\textbf{componentAmpN[]}} for $\mathcal{A}^{(1)}_{6}(-,+,-,+,-,+)$ required \textbf{6 ms, 58 $\mu$s} to complete.}\\[28pt]
\mathematicaSequence{\funL[]{timed}\brace{\funL[]{evaluate}\brace{\funL[]{ymLocalCovariantAmpData}\brace{6}/\!\!.\funL[]{helicityPolarizationRules}\brace{\stateYM{m},\stateYM{p},\stateYM{m},\stateYM{p},\stateYM{m},\stateYM{p}}}}}{\rule{0pt}{28pt}$\displaystyle{-}\frac{303905478694851114256700324213831390323258883}{7984839395151622586850827651481600}\rule{0pt}{34pt}$}
\\[-60pt]&\scalebox{0.6}{\hspace{-30pt}Evaluation of the function \texttt{\textbf{ymLocalCovariantAmp[]}} for $\mathcal{A}_{6}$ required \textbf{687 ms, 117 $\mu$s} to complete.}\\[28pt]
\mathematicaSequence{egId=\funL[1]{egKKIdentity}\brace{6}}{\{\funL{amp}\brace{1,{2},5,4,6,3},\mbox{\funL{amp}\brace{1,2,3,5,4,6}\,,}\mbox{\funL{amp}\brace{1,2,5,3,4,6}\,,}\\
\phantom{\{}\funL{amp}\brace{1,2,5,4,3,6},\funL{amp}\brace{1,3,2,5,4,6}\,\}}
\mathematicaSequence{egId/\!\!.\{amp\brace{\var{$x$}\patternTwo}\,$\mapsto$\funL[]{componentAmpN}\brace{\{\stateYM{m},\stateYM{p},\stateYM{m},\stateYM{p},\stateYM{m},\stateYM{p}\},\{\var{$x$}\}}\}}{\scalebox{0.575}{$\left\{\rule{0pt}{20pt}\right.\!\!\displaystyle\frac{90605959390952932033258346247194648329190445391108830889}{44995052808218374491562368697394236025059737600},\frac{115313045059750797255849964974626893280557908637}{10265827900086618516591888826740326400},$}\\
\scalebox{0.575}{$\displaystyle\frac{8017008759303048263064918003435046939715340554361277919}{1386741989350168168842555125532439632862003200},\text{-}\frac{100312319046513826881640469979335695624289102650989093653099}{19646728861596632293406512160386291038310224691200},$}\\
\scalebox{0.575}{$\displaystyle\text{-}\frac{827847812589080770401792079863041693216845524453441667}{59464242974926736245535394219262628268441600}\!\!\left.\rule{0pt}{20pt}\right\}$}}
\mathematicaSequence{Total\brace{\textnormal{\%}}}{0}
}
In the above example, even though the denominator in \funL{componentAmpN}\brace{\stateYM{m},\stateYM{p},\stateYM{m},\stateYM{p},\stateYM{m},\stateYM{p}} is merely 34 digits, 78-digits of precision would be required to reconstruct this rational number from its decimal expansion; for some terms in the KK identity, more than 100-digits would be required. 

Regardless of the author's preferences, however, the package does make it easy to evaluate amplitudes using real external momenta in $\mathbb{R}^{3,1}$. Moreover, randomly-generated on-shell, momentum-conserving four-momenta in $\mathbb{Q}^{1,3}$ can be generated using \funL{useRandomKinematics}\brace{\var{$n$},\built{True}}---for which no more than two particles' spinors involve square-root field extensions beyond $\mathbb{Q}(\sqrt{\text{-}1})$.

\newpage
\subsubsection{Lorentz-Invariant Spinor `Brackets' and their Generalizations}

When discussing the momentum of the $a$th (massless) particle, it is common to denote its spinors $\lambda_a,\tilde\lambda_a$ using a suggestive bra-ket notation: 
\eq{p_a\equivR\lambda_a\tilde\lambda_a\equivL \text{`}a\rangle[a\text{'}\,.}
One motivation for this is that Lorentz invariants must involve at least two momenta (if they are massless); and because the local (complexified) Lorentz group is $SU(2)_L\!\times\!SU(2)_R$, these invariants must correspond to $2\!\times\!2$ \emph{determinants}---contractions with the Levi-Cevita symbol: 
\eq{\begin{split}&\funL{ab}\brace{\var{a},\var{b}}\equivR\ab{\var{a\,b}}\equivR\det(\lambda_\var{a},\lambda_\var{b})\equivR\epsilon_{\alpha\,\beta}\lambda_\var{a}^{\alpha}\lambda_\var{b}^{\beta}\\
&\funL{sb}\brace{\var{a},\var{b}}\equivR\sb{\var{a}\,\var{b}}\equivR\det(\tilde\lambda_{\var{a}},\tilde\lambda_\var{b})\equivR\epsilon_{\dot\alpha\,\dot\beta}\tilde\lambda_\var{a}^{\dot\alpha}\tilde\lambda_\var{b}^{\dot\beta}\end{split}\label{spinor_brackets}}
The antisymmetry of $\ab{a\,b}{=}{-}\ab{b\,a}$ is more manifest (or harder to overlook) in this notation than in other variants that may be more familiar in the context of Weyl spinors (e.g.~statements that `$(\lambda_a)_\alpha(\lambda_b)^\alpha{=}{-}(\lambda_a)^\alpha(\lambda_b)_\alpha$' or similarly confusing constructions.)

Notice that under little group transformations---by $t_a$ for particle $a$ and by $t_b$ for particle $b$---$\ab{a\,b}\mapsto t_a\,t_b\ab{a\,b}$, while $\sb{a\,b}\mapsto\!t_a^{\text{-}1}\,t_b^{\text{-}1}\sb{a\,b}$. That is, $\ab{a\,b}$ and $\sb{a\,b}$ carry the same \emph{little group weights} as $\lambda_a\lambda_b$ and $\tilde\lambda_a\tilde\lambda_b$, respectively. 

In terms of these basic spinor brackets (\ref{spinor_brackets}), all other Lorentz-invariant quantities can be constructed. For example, the 2-particle Mandelstam involving massless particles labelled by $\{a,b\}\!\in\!\binom{[n]}{2}$ would be given by
\eq{\funL{s}\brace{\var{a},\var{b}}\equivR s_{\var{a\,b}}\equivR(p_\var{a}{+}p_\var{b})^2\equivR(p_\var{a}{+}p_\var{b})^\mu(p_\var{a}{+}p_\var{b})_\mu{=}2\,p_\var{a}\!\cdot\!p_\var{b}=\ab{\var{a\,b}}\sb{\var{a\,b}}\label{mandelstam_defn}}
This notation can be generalized simply enough:
\eq{\funL{s}\brace{\var{a},\var{b},\ldots,\var{c}}\equivR s_{\var{a\,b\cdots c}}\equivR(p_\var{a}{+}p_\var{b}{+}\ldots{+}p_{\var{c}})^2=\sum_{\substack{\r{i}<\r{j}\\
\fwboxL{30pt}{\{\r{i},\r{j}\}\!\subset\!\{\var{a},\var{b},\ldots,\var{c}\}}}}\ab{\r{i\,j}}\sb{\r{i\,j}}}
Clearly, Mandelstam invariants are invariant under little group transformations. 

Another important class of Lorentz-invariant function of spinors is the combination
\eq{\funL{asb}\brace{\var{a},\funL{p}\brace{\var{b},\ldots,\var{c}}\,,\var{d}}\equivR \langle\var{a}|(\var{b}\cdots\var{c})|\var{d}]\equivR\sum_{\fwboxL{20pt}{\r{i}\!\in\!\{\var{b},\ldots,\var{c}\}}}\ab{\var{a}\,\r{i}}\sb{\r{i}\,\var{d}}\,,}
which can be equivalently expressed as
\eq{\funL{sab}\brace{\var{d},\funL{p}\brace{\var{c},\ldots,\var{b}}\,,\var{a}}\equivR [\var{d}|(\var{c}\cdots\var{b})|\var{a}\rangle\equivR\sum_{\fwboxL{20pt}{\r{i}\!\in\!\{\var{c},\ldots,\var{b}\}}}\sb{\var{d}\,\r{i}}\ab{\r{i}\,\var{a}}\,.}
Notice that object only transforms under the little group for particles $\{\var{a},\var{d}\}\!\subset\![n]$ (and is little-group \emph{neutral} with respect to all other particles' little group transformations).

The notation in `$\langle1|(2\,3\,4)|5]$', say, may seem somewhat peculiar; it may help to recall the notation of (\ref{p_as_lambdas}):
\eq{\begin{split}\langle1|(\r{2\,3\,4})|5]\Leftrightarrow\langle1|(p_\r{2}{+}p_\r{3}{+}p_\r{4})|5]&\Leftrightarrow\langle1|\big(\r{2}\rangle[\r{2}{+}\r{3}\rangle[\r{3}{+}\r{4}\rangle[\r{4}\big)|5]\\
&=\big(\ab{1\,\r{2}}[\r{2}{+}\ab{1\,\r{3}}[\r{3}{+}\ab{1\,\r{4}}[\r{4}\big)|5]\\
&=\ab{1\,\r{2}}\sb{\r{2}\,5}{+}\ab{1\r{3}}\sb{\r{3}\,5}{+}\ab{1\,\r{4}}\sb{\r{4}\,5}\,.\end{split}}

This notation generalizes in a few different ways (that prove useful in the representation of amplitudes via on-shell recursion, for example). In particular, it is useful to define 
\eq{\begin{split}\ab{\var{a}|(\r{b_1\cdots b_{\text{-}1}})|(\b{c_1\cdots c_{\text{-}1}})|\var{d}}&\equivR\sum_{\fwboxL{20pt}{\r{i}\!\in\!\{\r{b_1},\ldots,\r{b_{\text{-}1}}\}}}\ab{\var{a}\,\r{i}}[\r{i}|(\b{c_1\cdots c_{\text{-}1}})|\var{d}\rangle=\sum_{\fwboxL{20pt}{\b{j}\!\in\!\{\b{c_1},\ldots,\b{c_{\text{-}1}}\}}}\langle\var{a}|(\r{b_1\cdots b_{\text{-}1}})|\b{j}]\ab{\b{j}\,\var{d}}\\[-4pt]
&\equivL\funL{ab}\brace{\var{a},\funL{p}\brace{\r{b_1},\ldots,\r{b_{\text{-}1}}}\,,\funL{p}\brace{\b{c_1},\ldots,\b{c_{\text{-}1}}}\,,\var{d}}\end{split}\label{general_ab_bracket}}
and similarly for 
\eq{\begin{split}
\sb{\var{a}|(\r{b_1\cdots b_{\text{-}1}})|(\b{c_1\cdots c_{\text{-}1}})|\var{d}}&\equivR\sum_{\fwboxL{20pt}{\r{i}\!\in\!\{\r{b_1},\ldots,\r{b_{\text{-}1}}\}}}\sb{\var{a}\,\r{i}}\langle\r{i}|(\b{c_1\cdots c_{\text{-}1}})|\var{d}]=\sum_{\fwboxL{20pt}{\b{j}\!\in\!\{\b{c_1},\ldots,\b{c_{\text{-}1}}\}}}[\var{a}|(\r{b_1\cdots b_{\text{-}1}})|\b{j}\rangle\sb{\b{j}\,\var{d}}\\[-4pt]
&\equivL\funL{sb}\brace{\var{a},\funL{p}\brace{\r{b_1},\ldots,\r{b_{\text{-}1}}}\,,\funL{p}\brace{\b{c_1},\ldots,\b{c_{\text{-}1}}}\,,\var{d}}\,.\end{split}\label{general_sb_bracket}}

Continuing this in the obvious way, we find angle-square-brackets defined with arbitrary \emph{odd-length} sequences of \funL{p}\brace{$\cdots$}\,'s
\eq{\begin{split}\hspace{-10pt}\langle\var{a}|(\b{b_1\cdots b_{\text{-}1}})|\cdots|({\color{hgreen}c_1\cdots c_{\text{-}1}})|(\r{d_1\cdots d_{\text{-}1}})|\var{e}]&\equivR\sum_{\fwboxL{20pt}{\r{j}\!\in\!\{\r{d_1},\ldots,\r{d_{\text{-}1}}\}}}\ab{\var{a}|(\b{b_1\cdots b_{\text{-}1}})|\cdots|({\color{hgreen}c_1\cdots c_{\text{-}1}})|\r{j}}\sb{\r{j}\,\var{e}}\\
&\equivL\funL{asb}\brace{\var{a},\funL{p}\brace{\b{b_1},\ldots,\b{b_{\text{-}1}}}\,,\ldots,\funL{p}\brace{\r{d_1},\ldots,\r{d_{\text{-}1}}}\,,\var{e}}\end{split}\label{general_asb_bracket}}
and angle/square-brackets defined for arbitrary \emph{even-length} sequences of \funL{p}\brace{$\cdots$}\,'s
\eq{\begin{split}\hspace{-10pt}\ab{\var{a}|(\b{b_1\cdots b_{\text{-}1}})|\cdots|({\color{hgreen}d_1\cdots d_{\text{-}1}})|(\r{e_1\cdots e_{\text{-}1}})|\var{f}}&\equivR\sum_{\fwboxL{20pt}{\r{j}\!\in\!\{\r{e_1},\ldots,\r{e_{\text{-}1}}\}}}\langle\var{a}|(\b{b_1\cdots b_{\text{-}1}})|\cdots|({\color{hgreen}d_1\cdots d_{\text{-}1}})|\r{j}]\ab{\r{j}\,\var{f}}\\[-4pt]
&\equivL\funL{ab}\brace{\var{a},\funL{p}\brace{\b{b_1},\ldots,\b{b_{\text{-}1}}}\,,\ldots,\funL{p}\brace{\r{e_1},\ldots,\r{e_{\text{-}1}}}\,,\var{f}}\\[4pt]
\hspace{-10pt}\sb{\var{a}|(\b{b_1\cdots b_{\text{-}1}})|\cdots|({\color{hgreen}d_1\cdots d_{\text{-}1}})|(\r{e_1\cdots e_{\text{-}1}})|\var{f}}&\equivR\sum_{\fwboxL{20pt}{\r{j}\!\in\!\{\r{e_1},\ldots,\r{e_{\text{-}1}}\}}}[\var{a}|(\b{b_1\cdots b_{\text{-}1}})|\cdots|({\color{hgreen}d_1\cdots d_{\text{-}1}})|\r{j}\rangle\sb{\r{j}\,\var{f}}\\[-4pt]
&\equivL\funL{sb}\brace{\var{a},\funL{p}\brace{\b{b_1},\ldots,\b{b_{\text{-}1}}}\,,\ldots,\funL{p}\brace{\r{e_1},\ldots,\r{e_{\text{-}1}}}\,,\var{f}}\,.\end{split}\label{generalized_ab_and_sb}}

Notice that in all the definitions above, expressions involving sums of $p$'s are additive.

\vspace{-0pt}\subsubsection{Relations and Redundancies of Functions of Spinors}\vspace{-0pt}

Of course, these functions of spinor variables are far from unique. For one thing, momentum conservation for $n$ particles always implies, for example that 
\eq{\funL{p}\brace{\var{a}}\,\,={-}\sum_{\r{j}\neq\var{a}}\funL{p}\brace{\r{j}}\,.}
This allows for many sometimes surprising identities. For example, for 10 particles
\eq{\ab{10|(1\,2)|(\r{3\,4})|(5\,6)|(\b{7\,8\,9})|10}={-}\ab{10|(1\,2)|(\r{3\,4\,5\,6})|(5\,6)|(\b{1\,2\,3\,4})|10}\,.\label{eg_10_pt_spinor_identity}}
This is not a hard identity to prove (try it!).\footnote{A useful, easier to prove identity is that $\ab{\var{x}\patternTwo,(\r{a\rule[-1.05pt]{12pt}{.65pt}})|(\r{a\rule[-1.05pt]{12pt}{.65pt}}),\b{y\rule[-1.05pt]{12pt}{.65pt}}}{=}{-}s_{\r{a}}\ab{\var{x}\,,\b{y}}$ for any repeated sequence $(\r{a\rule[-1.05pt]{12pt}{.65pt}})$. (The same identity holds also for square brackets or angle-square brackets.)} 
Using the package \package, this identity is easy to verify directly:
\mathematicaBox{
\mathematicaSequence{\funL[]{explicify}\brace{\\
\phantom{+}\funL{ab}\brace{10,\funL{p}\brace{1,2}\,,\funL{p}\brace{3,4}\,,\funL{p}\brace{5,6}\,,\funL{p}\brace{7,8,9}\,,10}\\
{+}\funL{ab}\brace{10,\funL{p}\brace{1,2}\,,\funL{p}\brace{3,4,5,6}\,,\funL{p}\brace{5,6}\,,\funL{p}\brace{1,2,3,4}\,,10}\,}}{0}
}
(As explained in \mbox{appendix~\ref{}}, the function \funL{explicify} evaluates expressions for random choices of their variables.)

But there are also identities among spinors expressions following from the dimensionality of their variables. In particular, as any spinor $\r{\lambda_a}$, say, is simply some two-component object, it can be expanded into any linearly independent pair of spinors---$\{\lambda_\b{b},\lambda_{\b{c}}\}$, say:
\eq{\r{\lambda_a}=\b{\lambda_{b}}\frac{\ab{\b{c}\,\r{a}}}{\ab{\b{b\,c}}}{+}\b{\lambda_{c}}\frac{\ab{\r{a}\,\b{b}}}{\ab{\b{b\,c}}}\,.}
This is an instance of what is known as \emph{Cramer}'s rule---or simply `the general solution to a $n$ linear equations':
\eq{\lambda_a\,\ab{b\,c}{+}\lambda_b\,\ab{c\,a}{+}\lambda_c\,\ab{a\,b}{=}0\,.\label{cramers_rule_spinors}}
If we view the collection of all $\binom{n}{2}$ angle-brackets $\ab{a\,b}$ as abstract \emph{symbols} disembodied from their meaning to us above, then the `Pl\"ucker relations' or \emph{Schouten identities}
\eq{\ab{a\,b}\,\ab{c\,d}{+}\ab{b\,c}\,\ab{a\,d}{+}\ab{c\,a}\,\ab{b\,d}{=}0\label{schouten_relation}}
(which follows by contracting (\ref{cramers_rule_spinors}) with $\lambda_d$) for all sets $\{a,b,c,d\}$ can be viewed as a collection of degree-two constraints on the space of brackets; satisfying these constraints is equivalent to the \emph{interpretation} of these symbols as \emph{meaning} the determinants of two-dimensional objects.

At any rate, the Schouten relations (\ref{schouten_relation}) can lead to tremendous simplifications among spinor expressions. We know of no universal algorithm that will ensure such simplifications---nor any universal algorithm that will canonically eliminate identities such as (\ref{eg_10_pt_spinor_identity}); but \emph{verifying} any claimed simplification is as trivial as evaluation for explicitly chosen sets of kinematic data; and verification at sufficiently random choices can rapidly exceed the Bayesian confidence threshold set by any Boltzmann brane. 

\vspace{-0pt}\subsubsection{\textbf{Momentum Twistors} (Mostly) for Ordered Amplitudes}\label{subsubsec:momentum_twistors_big_picture}\vspace{-0pt}

As described above, although spinor variables trivialize the on-shell condition for massless particles, momentum conservation (\ref{spinor_p_conserve}) remains a non-trivial constraint which must be imposed upon the space of allowed $\{\lambda,\tilde\lambda\}$ variables. \emph{Momentum-twistor} variables were introduced by Andrew Hodges in \cite{Hodges:2009hk} in part to simultaneously trivialize both the on-shell and momentum-conservation constraints (see also \cite{Mason:2009qx,ArkaniHamed:2009vw}). Although momentum-twistor variables are primarily used internally by the package \package, it is worthwhile to briefly note their definition (and our conventions for them) here. 

Momentum twistors are ordinary twistors, but defined on momentum-space (hence their name). Under the twistor map (see e.g.~\cite{Penrose:1962ij,Penrose:1967wn,Penrose:1968me,Penrose:1972ia}), every point in `momentum-$x$-space' \mbox{$x_a\!\in\!\mathbb{R}^{4}$} is mapped to a \emph{line} in twistor space---$\mathrm{span}\{z_A,z_a\}\equivL\text{`}(A\,a)\text{'}$ with $z_A,z_a\!\in\!\mathbb{P}^3$ corresponding to any pair of independent points along this particular line.\footnote{We will always consider twistors $z\!\in\!\mathbb{P}^3$ as being \emph{represented} by their \emph{four} homogeneous coordinates.} Two points in momentum space are light-like separated (that is, $(x_a{-}x_b)^2{=}0$) iff their lines in twistor space intersect at a point. When this happens, the point at their intersection defines a particular twistor---which is the image of \emph{that} `light ray' in spacetime.

Notice that for $n$ massless particles, each $p_a$ corresponds to some `light ray' in momentum-$x$-space because simply because $p_a^2{=}0$. Hence, there should be some particular twistor $z_a$ associated with every null momentum $p_a$. The linear span of any pair of such twistors $(a\,b)$ corresponds to some point in spacetime. Is there any meaning to these $\binom{n}{2}$ points? Well, clearly the spacetime points corresponding to the lines $(a\,b)$ and $(b\,c)$ in twistor space intersect at $z_b$; and so these lines correspond to to light-like separated points in spacetime---points whose light-like separation is exactly the light ray $p_b$. 

We have so far been careful to avoid any reference to how the labels $a\!\in\![n]$ which identify the momenta have been ordered. But let's use this ordering (however arbitrary) to construct a polygon through momentum-twistor space built out of consecutive pairs of twistors $(n\,1),(1\,2),\ldots,(n{-}1,n)$. Each consecutive pair of these lines intersects at a given twistor: e.g. the line $(n,1)$ and $(1,2)$ intersect at $z_1$ \emph{by construction}. Thus, these correspond to points light-like separated. If we denote the spacetime points $x$ corresponding to these lines as $x_a\Leftrightarrow(a\mi1\,a)$, then this sequence of points is pair-wise null-separated, with  $p_a\equivR(x_{a{+}1}{-}x_a)$ (with $x_{n{+}1}{=}x_{1}$ understood). Moreover, momentum conservation is manifested by the fact that the polygon \emph{closes}. 

The upshot of this story is that \emph{any arbitrary} set of $n$ momentum-twistors can be used in this way to define a set of \emph{massless} momenta $\{p_a\}$ that satisfy momentum conservation. In truth, this is little more than a different way to tell the story mentioned above: where we used $\lambda^{\perp}$ defined in (\ref{naive_generator_of_momentum_conserving_spinors}) to find sets of $\tilde{\lambda}$ which obey $\lambda\!\cdot\!\tilde\lambda{=}0$; but momentum-twistors formalize this procedure in a way that naturally ties together the meaning of the `random coefficients' $\rho$ to a geometric picture of twistor space geometry that proves very powerful.

Momentum-twistor kinematics is therefore described by a set of spinors $z\equivR\{z_1,\ldots,z_n\}$ with each $z_a$ being represented by a $4$-tuple of homogeneous coordinates. The only natural kinematic invariant associated with momentum twistors are the four-bracket which (slightly abusing notation) we denote by
\eq{\fun{ab}\brace{\var{a},\var{b},\var{c},\var{d}}\equivR\ab{\var{a\,b\,c\,d}}\equivR\det\{z_{\var{a}},z_{\var{b}},z_{\var{c}},z_{\var{d}},\}\,.}

Just as the natural invariants of spinors were generalizable to more involved versions (which prove useful in the context of on-shell recursion, for example), it turns out to be very useful to discuss twistors defined by geometric criteria. For example, we would like to allow for invariants involving not merely the twistors $z_a$ with $a\!\in\![n]$, but also perhaps the twistor defined as `where the line $(a\,b)$ intersects the plane $(c\,d\,e)$':\footnote{Here we have named the object `\funL{cap}' because of the \LaTeX command for the symbol `$\tncap$'.}
\eq{\funL{cap}\brace{\{\r{a},\r{b}\},\{\b{c},\b{d},\b{e}\}}\equivR z_{(\r{a\,b})\tcap(\b{c\,d\,e})}\equivR\text{`}(\r{a\,b})\tcap(\b{c\,d\,e})\text{'}\equivR\mathrm{span}\{z_\r{a},z_\r{b}\}\tcap\mathrm{span}\{z_\b{c},z_\b{d},z_\b{e}\}\,.}
This twistor is easy to describe concretely via the four-dimensional version of Cramer's rule:
\eq{z_a\ab{b\,c\,d\,e}{+}z_b\ab{c\,d\,e\,a}{+}z_{c}\ab{d\,e\,a\,b}{+}z_{d}\ab{e\,a\,b\,c}{+}z_{e}\ab{a\,b\,c\,d}{=}0\,}
for any set of twistors labeled by $\{a,b,c,d,e\}$. Moving the final three terms to the right hand side allows us to identify precisely a point simultaneously spanned by $\{z_\r{a},z_\r{b}\}$ and $\{z_\b{c},z_\b{d},z_\b{e}\}$:
\eq{\funL{cap}\brace{\{\r{a},\r{b}\},\{\b{c},\b{d},\b{e}\}}\equivR(\r{a\,b})\tcap(\b{c\,d\,e})=z_\r{a}\ab{\r{b}\,\b{c\,d\,e}}{+}z_{\r{b}}\ab{\b{c\,d\,e}\,\r{a}}\,.}
The translation from these kinematic variables to ordinary spinors is achieved through the identification of each twistor's four homogeneous coordinates 
\eq{z_a\equivL(\lambda_a^{1}\,,\lambda_a^2\,,\mu_a^1\,,\mu_a^2)}
which requires that the symmetry of the four-dimensional space of homogeneous twistor coordinates $z_a^I$ be explicitly broken\footnote{This can be seen as a manifest reference to the `infinity' bi-twistor $I_{\infty}$, used to identify the 2-dimensional subspace of each twistor's components corresponding to $\lambda_a$.} and then defining the $\tilde\lambda$ variables by the map
\eq{\tilde\lambda_\r{a}\equivR\frac{\ab{\r{a}\,\r{a}{+}1}\mu_{\r{a}{-}1}{+}\ab{\r{a}{+}1\,\r{a}{-}1}\mu_{\r{a}}{+}\ab{\r{a}{-}1\,\r{a}}\mu_{\r{a}{+}1}}{\ab{\r{a}{-}1\,\r{a}}\ab{\r{a}\,\r{a}{+}1}},}
which makes \emph{explicit} reference to the cyclic ordering of the particles' labels. It is worthwhile to bear in mind that although momentum twistors are defined for any arbitrary ordering, the \emph{map} between spinors and momentum-twistors breaks both conformal invariance and implicitly makes use of the ordering of these labels. This translation is very well suited for the \emph{ordered, partial} (a.k.a.~`planar') amplitudes of general field theories.

\vspace{-0pt}\subsection{Labeling External Sates of Scattering Amplitudes}\label{subsec:labeling_states}\vspace{-0pt}

\subsubsection{Helicity Amplitudes and Polarization Vectors/Tensors}

In pure gauge theory or gravity, we often think of labeling states (\emph{non-uniquely}) by their polarization vectors/tensors: $\epsilon_a^\pm$---which are defined only up to a choice of reference frame or, equivalently, a gauge choice. Of course, the truly \emph{invariant} information about the state is simply the sign `$\pm$' of the particle's helicity. From the Feynman rules, there is little distinction between amplitudes involving different distributions of helicity. 

Although the \package\, package is primarily concerned with states identified directly by their helicity labels, we have implemented a manifestly color-kinematic satisfying, local and covariant form of amplitudes in pure Yang-Mills. These are necessarily expressed in terms of polarization vectors. Specifically, we have provided a function \funL{ymLocalCovariantAmp}\brace{\var{$n$}} that gives the `color-striped' \emph{partial} or \emph{ordered} tree amplitude involving $\var{n}$ gluons; and also \funL{ymLocalCovariantColorDressedAmp}\brace{\var{$n$}} gives the full amplitude with color-dressing. These expressions are given in terms of manifestly dimensionally-agnostic Mandelstam invariants \funL{s}\brace{\var{$a$},\ldots,\var{$b$}} defined above in (\ref{mandelstam_defn}) and the ``squared sums'' 
\eq{\funL{ss}\brace{\funL{$\epsilon$}\brace{\var{a}}\,,\funL{$\epsilon$}\brace{\var{b}}}\,\,\equivR(\epsilon_{\var{a}}{+}\epsilon_{\var{b}})^2\quad\text{and}\quad\funL{ss}\brace{\funL{$\epsilon$}\brace{\var{a}}\,,\funL{p}\brace{\var{b}}}\,\,\equivR(\epsilon_{\var{a}}{+}p_{\var{b}})^2\,.}
(We choose to use squared-sums as opposed to dot products because it allows us to fully avoid any factors of 2 from appearing in our expressions.)

To evaluate these expressions, the polarization tensors must be specified, which can be done using \funL{helicityPolarizationRules}, as in:
\mathematicaBox{
\mathematicaSequence[1]{localAmp=\funL[]{timed}\brace{\funL[]{ymLocalCovariantAmp}\brace{6}};\\
LeafCount[localAmp]}{\rule{0pt}{20pt}634387}
\\[-45pt]&\scalebox{0.6}{\hspace{-30pt}Evaluation of the function \texttt{\textbf{ymLocalCovariantAmp[]}} for $\mathcal{A}^{(6)}$ required \textbf{341 ms, 284 $\mu$s} to complete.}\\[10pt]
\mathematicaSequence[1]{helAmp=localAmp/\!\!.\funL[]{helicityPolarizationRules}\brace{\stateYM{m},\stateYM{p},\stateYM{m},\stateYM{p},\stateYM{m},\stateYM{p}}\,;\\
\funL[]{timed}\brace{\funL[]{evaluate}\brace{helAmp}}}{$\rule{0pt}{26pt}\displaystyle\text{-}\frac{464575984240415}{164913754335264}$}
\\[-55pt]&\scalebox{0.6}{\hspace{-30pt}Evaluation of the function \texttt{\textbf{evaluate}} required \textbf{152 ms, 761 $\mu$s} to complete.}\\[25pt]
}
It is interesting to compare this with the more specific helicity amplitude
\mathematicaBox{
\mathematicaSequence[1]{recursedAmp=\funL[]{timed}\brace{\funL[]{componentAmp}\brace{\stateYM{m},\stateYM{p},\stateYM{m},\stateYM{p},\stateYM{m},\stateYM{p}}};
LeafCount[recursedAmp]}{\rule{0pt}{20pt}161}
\\[-45pt]&\scalebox{0.6}{\hspace{-30pt}Evaluation of the function \texttt{\textbf{componentAmp[]}} for $\mathcal{A}^{(6)}\!(-,+,-,+,-,+)$ required \textbf{11 ms, 367 $\mu$s} to complete.}\\[10pt]
\mathematicaSequence{\funL[]{nice}\brace{recursedAmp}}{\scalebox{0.925}{$\phantom{{-}}\displaystyle\frac{\sb{4\,6}^4\ab{3\,1}^4}{\sb{4\,5}\sb{5\,6}\ab{1\,2}\ab{2\,3}\langle1|(2\,3)|4]\langle3|(1\,2)|6]s_{1\,2\,3}}{+}
\frac{\sb{2\,4}^4\ab{5\,1}^4}{\sb{2\,3}\sb{3\,4}\ab{5\,6}\ab{6\,1}\langle1|(2\,3)|4]\langle5|(3\,4)|2]s_{2\,3\,4}}$}\\[5pt]
\scalebox{0.925}{$\displaystyle{-}\frac{\sb{2\,6}^4\ab{3\,5}^4}{\sb{1\,2}\sb{6\,1}\ab{3\,4}\ab{4\,5}\langle3|(1\,2)|6]\langle5|(3\,4)|2]s_{6\,1\,2}}$}}
\mathematicaSequence{\funL[]{timed}\brace{\funL[]{evaluate}\brace{recursedAmp}}}{$\rule{0pt}{26pt}\displaystyle\text{-}\frac{464575984240415}{164913754335264}$}
\\[-58pt]&\scalebox{0.6}{\hspace{-30pt}Evaluation of the function \texttt{\textbf{evaluate}} required \textbf{429 $\mu$s} to complete.}\\[27pt]
}

One complaint about the use of polarization vectors and the Feynman rules is that there is very little reason to suggest that there much to distinguish different choices of external helicities. Consider for example the following choice of helicities, for which the amplitude \emph{must} vanish:
\mathematicaBox{
\mathematicaSequence[2]{egAmp=\funL[]{ymLocalCovariantAmp}\brace{4};\\
cmptAmp=egAmp/\!/\!\!.\funL[]{helicityPolarizationRules}\brace{\stateYM{m},\stateYM{p},\stateYM{p},\stateYM{p}};\\
\funL[]{nice}@FullSimplify[cmptAmp}{$\rule{0pt}{20pt}\displaystyle\frac{1}{s_{1\,2}s_{1\,4}}\Big[(\epsilon_1^{-}\!{+}\epsilon_3^{+})^2\big(s_{1\,4}(\epsilon_2^+\!{+}p_1)^2(\epsilon_4^{+}\!{+}p_3)^2{-}s_{1\,2}(\epsilon_2^+\!{+}p_3)^2\big((\epsilon_4^+\!{+}p_2)^2\!{+}(\epsilon_4^{+}\!{+}p_3)^2\big)\big)$\\
$\displaystyle{+}(\epsilon_1^-\!{+}\epsilon_2^+)^2\big(s_{1\,4}(\epsilon_3^+\!{+}p_1)^2(\epsilon_4^+\!{+}p_2)^2\!{+}(s_{1\,2}{+}s_{1\,4})(\epsilon_3^+\!{+}p_2)^2\big((\epsilon_4^+\!{+}p_2)^2\!{+}(\epsilon_4^+\!{+}p_3)^2\big)\big)$\\
$\displaystyle{+}(\epsilon_1^-\!{+}\epsilon_4^+)^2\big(s_{1\,4}(\epsilon_2^+\!{+}p_1)^2\big(\!(\epsilon_3^+\!{+}p_1)^2\!{+}(\epsilon_3^+\!{+}p_2)^2\big)\!$\\
$\phantom{+}\hspace{60pt}\displaystyle{+}s_{1\,2}\big((\epsilon_2^+\!{+}p_1)^2(\epsilon_3^+\!\!\!{+}p_2)^2\!{-}(\epsilon_2^+\!\!\!{+}p_3)^2(\epsilon_3^+\!{+}p_1)^2\big)\big)\big)\Big]$}
\mathematicaSequence{\funL[]{evaluate}\brace{cmptAmp}}{0}
}

The vanishing of the amplitude for such a helicity configuration is obvious in the context of supersymmetric Yang-Mills theory (`sYM') and gravity (`sGR'). Interestingly, all \emph{tree-level} scattering amplitudes of $\mathcal{N}$-supersymmetric sYM or sGR (including \emph{non}-supersymmetric YM and GR) are identical to those of \emph{maximally} supersymmetric sYM and sGR involving whatever external states exist for the less-supersymmetric theory. This is a stronger version of the well-known fact that tree-level `helicity' amplitudes involving gluons or gravitons in maximally supersymmetric sYM and sGR are identical to those of pure YM and pure GR (a fact which follows trivially from the Feynman rules). 

As such, it is worthwhile to choose to label states in a way which identifies the states of $\mathcal{N}$-supersymmetric sYM or sGR manifestly as subsets of those states of the maximally supersymmetric theories. 

\subsubsection{The Spectra of States in Pure ($\mathcal{N}$-)supersymmetric sYM and sGR}\label{subsec:labeling_states}

\begin{table}[t]$$\scalebox{1}{\fwbox{0pt}{\hspace{0pt}\begin{array}{|lc@{}l@{$\!\!$}lc@{}|}
\multicolumn{5}{c}{\fwbox{450pt}{\text{\textbf{External States of \emph{Pure} $\mathcal{N}$-Supersymmetric superYang-Mills Theory (`sYM')}}}}\\
\hline\multicolumn{1}{|c}{\begin{array}{c}\text{(stylized)}\\[-4pt]\text{\textbf{state}}\\\end{array}}&\hspace{-2pt}\begin{array}{c}~\\[-4pt]\text{\textbf{helicity}}\end{array}\hspace{-2pt}&\multicolumn{1}{c}{\hspace{-25pt}\begin{array}{c}\textbf{`$R$-charge'}\\[-4pt]\text{{State-Label}}\end{array}\hspace{25pt}}&\multicolumn{1}{c}{\hspace{-20pt}\begin{array}{c}\package\\[-4pt]\text{\textbf{syntax}}\end{array}\hspace{20pt}}&\multicolumn{1}{c@{}|}{\hspace{-10pt}\begin{array}{c}\text{number}\\[-4pt]\text{{of states}}\end{array}}\\
\hline
|g\rangle^{\text{+}1}&1&\{\}\!\in\!\binom{[\mathcal{N}]}{0}&\stateYM[1]{p}&1\\[4pt]
|\psi\rangle^{\text{+}1/2}_{(I)}&\frac{1}{2}&\{I\}\!\in\!\binom{[\mathcal{N}]}{1}&\stateYM[1]{psi}\brace{\t{I}}&\binom{\mathcal{N}}{1}\\[4pt]
|\phi\rangle^{0}_{(I\,J)}\simeq|\bar{\phi}\rangle^0_{(\bar{K\,L})}&0&\{\!I\!,\!J\!\}\!\equivR\![4]\backslash\{\!K\!,\!L\!\}\!\in\!\binom{[\mathcal{N}]}{2}&\stateYM[1]{phi}\brace{\t{I\!,J}}\simeq\fun{phiBar}\brace{\t{K\!,L}}&\binom{\mathcal{N}}{2}\\[4pt]
|\bar{\phi}\rangle^{0}_{(\bar{I\,J})}\simeq|{\phi}\rangle^0_{({K\,L})}&0&\{\!K\!,\!L\!\}\!\!\equivL\bar{\{\!I\!,\!J\!\}}\!\in\![4]\backslash\binom{[\mathcal{N}]}{2}&\stateYM[1]{phiBar}\brace{\t{I\!,J}}\simeq\fun{phi}\brace{\t{K\!,L}}&\binom{\mathcal{N}}{2}\\[4pt]
|\bar{\psi}\rangle^{\text{-}1/2}_{(\bar{I})}&\text{-}\frac{1}{2}\phantom{\text{-}}&\{\!J\!,\!K\!,\!L\!\}\!\!\equivL\bar{\{I\}}\!\in\![4]\backslash\binom{[\mathcal{N}]}{1}&\stateYM[1]{psiBar}\brace{\t{I}}&\binom{\mathcal{N}}{1}\\[4pt]
|g\rangle^{\text{-}1}&\text{-}1\phantom{\text{-}}&\fwboxL{175pt}{\{1,2,3,4\}\!\!\equivL\bar{\{\}}\in\![4]\backslash\binom{[\mathcal{N}]}{0}}&\stateYM[1]{m}&1\\\hline
\end{array}
}}$$\vspace{-20pt}\caption{Conventions for labeling states of pure $\mathcal{N}$-supersymmetric superYang-Mills (`sYM'). Notice that for $\mathcal{N}\!=\!4$, the scalars are duplicated between $\phi$ and $\bar{\phi}$. \label{sym_state_table}}\vspace{-15pt}
\end{table}

In $\mathcal{N}$-supersymmetric Yang-Mills theory (`sYM') or gravity (`sGR'), the external states consist of gluons $|g\rangle^{\pm1}$ (in the adjoint representation of some Lie algebra (color labels suppressed)) or gravitons $|g\rangle^{\pm2}$, and all the other states related to these by supersymmetry. In \emph{pure} sYM and sGR, these are the \emph{only} external states of the theory. 

The spectrum of states in $\mathcal{N}$-supersymmetric sYM is summarized in \mbox{table~\ref{sym_state_table}}, and the spectrum of $\mathcal{N}$-supersymemtric sGR is given in \mbox{table~\ref{sgr_state_table}}.

\begin{table}[t]$$\scalebox{1}{\fwbox{0pt}{\hspace{0pt}\begin{array}{|lc@{}l@{$\!\!$}lc@{}|}
\multicolumn{5}{c}{\fwbox{450pt}{\text{\textbf{External States of \emph{Pure} $\mathcal{N}$-Supersymmetric superGravity Theory (`sGR')}}}}\\
\hline\multicolumn{1}{|c}{\begin{array}{c}\text{(stylized)}\\[-4pt]\text{\textbf{state}}\\\end{array}}&\hspace{-10pt}\fwbox{30pt}{\begin{array}{c}~\\[-4pt]\text{\textbf{helicity}}\end{array}}\hspace{-10pt}&\multicolumn{1}{c}{\hspace{-25pt}\begin{array}{c}\textbf{`$R$-charge'}\\[-4pt]\text{{State-Label}}\end{array}\hspace{25pt}}&\multicolumn{1}{c}{\hspace{-30pt}\begin{array}{c}\package\\[-4pt]\text{\textbf{syntax}}\end{array}\hspace{10pt}}&\multicolumn{1}{c@{}|}{\hspace{-10pt}\begin{array}{c}\text{number}\\[-4pt]\text{{of states}}\end{array}\hspace{0pt}}\\
\hline
|g\rangle^{\text{+}2}&2&\{\}\!\in\!\binom{[\mathcal{N}]}{0}&\stateGR[1]{pp}&1\\[4pt]
|\psi\rangle^{\text{+}3/2}_{(I)}&\frac{3}{2}&\{I\}\!\in\!\binom{[\mathcal{N}]}{1}&\stateGR[1]{gravitino}\brace{\t{I}}&\binom{\mathcal{N}}{1}\\[4pt]
|\gamma\rangle^{\text{+}1}_{(I\,J)}&1&\{\!I\!,\!J\!\}\!\in\!\binom{[\mathcal{N}]}{2}&\stateGR[1]{graviPhoton}\brace{\t{I\!,J}}&\binom{\mathcal{N}}{2}\\[4pt]
|\psi\rangle^{\text{+}1/2}_{(I\,J\,K)}&\frac{1}{2}&\{\!I\!,\!J\!,\!K\!\}\!\in\!\binom{[\mathcal{N}]}{3}&\stateGR[1]{psi}\brace{\t{I\!,J\!,K}}&\binom{\mathcal{N}}{3}\\[4pt]
\fwboxL{100pt}{|\phi\rangle^{0}_{\!(I\!J\!K\!L)}\!\simeq\!|\bar{\phi}\rangle^{0}_{\!(\bar{M\!N\!O\!P})}}&0&\{\!I\!,\!J\!,\!K\!,\!L\!\}\equivR\![8]\backslash\{\!M\!,\!N\!,\!O\!P\!\}\!\in\!\binom{[\mathcal{N}]}{4}&\fwboxL{145pt}{\stateGR[1]{phi}\brace{\t{I\!,J\!,K\!,L}}\!\simeq\!\fun{phiBar}\brace{\t{M\!,N\!,O\!,P}}}&\binom{\mathcal{N}}{4}\\[4pt]
\fwboxL{90pt}{|\bar{\phi}\rangle^{0}_{\!(\bar{I\!J\!K\!L})}\!\simeq\!|{\phi}\rangle^{0}_{\!({M\!N\!O\!P})}}&0&\{\!M\!,\!N\!,\!O\!,\!P\!\}\!\!\equivL\bar{\{\!I\!,\!J\!,\!K\!,\!L\!\}}\!\in\![8]\backslash\binom{[\mathcal{N}]}{4}&\fwboxL{145pt}{\stateGR[1]{phiBar}\brace{\t{I\!,J\!,K\!,L}}\!\simeq\!\fun{phi}\brace{\t{M\!,N\!,O\!,P}}}&\binom{\mathcal{N}}{4}\\[4pt]
|\bar{\psi}\rangle^{\text{-}1/2}_{(\bar{I\,J\,K})}&\text{-}\frac{1}{2}\phantom{\text{-}}&\{\!L\!,\!M\!,\!N\!,\!O\!,\!P\!\}\!\!\equivL\bar{\{\!I\!,\!J\!,\!K\!\}}\!\in\![8]\backslash\binom{[\mathcal{N}]}{3}&\stateGR[1]{psiBar}\brace{\t{I\!,J\!,K}}&\binom{\mathcal{N}}{3}\\[4pt]
|\bar{\gamma}\rangle^{\text{-}1}_{(\bar{I\,J})}&\text{-}1\phantom{\text{-}}&\{\!K\!,\!L\!,\!M\!,\!N\!,\!O\!,\!P\!\}\!\!\equivL\bar{\{\!I\!,\!J\!\}}\!\in\![8]\backslash\binom{[\mathcal{N}]}{2}&\stateGR[1]{graviPhotonBar}\brace{\t{I\!,J}}\hspace{-4pt}&\binom{\mathcal{N}}{2}\\[4pt]
|\bar{\psi}\rangle^{\text{-}3/2}_{(I)}&\text{-}\frac{3}{2}\phantom{\text{-}}&\{\!J\!,\!K\!,\!L\!,\!M\!,\!N\!,\!O\!,\!P\!\}\!\!\equivL\bar{\{I\}}\!\in\![8]\backslash\binom{[\mathcal{N}]}{1}&\stateGR[1]{gravitinoBar}\brace{\t{I}}&\binom{\mathcal{N}}{1}\\[4pt]
|g\rangle^{\text{-}2}&\text{-}2\phantom{\text{-}}&\{\,1,2,3,4,5,6,7,8\,\}\!\!\equivL\bar{\{\}}\in\![8]\backslash\binom{[\mathcal{N}]}{0}\hspace{5pt}&\stateGR[1]{mm}&1\\\hline
\end{array}
}}$$\vspace{-20pt}\caption{Conventions for labeling states of pure $\mathcal{N}$-supersymmetric superGravity (`sGR'). Notice that for $\mathcal{N}\!=\!8$, the scalars are duplicated between $\phi$ and $\bar{\phi}$.\label{sgr_state_table}}
\vspace{-15pt}\end{table}

One advantage of labeling the states as we have by `$R$-charge' labels (even when there is no $R$-symmetry) is that these labels are independent of the amount of supersymmetry. That is, we chose to associate the (-)-helicity gluon \stateYM{m} as the state labelled by `$\{1,2,3,4\}$' and the sYM state \stateYM{psiBar}\brace{1}\, by `$\{2,3,4\}$' regardless of whether we have $\mathcal{N}{=}1$ supersymmetry generators or more. One convenience of this convention is that it makes it easy to define  \mbox{\emph{(super)selection rules}} which determine which collections of external states have non-trivial interactions via the Feynman rules:\\
\eq{\hspace{-20pt}\fbox{\begin{minipage}{0.675\textwidth}The external states of any non-vanishing amplitude must have the \emph{same} number, $k{+}2$, of instances of each index $I\!\in\![4]$ ($I\!\in\![8]$ for sGR) among the states' `$R$-charge' labels.\\
\mbox{~}\hspace{5pt} Such a set of states is said to be of `N${}^k$MHV' degree.\end{minipage}}\hspace{-20pt}\label{superselection}}
We will soon see that $n$-particle amplitudes with $k\!\notin[0,n{-}4]$ must vanish by super-momentum-conservation. 

In the package \package, we have implemented a `random' choice of external states for $\mathcal{N}$-supersymmetric sYM or sGR, and it is easy to verify that these satisfy super-selection rules. 
\mathematicaBox{
\mathematicaSequence{egStates=\funL[]{randomComponentStates}\brace{6,1}}{\{\fun{psi}\![2],\fun{phiBar}\![1,2],\fun{psi}[1],\fun{psiBar}\![1],\fun{psi}\![1],\fun{m}\}}
\mathematicaSequence{\funL[]{rChargeLabels}\brace{egStates}}{\{\{2\},\{3,4\},\{1\},\{2,3,4\},\{1\},\{1,2,3,4\}\}}
\mathematicaSequence{Count[\textnormal{\%},\#,2]\&/@Range[4]}{\{3,3,3,3\}}
\mathematicaSequence{\funL[]{nmhvDegree}\brace{egStates}}{1}
}\vspace{-5pt}

\vspace{-0pt}\subsubsection{Unification: Supersymmetric Coherent States and \fun{superFunctions}}\vspace{-0pt}

In maximally supersymmetric sYM or sGR, all states are connected to one another by supersymmetry, and therefore it is convenient to unite into a more coherent framework. As described in \cite{ArkaniHamed:2008gz}, it is natural to group the states connected to gluons/gravitons into \emph{coherent states} by introducing a $\mathcal{N}$ Grassmann parameters $\tilde\eta_a^{I}$ for each particle indexed by $a\!\in\![n]$
\begin{align}\text{(sYM:)}\;|a\rangle&\equivR e^{\tilde{Q}_I\tilde\eta_a}|a\rangle^{\text{+}1}\\
&=|a\rangle^{\text{+}1}\!\!{+}\tilde\eta_a^I|a\rangle^{\text{+}1/2}_{\{I\}}\!\!{+}\frac{1}{2!}\tilde\eta_a^I\tilde\eta_a^J|a\rangle^{0}_{\{I,J\}}\!{+}\frac{1}{3!}\tilde\eta_a^I\tilde\eta_a^J\tilde\eta_a^K|a\rangle^{\text{-}1/2}_{\{I,J,K\}}\!{+}\frac{1}{4!}\tilde\eta_a^I\tilde\eta_a^J\tilde\eta_a^K\tilde\eta_a^{L}|a\rangle^{\text{-}1}_{\{I,J,K,L\}}\nonumber\\
\text{(sGR:)}\;|a\rangle&\equivR e^{\tilde{Q}_I\tilde\eta_a}|a\rangle^{\text{+}2}\\
&=|a\rangle^{\text{+}2}\!\!{+}\tilde\eta_a^I|a\rangle^{\text{+}3/2}_{\{I\}}\!\!{+}\frac{1}{2!}\tilde\eta_a^I\tilde\eta_a^J|a\rangle^{\text{+}1}_{\{I,J\}}\!{+}\ldots{+}(\tilde{\eta}^{1}\cdots\tilde{\eta}_a^{8})|a\rangle^{\text{-}2}_{\{1,2,3,4,5,6,7,8\}}\nonumber
\end{align}
In terms of these auxiliary book-keeping variables $\tilde\eta$ we may discuss \emph{superFunctions} and superAmplitudes which depend on the variables $\{\lambda,\tilde\lambda,\tilde\eta\}$ for all the particles collectively. To access or project-out a given state from the coherent state, one need only integrate over the appropriate Grassmann parameters on any expression involving the coherent state $|a\rangle$. For example, the $a$th particle's \stateYM{psi}\brace{1}\, \emph{component state} could be projected out from $|a\rangle$ via 
\eq{|a\rangle^{{+}1/2}_{\{1\}}=\int\!\!d\tilde\eta_a^{1}\,\,|a\rangle}
or its (-)-helicity gluon state projected out via
\vspace{-4pt}\eq{|a\rangle^{\text{-}1}_{\{1,2,3,4\}}=\int\!\!d\tilde\eta_a^1d\tilde\eta_a^2d\tilde\eta_a^3d\tilde\eta_a^4\,\,|a\rangle\,.\vspace{-4pt}}

A particularly important role is played by superFunctions that can be expressed as sums of \funL{superFunction} objects defined as follows:
\eq{\fun{superFunction}\brace{\r{f},\var{{C}}}\,\,\Leftrightarrow\,\, \r{f}(\funL{$\lambda$},\funL{$\lambda$b})\,\,\delta^{k\!\times\!\mathcal{N}}\!\big(\var{{C}}(\funL{$\lambda$},\funL{$\lambda$b})\!\cdot\!\tilde\eta\big)\delta^{2\!\times\!\mathcal{N}}\!\big(\funL{$\lambda$}\!\cdot\!\tilde\eta\big)\delta^{2\!\times\!2}\!\big(\funL{$\lambda$}\!\cdot\!\funL{$\lambda$b}\big)\,,}
where above, $\mathcal{N}{=}4,8$ for sYM and sGR, respectively. Notice that we have \emph{manifestly} factored-out super-momentum conservation from the above.  

\vspace{-0pt}\subsubsection{Extracting Component Functions from \fun{superFunctions}}\vspace{-5pt}

As mentioned above, particular \emph{component} functions (which are ordinary functions) can be extracted from \funL{superFunction} objects by the appropriate $\tilde\eta$-integrations. In the package \package, this is achieved using \mbox{\funL{component}\brace{\var{externalStates}}\,\brace{\var{expression}}\,,} which extracts from any \funL{superFunction} object the corresponding particular function. Due to the antisymmetry of the Grassmann $\tilde\eta$ parameters labeling the components of coherent states, these $\tilde\eta$-integrations result in\vspace{-2pt}
\eq{\begin{split}\hspace{00pt}&\hspace{-80pt}\fun{component}\brace{\var{\text{\textsl{componentStates}}}}\,\brace{\funL{superFunction}\brace{f,C}\,}\\\hspace{00pt}&\hspace{-80pt}\fwboxR{0pt}{\equivR}\fun{component}\brace{\var{\text{\textsl{componentStates}}}}\,\brace{f(\lambda,\tilde\lambda)\,\,\delta^{k\!\times\!\mathcal{N}}\!\big(C\!\cdot\!\tilde\eta\big)\delta^{2\!\times\!\mathcal{N}}\!\big(\lambda\!\cdot\!\tilde\eta\big)}\\[-5pt]&\equivR\int\!\!\!\prod_{I=1}^{\mathcal{N}}\!\!\!d\tilde{\eta}_{I}^{a^I_1}\cdots d\tilde\eta_I^{a^I_{k+2}}\left(f(\lambda,\tilde\lambda)\,\,\delta^{k\!\times\!\mathcal{N}}\!\big(C\!\cdot\!\tilde\eta\big)\delta^{2\!\times\!\mathcal{N}}\!\big(\lambda\!\cdot\!\tilde\eta\big)\right)\,\,\,\\[-10pt]
&\,=f(\lambda,\tilde\lambda)\prod_{I=1}^{\mathcal{N}}\det\{\hat{c}_{a^I_1},\ldots,\hat{c}_{a^I_{k{+}2}}\}\quad\text{with}\quad\hat{C}\equivR(\hat{c}_1,\ldots,\hat{c}_n)\equivR\left(\rule{0pt}{12pt}\right.\begin{array}{@{}c@{}}\funL{$\lambda$}\\[-5pt]C\end{array}\left.\rule{0pt}{12pt}\right)\,.\hspace{-80pt}\\[-10pt]\end{split}\label{component_functions_defined}}
where the particular determinants involve columns labelled by the $\mathcal{N}\!\in\!\{4,8\}$ $(k{+}2)$-tuples $\{\{a^1_1,\ldots,a^1_{k+2}\},\ldots,\{a^{\mathcal{N}}_1,\ldots,a^{\mathcal{N}}_{k+2}\}\}$ as determined from the list of \mbox{\{\var{componentStates}\}}\,via the function \mbox{\fun{stateListToMinorList}\brace{\var{componentStates}}\,.} More detail on the functioning of this organization of component amplitudes is described in the \mbox{\hyperlink{context_organization_of_appendix}{appendix}}.\\[-20pt]
\\[-8pt]

\newpage\addtocontents{toc}{\protect\newpage}
\vspace{-0pt}\section{\texorpdfstring{Using the \package\, Package}{Using the tree\_amplitudes Package}}\label{sec:using_the_package}\vspace{-0pt}
%

The \textsc{Mathematica} \package\, package, together with a demonstration notebook demonstrating its primary functionality through illustrative examples, is available from the abstract page of this work on the \t{arXiv}.

\vspace{-0pt}\subsection{Getting Started: Loading and Installing the \texorpdfstring{\package}{tree\_amplitudes}\, Package}\label{subsec:getting_started}\vspace{-0pt}
%

To load the package, simply make sure the file \package\!\t{\textbf{.m}} is the same directory as any (saved) \built{Notebook} that you are using, and evaluate the following:
\mathematicaBox{
\mathematicaSequence{SetDirectory[NotebookDirectory[]];\\
<\hspace{0pt}<\,\package\!\textbf{.m}}{}\\[-10pt]
&\fig{-20pt}{0.4}{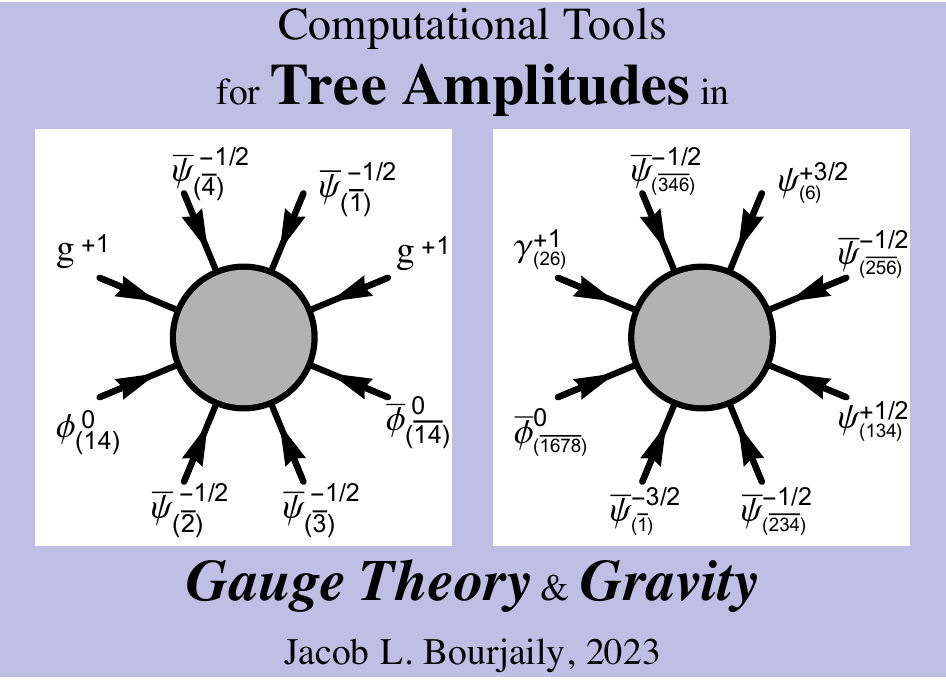}
}
(These are the first lines called by the demonstration notebook.)

If you would like the \package\, package accessible to future sessions---or to packages not in the directory of the file \package\!\t{\textbf{.m}}---you may call the function \builtL{installTreeAmplitudesPackage}, which makes a copy of the file \package\!\t{\textbf{.m}} and places it into one of the directories of \built{\$Path}. Once `installed', the package can be loaded into any new notebook by simply calling ``\t{<\hspace{0.pt}<\,\package\!\textbf{.m}}''.

\subsection{Documentation of Functions Provided by the Package}

The package's code is organized in parallel to that of this work's appendix, which represents the primary documentation for the package.

If inspecting the code, the user that many functions have been placed within the \built{Context} categories `\textbf{\t{treeTools\`{}}}' or `\textbf{\t{positroids\`{}}}'; these represent functions and code used primarily internally by the package, with less obvious use-cases for the user.

\newpage
%
\vspace{-0pt}\section{Conclusions and Potential, Additional Functionality}\label{sec:conclusions}\vspace{-5pt}

\vspace{-0pt}\subsection{Potentially Useful, Additional Functionality at Tree-Level}\vspace{-5pt}

\vspace{-0pt}\subsubsection{Other Forms of Recursion that Remain to be Implemented}\vspace{-5pt}

While we have endeavored to be extremely general in the variety of approaches to representing (tree-level) scattering amplitudes in sYM and sGR, there are a few obviously useful and important formalisms that we have left for future work. Among the most important and obvious of such oversights are the Berend-Giele (see e.g.~\cite{Berends:1987me}), the recursion relations described by Cachazo, Svrcek, and Witten in \cite{Cachazo:2004kj} (see also~\mbox{\cite{Risager:2005vk,Elvang:2008vz,Kiermaier:2009yu,ArkaniHamed:2009sx}}), and various related algorithms (see e.g.~\cite{Risager:2008yz}). 

For the most part, we have left these formalisms for future work due to the complications raised by explicit reference to auxiliary frames/spinors/twistors/etc.---which would require some increased flexibility of our implementation of \funL{evaluate}, for example. We have no doubt in the value of and interest in such representations of scattering amplitudes, but we must leave their implementation to future work.

\vspace{-0pt}\subsubsection{Explicit Color Tensors for Specific Gauge Groups}\vspace{-5pt}

Only (surprisingly) recently has it become relatively easy to generate explicit color tensors \funL{colorFactor} for arbitrary simply-laced Lie algebras. This is in part thanks to the implementation of structure constants in \t{SageMath} \cite{sage}. We have noticed several intriguing cancellations for particular color-dressed amplitudes with particular choices of gauge group (e.g.~$\mathfrak{g}_2$ is especially interesting), and it would seem worthwhile to make this functionality more widely available. We have much of this prepared, but it remains unclear how best to package this data in its most useable form for investigating amplitudes (nor is it especially clear that there is much interest to the physics community beyond $\mathfrak{su}_n$ gauge groups).

\vspace{-0pt}\subsubsection{Massless `QCD' Amplitudes: Including non-Adjoint Matter Fields}\vspace{-5pt}

As noticed in \cite{Dixon:2010ik}, the color-stripped amplitudes of sYM naturally encode the color-stripped amplitudes of massless quarks and gluons for arbitrary numbers of species; once these partial/primitive amplitudes are `dressed' with appropriate color-tensors (which depend on the representation $\mathbf{R}$ under which the quarks transform), tree amplitudes involving arbitrary numbers of quark species in arbitrary representations can be built from the partial amplitudes of sYM. This fact was generalized and considerably strengthened by Melia in a series of excellent papers  \cite{Melia:2013bta,Melia:2013epa,Melia:2015ika}, where it was shown moreover that the partial amplitudes involving \emph{distinct} quark-lines were entirely spanned by those consisting of only a single quark flavor.

More recently, the color-tensors required to dress these partial/primitive amplitudes for quarks transforming under arbitrary representations $\mathbf{R}$ of an arbitrary gauge group $\mathfrak{g}$ were studied by \cite{Johansson:2015oia,Ochirov:2019mtf}. Their results are surprisingly reminiscent of the \funL{colorFactor} tensors arising in the work of DDM  \cite{DelDuca:1999rs} for adjoint color-tensors. 

As such, it would seem that all ingredients exist to generate fully color-dressed amplitudes for (massless) quarks and gluons in `QCD' for arbitrary quark representations. It would seem incredibly worthwhile to automate this connection in future computational tools. However, it remains unclear how best such results should be encoded so that they would be maximally useful to those for whom they would be most interesting. Even in the case of fundamental quarks with gauge group $\mathfrak{su}_n$, it is unclear how best these results could be encoded for use by an interested user.)

\vspace{-0pt}\subsubsection{Amplitudes with Massive Particles via Massive Spinor Helicity}\vspace{-5pt}

While there is a great deal of simplicity to be found for massless scattering amplitudes in four dimensions, there are many reasons for interest in amplitudes involving massive particles. And there now exists the promising massive spinor-helicity formalism described in \cite{Arkani-Hamed:2017jhn}. 

Tools for working with massive spinor-helicity variables in \textsc{Mathematica} have been prepared in \cite{Kuczmarski:2014ara} as an addition to the codebase prepared by \cite{Maitre:2007jq}, and also more recently \cite{AccettulliHuber:2023ldr}. It would be worthwhile to strengthen those tools systematically---for example, to implement automated, on-shell recursion of tree amplitudes amplitudes involving both massive and massless particles. It would be worthwhile to have the results for scattering amplitudes even if limited to the case of massive QED, or dimensionally-regulated Yang-Mills theory.

\vspace{-0pt}\subsection{Tools for Amplitudes Beyond Tree-Level}\vspace{-5pt}

Of course, progress in our understanding of scattering amplitudes extends well beyond the leading order of perturbation. (Indeed, many of the most important discoveries about \emph{trees} were made while studying \emph{loops}.) This understanding includes all-loop recursion relations for maximally supersymmetric Yang-Mills theory in the planar limit \cite{ArkaniHamed:2010kv,Bourjaily:2023apy} (see also~\cite{CaronHuot:2010zt,Boels:2011tp}), and powerful, closed-formulae and code for amplitudes at the integrand-level (see e.g.~\cite{Bourjaily:2013mma,Bourjaily:2015jna,Bourjaily:2017wjl,Bourjaily:2021ewt}). There has also been important progress on recursion relations for amplitudes with less than maximal supersymmetry (see e.g.~\cite{Benincasa:2015zna,Benincasa:2016awv,Bourjaily:2021ujs}) and in maximally supersymmetric theories beyond the planar limit (see e.g.~\cite{Bourjaily:2018omh,Bourjaily:2019gqu,Bourjaily:2019iqr,Bourjaily:2021hcp,Bourjaily:2021iyq}).

However, starting at two loops (and honestly, even at one loop) the variety of options for master integrand bases (see e.g.~\cite{Bourjaily:2020qca}) is such that these results are still evolving too rapidly to seem ripe for stable consolidation.

That being said, it would be incredibly worthwhile to begin to build a consolidated resource for the results that do exist---and have this resource flexible enough to accommodate new discoveries as they are found. This is something that will require the efforts of many people, and we must leave this to future work. 

\vspace{\fill}
\vspace{6pt}
\section*{Acknowledgements}%
\vspace{-4pt}
\noindent The author is grateful for the valuable input, feedback, and contributions of Nikhil Kalyanapuram, Kokkimidis Patatoukos, Michael Plesser, Philip Velie, and Yaqi Zhang during the development of this package, and for helpful feedback and encouragement of Nima Arkani-Hamed, Carolina Figueiredo, Hofie Hannesdottir, Sebastian Mizera, and Jaroslav Trnka. The author is grateful also for the hospitality of the Institute for Advanced Study, Princeton, where many of these tools were developed. This work was supported in part by the US Department of Energy under contract DE-SC00019066.

\newpage
\addtocontents{toc}{\protect~\\[-10pt]\mbox{\vspace{0pt}}\protect\hrulefill\par}
\appendix
%
\setcounter{section}{-1}
\hypertarget{context_organization_of_appendix}{}\vspace{-0pt}\section[\mbox{\hspace{-18pt}Context-Organized List of Functions Provided by the Package}]{\hspace{-17.5pt}Context-Organized List of Functions Provided}\vspace{-0pt}

The primary functionality provided by the package can be found in the package using the function \fun{definedFunctionCategories}\brace{\var{section}\pattern\optArg{},\var{subSection}\pattern\optArg{}}:

\defnBox{definedFunctionCategories}{\var{section}\pattern\optArg{},\var{subSection}\pattern\optArg{}}{returns an \emph{formatted} table of the contents of this Appendix:
\mathematicaBox{\mathematicaSequence{\funL[1]{definedFunctionCategories}[]}{\textnormal{\begin{tabular}{@{}l@{$\,\,$}l}
\textbf{A}.&\textbf{\r{Abstract(/Protected) Symbols}}\\[-4pt]
\textbf{B}.&\textbf{\r{General Utilities and Mathematical Tools}}\\[-4pt]
\textbf{C}.&\textbf{\r{Aspects of External Component States}}\\[-4pt]
\textbf{D}.&\textbf{\r{Analytic Expressions of External Kinematics}}\\[-4pt]
\textbf{E}.&\textbf{\r{General Aspects of On-Shell Recursion Relations}}\\[-4pt]
\textbf{F}.&\textbf{\r{Ordered Partial Amplitudes of (pure) superYang-Mills}}\\[-4pt]
\textbf{G}.&\textbf{\r{Color-Dressed Amplitudes of (pure) superYang-Mills}}\\[-4pt]
\textbf{H}.&\textbf{\r{Tree Amplitudes of (pure) superGravity}}\\[-4pt]
\textbf{I}.&\textbf{\r{Tree Amplitudes of Other Theories}}\\[-4pt]
\textbf{J}.&\textbf{\r{Twistor String Theory and the Scattering Equation Formalism}}\\[-4pt]
\textbf{K}.&\textbf{\r{Positroids: Grassmannian Geometry of On-Shell Functions}}\\[-4pt]
\textbf{L}.&\textbf{\r{Kinematic Data: Specification and Evaluation}}\\[-4pt]
\textbf{M}.&\textbf{\r{Information and Syntax for TreeTools Package}}
\end{tabular}}}}
If the optional arguments \var{section} or \var{subSection} are specified, then a formatted table of categories is provided:
\mathematicaBox{\mathematicaSequence{\funL[1]{definedFunctionCategories}[F]}{\textnormal{\begin{tabular}{@{}l@{$\,\,$}l}\textbf{F}.&\textbf{\r{Ordered Partial Amplitudes of (pure) superYang-Mills}}\\[-4pt]
\textbf{F.1}&\textbf{\b{\;\;Ordered Partial Amplitudes: Bases, Identities, and Reduction}}\\[-4pt]
\textbf{F.2}&\textbf{\b{\;\;Analytic Expressions for Ordered superAmplitudes in sYM}}\\[-4pt]
\textbf{F.3}&\textbf{\b{\;\;Numeric Expressions for Ordered superAmplitudes in sYM}}\\[-4pt]
\textbf{F.4}&\textbf{\b{\;\;Analytic Component Amplitudes of sYM}}\\[-4pt]
\textbf{F.5}&\textbf{\b{\;\;Numerically Determined Component Amplitudes of sYM}}\\[-4pt]
\textbf{F.6}&\textbf{\b{\;\;Relations Among sYM Component Amplitudes}}\end{tabular}}}
\mathematicaSequence{\funL[1]{definedFunctionCategories}[F,2]}{\textnormal{\begin{tabular}{@{}l@{$\,\,$}l}\textbf{F}.&\textbf{\r{Ordered Partial Amplitudes of (pure) superYang-Mills}}\\[-4pt]
\textbf{F.2}&\textbf{\b{\;\;Analytic Expressions for Ordered superAmplitudes in sYM}}\\[-4pt]
\textbf{}&\textbf{{\;\;\funL{ymAmp}\brace{\var{$n$}\pattern,\var{$k$}\pattern,\var{legOrdering}\pattern\optArg{}\,}}}\\[-4pt]
\textbf{}&\textbf{{\;\;\funL{ymAmpRandom}\brace{\var{$n$}\pattern,\var{$k$}\pattern,\var{legOrdering}\pattern\optArg{}\,}}}\\[-4pt]~\\[-4pt]
\textbf{}&\textbf{{\;\;\funL{parkeTaylorAmp}\brace{\var{$n$}\pattern}}}\\[-4pt]
\textbf{}&\textbf{{\;\;\funL{parkeTaylorAmpBar}\brace{\var{$n$}\pattern}}}\\[-4pt]~\\[-4pt]
\textbf{}&\textbf{{\;\;\funL{ymTwistorAmp}\brace{\var{$n$}\pattern,\var{$k$}\pattern,\var{legOrdering}\pattern\optArg{}\,}}}\\[-4pt]
\textbf{}&\textbf{{\;\;\funL{ymTwistorAmpRandom}\brace{\var{$n$}\pattern,\var{$k$}\pattern,\var{legOrdering}\pattern\optArg{}\,}}}\\[-4pt]
\textbf{}&\textbf{{\;\;\funL{ymSuperTwistorAmp}\brace{\var{$n$}\pattern,\var{$k$}\pattern,\var{legOrdering}\pattern\optArg{}\,}}}\\[-4pt]
\textbf{}&\textbf{{\;\;\funL{ymSuperTwistorAmpRandom}\brace{\var{$n$}\pattern,\var{$k$}\pattern,\var{legOrdering}\pattern\optArg{}\,}}}
\end{tabular}}}}
}

\newpage
\vspace{-0pt}\sectionAppendix{Abstract Symbols and (Protected) Namespace}{appendix:abstract_protected_symbols}\vspace{-0pt}
%

\vspace{0pt}\subsectionAppendix{Protected Symbols for Labeling External, Component States}{appendix:protected_symbols_for_components}\vspace{-0pt}

As described in \mbox{section \ref{}}, the external component-states of sYM and sGR for any degree of supersymmetry are labelled according to the uniform description as truncations of the maximally supersymmetric theories. In particular, the labels for states in sYM and sGR are, respectively:

\boxedList{\stateYM{m}, \stateYM{p}, \stateYM{psi}\brace{\var{$i$}\pattern,\var{$jk$\patternTwo}\optArg{}}\,, \stateYM{psiBar}\brace{\var{$i$}\pattern,\var{$jk$\patternTwo}\optArg{}}\,, \stateYM{phi}\brace{\var{$i$\pattern},\var{$j$\pattern},\var{$kl$\patternTwo}\optArg{}}\,, \fun{phiBar}\brace{\var{$i$\pattern},\var{$j$\pattern},\var{$kl$\patternTwo}\optArg{}}}

\boxedList{\mbox{\stateGR{mm}, \stateGR{pp}, \stateGR{gravitino}\brace{\var{$i$}\pattern}, \mbox{\stateGR{gravitinoBar}\brace{\var{$i$}\pattern}}, \mbox{\stateGR{graviPhoton}\brace{\var{$i$\pattern},\var{$j$\pattern}}}, \mbox{\stateGR{graviPhotonBar}\brace{\var{$i$\pattern},\var{$j$\pattern}}}}}

\vspace{5pt}\subsectionAppendix{Protected Symbols Related to Kinematic Invariants}{appendix:protected_kinematic_symbols}\vspace{-0pt}
\vspace{0pt}\subsubsectionAppendix{Kinematic Functions of Spinor Variables and Momenta}{appendix:kinematic_involving_spinors}\vspace{-0pt}

\vspace{-10pt}

\defnBox{$\lambda$}{\var{$a$\pattern},\var{$\alpha$}\pattern\optArg{}}{an \emph{abstract} symbol representing the spinor $\lambda_{\text{\var{$a$}}}$ for the \var{$a$}th particle, or the \emph{component} $\lambda_{\text{\var{$a$}}}^{\text{\var{$\alpha$}}}$ of the spinor $\lambda_{\text{\var{$a$}}}$ if the optional argument \var{$\alpha$} is also provided.
\mathematicaBox{\mathematicaSequence[1]{\funL[]{useReferenceKinematics}[9];\\
\funL[]{evaluate}[\fun{$\lambda$}[5,2]]===\builtL[]{Ls}[[5,2]]}{True}
\mathematicaSequence{\funL[]{evaluate}[\fun{$\lambda$}/@Range[9]]}{\{\{1,0\},\{1,1\},\{1,3\},\{1,6\},\{1,10\},\{1,15\},\{1,21\},\{1,28\},\{1,36\}\}}
}}

\defnBox{$\lambda$b}{\var{$a$}\pattern,\var{$\dot{\alpha}$}\pattern}{an \emph{abstract} symbol representing the spinor $\tilde{\lambda}_{\text{\var{$a$}}}$ for the \var{$a$}th particle, or the \emph{component} $\tilde{\lambda}_{\text{\var{$a$}}}^{\text{\var{$\dot{\alpha}$}}}$ of the spinor $\tilde{\lambda}_{\text{\var{$a$}}}$ if the optional argument \var{$\dot\alpha$} is also provided. Similar to \fun{$\lambda$}.}

\defnBox{p}{\var{$a$}\pattern,\var{$b$}\patternTwo\optArg{}}{\emph{abstractly} represents the four-momentum of particle indexed by \var{$a$} in terms of the $(2\!\times\!2)$ matrix
\vspace{-0pt}\eq{\fun{p}\texttt{[}\var{a}\texttt{]}\,\,\Leftrightarrow\,\, p_{\var{a}}^{\alpha\,\dot{\alpha}}\equivR p_\var{a}^{\mu}\sigma_{\mu}^{\alpha\,\dot\alpha}=\left(\begin{array}{@{}cc@{}}p_\var{a}^0{+}p_\var{a}^3&p_\var{a}^1{-}i\,p_\var{a}^2\\
p_\var{a}^1{+}i\,p_\var{a}^2&p_\var{a}^0{-}p_{\var{a}}^3\end{array}\right)\,.\vspace{-0pt}}
When provided multiple indices (as an \built{Integer} \built{Sequence}), 
\vspace{-0pt}\eq{\fun{p}\texttt{[}\var{a},\var{b},\ldots,\var{c}\texttt{]}\,\Leftrightarrow\,\,(\fun{p}\texttt{[}\var{a}\texttt{]}{+}\fun{p}\texttt{[}\var{b}\texttt{]}{+}\cdots\fun{p}\texttt{[}\var{c}\texttt{]}).\vspace{-20pt}}
}

\defnBox{s}{\var{$a$}\pattern,\var{$b$}\patternTwo\optArg{}}{\emph{abstractly} represents the Mandelstam invariant
\vspace{-0pt}\eq{\fun{s}\texttt{[}\var{a},\var{b},\ldots\texttt{]}\,\Leftrightarrow\,\,\left(\fun{p}\texttt{[}\var{a}\texttt{]}{+}\fun{p}\texttt{[}\var{b}\texttt{]}{+}\ldots\right)^2\,.\vspace{-20pt}}
}

\defnBox{ab}{\var{$a$}\pattern\built{Integer},\var{$b$}\pattern\built{Integer}}{\emph{abstractly} represents the `\textbf{\texttt{a}}ngle-\textbf{\texttt{b}}racket' invariant 
\vspace{-0pt}\eq{\fun{ab}\texttt{[}\var{a},\var{b}\texttt{]}\,\Leftrightarrow\,\,\ab{\var{a\,b}}\equivR\det(\lambda_{\var{a}},\lambda_{\var{b}})\equivR\epsilon_{\alpha\,\beta}\lambda_{\var{a}}^{\alpha}\lambda_{\var{b}}^{\beta}\,.\vspace{-20pt}}}

\defnBox{sb}{\var{$a$}\pattern\built{Integer},\var{$b$}\pattern\built{Integer}}{\emph{abstractly} represents the `\textbf{\texttt{s}}quare-\textbf{\texttt{b}}racket' invariant 
\vspace{-0pt}\eq{\fun{sb}\texttt{[}\var{a},\var{b}\texttt{]}\,\Leftrightarrow\,\,\sb{\var{a\,b}}\equivR\det(\tilde\lambda_{\var{a}},\tilde\lambda_{\var{b}})\equivR\epsilon_{\dot\alpha\,\dot\beta}\tilde\lambda_{\var{a}}^{\dot\alpha}\tilde\lambda_{\var{b}}^{\dot\beta}\,.\vspace{-20pt}}}

\defnBox{asb}{\var{$a$}\pattern\built{Integer},\var{pSequence}\patternTwo\head{\fun{p}},\var{$b$}\pattern\built{Integer}}{\emph{abstractly} represents the kinematic invariant `\textbf{\t{a}}ngle-\textbf{\t{s}}quare-\textbf{\t{b}}racket'
\vspace{-4pt}\eq{\fun{asb}\t{[}\var{a},\fun{p}\t{[}\var{c},\ldots,\var{d}\t{]},\var{b}\t{]}\Leftrightarrow\,\,\langle\var{a}|(\var{c}\cdots\var{d})|\var{b}]\equivR\ab{\var{a\,c}}\sb{\var{c\,b}}{+}\ldots{+}\ab{\var{a\,d}}\sb{\var{d\,b}}\,.\vspace{-4pt}}
It generalizes to cases involving any \emph{odd}-length \built{Sequence} of \funL{p}'s (\ref{general_asb_bracket}):
\mathematicaBox{\mathematicaSequence{\funL[]{expandSpinorBrackets}[\fun{asb}[1,\funL{p}[2,3],4]]}{\funL{ab}[1,2]\funL{sb}[2,4]{+}\funL{ab}[1,3]\funL{ab}[3,4]}
\mathematicaSequence{\funL[]{expandSpinorBrackets}[\fun{asb}[1,\funL{p}[2],\funL{p}[3],\funL{p}[4],5]]}{\funL{ab}[1,2]\funL{sb}[2,3]\funL{ab}[3,4]\funL{sb}[4,5]}}
}

\defnBox{sab}{\var{$a$}\pattern\built{Integer},\var{pSequence}\patternTwo\head{\fun{p}},\var{$b$}\pattern\built{Integer}}{\emph{abstractly} represents the kinematic invariant `\textbf{\t{s}}quare-\textbf{\t{a}}ngle-\textbf{\t{b}}racket'
\vspace{-4pt}\eq{\fun{sab}\t{[}\var{a},\fun{p}\t{[}\var{c},\ldots,\var{d}\t{]},\var{b}\t{]}\,\Leftrightarrow\,\,[\var{a}|(\var{c}\cdots\var{d})|\var{b}\rangle\equivR\sb{\var{a\,c}}\ab{\var{c\,b}}{+}\ldots{+}\sb{\var{a\,d}}\ab{\var{d\,b}}\,.\vspace{-4pt}}
It generalizes to cases involving any \emph{odd}-length \built{Sequence} of \funL{p}'s (\ref{general_asb_bracket}):
\mathematicaBox{\mathematicaSequence{\funL[]{expandSpinorBrackets}[\funL[1]{sab}[1,\funL{p}[2,3],4]]}{\funL{sb}[1,2]\funL{ab}[2,4]{+}\funL{sb}[1,3]\funL{ab}[3,4]}
\mathematicaSequence{\funL[]{expandSpinorBrackets}[\funL[1]{sab}[1,\funL{p}[2],\funL{p}[3],\funL{p}[4],5]]}{\funL{sb}[1,2]\funL{ab}[2,3]\funL{sb}[3,4]\funL{ab}[4,5]}}
}

\defnBox[]{ab}{\var{$a$}\pattern\built{Integer},\var{pSequence}\patternTwo\head{\fun{p}},\var{$b$}\pattern\built{Integer}}{\emph{abstractly} represents the generalized momentum-space `\textbf{\t{a}}ngle-\textbf{\t{b}}racket'
\vspace{-4pt}\eq{\begin{split}\fun{ab}\t{[}\var{a},\fun{p}\t{[}\var{c_1},\ldots,\var{c_{{-}1}}\t{]},\fun{p}\brace{\var{d_1},\ldots,\var{d_{{-}1}}}\,,\var{b}\t{]}\Leftrightarrow&
\langle\var{a}|(\var{c_1}\cdots\var{c_{{-}1}})|(\var{d_1}\cdots\var{d_{{-}1}})|\var{b}\rangle\\
\equivR\sum_{\var{c},\var{d}}\ab{\var{a\,c}}\sb{\var{c\,d}}\ab{\var{d\,b}}\\[-14pt]\end{split}\vspace{-4pt}}
It generalizes to cases involving any \emph{even}-length \built{Sequence} of \fun{p}'s (\ref{generalized_ab_and_sb}).
}

~

\defnBox[]{sb}{\var{$a$}\pattern\built{Integer},\var{pSequence}\patternTwo\head{\fun{p}},\var{$b$}\pattern\built{Integer}}{\emph{abstractly} represents the generalized momentum-space `\textbf{\t{s}}quare-\textbf{\t{b}}racket'
\vspace{-4pt}\eq{\begin{split}\fun{sb}\t{[}\var{a},\fun{p}\t{[}\var{c_1},\ldots,\var{c_{{-}1}}\t{]},\fun{p}\brace{\var{d_1},\ldots,\var{d_{{-}1}}}\,,\var{b}\t{]}\Leftrightarrow&
[\var{a}|(\var{c_1}\cdots\var{c_{{-}1}})|(\var{d_1}\cdots\var{d_{{-}1}})|\var{b}]\\
\equivR\sum_{\var{c},\var{d}}\sb{\var{a\,c}}\ab{\var{c\,d}}\sb{\var{d\,b}}\\[-14pt]\end{split}\vspace{-4pt}}
It generalizes to cases involving any \emph{even}-length \built{Sequence} of \funL{p}'s (\ref{generalized_ab_and_sb}).
}

\newpage
\vspace{0pt}\subsubsectionAppendix{Kinematic Functions of Momentum-Twistors}{appendix:kinematic_involving_twistors}\vspace{-5pt}

\defnBox[5]{ab}{\var{$a$}\pattern\built{Integer},\var{$b$}\pattern\built{Integer},\var{$c$}\pattern\built{Integer},\var{$d$}\pattern\built{Integer}}{\emph{abstractly} denotes the  `\texttt{a}ngle-\texttt{b}racket'  (or  \emph{momentum-twistor} `four-bracket') invariant (see e.g.~\cite{Hodges:2009hk})
\vspace{-6pt}\eq{\fun{ab}\texttt{[}\var{a},\var{b},\var{c},\var{d}\texttt{]}\Leftrightarrow\,\,\ab{\var{a}\,\var{b}\,\var{c}\,\var{d}}\equivR\det(z_{\var{a}},z_{\var{b}},z_{\var{c}},z_{\var{d}})\equivR\epsilon_{I\,J\,K\,L}z_{\var{a}}^{I}z_{\var{b}}^{J}z_{\var{c}}^{K}z_{\var{d}}^{L}\,.\vspace{-0pt}}
\textbf{\emph{Note}}:\! the arguments of the \fun{ab} can include arbitrarily-nested references to twistors defined geometrically using \fun{cap}.
\vspace{0pt}}

\defnBox{cap}{\var{line}\pattern,\var{plane}\pattern}{\emph{abstractly} represents the geometric intersection (encoded by `$\backslash$cap' in \LaTeX) of two sets of twistors indexed by the \built{List}s \var{line} and \var{plane}. Specifically, 
\vspace{-6pt}\begin{align}\fun{cap}\texttt{[\{}\var{a},\var{b}\texttt{\}},\texttt{\{}\var{c},\var{d},\var{e}\texttt{\}]}&\Leftrightarrow\,\,(\var{a}\,\var{b})\tcap(\var{c}\,\var{d}\,\var{e})=\text{span}\{z_{\var{a}},z_{\var{b}}\}\tcap\,\text{span}\{z_\var{c},z_\var{d},z_\var{e}\}\label{defn_of_cap}\\&\equivR z_{\var{a}}\ab{\var{b\,c\,d\,e}}{+}z_{\var{b}}\ab{\var{c\,d\,e\,a}}=z_{\var{c}}\ab{\var{d\,e\,b\,a}}{+}z_{\var{d}}\ab{\var{e\,b\,a\,c}}{+}z_{\var{e}}\ab{\var{b\,a\,c\,d}}\,.\nonumber\end{align}

The ordering of the \var{line} and the \var{plane} can be interchanged at the cost of a minus sign:
\vspace{-6pt}\eq{\fun{cap}\texttt{[\{}\var{a},\var{b}\texttt{\}},\texttt{\{}\var{c},\var{d},\var{e}\texttt{\}]}\simeq{-}\fun{cap}\texttt{[\{}\var{c},\var{d},\var{e}\texttt{\}},\texttt{\{}\var{a},\var{b}\texttt{\}]}\,.}
In addition, the intersection of two planes is also supported:
\vspace{-6pt}\begin{align}\fun{cap}\texttt{[\{}\r{a},\r{b},\r{c}\texttt{\}},\texttt{\{}\var{d},\var{e},\var{f}\texttt{\}]}&\Leftrightarrow\,\,(\r{a}\,\r{b}\,\r{c})\tcap(\var{d}\,\var{e}\,\var{f})=\text{span}\{z_{\r{a}},z_{\r{b}},z_\r{c}\}\tcap\,\text{span}\{z_\var{d},z_\var{e},z_\var{f}\}\label{defn_of_cap2}\\[-15pt]&\equivR (\r{a\,b})\ab{\var{\r{c}\,d\,e\,f}}{+}(\r{b\,c})\ab{\var{\r{a}\,d\,e\,f}}{+}(\r{c\,a})\ab{\var{\r{b}\,d\,e\,f}}\,.\nonumber\\[-40pt]\nonumber\end{align}
\mathematicaBox{\mathematicaSequence{\funL[]{expandTwistorBrackets}[\fun{ab}[1,2,3,\fun{cap}[\{4,5\},\{6,7,8\}]]]}{\fun{ab}[1,2,3,4]\fun{ab}[5,6,7,8]{+}\fun{ab}[1,2,3,5]\fun{ab}[6,7,8,4]}
\mathematicaSequence{\funL[]{expandTwistorBrackets}[\fun{ab}[1,2,\fun{cap}[\{3,4,5\},\{6,7,8\}]]]}{\fun{ab}[1,2,4,5]\fun{ab}[3,6,7,8]{+}\fun{ab}[1,2,5,3]\fun{ab}[4,6,7,8] {+}\fun{ab}[1,2,3,4]\fun{ab}[5,6,7,8]}}

\textbf{\emph{Note}}:\! the arguments of the \fun{cap} can include arbitrarily-nested references to twistors defined geometrically through other instances of \fun{cap}. 
}

\newpage
\vspace{0pt}\subsubsectionAppendix{Polarization Vectors and Covariant Squared-Sums}{appendix:polarization_rules_for_gluons}\vspace{-10pt}

\defnBox{$\epsilon$}{\var{$a$}\pattern}{\emph{abstractly} denotes the polarization tensor $\epsilon_\var{a}$ for the \var{$a$}th particle.}

\defnBox{$\epsilon$m}{\var{$a$}\pattern}{\emph{abstractly} denotes the (-)-polarization tensor $\epsilon^{-}_\var{a}$ for the \var{$a$}th particle. When used in evaluation, this is defined via 
\vspace{-4pt}\eq{\fun{$\epsilon$m}\brace{\var{a}}\,\,\Leftrightarrow\,\, \epsilon^{-}_{\var{a}}\equivR\frac{\var{a}\rangle[\rho}{\sb{\rho\,\var{a}}}\,}
where the reference frame is defined by the light-like momentum 
\eq{\rho^{\alpha\,\dot\alpha}\equivR\left(\begin{array}{@{}cc@{}}\fwboxR{10pt}{1}&\fwboxR{14pt}{\text{-}1}\\\fwboxR{10pt}{\text{-}1}&\fwboxR{14pt}{1}\end{array}\right)\equivL\lambda_\rho^{\alpha}\tilde\lambda_\rho^{\dot\alpha}\equivL\rho\rangle[\rho\;\;\; \Leftrightarrow\;\;\; \lambda_\rho=\tilde{\lambda}_{\rho}=(1\,\,\text{-}1).\vspace{-14pt}}
}

\defnBox{$\epsilon$p}{\var{$a$}\pattern}{\emph{abstractly} denotes the (+)-polarization tensor $\epsilon^{+}_\var{a}$ for the \var{$a$}th particle. When used in evaluation, this is defined via 
\vspace{-4pt}\eq{\fun{$\epsilon$p}\brace{\var{a}}\,\,\Leftrightarrow\,\, \epsilon^{+}_{\var{a}}\equivR\frac{\rho\rangle[\var{a}}{\ab{\rho\,\var{a}}}\,}
where the reference frame is defined by the light-like momentum 
\eq{\rho^{\alpha\,\dot\alpha}\equivR\left(\begin{array}{@{}cc@{}}\fwboxR{10pt}{1}&\fwboxR{14pt}{\text{-}1}\\\fwboxR{10pt}{\text{-}1}&\fwboxR{14pt}{1}\end{array}\right)\equivL\lambda_\rho^{\alpha}\tilde\lambda_\rho^{\dot\alpha}\equivL\rho\rangle[\rho\;\;\; \Leftrightarrow\;\;\; \lambda_\rho=\tilde{\lambda}_{\rho}=(1\,\,\text{-}1).\vspace{-14pt}}
}

\defnBox{ss}{\fun{$\epsilon$}\brace{\var{a}\pattern},\fun{$\epsilon$}\brace{\var{$b$}\pattern}}{\emph{abstractly} represents the `\built{s}quared-\built{s}um'
\eq{\fun{ss}\brace{\fun{$\epsilon$}\brace{\var{a}},\,\fun{$\epsilon$}\brace{\var{b}}}\,\,\Leftrightarrow\,\, (\epsilon_{\var{a}}{+}\epsilon_{\var{b}})^2\,.}
The polarization-tensor arguments can be replaced by \fun{$\epsilon$p} or \fun{$\epsilon$m} (and in fact \emph{must} be replaced by one of these for the sake of numeric evaluation).\\[-10pt]

\textbf{\emph{Aside}}: we prefer to use `squared-sums' as opposed to the more familiar dot products because this avoids any appearance of factors of 2 from our results.
}

~\vspace{10pt}
\defnBox[]{ss}{\fun{$\epsilon$}\brace{\var{a}\pattern},\fun{p}\brace{\var{$b$}\pattern}}{\emph{abstractly} represents the `\built{s}quared-\built{s}um'
\eq{\fun{ss}\brace{\fun{$\epsilon$}\brace{\var{a}},\,\fun{p}\brace{\var{b}}}\,\,\Leftrightarrow\,\, (\epsilon_{\var{a}}{+}p_{\var{b}})^2\,.}
The polarization-tensor argument can be replaced by \fun{$\epsilon$p} or \fun{$\epsilon$m} (and in fact \emph{must} be replaced by one of these for the sake of numeric evaluation).
}

~\vspace{10pt}
\defnBox[]{ss}{\fun{p}\brace{\var{a}\pattern},\fun{p}\brace{\var{$b$}\pattern}}{\emph{abstractly} represents the `\built{s}quared-\built{s}um'
\eq{\fun{ss}\brace{\fun{p}\brace{\var{a}},\,\fun{p}\brace{\var{b}}}\equivR\fun{s}\brace{\var{a},\var{b}}\,\,\Leftrightarrow\,\, (p_{\var{a}}{+}p_{\var{b}})^2\,.\vspace{-20pt}}
}

\newpage
\vspace{5pt}\subsectionAppendix{Protected Symbols Related to Amplitudes}{appendix:protected_amplitude_symbols}\vspace{-10pt}

\defnBox{amp}{\var{legOrderingSequence\patternTwo}}{\emph{abstractly} represents the color-stripped \emph{ordered} sYM amplitude where the external legs are labelled by the \built{Sequence} of \built{Integer}'s \var{legOrderingSequence}.\\[-10pt]

\fun{amp} objects are used in the representation of various identities, for example:
\mathematicaBox{
\mathematicaSequence{\fun{amp}@@\funL[]{randomPerm}\brace{7}}{\fun{amp}[1,6,7,5,2,3,4]}
\mathematicaSequence{\textnormal{\%}/\!\!.\builtL[]{kkAmpReductionRule}}{\phantom{+}\fun{amp}[1,4,3,2,5,6,7]{+}\fun{amp}[1,4,3,2,6,5,7]+\fun{amp}[1,4,3,6,2,5,7] +\fun{amp}[1,4,6,3,2,5,7]+\fun{amp}[1,6,4,3,2,5,7]}
}
It also appears in the abstract representation of GR amplitudes via KLT, for example:
\mathematicaBox{
\mathematicaSequence{Grid\brace{\funL[]{kltRepresentationSeeds}\brace{6}}}{\begin{tabular}{@{}ccc@{}}\fun{amp}[1,2,3,4,5,6]&-\funL{s}[1,2](\funL{s}[1,3]+\funL{s}[2,3])\funL{s}[4,5]&\fun{amp}[1,2,3,6,4,5]\\
\fun{amp}[1,2,3,4,5,6]&-\funL{s}[1,2]\funL{s}[1,3]\funL{s}[4,5]&\fun{amp}[1,3,2,6,4,5]\end{tabular}}
}
\textbf{\emph{Note}}: in the context of \emph{e.g.}~supersymmetric \emph{Ward identities}, the arguments of \fun{amp} may be external \emph{states}, the meaning of which should be obvious.
}

\defnBox{superFunction}{\var{function}\pattern,\var{cMatrix}\pattern}{\emph{abstractly} represents
\eq{\fun{superFunction}\brace{\r{f},\var{{C}}}\,\,\Leftrightarrow\,\, \r{f}(\funL{$\lambda$},\funL{$\lambda$b})\,\,\delta^{k\!\times\!\mathcal{N}}\!\big(\var{{C}}(\funL{$\lambda$},\funL{$\lambda$b})\!\cdot\!\tilde\eta\big)\delta^{2\!\times\!\mathcal{N}}\!\big(\funL{$\lambda$}\!\cdot\!\tilde\eta\big)\delta^{2\!\times\!2}\!\big(\funL{$\lambda$}\!\cdot\!\funL{$\lambda$b}\big)\,,}
where \r{$f$}$(\funL{$\lambda$},\!\funL{$\lambda$b})$ is an ordinary (`Bosonic') function of the external kinematics represented by the spinors \funL{$\lambda$},\funL{$\lambda$b} encoding the external momenta of $n$ particles and $\var{{C}}$ is a $(k\!\times\!n)$ \built{Array} of ordinary functions $\{\b{\var{c}_{\var{a}}^{\var{\alpha}}}(\funL{$\lambda$},\funL{$\lambda$b})\}$ indexed by $\var{a}\!\in\![n]$ and $\var{\alpha}\!\in\![k]$. Here, $k$ refers to the N${}^k$MHV-degree for which the \fun{superFunction} has support and $\mathcal{N}\!\in\!\{4,8\}$ for sYM and sGR \fun{superFunction} objects, respectively.\\[-10pt]

\emph{\textbf{Note}}: to (internally) distinguish between a \fun{superFunction} relevant to sYM from one for sGR, the code represents those of sGR \emph{always} as involving a $(2k)\!\times\!n$ matrix ${C}\equivR\built{Join}\brace{C_1,C_2\,}$ (where $C_1$ and ${C_2}$ need not be distinct) with 
\eq{\delta^{k\!\times\!8}\!\left({C}\!\cdot\!\tilde\eta\right)\equivL\delta^{k\!\times\!4}\!\left(C_1\!\cdot\!\tilde\eta\right)\delta^{k\!\times\!4}\!\left(C_2\!\cdot\!\tilde\eta\right)\,.}
This allows \fun{superFunction} objects arising from KLT (as products of those of sYM) to be treated internally in the same way as those arising in BCFW (as powers of those of sYM). (This possible duplication of rows of the coefficient matrix is suppressed in the output of \funL{nice}.)

}

\defnBox{R}{\var{twistorArgumentSequence}\patternTwo}{\emph{abstractly} encodes the momentum-(super-)twistor `five-bracket' defined by
\vspace{5pt}\eq{\fwbox{0pt}{\hspace{-45pt}\fun{R}\brace{\var{a},\var{b},\var{c},\var{d},\var{e}}\,\,\Leftrightarrow 
\frac{\delta^{1\!\times\!4}\!\big(\eta_{\var{a}}\ab{\var{b\,c\,d\,e}}{+}\eta_\var{b}\ab{\var{c\,d\,e\,a}}{+}\eta_{\var{c}}\ab{\var{d\,e\,a\,b}}{+}\eta_{\var{d}}\ab{\var{e\,a\,b\,c}}{+}\eta_{\var{e}}\ab{\var{a\,b\,c\,d}}\big)}{\ab{\var{a\,b\,c\,d}}\ab{\var{b\,c\,d\,e}}\ab{\var{c\,d\,e\,a}}\ab{\var{d\,e\,a\,b}}\ab{\var{e\,a\,b\,c}}}\!.}\vspace{-6pt}}
}

\defnBox{colorFactor}{\var{indexSequence}\patternTwo\built{Integer}}{\emph{abstractly} represents the `color tensor'
\eq{\text{\fun{colorFactor}\texttt{[}\r{$a$},\var{$b_{1}$},\ldots,\var{$b_{\text{-}1}$},\r{$c$}\texttt{]}}\equivR\sum_{\b{e_i}}f^{\text{\r{$a$}\,\var{$b{}_{1}$}}\,\b{e_1}}f^{\b{e_1}\,\text{\var{$b{}_{2}$}}\,\b{e_2}}\cdots f^{\b{e_{\text{-}3}}\,\text{\var{$b{}_{\text{-}2}$}}\,\b{e_{\text{-}2}}}f^{\b{e_{\text{-}1}}\,\text{\var{$b{}_{\text{-}1}$}\,\r{$c$}}},\label{tree_like_color_tensors_defined}\vspace{-10pt}}
where $f^{\var{c_1\,c_2\,c_3}}$ are the structure constants of some Lie algebra $\mathfrak{g}$ and $\var{c_a}\!\in\![\dim(\mathfrak{g})]$ indexes the `color' of the \var{$a$}th gluon (in the adjoint representation of the algebra). 
}

\defnBox{tr}{\var{indexSequence}\patternTwo\built{Integer}}{\emph{abstractly} represents the `color trace'
\eq{\fun{tr}\brace{\var{a_1},\ldots,\var{a_{\text{-}1}}}\equivR\mathrm{tr}\big(T^{\var{a_1}}_{\mathbf{R}}\ldots T^{\var{a_{\text{-}1}}}_{\mathbf{R}}\big)}
where $T_{\mathbf{R}}^{\text{\var{$c{}_{a}$}}}$ are the $\dim(\mathbf{R})\!\times\!\dim(\mathbf{R})$ matrices encoding the \emph{generators} of some particular (matrix) representation $\mathbf{R}$ of the Lie algebra $\mathfrak{g}$ and \var{$c{}_{a}$}$\,\in\![\dim(\mathfrak{g})]$ is an index encoding the `color' of the \var{$a$}th gluon; the normalization of these tensors are chosen so that they are related to the \emph{structure constants} defining the Lie algebra $\mathfrak{g}$ via
\eq{f^{c_1\,c_2\,c_3}\equivR\mathrm{tr}\big(T_{\mathbf{R}}^{c_1}T_{\mathbf{R}}^{c_2}T_{\mathbf{R}}^{c_3}\big){-}\mathrm{tr}\big(T_{\mathbf{R}}^{c_1}T_{\mathbf{R}}^{c_3}T_{\mathbf{R}}^{c_2}\big)\,.\vspace{-20pt}}
}

\vspace{10pt}\subsectionAppendix{Protected Symbols Related to Positroids}{appendix:positroid_symbols}\vspace{-10pt}

\defnBox{positroid}{\var{cHatMatrix}\pattern}{\emph{abstractly} defines the Grassmannian integral
\eq{\hspace{-20pt}\fwbox{0pt}{\fun{postiroid}\brace{\var{\hat{C}}(\vec{\fun{$\alpha$}})}\,\,\Leftrightarrow\int\!\!\!\left(\prod_{i}\frac{d\fun{$\alpha_i$}}{\fun{$\alpha_i$}}\right)\,\delta^{(k+2)\!\times\!2}\!\big(\var{\hat{C}}(\vec{\fun{$\alpha$}})\!\cdot\!\funL{$\lambda$b}\big)\delta^{(n{-}k{-}2)\!\times\!2}\!\big(\funL{$\lambda$}\!\cdot\!\var{\hat{C}}^{\perp}\!\!(\vec{\fun{$\alpha$}})\big)\delta^{(k+2)\!\times\!4}\!\big(\var{\hat{C}}(\vec{\fun{$\alpha$}})\!\cdot\!\tilde\eta\big)\, ,}\nonumber}
where $\var{\hat{C}}(\vec{\fun{$\alpha$}})$ is a $(k{+}2)\!\times\!n$ matrix parameterized by $\vec{\fun{$\alpha$}}$ representing a particular positroid variety labeled by some decorated permutation. As described in much more detail in \cite{ArkaniHamed:book} (see also \cite{Bourjaily:2012gy}), this integral imposes $2n{-}4$ excess constraints on the parameters $\{\vec{\fun{$\alpha$}}\}$ via the Bosonic $\delta$ functions involving \funL{$\lambda$} and \funL{$\lambda$b}. (There are $2n$ of these constraints, but 4 are equivalent to $\delta^{2\!\times\!2}\!(\funL{$\lambda$}\!\cdot\!\funL{$\lambda$b})$---momentum conservation.)\\[-10pt]

Notice that this is not directly presented as a \funL{superFunction}, because while the constraints \emph{require} that $\var{\hat{C}}$ \emph{contains} \funL{$\lambda$}, it does not define the part $\var{C}\equivR \var{\hat{C}}\,\tcap\,\funL{$\lambda$}^\perp$. 
}

\defnBox{$\sigma$}{\var{permutation}\patternTwo}{\emph{abstractly} denotes the positroid variety labeled by the decorated permutation \cite{ArkaniHamed:book,Bourjaily:2012gy} whose \emph{images} are listed by the \built{Integer} \built{Sequence} \var{permutation}:
\eq{\fun{$\sigma$}\brace{\var{\sigma(1)},\var{\sigma(2)},\ldots,\var{\sigma(n)}}\,\,\Leftrightarrow\,\,\sigma\!:\!a\mapsto\sigma(a)\quad\text{for}\quad a\!\in\![n]\quad\text{with}\quad\sigma(a)\geq\,a\,}
where the final requirement is merely a convention related to the unambiguously labeling decorated permutations. 
}

\defnBox{m}{\var{indices}\patternTwo}{\emph{abstractly} the Pl\"ucker coordinate (or `\textbf{m}inor') with \{\var{indices}\}$\in\!\binom{[n]}{k}$:
\eq{\fun{m}\brace{\var{a},\var{\ldots},\var{c}}\,\,\Leftrightarrow(\var{a}\cdots\var{c})\equivR\det\{c_{\var{a}},\ldots,c_{\var{c}}\}.\vspace{-20pt}}
}

\defnBox{$\alpha$}{\var{index}\pattern}{\emph{abstractly} represents a \emph{canonical coordinate} \fun{$\alpha$}\brace{\var{index}}\, parameterizing some positroid subvariety in the Grassmannian.}

\defnBox{$\tau$}{\var{$a$}\pattern,\var{$b$}\pattern}{\emph{abstractly} represents the \emph{transposition} which swaps the images \emph{of} $\{\var{a},\var{b}\}$ of a decorated permutation:
\eq{\fun{$\tau$}\brace{\var{a},\var{b}}:\!\fun{$\sigma$}\brace{\var{\sigma(1)},\var{\ldots},\var{\sigma(a)},\var{\ldots},\var{\sigma(b)},\var{\ldots},\var{\sigma(n)}}\mapsto\!\fun{$\sigma$}\brace{\var{\ldots},\var{\sigma(b)},\var{\ldots},\var{\sigma(a)},\var{\ldots}}\vspace{-20pt}}
}

\defnBox{$\pi$}{\var{$a$}\pattern,\var{$b$}\pattern}{\emph{abstractly} represents the \emph{inversion} which swaps the \emph{images} $\{\var{a},\var{b}\}$ of a decorated permutation:
\eq{\fun{$\pi$}\brace{\var{a},\var{b}}:\!\fun{$\sigma$}\brace{\var{\ldots},\var{a},\var{\ldots},\var{b},\var{\ldots}}\mapsto\!\fun{$\sigma$}\brace{\var{\ldots},\var{b},\var{\ldots},\var{a},\var{\ldots}}\vspace{-20pt}}
}


\vspace{5pt}\subsectionAppendix{Symbols Related to Scattering Equations and Twistor String}{appendix:symbols_for_chy}\vspace{-0pt}

\defnBox[]{$\sigma$}{\var{a}\pattern,\var{b}\pattern\optArg{}}{in the context of the \emph{scattering equations}, \fun{$\sigma$}\brace{\var{$a$}}$\,\in\!\mathbb{P}^1$ represents an auxiliary variable associated with the particle indexed by $\var{a}\!\in\![n]$. When two arguments are provided, \fun{$\sigma$}\brace{\var{a},\var{b}}\, merely denotes the difference \mbox{\fun{$\sigma$}\brace{\var{a},\var{b}}$\,\equivR$\fun{$\sigma$}\brace{\var{$a$}}\,$-$\mbox{\fun{$\sigma$}\brace{\var{$b$}}}}\,.
}

~

\defnBox[]{$\sigma$}{\var{$a$}\pattern,\var{$b$}\pattern}{\emph{abstractly} denotes \fun{$\sigma$}\brace{\var{$a$}}$-$\fun{$\sigma$}\brace{\var{$b$}}\,.}

\defnBox{$\xi$}{\var{$a$}\pattern}{in the context of \emph{twistor string theory} (as described by RSV \cite{RSV,Roiban:2004yf,ArkaniHamed:2009dg}),  \fun{$\xi$}${}_{\var{a}}\!\!\in\!\mathbb{P}^1$ represents an auxiliary variable associated with the particle indexed by $\var{a}\!\in\![n]$.}

\newpage
\vspace{-0pt}\sectionAppendix{General-Purpose Functions and Miscellaneous Tools}{appendix:general_tools}\vspace{-0pt}
%
\vspace{0pt}\subsectionAppendix{Mathematical Functions and Operations}{appendix:mathematical_tools}\vspace{-10pt}

\defnBox{complement}{\var{aList}\pattern,\var{bList}\patternTwo}{very similar to \textsc{Mathematica}'s \mbox{\built{Complement}}, but preserves the ordering of the elements of \var{aList}:
\mathematicaBox{
\mathematicaSequence{Complement\brace{\{5,2,3,1,4\},\{1,2\}}}{\{3,4,5\}}
\mathematicaSequence{\funL[1]{complement}\brace{\{5,2,3,1,4\},\{1,2\}}}{\{5,3,4\}}
}
}

\defnBox{pfaffian}{\var{matrix}\pattern}{returns the \emph{Pfaffian} of \var{matrix}, $\text{Pf}($\var{matrix}). If \var{$\Psi$} is an even-sized skew-symmetric square matrix, then $\text{Pf}(\var{\Psi})^2\!=\!\det(\var{\Psi})$.
}

\defnBox{eulerianNumber}{\var{$n$}\pattern,\var{$m$}\pattern}{provides the \emph{eulerian number} 
\eq{\fun{eulerianNumber}[\var{n},\var{m}]\,\,\Leftrightarrow\left\langle\begin{array}{@{}c@{}}\var{n}\\\var{m}\end{array}\right\rangle\equivR\sum_{k=0}^{\var{m}}({-}1)^{k}\binom{\var{n}{+}1}{k}(1{+}\var{m}{-}k)^{\var{n}}\,.\vspace{-14pt}}
}

\defnBox{randomPerm}{\var{$n$}\pattern}{returns a random permutation of the \built{List} $[\var{n}]\equivR\!\{1,\ldots,\var{n}\}$:
\mathematicaBox{\mathematicaSequence{\funL[1]{randomPerm}[6]}{\{4,2,3,6,1,5\}}}
}

\defnBox{randomSubset}{\var{$n$}\pattern,\var{$k$}\pattern}{when \var{$n$} is an \built{Integer}, it returns a random $\var{k}$-element subset of $[\var{n}]\equivR\!\{1,\ldots,\var{n}\}$:
\mathematicaBox{\mathematicaSequence{\funL[1]{randomSubset}[6,3]}{\{2,3,6\}}}
Alternatively, if \var{$n$} is a \built{List} it will return a random \var{$k$}-element subset of $\var{n}$.
}

\defnBox{shuffle}{\var{$a$}\pattern\built{Integer},\var{$b$}\pattern\built{Integer}}{returns the elements of the shuffle product
\eq{\fun{shuffle}\brace{\var{a},\var{b}}\,\Leftrightarrow [\var{a}]\bigger{\shuffle}[\var{b}]\subset\mathfrak{S}([\var{a}{+}\var{b}])}
which consists of all permutations of the set $[\var{a}{+}\var{b}]$ for which the elements $\{1,\ldots,\var{a}\}$ and $\{\var{a}{+}1\,\ldots,\var{a}{+}\var{b}\}$ each preserve their ordering.
\mathematicaBox{\mathematicaSequence{\funL[1]{shuffle}[2,3]/\!\!.\{\var{q}\pattern:>Style[\var{q},If[\var{q}>2,Blue,Red]]\}}{\{\{\r{1},\r{2},\b{3},\b{4},\b{5}\},\{\r{1},\b{3},\r{2},\b{4},\b{5}\},\{\r{1},\b3,\b4,\r{2},\b5\},\{\r{1},\b3,\b4,\b5,\r{2}\},\{\b3,\r{1},\r{2},\b4,\b5\}, \phantom{\{}\{\b3,\r{1},\b4,\r{2},\b5\},\{\b3,\r{1},\b4,\b5,\r{2}\},\{\b3,\b4,\r1,\r2,\b5\},\{\b3,\b4,\r1,\b5,\r2\},\{\b3,\b4,\b5,\r1,\r2\}\}}}
}

\defnBox{shuffle}{\var{$A$}\pattern\built{List},\var{$B$}\pattern\built{List}}{returns a \built{List} of the elements of the shuffle product of the \built{List}s $\var{A},\var{B}$.
}

\vspace{10pt}\subsectionAppendix{Generally Useful \textsc{Mathematica}-Like Functions}{appendix:mathematica_like_misc}\vspace{-10pt}

\defnBox{rationalize}{\var{number}\pattern}{behaves like \textsc{Mathematica}'s \built{Rationalize}, but only if the result is stable within the degree of \built{Precision} of the \var{number}
\mathematicaBox{
\mathematicaSequence[1]{\funL[]{useReferenceKinematics}\brace{12};\\
egCmptN=\funL[]{componentAmpN}\brace{\stateYM{p},\stateYM{m},\stateYM{p},\stateYM{p},\stateYM{p},\stateYM{p},\stateYM{m},\stateYM{m},\stateYM{m},\stateYM{m},\stateYM{p},\stateYM{m}};}{}
\mathematicaSequence{\funL[1]{rationalize}\brace{N\brace{egCmptN,200}}}{\scalebox{0.85}{${-}3.8785559513223388133705062343428668499616846071829398563170738061920041211\raisebox{3pt}{\rotatebox{-35}{\scalebox{0.55}{$\ddots$}}}$}\\
\scalebox{0.85}{$5855076317048164834493008764502236264358005013900342112883717010759767366431\raisebox{3pt}{\rotatebox{-35}{\scalebox{0.55}{$\ddots$}}}$}\\
\scalebox{0.85}{$51039930500725089626152736096793692143024404263628\!\times\!10^{-6}$}}
\mathematicaSequence{\funL[1]{rationalize}\brace{N\brace{egCmptN,240}}}{\scalebox{0.85}{$\text{-}\!\frac{27645335319948778562212486355454478754249849140149534616844538844093116149731551812150851420059444728717}{7127739206784806690102854315546372372524783316691980030837844052674950242430805728775854915363781568000000000}$}}
\mathematicaSequence{\textnormal{\%}===egCmptN}{True}
}
As exemplified by the example above, for which the amplitude component in question is a \built{Rational} number---a by a 104-digit \built{Integer} divided by a 109-digit \built{Integer}---in order to `reconstruct' the result from a decimal approximation would require that the answer be known to approximately 240(!) digits. This is one of the key advantages of using \built{Rational} kinematics, as opposed to \built{Real}-valued ones: results are automatically valid to \emph{infinite} precision. 
}

\defnBox{variables}{\var{expression}\pattern}{similar to \textsc{Mathematica}'s \mbox{\built{Variables}\brace{}\,}, but strips any protected \built{Head}s defined by the package---e.g.
\mathematicaBox{\mathematicaSequence{Variables[\funL{positroid}[\{\{1\!,\funL{$\alpha$}\![2]{+}\funL{$\alpha$}\![4],\funL{$\alpha$}\![2]\funL{$\alpha$}\![3],0\},\{0,1,\funL{$\alpha$}\![3],\funL{$\alpha$}\![1]\}\}]\!]}{\funL{positroid}[\{\{1,\funL{$\alpha$}\![2]{+}\funL{$\alpha$}\![4],\funL{$\alpha$}\![2]\funL{$\alpha$}\![3],0\},\{0,1,\funL{$\alpha$}\![3],\funL{$\alpha$}\![1]\}\}]}
\mathematicaSequence{\funL[1]{variables}[\funL{positroid}[\{\{1\!,\funL{$\alpha$}\![2]{+}\funL{$\alpha$}\![4],\funL{$\alpha$}\![2]\funL{$\alpha$}\![3],0\},\{0,1,\funL{$\alpha$}\![3],\funL{$\alpha$}\![1]\}\}]\!]}{\{\funL{$\alpha$}\![1],\funL{$\alpha$}\![2],\funL{$\alpha$}\![3],\funL{$\alpha$}\![4]\}}}
}

\defnBox{explicify}{\var{expression}\pattern}{\built{Replace}s all variables in \var{expression} to random values. For example:
\mathematicaBox{\mathematicaSequence{\funL[]{nice}@\funL[1]{explicify}[\{\{1,\funL{$\alpha$}\![2]{+}\funL{$\alpha$}\![4],\funL{$\alpha$}\![2]\funL{$\alpha$}\![3],0\},\{0,1,\funL{$\alpha$}\![3],\funL{$\alpha$}\![1]\}\}]}{$\left(\begin{array}{@{}cccc@{}}1&471&33552&0\\0&1&233&404\end{array}\right)$}}

\textbf{\emph{Note}}: for any variables associated with \emph{kinematics} (e.g.~\funL{ab} or \funL{asb}), randomly-chosen, on-shell and momentum-conserving kinematics will be used for their evaluation. This ensures that any symbolic consequences of momentum conservation or related identities are always observed: 
\mathematicaBox{\mathematicaSequence{\funL[1]{explicify}\brace{\funL{ab}\brace{1,2}\,\funL{ab}\brace{3,4}\,+\funL{ab}\brace{1,3}\,\funL{ab}\brace{4,2}\,+\funL{ab}\brace{1,4}\,\funL{ab}\brace{2,3}\,}}{0}}
}

\defnBox{fuzzify}{\var{expression}\pattern}{first applies \fun{explicify} to \var{expression}, and then replaces any \built{Matrix} expression with a `fuzzy' version thereof, where all 0 entries are dimmed and all non-vanishing quantities are replaced by `$\b{\star}$' for ease of visual recognition of the sparsity of data. 
\mathematicaBox{
\mathematicaSequence{\funL[1]{fuzzify}\brace{\funL[]{kltKernel}\brace{\funL[]{bcjBasisAmps}\brace{6},\funL[]{kltBasisAmpsBeta}\brace{6}}}}{$\displaystyle\left(\begin{array}{@{}cccccc@{}}
\b{\rstar}&\deemph{0}&\b{\rstar}&\deemph{0}&\deemph{0}&\deemph{0}\\[-6pt]
\b{\rstar}&\deemph{0}&\b{\rstar}&\deemph{0}&\deemph{0}&\deemph{0}\\[-6pt]
\deemph{0}&\b{\rstar}&\deemph{0}&\deemph{0}&\b{\rstar}&\deemph{0}\\[-6pt]
\deemph{0}&\b{\rstar}&\deemph{0}&\deemph{0}&\b{\rstar}&\deemph{0}\\[-6pt]
\deemph{0}&\deemph{0}&\deemph{0}&\b{\rstar}&\deemph{0}&\b{\rstar}\\[-6pt]
\deemph{0}&\deemph{0}&\deemph{0}&\b{\rstar}&\deemph{0}&\b{\rstar}
\end{array}\right)$}
\mathematicaSequence{\funL[1]{fuzzify}\brace{\funL[]{kltKernel}\brace{\funL[]{kltBasisAmpsAlpha}\brace{6},\funL[]{kltBasisAmpsBeta}\brace{6}}}}{$\displaystyle\left(\begin{array}{@{}cccccc@{}}\b{\rstar}&\b{\rstar}&\deemph{0}&\deemph{0}&\deemph{0}&\deemph{0}\\[-6pt]
\b{\rstar}&\b{\rstar}&\deemph{0}&\deemph{0}&\deemph{0}&\deemph{0}\\[-6pt]
\deemph{0}&\deemph{0}&\b{\rstar}&\b{\rstar}&\deemph{0}&\deemph{0}\\[-6pt]
\deemph{0}&\deemph{0}&\b{\rstar}&\b{\rstar}&\deemph{0}&\deemph{0}\\[-6pt]
\deemph{0}&\deemph{0}&\deemph{0}&\deemph{0}&\b{\rstar}&\b{\rstar}\\[-6pt]
\deemph{0}&\deemph{0}&\deemph{0}&\deemph{0}&\b{\rstar}&\b{\rstar}
\end{array}\right)$}
\mathematicaSequence{\funL[1]{fuzzify}\brace{\funL[]{kltKernel}\brace{\funL[]{bcjBasisAmps}\brace{6},\funL[]{bcjBasisAmps}\brace{6}}}}{$\displaystyle\left(\begin{array}{@{}cccccc@{}}\b{\rstar}&\b{\rstar}&\b{\rstar}&\b{\rstar}&\b{\rstar}&\b{\rstar}\\[-6pt]
\b{\rstar}&\b{\rstar}&\b{\rstar}&\b{\rstar}&\b{\rstar}&\b{\rstar}\\[-6pt]
\b{\rstar}&\b{\rstar}&\b{\rstar}&\b{\rstar}&\b{\rstar}&\b{\rstar}\\[-6pt]
\b{\rstar}&\b{\rstar}&\b{\rstar}&\b{\rstar}&\b{\rstar}&\b{\rstar}\\[-6pt]
\b{\rstar}&\b{\rstar}&\b{\rstar}&\b{\rstar}&\b{\rstar}&\b{\rstar}\\[-6pt]
\b{\rstar}&\b{\rstar}&\b{\rstar}&\b{\rstar}&\b{\rstar}&\b{\rstar}
\end{array}\right)$}
}
}

\newpage
\vspace{-0pt}\subsectionAppendix{Generally Useful Non-Mathematical, \textsc{Mathematica}-Like Functions}{appendix:non_math_mathematica_like_misc}\vspace{-10pt}

\defnBox{memory}{}{returns the amount of computer memory used in the current \textsc{Mathematica} session---differentiating between that used by the \built{Kernel} and the \built{FrontEnd}:\\[-12pt]
\mathematicaBox{\mathematicaSequence{\builtL[1]{memory}}{172.68 MB (1.11 GB for FrontEnd)}}
}

\defnBox{timed}{\var{expression}\pattern,\var{minimalTimeInSeconds}\pattern\optArg{0\,}}{provides information on the time required to \built{Evaluate} \var{expression}. Provided the time required exceeds the \emph{optional} value set by \var{minimalTimeInSeconds} (which has a \built{Default} value of 0), a message indicating the time required will be \built{Print} to screen.\\[-10pt]

Regardless of \var{minimalTimeInSeconds} which adjusts what is \built{Print} to the user, the function \fun{timed} will store the timing information (including the \built{Head} and time) to \fun{timingData}\brace{\built{\$Line}}\,, and also to \built{timingDataList}.\\[-10pt]

\emph{\textbf{Note}}: the function \fun{timed} automatically threads over certain types of \var{expression}s---for example, those whose \built{Head} is \built{Map}, \built{Table}, or \built{MapApply}.
\mathematicaBox{\mathematicaSequence{\funL[1]{timed}[\funL[]{componentAmpN}[\funL[]{randomComponentStates}[0][12,4]]\&/@Range[2]]\!;}{}
\\[-22pt]&\scalebox{0.6}{\hspace{-30pt}Evaluation of the function \texttt{\textbf{componentAmpN[]}} for $\mathcal{A}^{(4)}_{12}(-,+,-,+,+,-,+,-,-,+,-,+)$ required \textbf{1 seconds, 157 ms} to complete.}\\[-5pt]
&\scalebox{0.6}{\hspace{-30pt}Evaluation of the function \texttt{\textbf{componentAmpN[]}} for $\mathcal{A}^{(4)}_{12}(-,-,-,+,+,-,+,+,-,+,+,-)$ required \textbf{1 seconds, 833 $\mathbf{\mu}$s} to complete.}\\[-5pt]
&\scalebox{0.6}{\hspace{-40pt}Evaluation of the function \texttt{\textbf{Map}} required \textbf{2 seconds, 159 ms} to complete.}\\[-5pt]
\mathematicaSequence{\funL[1]{timingData}[1]}{\{\{Map,2.159886\},\\\{componentAmpN[m,p,m,p,p,m,p,m,m,p,m,p],1.157529\}, \{componentAmpN[m,m,m,p,p,m,p,p,m,p,p,m],1.000833\}\}}}
}

\defnBox{timingData}{\var{lineNumber}\pattern}{a function defined by \fun{timed}, with arguments given by the built-in variable \built{\$Line}. If multiple instances of \fun{timed} are called within a given \built{\$Line} of code, then this \fun{timingData} gives a \built{List} of pairs $\{\texttt{functionHead},\text{\var{timeInSeconds}}\}$.\\[-10pt]

If \var{lineNumber} is negative, then \fun{timingData}\brace{\var{${-}q$}} will return the data for the \var{$q$}th previous line.}

\defnBox{timingDataList}{}{a \built{List} of all the \built{DownValues} of the function \fun{timingData}\brace{}. This is useful, as not all \built{\$Line} values have timing data associated with them.}

\newpage
\vspace{-0pt}\subsectionAppendix{Formatting or Stylizing Expressions}{appendix:formatting_output}\vspace{-10pt}

\defnBox{nice}{\var{expression}\pattern}{displays the contents of \var{expression} in a way that is much more human-readable. For example, \fun{nice} will automatically display matrices in \built{MatrixForm}, and stylize/format the contents of many protected symbolic functions outlined in \mbox{Appendix \ref{appendix:abstract_protected_symbols}}.\\[-10pt]

To illustrate how \fun{nice} formats expressions, consider the following examples:
\begin{tabular}{|@{$\;$}l@{$\;$}|@{$\;\;$}l@{$\,$}|}\multicolumn{1}{c}{\textbf{Image under }\fun{nice}\brace{}}&\multicolumn{1}{c}{\textsc{\textbf{Mathematica}} \textbf{syntax}}\\
\hline
$\displaystyle|\psi\rangle_{(3)}^{\text{+}1/2}$&\stateYM{psi}\brace{\t{3}}\\\hline
$\displaystyle\lambda_1^2$&\funL{$\lambda$}\brace{\t{1,2}}\\
$\displaystyle\ab{1\,2}$&\funL{ab}\brace{\t{1,2}}\\
$\displaystyle\sb{3\,4}$&\funL{sb}\brace{\t{3,4}}\\
$\displaystyle s_{1\,2\,3}$&\funL{s}\brace{\t{1,2,3}}\\
$\displaystyle\langle1|(2\,3\,4)|(5\,6\,7\,8)|(9\,10)|11]$&\funL{asb}\brace{\t{1,\funL{p}\brace{\t{2,3,4}}\,,\funL{p}\brace{\t{5,6,7,8}}\,,\funL{p}\brace{\t{9,10}}\,,11}}\\
$\displaystyle[1|(2\,3\,4)|(5)|6\rangle$&\funL{sab}\brace{\t{1,\funL{p}\brace{\t{2,3,4}}\,\funL{p}\brace{\t{5}}\,,6}}\\
$\displaystyle\langle1|(2\,3)|(4\,5\,6)|7\rangle$&\funL{ab}\brace{\t{1,\funL{p}\brace{\t{2,3}}\,,\funL{p}\brace{\t{4,5,6}}\,,7}}\\
$\displaystyle[1|(2\,3\,4)|(5\,6)|7]$&\funL{sb}\brace{\t{1,\funL{p}\brace{2,3,4}\,,\funL{p}\brace{5,6}\,,7}}\\\hline
$\displaystyle\ab{1\,2\,3\,4}$&\fourBr\brace{\t{1,2,3,4}}\\
$\displaystyle\ab{(1\,2)\tcap(3\,4\,5)\,6\,7\,8}$&\fourBr\brace{\t{\funL{cap}\brace{\{1,2\},\{3,4,5\}\,},6,7,8}}\\\hline
$\displaystyle(1\,3\,5\,7)$&\funL{m}\brace{\t{1,3,5,7}}\\
$\displaystyle(1\,3)$&\fun{$\sigma$}\brace{\t{1,3}}\\\hline
\fig{-15pt}{0.5}{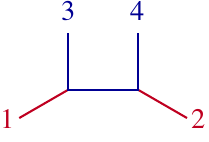}&\hspace{-1pt}\texttt{\funL{colorFactor}[1,3,4,2]}\\\hline
$\displaystyle\frac{1}{\alpha_1\,\alpha_2\,\alpha_3\,\alpha_4},\left(\begin{array}{@{}cccc@{}}1&\alpha_2\text{+}\alpha_4&\alpha_2\,\alpha_3&0\\
0&1&\alpha_3&\alpha_1\end{array}\right)\!\cdot\!\tilde\eta$&\begin{tabular}{@{}l@{}}\funL{positroid}[\t{\{\{1,\funL{$\alpha$}\![2]+\funL{$\alpha$}\![4],\funL{$\alpha$}\![2]\,\funL{$\alpha$}\![3],0\}\!,\!}\\\texttt{\{0,1,\funL{$\alpha$}\![3],\funL{$\alpha$}\![1]\}\}]}\end{tabular}\\\hline
$\displaystyle f,\left(\begin{array}{@{}ccc@{}}c[1,1]&c[1,2]&c[1,3]\\c[2,1]&c[2,2]&c[2,3]\end{array}\right)\!\cdot\!\tilde\eta$&\funL{superFunction}\brace{\t{f,Array\brace{c,\{2,3\}}\,}}\\\hline
\end{tabular}

\vspace{5pt}
\textbf{\emph{Note}}: the output of \fun{nice} is rarely (if ever!) useable for subsequent evaluation or algebraic manipulation. We have chosen to make users actively \emph{opt-in} using \fun{nice} to view `nice' expressions because we believe that the machine-readable (and user-manipulable) output is more valuable to most users. 
}

\defnBox{illustratingNice}{}{returns a table of examples similar those given above for \fun{nice}.}

\defnBox{niceTime}{\var{timeInSeconds}\pattern}{expresses  \var{timeInSeconds} in more appropriate units:
\mathematicaBox{\mathematicaSequence{\funL[1]{niceTime}[3.1 Power[10,\#]]\&/@Range[5]}{\{31 seconds, 400 ms,\\\phantom{\{}5 minutes, 14 seconds,\\\phantom{\{}52 minutes, 20 seconds,\\\phantom{\{}8 hours, 43 minutes,\\\phantom{\{}3 days, 15 hours\}}}\vspace{-100pt}}

\newpage
\vspace{-0pt}\sectionAppendix{Aspects of External, On-Shell Particles and States}{appendix:aspects_of_external_states}\vspace{-0pt}
%

\vspace{-0pt}\subsectionAppendix{\emph{Stylized} Listing of Component States for sYM and sGR}{appendix:stylized_component_lists}\vspace{-10pt}

\defnBox{componentStateList}{\var{$\mathcal{N}$}\pattern\optArg{4\,}}{equivalent to \fun{componentStateListYM}\brace{\var{$\mathcal{N}$}}.}

\defnBox{componentStateListYM}{\var{$\mathcal{N}$}\pattern\optArg{4\,}}{displays the coherent states (and the syntax by which they are named) of pure supersymemtric Yang-Mills theory with $\mathcal{N}$ supersymmetry generators.\\[-12pt]
\mathematicaBox{\mathematicaSequence{\funL[1]{componentStateListYM}[0]}{$\begin{array}{@{}ccc}\rule[-10pt]{0pt}{67.5pt}
\end{array}$}}
\mathematicaBox{\mathematicaSequence{\funL[1]{componentStateListYM}[1]}{$\begin{array}{@{}ccc}\rule[-10pt]{0pt}{105pt}
\end{array}$}}
\mathematicaBox{\mathematicaSequence{\funL[1]{componentStateListYM}[2]}{$\begin{array}{@{}ccc}\rule[-10pt]{0pt}{175pt}
\end{array}$}}
}
\raisebox{376pt}[000pt]{\scalebox{1}{\fwboxL{250pt}{\hspace{40pt}\begin{array}{|lclll|}
\hline\multicolumn{1}{|c}{\begin{array}{c}\text{(stylized)}\\\text{\textbf{state name}}\\\end{array}}&\begin{array}{c}~\\\text{\textbf{helicity}}\end{array}&\multicolumn{1}{c}{\begin{array}{c}~\\\text{\textbf{$R$-charge}}\end{array}}&\multicolumn{1}{c}{\begin{array}{c}~\\\text{\textbf{syntax}}\end{array}}&\multicolumn{1}{c|}{\begin{array}{c}\text{optional/}\\\text{{alt.\ syntax}}\end{array}}\\
\hline
|g\rangle^{\text{+}1}&1&\{\}&\stateYM{p}&\\
|g\rangle^{\text{-}1}&\text{-}1\phantom{\text{-}}&\bar{\{\}}\!\equivR\!\{1,2,3,4\}&\stateYM{m}&\\\hline
\end{array}
}}}\vspace{-10pt}\\
\raisebox{262pt}[000pt]{\scalebox{1}{\fwboxL{250pt}{\hspace{40pt}\begin{array}{|lclll|}
\hline\multicolumn{1}{|c}{\begin{array}{c}\text{(stylized)}\\\text{\textbf{state name}}\\\end{array}}&\begin{array}{c}~\\\text{\textbf{helicity}}\end{array}&\multicolumn{1}{c}{\begin{array}{c}~\\\text{\textbf{$R$-charge}}\end{array}}&\multicolumn{1}{c}{\begin{array}{c}~\\\text{\textbf{syntax}}\end{array}}&\multicolumn{1}{c|}{\begin{array}{c}\text{optional/}\\\text{{alt.\ syntax}}\end{array}}\\
\hline
|g\rangle^{\text{+}1}&1&\{\}&\stateYM{p}&\\
|\psi\rangle^{\text{+}1/2}_{(1)}&\frac{1}{2}&\{1\}&\stateYM{psi}\brace{1}&\\
|\bar{\psi}\rangle^{\text{-}1/2}_{(\bar{1})}&\text{-}\frac{1}{2}\phantom{\text{-}}&\bar{\{1\}}\!\equivR\!\{2,3,4\}&\stateYM{psiBar}\brace{1}&\stateYM{psiBar}\brace{2,3,4}\\
|g\rangle^{\text{-}1}&\text{-}1\phantom{\text{-}}&\bar{\{\}}\!\equivR\!\{1,2,3,4\}&\stateYM{m}&\\\hline
\end{array}
}}}\vspace{-10pt}\\
\raisebox{95pt}[000pt]{\scalebox{1}{\fwboxL{250pt}{\hspace{40pt}\begin{array}{|lclll|}
\hline\multicolumn{1}{|c}{\begin{array}{c}\text{(stylized)}\\\text{\textbf{state name}}\\\end{array}}&\begin{array}{c}~\\\text{\textbf{helicity}}\end{array}&\multicolumn{1}{c}{\begin{array}{c}~\\\text{\textbf{$R$-charge}}\end{array}}&\multicolumn{1}{c}{\begin{array}{c}~\\\text{\textbf{syntax}}\end{array}}&\multicolumn{1}{c|}{\begin{array}{c}\text{optional/}\\\text{{alt.\ syntax}}\end{array}}\\
\hline
|g\rangle^{\text{+}1}&1&\{\}&\stateYM{p}&\\
|\psi\rangle^{\text{+}1/2}_{(1)}&\frac{1}{2}&\{1\}&\stateYM{psi}\brace{1}&\\
|\psi\rangle^{\text{+}1/2}_{(2)}&\frac{1}{2}&\{2\}&\stateYM{psi}\brace{2}&\\
|\phi\rangle^{0}_{(1\,2)}&0&\{1,2\}&\stateYM{phi}\brace{1,2}&\\
|\bar{\phi}\rangle^{0}_{(\bar{1\,2})}&0&\bar{\{1,2\}}\!\equivR\!\{3,4\}&\stateYM{phiBar}\brace{1,2}&\stateYM{phi}\brace{3,4}\\
|\bar{\psi}\rangle^{\text{-}1/2}_{(\bar{2})}&\text{-}\frac{1}{2}\phantom{\text{-}}&\bar{\{2\}}\!\equivR\!\{1,3,4\}&\stateYM{psiBar}\brace{2}&\stateYM{psiBar}\brace{1,3,4}\\
|\bar{\psi}\rangle^{\text{-}1/2}_{(\bar{1})}&\text{-}\frac{1}{2}\phantom{\text{-}}&\bar{\{1\}}\!\equivR\!\{2,3,4\}&\stateYM{psiBar}\brace{1}&\stateYM{psiBar}\brace{2,3,4}\\
|g\rangle^{\text{-}1}&\text{-}1\phantom{\text{-}}&\bar{\{\}}\!\equivR\!\{1,2,3,4\}&\stateYM{m}&\\\hline
\end{array}
}}}\vspace{-10pt}

\defnBox{componentStateListGR}{\var{$\mathcal{N}$}\pattern\optArg{8\,}}{displays the coherent states (and their relevant syntax) of pure supersymmemtric gravitational theories with $\mathcal{N}$ supersymmetry generators.\\[-12pt]
\mathematicaBox{\mathematicaSequence{\funL[1]{componentStateListGR}[4]}{$\begin{array}{@{}ccc}\rule[-10pt]{0pt}{530pt}
\end{array}$}}
\raisebox{272pt}[000pt]{\scalebox{1.02}{\fwboxL{250pt}{\hspace{16pt}\begin{array}{|l@{$\!\!$}c@{$\,\,$}lll|}
\hline\multicolumn{1}{|c}{\fwbox{45pt}{\begin{array}{c}\text{(stylized)}\\[-5pt]\text{\hspace{-10pt}\textbf{state}}\\\end{array}}}&\fwbox{20pt}{\begin{array}{c}~\\[-5pt]\text{\textbf{helicity}}\end{array}}&\multicolumn{1}{c}{\begin{array}{c}~\\[-5pt]\text{\textbf{$R$-charge}}\end{array}}&\multicolumn{1}{c}{\begin{array}{c}~\\[-5pt]\text{\textbf{syntax}}\end{array}}&\multicolumn{1}{c|}{\begin{array}{c}\text{optional/}\\[-5pt]\text{{alt.\ syntax}}\end{array}}\\
\hline
\scalebox{0.8}{$|g\rangle^{\text{+}2}$}&2&\scalebox{0.9}{$\{\}$}&\scalebox{0.8}{\stateGR{pp}}&\\[-1pt]
\scalebox{0.8}{$|\psi\rangle^{\text{+}3/2}_{(1)}$}&\frac{3}{2}&\scalebox{0.9}{$\{1\}$}&\scalebox{0.8}{\stateGR{gravitino}\brace{1}}&\\[-1pt]
\scalebox{0.8}{$|\psi\rangle^{\text{+}3/2}_{(2)}$}&\frac{3}{2}&\scalebox{0.9}{$\{2\}$}&\scalebox{0.8}{\stateGR{gravitino}\brace{2}}&\\[-1pt]
\scalebox{0.8}{$|\psi\rangle^{\text{+}3/2}_{(3)}$}&\frac{3}{2}&\scalebox{0.9}{$\{3\}$}&\scalebox{0.8}{\stateGR{gravitino}\brace{3}}&\\[-1pt]
\scalebox{0.8}{$|\psi\rangle^{\text{+}3/2}_{(4)}$}&\frac{3}{2}&\scalebox{0.9}{$\{4\}$}&\scalebox{0.8}{\stateGR{gravitino}\brace{4}}&\\[-1pt]
\scalebox{0.8}{$|\gamma\rangle^{\text{+}1}_{(1\,2)}$}&1&\scalebox{0.9}{$\{1\,2\}$}&\scalebox{0.8}{\stateGR{graviPhoton}\brace{\t{1\!,\!2}}}&\\[-1pt]
\scalebox{0.8}{$|\gamma\rangle^{\text{+}1}_{(1\,3)}$}&1&\scalebox{0.9}{$\{1\,3\}$}&\scalebox{0.8}{\stateGR{graviPhoton}\brace{\t{1\!,\!3}}}&\\[-1pt]
\scalebox{0.8}{$|\gamma\rangle^{\text{+}1}_{(1\,4)}$}&1&\scalebox{0.9}{$\{1\,4\}$}&\scalebox{0.8}{\stateGR{graviPhoton}\brace{\t{1\!,\!4}}}&\\[-1pt]
\scalebox{0.8}{$|\gamma\rangle^{\text{+}1}_{(2\,3)}$}&1&\scalebox{0.9}{$\{2\,3\}$}&\scalebox{0.8}{\stateGR{graviPhoton}\brace{\t{2\!,\!3}}}&\\[-1pt]
\scalebox{0.8}{$|\gamma\rangle^{\text{+}1}_{(2\,4)}$}&1&\scalebox{0.9}{$\{2\,4\}$}&\scalebox{0.8}{\stateGR{graviPhoton}\brace{\t{2\!,\!4}}}&\\[-1pt]
\scalebox{0.8}{$|\gamma\rangle^{\text{+}1}_{(3\,4)}$}&1&\scalebox{0.9}{$\{3\,4\}$}&\scalebox{0.8}{\stateGR{graviPhoton}\brace{\t{3\!,\!4}}}&\\[-1pt]
\scalebox{0.8}{$|\psi\rangle^{\text{+}1/2}_{(1\,2\,3)}$}&\frac{1}{2}&\scalebox{0.9}{$\{1,\!2,\!3\}$}&\scalebox{0.8}{\stateGR{psi}\brace{\t{1\!,\!2\!,\!3}}}&\\[-1pt]
\scalebox{0.8}{$|\psi\rangle^{\text{+}1/2}_{(1\,2\,4)}$}&\frac{1}{2}&\scalebox{0.9}{$\{1,\!2,\!4\}$}&\scalebox{0.8}{\stateGR{psi}\brace{\t{1\!,\!2\!,\!4}}}&\\[-1pt]
\scalebox{0.8}{$|\psi\rangle^{\text{+}1/2}_{(1\,3\,4)}$}&\frac{1}{2}&\scalebox{0.9}{$\{1,\!3,\!4\}$}&\scalebox{0.8}{\stateGR{psi}\brace{\t{1\!,\!3\!,\!4}}}&\\[-1pt]
\scalebox{0.8}{$|\psi\rangle^{\text{+}1/2}_{(2\,3\,4)}$}&\frac{1}{2}&\scalebox{0.9}{$\{2,\!3,\!4\}$}&\scalebox{0.8}{\stateGR{psi}\brace{\t{2\!,\!3\!,\!4}}}&\\[-1pt]
\scalebox{0.8}{$|\phi\rangle^{0}_{(1\,2\,3\,4)}$}&0&\scalebox{0.9}{$\{1,\!2,\!3,\!4\}$}&\scalebox{0.8}{\stateGR{phi}\brace{\t{1\!,\!2\!,\!3\!,\!4}}}&\\[-1pt]
\scalebox{0.8}{$|\bar{\phi}\rangle^{0}_{(\bar{1\,2\,3\,4})}$}&0&\scalebox{0.9}{$\bar{\{1,\!2,\!3,\!4\}}\!\equivR\!\{5,\!6,\!7,\!8\}$}&\scalebox{0.8}{\stateGR{phiBar}\brace{\t{1\!,\!2\!,\!3\!,\!4}}}&\scalebox{0.8}{\stateGR{phi}\brace{\t{5\!,\!6\!,\!7\!,\!8}}}\\[-1pt]
\scalebox{0.8}{$|\bar{\psi}\rangle^{\text{-}1/2}_{(\bar{2\,3\,4})}$}&\text{-}\frac{1}{2}\phantom{\text{-}}&\scalebox{0.9}{$\bar{\{2,\!3,\!4\}}\!\equivR\!\{1,\!5,\!6,\!7,\!8\}$}&\scalebox{0.8}{\stateGR{psiBar}\brace{\t{2\!,\!3\!,\!4}}}&\scalebox{0.8}{\stateGR{psiBar}\brace{\t{1\!,\!5\!,\!6\!,\!7\!,\!8}}}\\[-1pt]
\scalebox{0.8}{$|\bar{\psi}\rangle^{\text{-}1/2}_{(\bar{1\,3\,4})}$}&\text{-}\frac{1}{2}\phantom{\text{-}}&\scalebox{0.9}{$\bar{\{1,\!3,\!4\}}\!\equivR\!\{2,\!5,\!6,\!7,\!8\}$}&\scalebox{0.8}{\stateGR{psiBar}\brace{\t{1\!,\!3\!,\!4}}}&\scalebox{0.8}{\stateGR{psiBar}\brace{\t{2\!,\!5\!,\!6\!,\!7\!,\!8}}}\\[-1pt]
\scalebox{0.8}{$|\bar{\psi}\rangle^{\text{-}1/2}_{(\bar{1\,2\,4})}$}&\text{-}\frac{1}{2}\phantom{\text{-}}&\scalebox{0.9}{$\bar{\{1,\!2,\!4\}}\!\equivR\!\{3,\!5,\!6,\!7,\!8\}$}&\scalebox{0.8}{\stateGR{psiBar}\brace{\t{1\!,\!2\!,\!4}}}&\scalebox{0.8}{\stateGR{psiBar}\brace{\t{3\!,\!5\!,\!6\!,\!7\!,\!8}}}\\[-1pt]
\scalebox{0.8}{$|\bar{\psi}\rangle^{\text{-}1/2}_{(\bar{1\,2\,3})}$}&\text{-}\frac{1}{2}\phantom{\text{-}}&\scalebox{0.9}{$\bar{\{1,\!2,\!3\}}\!\equivR\!\{4,\!5,\!6,\!7,\!8\}$}&\scalebox{0.8}{\stateGR{psiBar}\brace{\t{1\!,\!2\!,\!3}}}&\scalebox{0.8}{\stateGR{psiBar}\brace{\t{4\!,\!5\!,\!6\!,\!7\!,\!8}}}\\[-1pt]
\scalebox{0.8}{$|\gamma\rangle^{\text{-}1}_{(\bar{3\,4})}$}&\text{-}1\phantom{\text{-}}&\scalebox{0.9}{$\bar{\{3\,4\}}\!\equivR\!\{1,\!2,\!5,\!6,\!7,\!8\}$}&\scalebox{0.8}{\stateGR{graviPhotonBar}\brace{\t{3\!,\!4}}}&\scalebox{0.8}{\stateGR{graviPhotonBar}\brace{\t{1\!,\!2\!,\!5\!,\!6\!,\!7\!,\!8}}}\\[-1pt]
\scalebox{0.8}{$|\gamma\rangle^{\text{-}1}_{(\bar{2\,4})}$}&\text{-}1\phantom{\text{-}}&\scalebox{0.9}{$\bar{\{2\,4\}}\!\equivR\!\{1,\!3,\!5,\!6,\!7,\!8\}$}&\scalebox{0.8}{\stateGR{graviPhotonBar}\brace{\t{2\!,\!4}}}&\scalebox{0.8}{\stateGR{graviPhotonBar}\brace{\t{1\!,\!3\!,\!5\!,\!6\!,\!7\!,\!8}}}\\[-1pt]
\scalebox{0.8}{$|\gamma\rangle^{\text{-}1}_{(\bar{2\,3})}$}&\text{-}1\phantom{\text{-}}&\scalebox{0.9}{$\bar{\{2\,3\}}\!\equivR\!\{1,\!4,\!5,\!6,\!7,\!8\}$}&\scalebox{0.8}{\stateGR{graviPhotonBar}\brace{\t{2\!,\!3}}}&\scalebox{0.8}{\stateGR{graviPhotonBar}\brace{\t{1\!,\!4\!,\!5\!,\!6\!,\!7\!,\!8}}}\\[-1pt]
\scalebox{0.8}{$|\gamma\rangle^{\text{-}1}_{(\bar{1\,4})}$}&\text{-}1\phantom{\text{-}}&\scalebox{0.9}{$\bar{\{1\,4\}}\!\equivR\!\{2,\!3,\!5,\!6,\!7,\!8\}$}&\scalebox{0.8}{\stateGR{graviPhotonBar}\brace{\t{1\!,\!4}}}&\scalebox{0.8}{\stateGR{graviPhotonBar}\brace{\t{2\!,\!3\!,\!5\!,\!6\!,\!7\!,\!8}}}\\[-1pt]
\scalebox{0.8}{$|\gamma\rangle^{\text{-}1}_{(\bar{1\,3})}$}&\text{-}1\phantom{\text{-}}&\scalebox{0.9}{$\bar{\{1\,3\}}\!\equivR\!\{2,\!4,\!5,\!6,\!7,\!8\}$}&\scalebox{0.8}{\stateGR{graviPhotonBar}\brace{\t{1\!,\!3}}}&\scalebox{0.8}{\stateGR{graviPhotonBar}\brace{\t{2\!,\!4\!,\!5\!,\!6\!,\!7\!,\!8}}}\\[-1pt]
\scalebox{0.8}{$|\gamma\rangle^{\text{-}1}_{(\bar{1\,2})}$}&\text{-}1\phantom{\text{-}}&\scalebox{0.9}{$\bar{\{1\,2\}}\!\equivR\!\{3,\!4,\!5,\!6,\!7,\!8\}$}&\scalebox{0.8}{\stateGR{graviPhotonBar}\brace{\t{1\!,\!2}}}&\scalebox{0.8}{\stateGR{graviPhotonBar}\brace{\t{3\!,\!4\!,\!5\!,\!6\!,\!7\!,\!8}}}\\[-1pt]
\scalebox{0.8}{$|\bar{\psi}\rangle^{\text{-}3/2}_{(\bar{4})}$}&\text{-}\frac{3}{2}\phantom{\text{-}}&\scalebox{0.9}{$\bar{\{4\}}\!\equivR\!\{1,\!2,\!3,\!5,\!6,\!7,\!8\}$}&\scalebox{0.8}{\stateGR{gravitinoBar}\brace{\t{4}}}&\scalebox{0.8}{\stateGR{gravitinoBar}\brace{\t{1\!,\!2\!,\!3\!,\!5\!,\!6\!,\!7\!,\!8}}}\\[-1pt]
\scalebox{0.8}{$|\bar{\psi}\rangle^{\text{-}3/2}_{(\bar{3})}$}&\text{-}\frac{3}{2}\phantom{\text{-}}&\scalebox{0.9}{$\bar{\{3\}}\!\equivR\!\{1,\!2,\!4,\!5,\!6,\!7,\!8\}$}&\scalebox{0.8}{\stateGR{gravitinoBar}\brace{\t{3}}}&\scalebox{0.8}{\stateGR{gravitinoBar}\brace{\t{1\!,\!2\!,\!4\!,\!5\!,\!6\!,\!7\!,\!8}}}\\[-1pt]
\scalebox{0.8}{$|\bar{\psi}\rangle^{\text{-}3/2}_{(\bar{2})}$}&\text{-}\frac{3}{2}\phantom{\text{-}}&\scalebox{0.9}{$\bar{\{2\}}\!\equivR\!\{1,\!3,\!4,\!5,\!6,\!7,\!8\}$}&\scalebox{0.8}{\stateGR{gravitinoBar}\brace{\t{2}}}&\scalebox{0.8}{\stateGR{gravitinoBar}\brace{\t{1\!,\!3\!,\!4\!,\!5\!,\!6\!,\!7\!,\!8}}}\\[-1pt]
\scalebox{0.8}{$|\bar{\psi}\rangle^{\text{-}3/2}_{(\bar{1})}$}&\text{-}\frac{3}{2}\phantom{\text{-}}&\scalebox{0.9}{$\bar{\{1\}}\!\equivR\!\{2,\!3,\!4,\!5,\!6,\!7,\!8\}$}&\scalebox{0.8}{\stateGR{gravitinoBar}\brace{\t{1}}}&\scalebox{0.8}{\stateGR{gravitinoBar}\brace{\t{2\!,\!3\!,\!4\!,\!5\!,\!6\!,\!7\!,\!8}}}\\[-1pt]
\scalebox{0.8}{$|g\rangle^{\text{-}2}$}&\text{-}2\phantom{\text{-}}&\scalebox{0.9}{$\bar{\{\}}\!\equivR\!\{1,\!2,\!3,\!4,\!5,\!6,\!7,\!8\}$}&\scalebox{0.8}{\stateGR{mm}}&\\\hline
\end{array}}}}\vspace{-20pt}
}

\newpage
\vspace{-0pt}\subsectionAppendix{Generating \emph{Random} Sets of External States}{appendix:random_states}\vspace{-10pt}


\defnBox{randomHelicityComponent}{\var{$n$}\pattern,\var{$k$}\pattern\optArg{}}{same as \mbox{\fun{randomComponentStatesYM}\brace{0}\,\brace{\var{$n$},\var{$k$}}\,.} When the optional argument \var{$k$} is not specified, it is taken to be a \built{RandomChoice} of $[0,\var{n}{-}4]$.}

\defnBox{randomHelicityComponentGR}{\var{$n$}\pattern,\var{$k$}\pattern\optArg{}}{same as \mbox{\fun{randomComponentStatesGR}\brace{0}\,\brace{\var{$n$},\var{$k$}}\,.} When the optional argument \var{$k$} is not specified, it is taken to be a \built{RandomChoice} of $[0,\var{n}{-}4]$.}

\defnBox{randomComponentStates}{\var{$n$}\pattern,\var{$k$}\pattern}{same as \mbox{\fun{randomComponentStatesYM}\brace{4}\,\brace{\var{$n$},\var{$k$}}\,.}}

\defnBox{randomComponentStatesGR}{\var{$n$}\pattern,\var{$k$}\pattern}{same as \mbox{\fun{randomComponentStatesGR}\brace{8}\,\brace{\var{$n$},\var{$k$}}\,.}}

\defnBoxTwo{randomComponentStatesYM}{\var{$\mathcal{N}$}\pattern\optArg{4}\,}{\var{$n$}\pattern,\var{$k$}\pattern}{gives a \emph{random} choice of \var{$n$} external states corresponding to a non-trivial N${}^{\var{k}}$MHV amplitude in pure Yang-Mills theory with \var{$\mathcal{N}$} supersymmetries.
\mathematicaBox{
\mathematicaSequence{\funL[1]{randomComponentStatesYM}\brace{\#}\,\brace{6,1}\,\&/\!@Range[4]}{\{\{\stateYM{m},\stateYM{p},\stateYM{p},\stateYM{m},\stateYM{m},\stateYM{p}\}, 
\{\stateYM{phi}[1,2],\stateYM{psiBar}[1],\stateYM{p},\stateYM{p},\stateYM{m},\stateYM{psiBar}[2]\},\\
\phantom{\{}\{\stateYM{phi}[1,3],\stateYM{m},\stateYM{psi}[2],\stateYM{psi}[4],\stateYM{psi}[4],\stateYM{psiBar}[4]\},\\
\phantom{\{}\{\stateYM{psiBar}[1],\stateYM{psi}[3],\stateYM{phi}[1,2],\stateYM{psiBar}[4],\stateYM{psi}[4],\stateYM{phi}[1,4]\}\}}
\mathematicaSequence{\funL[]{nice}/\!@\,\funL{amp}@@@\textnormal{\%}}{$\Big\{\mathcal{A}_6^{(1)}\!\big(-,+,+,-,-,+\big),$ $\mathcal{A}_6^{(1)}\!\big(|\phi\rangle^0_{(1\,2)},|\bar{\psi}\rangle^{\text{-}1/2}_{(\bar{1})},+,+,-,|\bar{\psi}\rangle^{\text{-}1/2}_{(\bar{2})}\big),$\\
\phantom{\{}$\mathcal{A}_6^{(1)}\!\big(|\phi\rangle^0_{(1\,3)},-,|{\psi}\rangle^{\text{+}1/2}_{({2})},|{\psi}\rangle^{\text{+}1/2}_{({4})},|{\psi}\rangle^{\text{+}1/2}_{({4})},|\bar{\psi}\rangle^{\text{-}1/2}_{(\bar{4})}\big),$\\
\phantom{\{}$\mathcal{A}_6^{(1)}\!\big(|\bar{\psi}\rangle^{\text{-}1/2}_{(\bar{1})},|{\psi}\rangle^{\text{+}1/2}_{({3})},|\phi\rangle^0_{(1\,2)},|\bar{\psi}\rangle^{\text{-}1/2}_{(\bar{4})},|{\psi}\rangle^{\text{+}1/2}_{({4})},|\phi\rangle^{0}_{({1\,4})}\big)\Big\}$
}
}
\\[-24pt]%
}

\defnBoxTwo{randomComponentStatesGR}{\var{$\mathcal{N}$}\pattern\optArg{8}\,}{\var{$n$}\pattern,\var{$k$}\pattern}{gives a \emph{random} choice of \var{$n$} external states corresponding to a non-trivial N${}^{\var{k}}$MHV amplitude in pure supersymmetric gravitational theory with \var{$\mathcal{N}$} supersymmetries.
\mathematicaBox{
\mathematicaSequence{\funL[1]{randomComponentStatesGR}\brace{\#}\,\brace{6,1}\,\&/\!@\{0,1,8\}}{\{\{\stateGR{mm},\stateGR{pp},\stateGR{mm},\stateGR{pp},\stateGR{pp},\stateGR{mm}\},\\
\phantom{\{}\{\stateGR{gravitino}\![1]\!,\hspace{-1pt}\stateGR{pp}\!,\hspace{-1pt}\stateGR{gravitino}\![1]\!,\hspace{-1pt}\stateGR{mm}\!,\hspace{-1pt}\stateGR{gravitinoBar}\![1]\!,\hspace{-1pt}\stateGR{gravitinoBar}\![1]\!\}\!,\\
\phantom{\{}\{\stateGR{psi}\![3,7,8]\!,\stateGR{phi}\![1,2,5,8]\!,\stateGR{psiBar}\![2,6,7]\!,\!\stateGR{phiBar}\![1,3,5,8]\!,\\
\phantom{\{\{}\stateGR{psiBar}\![4,7,8],\stateGR{psi}\![4,6,7]\}\}
}
\mathematicaSequence{\funL[]{nice}/\!@\,\funL{amp}@@@\textnormal{\%}}{$\Big\{\mathcal{A}_6^{(1)}\!\big(--,++,--,++,++,--\big),$ $\mathcal{A}_6^{(1)}\!\big(|\phi\rangle^0_{(1\,2)},|\bar{\psi}\rangle^{\text{-}1/2}_{(\bar{1})},+,+,-,|\bar{\psi}\rangle^{\text{-}1/2}_{(\bar{2})}\big),$\\
\phantom{\{}$\mathcal{A}_6^{(1)}\!\big(|\psi\rangle^{\text{+}3/2}_{(1)},++,|{\psi}\rangle^{\text{+}3/2}_{({1})},--,|\bar{\psi}\rangle^{\text{-}3/2}_{(\bar{1})},|\bar{\psi}\rangle^{\text{-}3/2}_{(\bar{1})}\big),$\\
\phantom{\{}$\mathcal{A}_6^{(1)}\!\big(|{\psi}\rangle^{\text{+}1/2}_{({3\,4\,8})},|{\phi}\rangle^{0}_{({1\,2\,5\,8})},|\bar{\psi}\rangle^{\text{-}1/2}_{(\bar{2\,6\,7})},|\bar{\phi}\rangle^{0}_{(\bar{1\,3\,5\,8})},|\bar{\psi}\rangle^{\text{-}1/2}_{(\bar{4\,7\,8})},|\psi\rangle^{\text{+}1/2}_{({4\,6\,7})}\big)\Big\}$
}
}\vspace{-40pt}
}

\defnBox{randomStateSplitting}{\var{componentStates}\pattern}{returns a \emph{random} choice of how to represent the superGravity \var{componentStates} as a `product' of those of superYang-Mills theory. This is achieved by \emph{randomly} partitioning the 8 minors (as generated by \funL{stateListToMinorList}, say) used to project out \var{componentStates} into two sets of 4. It is useful to verify that the $SU_8$ $R$-symmetry can be broken into any arbitrary pairs of $SU_4$ subgroups when using KLT, for example---which is why we have implemented the making of a `random' choice:
\mathematicaBox{
\mathematicaSequence{egStates=\funL[]{randomComponentStatesGR}\brace{6,1}}{\{\stateGR{psiBar}[2,3,7],\stateGR{phi}[1,4,6,7],\stateGR{psi}[2,3,8],\stateGR{psi}[1,2,7],\\
\phantom{\{}\stateGR{psi}[2,3,5],\stateGR{graviPhotonBar}[1,2]\}}
\mathematicaSequence{refAmpN=\funL[]{componentAmpN}\brace{egStates}}{$\displaystyle\text{-}\frac{286986179759783753}{12040553430847766}$}
\mathematicaSequence{\{statesL1,statesR1\}=\funL[1]{randomStateSplitting}\brace{egStates}}{\{\{\stateYM{m},\stateYM{phi}[1,3],\stateYM{psi}[4],\stateYM{p},\stateYM{psi}[2],\stateYM{m}\},\\
\phantom{\{}\{\stateYM{psi}[1],\stateYM{phi}[1,4],\stateYM{phiBar}[1,4]\stateYM{psiBar}[3],\stateYM{phiBar}[1,4],\stateYM{phiBar}[1,2]\}\}}
\mathematicaSequence[3]{\{ampsL,ampsR\}=\{\funL[]{kltBasisAmpsAlpha}\brace{6},\,\funL[]{kltBasisAmpsBeta}\brace{6}\,\};\\
leftTerms1=ampsL/\!\!.\{amp\brace{\var{x}\patternTwo}:\!>\mbox{\funL[]{componentAmp}\brace{statesL1,\{\var{x}\}};}\\
rightTerms1=ampsL/\!\!.\{amp\brace{\var{x}\patternTwo}:\!>\mbox{\funL[]{componentAmp}\brace{statesR1,\{\var{x}\}};}\\
\funL[]{evaluate}\brace{leftTerms1.\funL[]{kltKernel}\brace{6}\,.rightTerms1}}{$\rule{0pt}{24pt}\displaystyle\text{-}\frac{286986179759783753}{12040553430847766}$}
\mathematicaSequence{\{statesL2,statesR2\}=\funL[1]{randomStateSplitting}\brace{egStates}}{\{\{\stateYM{psiBar}[2],\stateYM{phi}[1,4],\stateYM{psi}[2],\stateYM{phi}[1,2],\stateYM{phiBar}[1,4],\stateYM{phiBar}[1,2]\},\\
\phantom{\{}\{\stateYM{phiBar}[1,3],\stateYM{phiBar}[1,4],\stateYM{phi}[1,4],\stateYM{psi}[3],\stateYM{psi}[1],\stateYM{m}\}\}}
\mathematicaSequence[2]{
leftTerms2=ampsL/\!\!.\{\funL{amp}\brace{\var{x}\patternTwo}:\!>\mbox{\funL[]{componentAmp}\brace{statesL2,\{\var{x}\}};}\\
rightTerms2=ampsL/\!\!.\{\funL{amp}\brace{\var{x}\patternTwo}:\!>\mbox{\funL[]{componentAmp}\brace{statesR2,\{\var{x}\}};}\\
\funL[]{evaluate}\brace{leftTerms2.\funL[]{kltKernel}\brace{6}\,.rightTerms2}}{$\rule{0pt}{24pt}\displaystyle\text{-}\frac{286986179759783753}{12040553430847766}$}
}
}

\newpage
\vspace{-0pt}\subsectionAppendix{Aspects of and Manipulations to External States}{appendix:manipulations_of_external_states}\vspace{-10pt}

\defnBox{rChargeLabels}{\var{componentStates}\patternTwo}{returns a \built{List} of `$R$-charge' labels associated with each of the states in \var{componentStates}. Superselection rules require that each index of $I\!\in\![\mathcal{N}]$ (with $\mathcal{N}\!\in\!\{4,8\}$) must occur exactly the same number of instances. See \mbox{Tables \ref{sym_state_table} and \ref{sgr_state_table}} for our conventions. 
\mathematicaBox{
\mathematicaSequence{egStates=\funL[]{randomComponentStates}\brace{6,1}}{\{\stateYM{psi}\![2],\stateYM{phiBar}\![1,2],\stateYM{psi}[1],\stateYM{psiBar}\![1],\stateYM{psi}\![1],\stateYM{m}\}}
\mathematicaSequence{\funL[1]{rChargeLabels}\brace{egStates}}{\{\{2\},\{3,4\},\{1\},\{2,3,4\},\{1\},\{1,2,3,4\}\}}
}\vspace{-10pt}
}

\defnBox{legalComponentStatesQ}{\var{componentStates}\pattern}{returns \built{True} or \built{False} if the \var{componentStates} correspond to those of pure sYM or sGR theories.}

\defnBox{nonVanishingComponentAmpQ}{\var{componentStates}\patternTwo}{returns \built{True} if the set of states of \mbox{\{\var{componentStates}\}} are \emph{consistent} with supersymmetry (for arbitrary $\mathcal{N}$)---that is, if the distribution of $R$-charges of the states have the property that each index $I\!\in\![\mathcal{N}]$ appears exactly $k{+}2$ times with $k\!\in\![0,n{-}4]$. If this is not the case, the function returns \built{False}.\\[-10pt]

\textbf{\emph{Note}}: this is \emph{not equivalent} to an assessment that the tree-level \emph{amplitude} for \{\var{componentStates}\} is `non-vanishing'---merely if it vanishes by \emph{superselection}:
\mathematicaBox{
\mathematicaSequence[1]{egCmpt=\{\stateYM{phi}\![1\!,\!2]\!,\stateYM{phi}\![1\!,\!3]\!,\stateYM{phi}\![1\!,\!4]\!,\stateYM{phiBar}\![1\!,\!2]\!,\stateYM{phiBar}\![1\!,\!3]\!,\stateYM{phiBar}\![1\!,\!4]\};\\
\funL[1]{nonVanishingComponentAmpQ}\brace{egCmpt}}{True}\vspace{-5pt}
\mathematicaSequence{\funL[]{componentAmp}[egCmpt]}{0}
}
\vspace{-10pt}}

\defnBox{nmhvDegree}{\var{componentStates}\patternTwo}{gives the `N${}^k$MHV-\emph{degree}' $k$ of \var{componentStates}.}

\defnBox{minimalSusyDegree}{\var{componentStates}\pattern}{returns the number of supersymmetries $\mathcal{N}$ for which an amplitude involving \var{componentStates} may exist.\\[-10pt]

\textbf{\emph{Note}}: this degree is chosen irrespective of what $R$-charge assignments have been chosen. For example, although we would not consider $\{|\psi\rangle^{\text{+}1/2}_{(4)},|\bar{\psi}\rangle^{\text{-}1/2}_{(\bar{4})}\}$ to be external states for $\mathcal{N}\!=\!1$ sYM, it is clear that an amplitude involving only these states and gluons would be `valid' in sYM with $\mathcal{N}\!=\!1$; thus,  
\mathematicaBox{
\mathematicaSequence{\funL[1]{minimalSusyDegree}\brace{\stateYM{m},\stateYM{p},\stateYM{psiBar}[4],\stateYM{p},\stateYM{m},\stateYM{psi}[4]}}{1}
}\vspace{-10pt}
}

\defnBox{conjugateComponentStates}{\var{componentStates}\patternTwo}{returns the \emph{parity} conjugated states relative to \var{componentStates}.
\vspace{-40pt}}

\newpage
\vspace{-0pt}\subsectionAppendix{Helicities and Polarization Rules}{appendix:helicities_and_polarization_rules}\vspace{-10pt}

\defnBox{helicityPolarizationRules}{\var{gluonHelicities}\pattern}{returns a \built{Rule} for assigning specific \emph{helicity} choices to otherwise abstract polarization tensors \funL{$\epsilon$}\brace{\var{$a$}} to each particle according to the specified \var{gluonHelicities}. Moreover, these rules assume that a single reference frame $\rho\rangle[\rho$ has been chosen for all polarizations, so that $\epsilon_\var{a}^{\pm}\!\cdot\!\epsilon_{\var{b}}^{\pm}\!=\!0$ for all pairs $\var{a},\var{b}$.
\mathematicaBox{
\mathematicaSequence{\funL[1]{helicityPolarizationRules}\brace{\stateYM{m},\stateYM{p},\stateYM{m},\stateYM{p},\stateYM{m},\stateYM{p}}}{\{\funL{$\epsilon$}\![1]$\!\rightarrow$\funL{$\epsilon$m}[1]\!,\!\funL{$\epsilon$}[2]$\!\rightarrow$\funL{$\epsilon$p}[2]\!,\!\funL{$\epsilon$}[3]$\!\rightarrow$\funL{$\epsilon$m}[3]\!,\!\funL{$\epsilon$}[4]$\!\rightarrow$\funL{$\epsilon$p}[4]\!,\!\funL{$\epsilon$}[5]$\!\rightarrow$\funL{$\epsilon$m}[5]\!,\!\funL{$\epsilon$}\![6]$\!\rightarrow$\funL{$\epsilon$p}\![6]\!,\\ 
\phantom{\{}\funL{ss}[\funL{$\epsilon$p}[\var{$x$\pattern}],\funL{$\epsilon$p}[\var{$y$\pattern}]]:$\!\!\rightarrow$0,\funL{ss}[\funL{$\epsilon$m}[\var{$x$\pattern}],\funL{$\epsilon$m}[\var{$y$\pattern}]]:$\!\!\rightarrow$0\}}
}
Such a specification of polarization tensors' helicities must be used before any expression (e.g.~for CHY representations of non-supersymmetric Yang-Mills amplitudes) involving polarization tensors may be \funL{evaluate}d. 
}

\newpage
\vspace{-0pt}\sectionAppendix{Analytic Functions of External Kinematics}{appendix:analytic_functions_of_kinematics}\vspace{-0pt}
%
\vspace{-0pt}\subsectionAppendix{General Aspects of Spinor Expressions}{appendix:general_aspects_of_spinor_expressions}\vspace{-10pt}

\defnBox{littleGroup}{\var{$n$}\pattern\built{Integer}\optArg{},\var{expression}\pattern}{returns a \built{List} of the little-group scaling weights $\{w_1,\ldots,w_\var{n}\}$ with respect to the little group transformation:
\eq{t_a\!:\!\left(\lambda_a,\tilde\lambda_a\right)\mapsto \left(t_a\,\lambda_a,t_a^{-1}\tilde\lambda_a\right),\qquad t_a\!:\!(\text{\var{expression}})\mapsto t_a^{w_a}(\text{\var{expression}})}
under which the \var{expression} is should transform with uniform weight $w_a$. 
\mathematicaBox{\mathematicaSequence{egStates=\funL[]{randomComponentStates}[6,1]}{\{\stateYM{psi}[3],\stateYM{psi}[4],\stateYM{psiBar}[1],\stateYM{phi}[1,2],\stateYM{psiBar}[3],\stateYM{phi}[1,3]\}}
\mathematicaSequence{\funL[]{nice}[egStates]}{$\left\{|\psi\rangle^{\text{+}1/2}_{(3)},|\psi\rangle^{\text{+}1/2}_{(4)},|\bar{\psi}\rangle^{\text{-}1/2}_{(\bar{1})},|\phi\rangle^{0}_{(1\,2)},|\bar{\psi}\rangle^{\text{-}1/2}_{(\bar{3})},|\phi\rangle^{0}_{(\bar{1\,3})}\right\}$}
\mathematicaSequence{\funL[1]{littleGroup}[egStates]}{\{-1,-1,1,0,1,0\}}
\mathematicaSequence{\funL[1]{littleGroup}[\funL[]{componentAmp}[egStates]]}{\{-1,-1,1,0,1,0\}}
\mathematicaSequence[2]{ReplacePart[\{\builtL[]{Ls},\builtL[]{Lbs}\},\{\{1,5\}$\rightarrow$2\,\builtL[]{Ls}[[5]],\{2,5\}$\rightarrow$1/2\,\builtL[]{Lbs}[[5]]\}]\!;\\
\funL[]{evaluateWithSpinors}[\textnormal{\%}][\funL[]{componentAmp}[egStates]]\!;\\
\textnormal{\%}/\funL[]{componentAmpN}[egStates]}{2}
}
The first argument $\var{n}$ is optional precisely because it often does not matter or may be inferred. This argument need only be called to `pad' the output to uniform length $\var{n}$ (in case some leg labels do not make any appearance).\\[-10pt]

\textbf{\emph{Note}}: when applied to an expression containing \funL{superFunction} arguments, the little group weights of $\tilde\eta$'s are taken into account:
\mathematicaBox{\mathematicaSequence{\funL[1]{littleGroup}[\funL[]{ymAmp}[6,1]]}{\{\{-2,-2,-2,-2,-2,-2\},\{-2,-2,-2,-2,-2,-2\},\{-2,-2,-2,-2,-2,-2\}\}}
\mathematicaSequence{\funL[1]{littleGroup}[\funL[]{ymSuperTwistorAmp}[6,1]]}{\{\{0,0,0,0,0,0\},\{0,0,0,0,0,0\},\{0,0,0,0,0,0\}\}}}

\textbf{\emph{Note}}: polarization vectors with \emph{unspecified} helicity do not have any defined little group weights.
}

\defnBox{massDimension}{\var{expression}\pattern}{returns the mass/energy \emph{scaling dimension} $d$ of \var{expression}. This is the scaling of 
\eq{m\!:\!p_a\!\mapsto\!m\,p_a\;\;\forall\,a\;\;\text{under which}\;\; m\!:\!\!(\text{\var{expression}})\!\mapsto\! m^{d}\,(\text{\var{expression}})\,.}
Thus, for example,
\mathematicaBox{
\mathematicaSequence{\funL[1]{massDimension}/@\{\funL{ab}\brace{1,2}\,,\funL{s}\brace{1,2}\,,\funL{asb}\brace{1,\funL{p}\brace{2}\,,4}\,,\funL{ab}\brace{1,\funL{p}\brace{2}\,,\funL{p}\brace{3}\,,4}\,\}}{\{1,2,2,3\}}
}
This function is especially useful for simplifying sub-expressions, especially when combined with \fun{littleGroup}; this is because any \built{Sum} of terms must carry identical little group weights and \fun{massDimension}---greatly limiting the scope of potential algebraic expressions.\\[-10pt]

Any $n$-particle component amplitude of sYM will carry a mass dimension of $(4{-}n)$, while \emph{all} $n$-particle amplitudes of sGR carry mass dimension of $\text{+}2$. 
\mathematicaBox{\mathematicaSequence{\funL[1]{massDimension}[\funL[]{componentAmp}[\funL[]{randomComponentStatesYM}[\#,0]]]\& /@Range[4,8]}{\{0,-1,-2,-3,-4\}}
\mathematicaSequence{\funL[1]{massDimension}[\funL[]{componentAmp}[\funL[]{randomComponentStatesGR}[\#,0]]]\& /@Range[4,8]}{\{2,2,2,2,2\}}}
}

\defnBox{parityConjugationRule}{}{a \built{Rule} that replaces all spinor arguments with their conjugated form and applies \funL{parityConjugatePositroid}\brace{} to any \funL{$\sigma$} objects.\\[-10pt]

\textbf{\emph{Note}}: this should \emph{\textbf{not} be applied to} \funL{superFunction} \emph{objects}, but only to bosonic functions (such as \emph{component} amplitudes): the conventions for \funL{superFunction} involving $\tilde\eta$ coefficients are not conjugated by \built{parityConjugationRule} and the resulting expressions therefore would be non-sensical/problematic. 
}

\newpage
\vspace{0pt}\subsectionAppendix{\fun{superFunctions} of Spinor-Helicity Variables}{appendix:superFunctions_of_spinors}\vspace{-5pt}

\defnBox[]{positroidLabel}{\funL{superFunction}\brace{\var{$f$}\pattern,\var{$C$}\pattern}}{will provide the decorated permutation label for the relevant \funL{superFunction}. This function \emph{assumes} that \fun{superFunction} corresponds to one of sYM---that is, that the matrix $\var{C}\equivL(\var{c}_1,\ldots,\var{c}_n)$ is full rank. To be clear, this label is actually defined for the \emph{entire} matrix of $\tilde\eta$ coefficients, $\var{\hat{C}}\equivR\built{Join}\brace{\var{\lambda},\var{C}}\equivL(\var{\hat{c}}_1,\ldots,\var{\hat{c}}_n)$.\\[-10pt]

As described in \cite{ArkaniHamed:book,Bourjaily:2012gy} (see also \cite{Franco:2012mm,Franco:2012wv}), this permutation is defined in the following way: for each $a\!\in\![n]$, there is a unique $\sigma(a)\geq a$ such that $\var{\hat{c}}_a\!\in\!\mathrm{span}\{\var{\hat{c}}_{a+1},\ldots,\var{\hat{c}}_{\sigma(a)}\}$; when $\var{\hat{c}}_a\!=\!\vec{0}\in\!\mathbb{C}^{k{+}2}$, $\var{\hat{c}}_a\!\in\!\mathrm{span}\{\}$, which should be understood as the case where $\sigma(a)\!=\!a$.\\[-12pt]
\mathematicaBox{\mathematicaSequence{\funL[1]{positroidLabel}/@\funL[]{ymAmp}[6,1]}{$\{\funL{$\sigma$}\brace{4,5,6,8,7,9}\,,\funL{$\sigma$}\brace{3,5,6,7,8,10}\,,\funL{$\sigma$}\brace{4,6,5,7,8,9}\,\}$}}\vspace{-10pt}
}\vspace{-0pt}

\defnBox{standardizeSuperFunctions}{\var{expression}\pattern}{because \funL{superFunction} objects relevant to N${}^k$MHV amplitudes are equivalent under 
\eq{\funL{superFunction}\brace{\var{f},\var{C}}\simeq\funL{superFunction}\brace{\var{f}/(\det(\mathbf{M})^4),\mathbf{M}.\var{C}{+}\mathbf{Q}\!.\funL{$\lambda$}}}
for any (non-singular) $(k\!\times\!k)$ matrix $\mathbf{M}$ and any $(k\!\times\!2)$ matrix $\mathbf{Q}$, there are many non-identical ways of expressing the same \fun{superFunction}. This equivalence class spans a $GL(k\text{+}2)$ redundancy in their representation.\\[-10pt]

A \emph{canonical} representative of this $GL(k\text{+}2)$ equivalence class (for each \fun{superFunction}) is chosen by \fun{standardizeSuperFunctions}. Thus, two \fun{superFunction} objects are \emph{equivalent} iff they are \emph{identical} after the application of \fun{standardizeSuperFunctions}. 
\mathematicaBox{
\mathematicaSequence{egFcn=RandomChoice\brace{\funL[]{ymTwistorAmp}\brace{7,2}\,}}{\funL{R}\brace{1,2,3,6,7}\,\funL{R}\brace{\funL{cap}\brace{\{2,3\},\{7,6,1\}}\,,3,4,5,6}}
\mathematicaSequence[2]{image1=\funL[]{fromTwistorsToSpinors}\brace{\funL[]{toSuperTwistorFunctions}\brace{7}\,\brace{egFcn}\,}\,;\\
image2=\funL[]{toSuperSpinorFunctions}\brace{7}\,\brace{egFcn}\,;\\
\funL[]{nice}/@\funL[]{evaluate}\brace{\{image1,image2\}\,}}{\scalebox{0.75}{$\left\{\rule{0pt}{20pt}\right.\!\!\displaystyle\text{-}\frac{1}{286826570032774276519460740205940999979008000},\left(\begin{array}{@{}ccccccc@{}}0&3408&\text{-}792&\text{-}540&\text{-}204&216&0\\0&0&2258280&\text{-}2252772&633420&0&0\end{array}\right)\!\!\cdot\!\tilde\eta$,}\\
\scalebox{0.75}{$\phantom{\Big\{}\displaystyle\text{-}\frac{945}{352166721556936},\left(\begin{array}{@{}ccccccc@{}}\text{-}\tfrac{29}{7}&\text{-}\tfrac{52}{3}&0&0&0&0&\tfrac{20}{7}\\0&0&\text{-}\tfrac{41}{12}&\tfrac{409}{120}&\text{-}\tfrac{23}{24}&0&0\end{array}\right)\!\!\cdot\!\tilde\eta\left.\rule{0pt}{20pt}\right\}$}
}
\mathematicaSequence{\funL[]{nice}/@\funL[1]{standardizeSuperFunctions}\brace{\{image1,image2\}\,}}{$\displaystyle\left\{\rule{0pt}{22pt}\right.\!\!\!\text{-}\frac{123337920155}{10083388552060824},\left(\begin{array}{@{}ccccccc@{}}0&1&0&\text{-}\frac{11361}{29110}&\frac{31}{5822}&\frac{9}{142}&0\\
0&0&1&\text{-}\frac{409}{410}&\frac{23}{82}&0&0\end{array}\right)\!\!\cdot\tilde\eta$\,,\\
$\displaystyle~\hspace{7.4pt}\text{-}\frac{123337920155}{10083388552060824},\left(\begin{array}{@{}ccccccc@{}}0&1&0&\text{-}\frac{11361}{29110}&\frac{31}{5822}&\frac{9}{142}&0\\
0&0&1&\text{-}\frac{409}{410}&\frac{23}{82}&0&0\end{array}\right)\!\!\cdot\tilde\eta\left.\rule{0pt}{22pt}\right\}$
}
}\vspace{-40pt}
}

\newpage
\defnBox{superFunctionProductRule}{}{a \built{Rule} which implements the facts that 
\eq{\fwbox{0pt}{\hspace{-10pt}\text{\funL{superFunction}\brace{\var{$f$},\var{$C$}}\,\funL{superFunction}\brace{\var{$g$},\var{$D$}}\,=\,\funL{superFunction}\brace{\var{$f\,g$},\built{Join}\brace{\var{$C$},\var{$D$}}}}\,,}\nonumber}
and $\text{(\var{$g$}\,\,\funL{superFunction}\brace{\var{$f$},\var{$C$}}\,)=\,\funL{superFunction}\brace{\var{$f\,g$},\var{$C$}}\,,}$ for all functions \var{$g$}.
}

\vspace{10pt}\subsectionAppendix{Component Functions of \fun{superFunction} Objects}{appendix:components_of_superFunctions}\vspace{-10pt}

\defnBoxTwo{component}{\var{componentStates}\patternTwo}{\var{expression}\pattern}{returns the particular \emph{component} function of an \var{expression} involving \funL{superFunction} objects via 
\eq{\begin{split}\hspace{00pt}&\hspace{-80pt}\fun{component}\brace{\var{\text{\textsl{componentStates}}}}\,\brace{\funL{superFunction}\brace{f,C}\,}\\\hspace{00pt}&\hspace{-80pt}\fwboxR{0pt}{\equivR}\fun{component}\brace{\var{\text{\textsl{componentStates}}}}\,\brace{f(\lambda,\tilde\lambda)\,\,\delta^{k\!\times\!\mathcal{N}}\!\big(C\!\cdot\!\tilde\eta\big)\delta^{2\!\times\!\mathcal{N}}\!\big(\lambda\!\cdot\!\tilde\eta\big)}\\[-5pt]&\equivR\int\!\!\!\prod_{I=1}^{\mathcal{N}}\!\!\!d\tilde{\eta}_{I}^{a^I_1}\cdots d\tilde\eta_I^{a^I_{k+2}}\left(f(\lambda,\tilde\lambda)\,\,\delta^{k\!\times\!\mathcal{N}}\!\big(C\!\cdot\!\tilde\eta\big)\delta^{2\!\times\!\mathcal{N}}\!\big(\lambda\!\cdot\!\tilde\eta\big)\right)\,\,\,\\[-10pt]
&\,=f(\lambda,\tilde\lambda)\prod_{I=1}^{\mathcal{N}}\det\{\hat{c}_{a^I_1},\ldots,\hat{c}_{a^I_{k{+}2}}\}\quad\text{with}\quad\hat{C}\equivR(\hat{c}_1,\ldots,\hat{c}_n)\equivR\left(\rule{0pt}{12pt}\right.\begin{array}{@{}c@{}}\funL{$\lambda$}\\[-5pt]C\end{array}\left.\rule{0pt}{12pt}\right)\,.\hspace{-80pt}\\[-10pt]\end{split}\label{component_functions_defined}}
where the particular determinants involve columns labelled by the $\mathcal{N}\!\in\!\{4,8\}$ $(k{+}2)$-tuples $\{\{a^1_1,\ldots,a^1_{k+2}\},\ldots,\{a^{\mathcal{N}}_1,\ldots,a^{\mathcal{N}}_{k+2}\}\}$ as determined from the list of \mbox{\{\var{componentStates}\}}\,via the function \mbox{\fun{stateListToMinorList}\brace{\var{componentStates}}\,.}\\[-12pt]
}

\defnBox{stateListToMinorList}{\var{componentStates}\patternTwo}{returns the list of `minors'---given as a \built{List} of \built{List}s of $(k\text{+}2)$-tuples of $\binom{[n]}{k\text{+}2}$ of the form $\{\{a^1_1,\ldots,a^1_{k+2}\},\ldots,\{a^{\mathcal{N}}_1,\ldots,a^{\mathcal{N}}_{k{+}2}\}\}$ (where $\mathcal{N}\!=\!4,8$ for sYM or sGR, respectively)---whose product will project out the component function corresponding to \mbox{\{\var{componentStates}\}} from any \funL{superFunction} \`a la (\ref{component_functions_defined}).
}

\defnBoxTwo{minorListToStateList}{\var{$n$}\pattern\optArg{}}{\var{minorList}\patternTwo}{essentially the inverse function of \fun{stateListToMinorList}, but where the multiplicity \var{$n$} must be specified or inferred (by the maximal index appearing in the array \var{minorList}) as the same list of minors used to project out \{\var{stateList}\} will also be used to project-out the component $\{\var{stateList},\funL{p},\funL{p},\funL{p},\ldots\}$, for example. 
\mathematicaBox{
\mathematicaSequence{egMinors=\funL[]{randomSubset}\brace{6,3}\,\&/@Range[4]}{\{\{2,4,5\},\{1,2,5\},\{2,3,5\},\{3,5,6\}\}}
\mathematicaSequence{\funL[]{nice}\brace{\funL[1]{minorListToStateList}\brace{6}\,\brace{egMinors}}}{$\{|\psi\rangle^{\text{+}1/2}_{(2)},|\bar{\psi}\rangle^{\text{-}1/2}_{(\bar{4})},|\bar{\phi}\rangle^0_{(\bar{1\,2})},|\psi\rangle^{\text{+}1/2}_{(1)},|g\rangle^{\text{-}1},|\psi\rangle^{\text{+}1/2}_{(4)}\}$}
\mathematicaSequence{egMinors=\funL[]{randomSubset}\brace{6,3}\,\&/@Range[8]}{\{\{4,5,6\},\{2,5,6\},\{1,2,5\},\{3,5,6\},\{3,4,6\},\{2,3,5\},\{2,4,6\},\{1,5,6\}\!\}}
\mathematicaSequence{\funL[]{nice}\brace{\funL[1]{minorListToStateList}\brace{6}\,\brace{egMinors}}}{$\{|\gamma\rangle^{\text{+}1}_{(3\,8)},|\bar{\phi}\rangle^{0}_{(\bar{1\,4\,5\,8})},|{\psi}\rangle^{\text{+}1/2}_{(4\,5\,6)},|\psi\rangle^{\text{+}1/2}_{(1\,5\,7)},|\gamma\rangle^{\text{-}1}_{(\bar{5\,7})},|\gamma\rangle^{\text{-}1}_{(\bar{3\,6})}\}$}
}
\vspace{-20pt}}

\defnBoxTwo{dressedMinor}{\{\var{columnNumbers}\patternTwo\}}{\var{$C$}\pattern}{for \mbox{$\{$\var{columnNumbers}$\}\equivL\{\var{a_1},\ldots,\var{a_{k+2}}\}$} a $(k\text{+}2)$-tuple of \built{Integer}s and $\var{C}\equivL(\var{c_1},\ldots,\var{c_n})$ a $(k\!\times\!n)$-matrix, \fun{dressedMinor} returns the  determinant/minor 
\eq{\det\{\var{\hat{c}}_{a_1},\ldots,\var{\hat{c}}_{a_{k{+}2}}\}\qquad\text{where}\qquad\var{\hat{C}}\equivR(\var{\hat{c}}_1,\ldots,\var{\hat{c}}_n)\equivR\left(\rule{0pt}{12pt}\right.\begin{array}{@{}c@{}}\funL{$\lambda$}\\[-5pt]\var{C}\end{array}\left.\rule{0pt}{12pt}\right)}
This is done through the standard decomposition of $(k\text{+}2)\!\times\!(k\text{+}2)$ determinants into a sum of products of $(2\!\times\!2)$ and $(k\!\times\!k)$ determinants---thus ensuring the result to be Lorentz invariant (provided the matrix \var{$C$} is). 
}

\defnBox{minimalNonVanishingComponent}{\fun{superFunction}\brace{\var{$f$}\pattern,\var{$C$}\pattern}}{returns the \emph{lexicographically first} choice of ({-})-helicity gluon/graviton positions for which the \fun{superFunction} is non-vanishing.\\[-10pt]

In terms of the matrix $\var{\hat{C}}\equivR$\built{Join}\brace{\built{Transpose}\brace{\fun{$\lambda$}}\,,\var{$C$}}$\equivL(\var{\hat{c}_1},\ldots,\var{\hat{c}_n})$---a $(k{+}2)\!\times\!n$ matrix for $n$-particle N${}^k$MHV---this corresponds to the lexicographically-minimal list \mbox{$\{m_1,\ldots,m_{k+2}\}\!\in\!\binom{[n]}{k\text{+}2}$} such that $\det\{\var{\hat{c}_{m_1}},\ldots,\var{\hat{c}_{m_{k+2}}}\}\neq0$.\\[-10pt]
}

\newpage
\vspace{-0pt}\subsectionAppendix{Manipulations of Spinor Expressions}{appendix:manipulations_of_spinor_expressions}\vspace{-10pt}

\defnBoxTwo{canonicalizeSpinors}{\var{$n$}\pattern}{\var{expression}\pattern}{applies `trivial' identities to `canonicalize' most spinor functions, resulting in a more canonical form of the \var{expression}. For example,
\mathematicaBox{
\mathematicaSequence{\funL[1]{canonicalizeSpinors}\brace{6}\,\brace{\{\funL{s}[1,\!3,\!2],\funL{ab}[5,\!4],\funL{s}[1,\!6],\funL{asb}[5,\!\funL{p}[1,2,3],\!4]\}}}{\{\funL{s}[1,2,3],-\funL{ab}[4,5],\funL{ab}[6,1]\funL{sb}[6,1],\funL{ab}[5,6]\funL{sb}[4,6]\}}
}
Notice that there is a preference for \emph{cyclic} ordering (as in \funL{ab}\brace{1,\var{$n$}}$\,\mapsto\text{-}$\funL{ab}\brace{\var{$n$},1}) and for writing any spinor expressions which factorize as being factorized. Moreover, the multiplicity \var{$n$} plays an important role in applying identities induced by momentum conservation as in the second step of
\eq{\begin{split}\fwbox{0pt}{\funL{asb}\brace{5,\funL{p}\brace{1,2,3}\,,4}\mapsto{-}\funL{asb}\brace{5,\funL{p}\brace{5,6}\,,4}\mapsto{-}\funL{asb}\brace{5,\funL{p}\brace{6}\,,4}}\\
\fwbox{0pt}{\mapsto{-}\funL{ab}\brace{5,6}\,\,\funL{sb}\brace{6,4}\,\mapsto\funL{ab}\brace{5,6}\,\,\funL{sb}\brace{4,6}\,.}\end{split}}
\textbf{\emph{Note}}: at sufficiently large `depth' (such as \funL{ab}\brace{$1$,\funL{p}\brace{$\cdots$}\,,\ldots,\funL{p}\brace{$\cdots$}\,,2} with many $\funL{p}$'s appearing), it is possible to encounter identical monomials which will not be fully `canonicalized'. For example, both 
\eq{\ab{10|(1\,2)|(3\,4)|(5\,6)|(7\,8\,9)|10}\qquad\text{and}\qquad\ab{10|(1\,2)(3\,4\,5\,6)|(5\,6)|(1\,2\,3\,4)|10}}
are left invariant by \fun{canonicalizeSpinors}\brace{10} and yet these are (non-trivially) related by a sign; but their equivalence is easy to verify numerically:
\mathematicaBox{
\mathematicaSequence[1]{\funL[]{useRandomKinematics}\brace{10};\\
\funL[]{evaluate}\big[\funL{ab}\brace{10,\funL{p}\brace{1,2}\,,\funL{p}\brace{3,4}\,,\funL{p}\brace{5,6}\,,\funL{p}\brace{7,8,9}\,,10}\\\mbox{~\hspace{50pt}}+\funL{ab}\brace{10,\funL{p}\brace{1,2}\,,\funL{p}\brace{3,4,5,6}\,,\funL{p}\brace{5,6}\,,\funL{p}\brace{1,2,3,4}\,,10}\,\big]}{0}
}
}

\defnBoxTwo{permuteLegs}{\var{newLegLabels}\patternTwo}{\var{expression}\pattern}{returns a version of \var{expression} for which all the `leg-labels' appearing in kinematic functions such as \funL{ab}, \funL{sb}, \funL{asb}, etc.\ have been transformed according to the \built{Rule}:
\eq{\text{\built{Thread}\brace{\built{Rule}\brace{\built{Sort}\brace{\{\var{newLegLabels}\}}\,,\{\var{newLegLabels}\}}}}\,.}
Notice that this does \emph{not} require that \var{newLegLabels} span the entire range of leg labels appearing in \var{expression}. 
\mathematicaBox{
\mathematicaSequence[1]{seedTerm=(\funL[]{parkeTaylorAmp}[6])${}^2$\funL[]{grAmpDenominators}\brace{6,0}[\![1]\!];\\
\funL[]{nice}\brace{seedTerm}}{$\rule{0pt}{24pt}\displaystyle\frac{\ab{1|(2)|(3)|4}\ab{1|(2\,3)|(4)|5}\,s_{5\,6}}{\ab{1\,2}^2\ab{1\,4}\ab{1\,5}\ab{2\,3}^2\ab{3\,4}^2\ab{4\,5}^2\ab{5\,6}^2\ab{6\,1}^2}$}
\mathematicaSequence[1]{permSum=
Total\brace{\funL[1]{permuteLegs}\brace{\#}\,\brace{seedTerm}\,\&/\!@Permutations\brace{Range\brace{2,5}}};\\
\funL[]{evaluate}\brace{permSum}}{$\rule{0pt}{20pt}\displaystyle\frac{104334383}{34720812000000}$}
\mathematicaSequence{\funL[]{evaluate}\brace{\funL[]{grAmpHodges}\brace{6}}}{$\rule{0pt}{20pt}\displaystyle\frac{104334383}{34720812000000}$}
}

\textbf{\emph{Note}}: for \funL{superFunction} objects, the functional arguments of the coefficients of the $\tilde\eta$'s are permuted, but \emph{the ordering of the $\tilde\eta$'s is conventionally \textbf{always} taken to be $\{1,\ldots,n\}$}---as these are reflected also in the labeling of external momenta, and the listing of component functions. For example, 
\mathematicaBox{
\mathematicaSequence[2]{seedFcn=\funL[]{parkeTaylorAmpBar}\brace{5};\\
permutedFcn=\funL[1]{permuteLegs}\brace{\{5,3\}}\,\brace{seedFcn};\\
\funL[]{nice}/@(\{seedFcn,permutedFcn\}/\!\!.\{5->Style[5,Blue]\!,3->Style[3,Red]\})}{
$\rule{0pt}{23pt}\left\{\rule{0pt}{20pt}\right.\hspace{-5pt}\displaystyle\frac{1}{\sb{1\,2}\sb{2\,\b{3}}\sb{\b{3}\,4}\sb{4\,\r{5}}\sb{\r{5}\,1}},\left(\fwbox{20pt}{0}\fwbox{20pt}{0}\fwbox{40pt}{\frac{\sb{4\,\b{5}}}{\ab{1\,2}}}\fwbox{40pt}{\frac{\sb{\b{5}\,\r{3}}}{\ab{1\,2}}}\fwbox{40pt}{\frac{\sb{\r{3}\,4}}{\ab{1\,2}}}\right)\!\cdot\!\tilde\eta$\,,\\
$\rule{0pt}{23pt}\hspace{7pt}\displaystyle\frac{1}{\sb{1\,2}\sb{2\,\r{5}}\sb{\r{5}\,4}\sb{4\,\b{3}}\sb{\b{3}\,1}},\left(\fwbox{20pt}{0}\fwbox{20pt}{0}\fwbox{40pt}{\frac{\sb{\b{5}\,4}}{\ab{1\,2}}}\fwbox{40pt}{\frac{\sb{\r{3}\,\b{5}}}{\ab{1\,2}}}\fwbox{40pt}{\frac{\sb{4\,\r{3}}}{\ab{1\,2}}}\right)\!\cdot\!\tilde\eta\left.\rule{0pt}{20pt}\right\}$}
}
In the above, $\big(c_{\r{3}}\tilde\eta_{\r{3}}\text{+}c_{\b{5}}\tilde\eta_\b{5}\big)\!=\!\big(\frac{\sb{4\,\b{5}}}{\ab{1\,2}}\tilde\eta_{\r{3}}\text{+}\frac{\sb{\r{3}\,4}}{\ab{1\,2}}\tilde\eta_{\b{5}}\big)\!\mapsto\!\big(\frac{\sb{4\,\r{3}}}{\ab{1\,2}}\tilde\eta_{\b{5}}\text{+}\frac{\sb{\b{5}\,4}}{\ab{1\,2}}\tilde\eta_{\r{3}}\big)\!=\!\big(\frac{\sb{\b{5}\,4}}{\ab{1\,2}}\tilde\eta_{\r{3}}\text{+}\frac{\sb{4\,\r{3}}}{\ab{1\,2}}\tilde\eta_{\b{5}}\big)$, and so the coefficient of $\tilde\eta_\r{3}$ appears to change minimally (only by a sign). This reflects the convention that supermomentum conservation, say, 
\eq{\delta^{2\!\times\!\mathcal{N}}\!\big(\lambda\!\cdot\!\tilde\eta\big)\equivR\delta^{2\!\times\!\mathcal{N}}\!\Big(\sum_{a=1}^n\lambda_a\tilde\eta_a^I\Big)}
should itself be viewed as \emph{permutation \textbf{invariant}}.
}

\defnBoxTwo{bcfwShift}{\{\var{$a$}\pattern,\var{$b$}\pattern\}}{\var{expression}\pattern}{modifies \var{expression} to actively transform it according to the `BCFW shift' 
\eq{\big(\tilde\lambda_\var{a},\lambda_{\var{b}},\tilde\eta_\var{a}\big)\mapsto\big(\tilde\lambda_\var{a}{-}\r{\alpha}\,\tilde{\lambda}_{\var{b}},\lambda_\var{b}{+}\r{\alpha}\,\lambda_\var{a},\tilde\eta_\var{a}{-}\r{\alpha}\,\tilde\eta_\var{b}\big).\label{general_bcfw_shift_map}}
The introduced `shift' parameter is denoted `$\r{\alpha}$'. For example,
\mathematicaBox{\mathematicaSequence{\funL[1]{bcfwShift}[\{1,6\}][\{\funL{ab}[1,2],\funL{sb}[1,2],\funL{ab}[5,6],\funL{sb}[5,6],\funL{s}[1,6]\}]}{\{\funL{ab}[1,2],\funL{sb}[1,2]-$\r{\alpha}$\,\funL{sb}[6,2],$\r{\alpha}$\,\funL{ab}[5\,1]+\funL{ab}[5,6],\funL{sb}[5,6],\funL{s}[2,3,4,5]\}}}
Notice that the shift applied to $s_{1,6}\mapsto s_{2,3,4,5}$, which is simply a relabeling. This is due to the fact that the first step performed by \fun{bcfwShift} is to call \fun{expandSpinorBrackets}\brace{\{\var{$a$},\var{$b$}\},\var{expression}}\,, which `exposes' the dependence of any expression on the leg labeled $\{\var{a},\var{b}\}$ (and manifestly deemphasizes any appearance of these labels which can be removed by momentum conservation).\\[-10pt]

The shifting defined by (\ref{general_bcfw_shift_map}) on $\tilde\eta_\var{a}$'s induces a change to \funL{superFunction} object in the obvious way:
\mathematicaBox{\mathematicaSequence{\funL[]{nice}[\funL[1]{bcfwShift}[\{2,6\}][\funL[]{ymAmp}[6,1][[1]]]]}{
$\displaystyle1\bigg/\Big((\alpha\,\ab{5\,2}\text{+}\ab{5\,6})(\alpha\,\ab{2\,1}\text{+}\ab{6\,1})(\langle5|(3\,4\,5)|2]\text{-}\r\alpha\,\langle5|(3\,4\,5)|6])(s_{5\,6\,1}\text{-}\langle2|(5\,1)|6])\sb{3\,4}$ $\displaystyle(\sb{2\,3}\text{-}\r\alpha\,\sb{6\,3})(\langle1|(3)|4]\text{+}\ab{1\,2}(\sb{2\,4}\text{-}\r{\alpha}\,\sb{6\,4}))\!\!\!\Big),$\\
$\displaystyle\big(\fwbox{25pt}{0}\fwbox{25pt}{\sb{3\,4}}\fwbox{70pt}{(\sb{4\,3}\text{+}\r\alpha\sb{6\,4})}\fwbox{75pt}{(\sb{2\,3}\text{+}\r\alpha\sb{3\,6})}\,\,\fwbox{35pt}{0}\fwbox{45pt}{\r\alpha\sb{4\,3}}\big)\!\cdot\!\tilde\eta\!\!$}}
\textbf{\emph{Note}}: the shift does \emph{not} introduce the factor of $1/\r{\alpha}$ into the measure (which would arise in the case of actual BCFW recursion).
}

\defnBox{expandSpinorBrackets}{\var{expression}\pattern}{returns a form of \var{expression} expressed in terms of two-argument \funL{ab}\brace{\var{$i$},\var{$j$}} and \funL{sb}\brace{\var{$i$},\var{$j$}} invariants. This is achieved by applying the definition of \funL{asb} etc.\ to express these in terms of sums of products of two-particle invariants. 
\mathematicaBox{
\mathematicaSequence[1]{egBrackets=\{\funL{asb}\brace{1,\funL{p}\brace{2,3}\,,4}\,,\funL{ab}\brace{1,\funL{p}\brace{2,3}\,,\funL{p}\brace{4,5}\,,6}\,\};\\[-2pt]
\funL[1]{expandSpinorBrackets}\brace{egBrackets}}{\{\funL{ab}\brace{1,2}\,\,\funL{sb}\brace{2,4}\,\,+\funL{ab}\brace{1,3}\,\,\funL{sb}\brace{3,4}\,,\\
\phantom{+}\funL{ab}\brace{1,2}\,\,\funL{sb}\brace{2,4}\,\,\funL{ab}\brace{4,6}\,\,+\funL{ab}\brace{1,2}\,\,\funL{sb}\brace{2,5}\,\,\funL{ab}\brace{5,6}\,\,\\
+\funL{ab}\brace{1,3}\,\,\funL{sb}\brace{3,4}\,\,\funL{ab}\brace{4,6}\,\,+\funL{ab}\brace{1,3}\,\,\funL{sb}\brace{3,5}\,\,\funL{ab}\brace{5,6}\,\}}
}
}

\defnBox[]{expandSpinorBrackets}{\{\var{$a$}\pattern,\var{$b$}\pattern\},\var{expression}\pattern}{essentially applies the function \funL{expandSpinorBrackets} as above, but \emph{only} to those sub-expressions which depend on legs labelled by $\{\var{a},\var{b}\}$ through \mbox{\funL{p}\brace{$\cdots,\var{a},\cdots$}}\, or  \mbox{\funL{p}\brace{$\cdots,\var{b},\cdots$}}\,; thus, it `exposes' the spinorial dependence on $(\lambda_\var{a},\tilde\lambda_\var{a})$ and $(\lambda_\var{b},\tilde\lambda_\var{b})$ while leaving all other expressions compact and un-expanded.
\mathematicaBox{
\mathematicaSequence[1]{egBrackets=\{\funL{ab}\brace{1,\funL{p}\brace{2,3}\,,\funL{p}\brace{4,5}\,,6}\,,\funL{ab}\brace{4,\funL{p}\brace{5,6}\,,\funL{p}\brace{1,2}\,,3}\,\}\};\\[-2pt]
\funL[1]{expandSpinorBrackets}\brace{\{1,6\},egBrackets}}{\{\funL{ab}\brace{1,\funL{p}\brace{2,3}\,,\funL{p}\brace{4,5}\,,6}\,,\\
\phantom{+}\funL{ab}\brace{4,\funL{p}\brace{5}\,,\funL{p}\brace{2}\,,3}\,+\funL{ab}\brace{4,5}\,\funL{asb}\brace{3,\funL{p}\brace{2}\,,6}\,\\
+\funL{ab}\brace{1,3}\,\funL{asb}\brace{4,\funL{p}\brace{5}\,,1}\,+\funL{ab}\brace{3,1}\,\funL{sb}\brace{1,6}\,\funL{ab}\brace{4,6}\,\}}
}
This is especially useful in the context of shifting some subset of spinors (as in BCFW, say)---where many compact subexpressions need not be expanded at all to manifest the dependence on legs $\{\var{a},\var{b}\}$. 
}

\vspace{10pt}\subsectionAppendix{Analytic Expressions of Momentum-Twistors}{appendix:analytic_twistor_expressions}\vspace{-0pt}

\defnBox[]{R}{\var{twistorLabels}\patternTwo}{a \emph{protected} and \emph{abstract} symbol representing the momentum-twistor `five bracket' or `	$R$-invariant'
\eq{\fwbox{0pt}{\hspace{-40pt}\fun{R}\brace{\var{a},\var{b},\var{c},\var{d},\var{e}}\;\Leftrightarrow\;\frac{\delta^{1\!\times\!4}\!\big(\ab{\var{b\,c\,d\,e}}\eta_{\var{a}}\text{+}\ab{\var{c\,d\,e\,a}}\eta_{\var{b}}\text{+}\ab{\var{d\,e\,a\,b}}\eta_{\var{c}}\text{+}\ab{\var{e\,a\,b\,c}}\eta_{\var{d}}\text{+}\ab{\var{a\,b\,c\,d}}\eta_{\var{e}}\big)}{\ab{\var{a\,b\,c\,d}}\ab{\var{b\,c\,d\,e}}\ab{\var{c\,d\,e\,a}}\ab{\var{d\,e\,a\,b}}\ab{\var{e\,a\,b\,c}}}}\label{r_invariant_defined}}
where $\eta_{\var{a}}$ is the momentum-twistor Grassmann parameter associated to \emph{momentum-twistor coherent} states. These are related to the spinor-space $\tilde\eta$'s in the same way that $\tilde\lambda$ is related to the $\mu$'s of momentum-twistors. See e.g.\ \cite{Hodges:2009hk,Mason:2009qx,ArkaniHamed:2009vw,ArkaniHamed:book} for more details.\\[-10pt]

\textbf{\emph{Note}}: The arguments of \fun{R} can involve arbitrarily-nested instances of `\funL{cap}'. 
}

\defnBoxTwo{toSuperTwistorFunctions}{\var{$n$}\pattern\optArg{}}{\var{expression}\pattern}{converts any (products) of \fun{R}'s to \funL{superFunction} objects according to (\ref{r_invariant_defined}).
\mathematicaBox{
\mathematicaSequence{egExprn=RandomChoice[\funL[]{ymTwistorAmp}[8,2]]}{\funL{R}[1,2,3,5,\funL{cap}[\{5,6\},\{8,7,1\}]]\funL{R}[1,5,6,7,8]}
\mathematicaSequence{\funL[]{nice}\brace{\funL[1]{toSuperTwistorFunctions}\brace{8}\,\brace{\textnormal{\%}}}}{\scalebox{0.7}{$\displaystyle\frac{1}{\ab{1235}\ab{1567}\ab{235(56)\tcap(871)}\ab{1\,5\,6\,7}\ab{35(56)\tcap(871)1}\ab{5678}\ab{5(56)\tcap(871)12}\ab{6781}\ab{7812}\ab{8156}\ab{(56)\tcap(871)123}},$}\\
\scalebox{0.7}{$\displaystyle\left(\begin{array}{@{}cccccccc@{}}\ab{5678}&0&0&0&\ab{6781}&\ab{7812}&\ab{8156}&\ab{1567}\\
\text{-}\ab{1578}\ab{2356}\phantom{\text{-}}&\ab{1356}\ab{1578}&\text{-}\ab{1256}\ab{1578}\phantom{\text{-}}&0&\ab{1236}\ab{1578}&\text{-}\ab{1235}\ab{1578}\phantom{\text{-}}&0&0\end{array}\right)\!\cdot\!\eta$}}
}
}

\defnBox{simplifyTwistorBrackets}{\var{expressions}\pattern}{returns an expression where many (but not all) geometric identities of \funL{cap}-objects have been used to simplify the expression. For example, one of the \fun{R} invariants appearing in the 6-point N${}^2$MHV amplitude is \fun{R}\brace{\funL{cap}\brace{\{2,3\},\{6,5,1\}}\,,3,4,5,\funL{cap}\brace{\{6,5\},\{3,2,1\}}\,} which gives a bosonic part
\mathematicaBox{
\mathematicaSequence[1]{egFcn=First[\funL[1]{toSuperTwistorFunctions}\brace{6}\,[\\ \funL{R}\brace{\funL{cap}\brace{\{2,3\},\{6,5,1\}}\,,3,4,5,\funL{cap}\brace{\{6,5\},\{3,2,1\}}\,}];\\
\funL[]{nice}\brace{egFcn}}{\scalebox{0.69}{$\displaystyle\rule{0pt}{22pt}\text{-}\frac{1}{\ab{345(23)\tcap(651)}\ab{45(23)\tcap(651)(65)\tcap(321)}\ab{5(23)\tcap(651)(65)\tcap(321)3}\ab{(23)\tcap(651)(65)\tcap(321)34}\ab{(65)\tcap(321)345}}$}}
\mathematicaSequence[1]{\funL[1]{simplifyTwistorBrackets}\brace{egFcn};\\
\funL[]{nice}\brace{\textnormal{\%}}}{$\displaystyle\frac{1}{\ab{1234}\ab{1235}^3\ab{1356}^3\ab{1456}\ab{2345}\ab{2356}^3\ab{3456}}$}
}
}

\defnBox{expandTwistorBrackets}{\var{expressions}\pattern}{returns an expression free of any geometrically-defined arguments involving \funL{cap} by recursively applying the definitions (\ref{defn_of_cap}) and (\ref{defn_of_cap2}). 
\mathematicaBox{
\mathematicaSequence[1]{egExprn=\funL{ab}\brace{\funL{cap}\brace{\{1,2\},\{3,4,\funL{cap}\brace{\{5,6\},\{7,8,9\}}\,\}},10,11,12}\,;\\
\funL[]{nice}\brace{egExprn}}{$\rule{0pt}{14pt}\ab{(12)\tcap(34(56)\tcap(789))\,10\,11\,12}$}
\mathematicaSequence{\funL[1]{expandTwistorBrackets}\brace{egExprn}}{\fourBr[1,10,11,12](\fourBr[2,3,4,5]\fun{ab}[6,7,8,9]+\fun{ab}[2,3,4,6]\fun{ab}[7,8,9,5])+ \fun{ab}[2,10,11,12](\fun{ab}[3,4,5,1]\fun{ab}[6,7,8,9]+\fun{ab}[3,4,6,1]\fun{ab}[7,8,9,5])}
}
}

\vspace{10pt}\subsectionAppendix{Translations from Twistors to Spinor Variables}{appendix:from_twistors_to_spinors}\vspace{-10pt}

\defnBox{fromTwistorsToSpinors}{\var{twistorExpression}\pattern}{converts all momentum-twistor four-brackets into ordinary spinor brackets. If there are \funL{superFunction} arguments appearing in \var{twistorExpression}, then \fun{fromTwistorsToSpinors} assumes that the expression involved momentum twistor $\eta$'s whose coefficients should be translated to those of the $\tilde\eta$'s in spinor space. 
\mathematicaBox{
\mathematicaSequence{\funL[1]{fromTwistorsToSpinors}\brace{\fourBr[1,2,3,\funL{cap}[\{5,6\},\{7,8,1\}]]}}{\funL{ab}[1,2]\funL{ab}[2,3]\funL{ab}[5,6]\funL{ab}[7,8]\funL{ab}[8,1]\funL{sb}[2,\funL{p}[6,7,8,1],\funL{p}[6,7],8]}
\mathematicaSequence{\funL[]{nice}\brace{\textnormal{\%}}}{$\ab{1\,2}\ab{2\,3}\ab{5\,6}\ab{7\,8}\ab{8\,1}\sb{2|(6\,7\,8\,1)|(6\,7)|8}$}
}
}

\defnBoxTwo{toSuperSpinorFunctions}{\var{$n$}\pattern\optArg{},\var{gaugeFixQ}\pattern\optArg{\built{True}}\,}{\var{expressionOfRs}\pattern}{converts all expressions involving (products of) \funL{R} into spinor-space \funL{superFunction} objects according to the usual map.
\mathematicaBox{
\mathematicaSequence{\funL[]{ymTwistorAmp}\brace{6,1}}{\{\funL{R}[1,2,3,5,6]\funL{R}[\funL{cap}[\{2,3\},\{6,5,1\}],3,4,5,\funL{cap}[\{6,5\},\{3,2,1\}]]\}}
\mathematicaSequence[1]{\funL[1]{toSuperSpinorFunctions}\brace{6}\,\brace{\textnormal{\%}};\\
\funL[]{nice}\brace{\textnormal{\%}}}{\Big\{$\rule{0pt}{24pt}\displaystyle\frac{1}{\sb{1\,2}\sb{2\,3}\sb{3\,4}\sb{4\,5}\sb{5\,6}\sb{6\,1}}\,,\left(\begin{array}{@{}cccccc@{}}0&0&\text{-}\frac{\sb{4\,5}}{s_{6\,1\,2}}\phantom{\text{-}}&\frac{\sb{3\,5}}{s_{6\,1\,2}}&\text{-}\frac{\sb{3\,4}}{s_{6\,1\,2}}\phantom{\text{-}}&0\\
\sb{2\,6}&\sb{6\,1}&0&0&\sb{1\,2}\end{array}\right)\!\cdot\!\tilde\eta\Big\}$}
}
\textbf{\emph{Note}}: the optional argument of \var{gaugeFixQ} instructs \fun{toSuperSpinorFunction} to remove all excessive powers generated by the conversion from momentum-twistors to spinors by rescaling the coefficients of $\tilde\eta$'s. 
\mathematicaBox{
\mathematicaSequence[1]{\funL[1]{toSuperSpinorFunctions}\brace{6,False}\,\brace{\funL[]{ymTwistorAmp}\brace{6,1}};\\
\funL[]{nice}\brace{\textnormal{\%}\![\![1,1]\!]}}{\scalebox{0.9}{$\rule{0pt}{24pt}\displaystyle\frac{1}{\sb{1\,2}\sb{2\,3}\sb{3\,4}\sb{4\,5}\sb{5\,6}\sb{6\,1}\ab{1\,2}^8\ab{2\,3}^{12}\ab{3\,4}^4\ab{4\,5}^4\ab{5\,6}^{12}\ab{6\,1}^{8}\langle3|(1\,2)|6]^4\langle5|(6\,1)|2]^4s_{6\,1\,2}^4}$}}
}
}

\newpage
\vspace{-0pt}\sectionAppendix{Aspects and Ingredients of On-Shell Recursion}{appendix:on_shell_recursion_fcns}\vspace{-0pt}
%
\vspace{-0pt}\subsectionAppendix{Factorizations of Amplitudes \& Bridge Partitions}{appendix:partitions_etc}\vspace{-10pt}

\defnBox{bcfwPartitions}{\var{$n$}\pattern,\var{$k$}\pattern,\var{conjugatedQ}\optArg{\built{False}}\,}{returns a \built{List} of amplitude indices \texttt{\{\{$n_L$,$k_L$\},\{$n_R$,$k_R$\}\}} which appear as a factorization channel of the \var{$n$}-particle, N${}^{\text{\var{$k$}}}$MHV tree amplitude according to the BCFW shift $(\b{a},\b{b})$ where $\tilde{\lambda}_{\b{a}}\!\mapsto\!\tilde{\lambda}_{\b{a}}{-}\r{\alpha}\tilde{\lambda}_{\b{b}}$, $\lambda_{\b{b}}\!\mapsto\!\lambda_{\b{b}}{+}\r{\alpha}\lambda_{\b{a}}$, and $p_{\b{a}},p_{\b{b}}$ are in the left and right sides of the partition, respectively.\\[-10pt]

If the optional argument \var{conjugatedQ} is set to \built{True}, then returns the partitions for the \emph{conjugated} BCFW shift $(\b{b},\b{a})$ (still with $p_{\b{a}},p_{\b{b}}$ on the left and right sides, respectively).
\mathematicaBox{\mathematicaSequence{\funL[1]{bcfwPartitions}[6,1]}{\{\{\{5,1\},\{3,{-}1\}\},\{\{4,0\},\{4,0\}\},\{\{3,0\},\{5,0\}\}\}}
\mathematicaSequence{\funL[1]{bcfwPartitions}[6,1,True]}{\{\{\{5,0\},\{3,0\}\},\{\{4,0\},\{4,0\}\},\{\{3,-1\},\{5,1\}\}\}}}
}

\defnBox{bcfwPartitions}{\var{gluonHelicitySequence}\patternTwo}{lorem ipsum\\[-14pt]

\textbf{\emph{Note}}: \fun{bcfwPartitions}\brace{} \emph{automatically} returns parity-\emph{conjugated} BCFW partitions when \texttt{$\{$}\var{gluonHelicitySequence}\texttt{$\}$} is \texttt{\{}\var{p},\patternTwo,\var{m}\texttt{\}}.}

\vspace{10pt}\subsectionAppendix{Naming and Enumerating Terms in On-Shell Recursion}{appendix:naming_and_counting}\vspace{-10pt}

\defnBox{numberOfBCFWterms}{\var{$n$}\pattern,\var{$k$}\pattern}{returns the number of terms in the BCFW recursion of color-ordered, tree-level \var{n}-particle, N${}^{\text{\var{k}}}$MHV tree amplitude in YM, given by
\eq{\fun{numberOfBCFWterms}\brace{\var{n},\var{k}}\,\,\equivR\frac{1}{\text{\var{$n$}}{-}3}\binom{\text{\var{$n$}}{-}3}{\text{\var{$k$}}}\binom{\text{\var{$n$}}{-}3}{\text{\var{$k$}}{+}1}\,.}
For example, the number of terms in the $20$-particle N$^{8}$MHV amplitude be:
\mathematicaBox{\mathematicaSequence{\funL[1]{numberOfBCFWterms}[20,8]}{34\,763\,300}}
}

\defnBox{bcfwTermNames}{\var{$n$}\pattern,\var{$k$}\pattern}{returns a \built{List} of \emph{formatted} `names' for terms appearing in the BCFW recursion of the \var{n}-particle, N${}^{\text{\var{k}}}$MHV tree amplitude in YM:
\mathematicaBox{\mathematicaSequence{\funL[1]{bcfwTermNames}[6,1]}{$\{\mathcal{A}^{1}_{5}\otimes\mathcal{A}^{{-}1}_3$\!\!,$\mathcal{A}^{0}_{4}\otimes\mathcal{A}^{0}_{4}$,$\mathcal{A}^{0}_{3}\otimes\mathcal{A}^0_5\}$}}
}

\newpage
\vspace{-0pt}\subsectionAppendix{On-Shell Cut Data for Shifted Moments}{appendix:on_shell_cut_data}\vspace{-10pt}

\defnBox{bcfwCutData}{\{\var{$n_L$}\pattern,\var{$n_R$}\pattern\},\var{conjugatedQ}\pattern\optArg{}}{in BCFW recursion, an $n$-particle amplitude is expressed as products of two amplitudes with lower multiplicity \var{$n_L$} and \var{$n_R$} involving `shifted' momenta. If the legs labelled $\{\r{1},\b{n}\}$ are chosen for the shift, these lower-point amplitudes must be evaluated at 
\vspace{-5pt}\eq{\mathcal{A}_{\var{n_L}}\!\big(\r{\hat{1}}(\g{z^*}),2,\ldots,\hat{I}(\g{z^*})\big)\quad\text{and}\quad\mathcal{A}_{\var{n_R}}\!\big(\text{-}\hat{I}(\g{z^*}),\ldots,\b{\hat{n}}(\g{z^*})\big)\vspace{-5pt}}
where 
\eq{p_\r{1}\equivL\lambda_\r{1}\tilde\lambda_\r{1}\mapsto p_{\hat{\r{1}}}\equivR\lambda_1\big(\tilde{\lambda}_{\r{1}}{-}\g{z}\tilde{\lambda}_{\b{n}}\big)\quad\text{and}\quad p_\b{n}\equivL\lambda_\b{n}\tilde\lambda_\b{n}\mapsto p_{\hat{\b{n}}}\equivR\big(\lambda_\b{n}\text{+}\g{z}\lambda_\r{1}\big)\tilde\lambda_{\b{n}}\label{bcfw_split_amps}}
and the pole location $\g{z^*}$ is chosen so that $(p_{\r{\hat{1}}}\,{+}p_2{+}\ldots,p_{\var{n_L}{-}1})\equivL\text{-} p_I$ is on-shell. When $p_I(\g{z^*})$ is on-shell, it may be represented in any basis of spinors---but some choices are better for analytic expressions than others. 

The analytic details of this set of on-shell momenta are given by \fun{bcfwCutData}, encoded in a particular (possibly peculiar) form. Consider for example:
\mathematicaBox{
\mathematicaSequence{\funL[]{nice}\brace{\funL[1]{bcfwCutData}\brace{\{7,4\}}\,}}{\begin{tabular}{@{}l@{$\;$}c@{$\;$}l}zStar&$\rightarrow$&$\displaystyle\frac{s_{7\,8\,9}}{\langle1|(7\,8)|9]}$\\[10pt]
intL&$\rightarrow$&$\displaystyle\big\{\{7,9\},\big\{\big\{\text{-}\frac{\langle9|(7\,8)|9]}{s_{7\,9}},\frac{\sb{8\,9}\ab{7\,8}}{s_{7\,9}}\big\},\big\{1,\text{-}\frac{\langle1|(8\,9)|7]}{\langle1|(7\,8)|9]}\big\}\big\}$\\[10pt]
intR&$\rightarrow$&$\displaystyle\big\{\{7,9\},\big\{\big\{\text{-}\frac{\langle9|(7\,8)|9]}{s_{7\,9}},\frac{\sb{8\,9}\ab{7\,8}}{s_{7\,9}}\big\},\big\{\text{-}1,\frac{\langle1|(8\,9)|7]}{\langle1|(7\,8)|9]}\big\}\big\}$\\[10pt]
oneHat&$\rightarrow$&$\displaystyle\big\{\{1,9\},\big\{\big\{1,0\big\},\big\{1,\text{-}\frac{s_{7\,8\,9}}{\langle1|(7\,8)|9]}\big\}\big\}$\\[10pt]
nHat&$\rightarrow$&$\displaystyle\big\{\{9,1\},\big\{\big\{1,\frac{s_{7\,8\,9}}{\langle1|(7\,8)|9]}\big\},\big\{1,0\big\}\big\}$\\[10pt]
\end{tabular}}
}
where the meaning of `$x\!\mapsto\!\{\{a,b\},\{\{x_{1\,a},x_{1\,b}\},\{x_{2\,a},x_{2\,b}\}\}$' is that the on-shell momentum $p_x$ can be written as
\eq{p_x=\lambda_x\,\tilde\lambda_x\equivR\big(x_{1\,a}\lambda_a{+}x_{1\,b}\lambda_b\big)\big(x_{2\,a}\tilde\lambda_a{+}x_{2\,b}\tilde\lambda_b\big)\,.}
Thus, all the spinor-data required by the amplitudes in (\ref{bcfw_split_amps}) is given conveniently by \fun{bcfwCutData}.
}

\defnBox{bcfwCutSpinors}{\{\var{$n_L$}\pattern,\var{$n_R$}\pattern\},\var{conjugatedQ}\pattern\optArg{}}{analogous to \fun{bcfwCutData}, this function returns
\vspace{-5pt}\eq{\begin{split}\built{List\big[}&\{\{\lambda_\r{\hat{1}},\tilde\lambda_{\r{\hat{1}}}\},\{\lambda_2,\tilde\lambda_2\},\ldots,\{\lambda_{I_L},\tilde{\lambda}_{I_L}\}\},\\
&\{\{\lambda_{I_R},\tilde{\lambda}_{I_R}\},\ldots,\ldots,\{\lambda_{\b{\hat{n}}},\tilde\lambda_{\b{\hat{n}}}\}\}\big]
\end{split}\vspace{-10pt}}
corresponding to the sets of spinors relevant to the left/right amplitude of (\ref{bcfw_split_amps}) respectively.\vspace{-40pt} 
}

\newpage\addtocontents{toc}{\protect\newpage}
\vspace{0pt}\sectionAppendix{Ordered Partial Amplitudes of (pure) superYang-Mills}{appendix:planar_amps_of_sym}\vspace{-0pt}
%
\vspace{-0pt}\subsectionAppendix{Bases of Partial Amplitudes: Identities and Reduction Rules}{appendix:abstract_amps}\vspace{-0pt}

\vspace{0pt}\subsubsectionAppendix{Cyclic and Dihedral Sets of Amplitudes and Reductions}{appendix:cyclic_and_dihedral}\vspace{-10pt}

\defnBox{cyclicAmps}{\var{$n$}\pattern}{provides a \built{List} of abstract \funL{amp} objects with specified leg-ordering for any multiplicity \var{$n$} for which the cyclic redundancy has been eliminated by setting the leg indexed $\r{1}$ to the first slot. 
\mathematicaBox{
\mathematicaSequence[1]{\funL[1]{cyclicAmps}\brace{4};\\
\funL[]{nice}\brace{\textnormal{\%}/\!\!.\{\funL{amp}[\var{$x$\pattern}]$\!\mapsto$\funL{amp}@@(\{\var{$x$}\}/\!\!.\{1->Style[1,Red]\}\}}}{$\rule{0pt}{14pt}\{\mathcal{A}_4\!(\r{1},2,3,4),\mathcal{A}_4\!(\r{1},2,4,3),\mathcal{A}_4\!(\r{1},3,2,4),\mathcal{A}_4\!(\r{1},3,4,2),\mathcal{A}_4\!(\r{1},4,2,3),\mathcal{A}_4\!(\r{1},4,3,2)\}$}
}
The cyclic basis consists of $(\var{n}{-}1)!$ elements.
}

\defnBox{cyclicAmpReductionRule}{}{a \built{Rule} which canonically rotates the arguments of any \funL{amp} according to the \built{Rule}
\eq{\big\{\funL{amp}\brace{\var{x}\patternThree,\r{1},\var{y}\patternThree}\mapsto\funL{amp}\brace{\r{1},\var{y},\var{x}}\,\big\}}
so any \funL{amp}'s image will be an element of \fun{cyclicAmps}\brace{\var{$n$}}.
}

\defnBox{dihedralAmps}{\var{$n$}\pattern}{provides a \built{List} of abstract \funL{amp} objects with specified leg-ordering for any multiplicity \var{$n$} for which the entire \emph{dihedral} symmetry has been eliminated by setting the leg indexed $\r{1}$ to the first slot and ensuring that for each \funL{amp}\brace{\r{1},\b{$a$\rule[-1.05pt]{7.5pt}{.65pt}},\var{$x$}\patternThree,\b{$b$\rule[-1.05pt]{7.5pt}{.65pt}}}\,, $\b{a}\!<\!\b{b}$. 
\mathematicaBox{
\mathematicaSequence[1]{\funL[1]{dihedralAmps}\brace{5};\\
\mbox{\textnormal{\%}/\!\!.\{\funL{amp}\![\!1,\!\var{$a$}\pattern\!,\var{$x$}\patternThree\!,\var{$b$\pattern}]\!$\mapsto$\funL{amp}\brace{Style\![1\!,Red]\!\!,Style\![\var{$a$}\!,Blue]\!\!,\var{$x$}\!,Style\![\var{$b$}\!,Blue]\!}\}\!;}\\
\funL[]{nice}\brace{\textnormal{\%}}}{$\rule{0pt}{14pt}\{\mathcal{A}_5\!(\r{1},\b2,3,4,\b5),\mathcal{A}_5\!(\r{1},\b2,3,5,\b4),\mathcal{A}_5\!(\r{1},\b2,4,3,\b5),\mathcal{A}_5\!(\r{1},\b2,4,5,\b3),$\\
$\phantom{\{}\mathcal{A}_5\!(\r{1},\b2,5,3,\b4),\mathcal{A}_5\!(\r{1},\b2,5,4,\b3),\mathcal{A}_5\!(\r{1},\b3,2,4,\b5),\mathcal{A}_5\!(\r{1},\b3,2,5,\b4),$\\
$\phantom{\{}\mathcal{A}_5\!(\r{1},\b3,4,2,\b5),\mathcal{A}_5\!(\r{1},\b3,5,2,\b4),\mathcal{A}_5\!(\r{1},\b4,2,3,\b5),\mathcal{A}_5\!(\r{1},\b4,3,2,\b5)\}$}
}
The dihedral basis consists of $(\var{n}{-}1)/2!$ elements.
}

\defnBox{dihedralAmpReductionRule}{}{a \built{Rule} which canonically rotates the arguments of any \funL{amp} according to the rule
\eq{\begin{split}\big\{\funL{amp}\brace{\var{x}\patternThree,\r{1},\var{y}\patternThree}\mapsto\texttt{If}&\texttt{\big[OrderedQ}\brace{\{\var{y},\var{x}\}[\![1,1]\!]}\,,\funL{amp}\brace{\r{1},\var{y},\var{x}}\,,\\
&\;\;\;(\text{-}1)^{|\var{x},\var{y}|\text{+}1}\funL{amp}\brace{1,\texttt{\#\#}}\texttt{\&@@}\texttt{Reverse}\brace{\{\var{y},\var{x}\}}\texttt{\big]}\}\\[-12pt]
\end{split}}
so any \funL{amp}'s image will be an element of \fun{dihedralAmps}\brace{\var{$n$}}.
\mathematicaBox{
\mathematicaSequence{\funL{amp}@@\funL[]{randomPerm}\brace{7}}{\funL{amp}\brace{5,1,6,3,7,4,2}}
\mathematicaSequence{\textnormal{\%}/\!\!.\builtL[1]{dihedralAmpReductionRule}}{-\funL{amp}\brace{1,5,2,4,7,3,6}}
}\vspace{-40pt}
}

\vspace{10pt}\subsubsectionAppendix{Kleiss-Kuijf Relations, Bases, and Reductions}{appendix:kk_relations}\vspace{-10pt}

\defnBox{kkBasisAmps}{\var{$n$}\pattern}{provides a \built{List} of abstract \funL{amp} objects with specified leg-ordering for any multiplicity \var{$n$} modulo all KK relations; this basis consists of all $(\var{n}\text{-}2)!$ ordered amplitudes of the form \funL{amp}\brace{\r{1},\b{$x$\rule[-1.05pt]{14.5pt}{.65pt}},\var{$n$}}.}

\defnBox{u1DecouplingIdentity}{\var{$n$}\pattern}{returns a \built{List} of \funL{amp} objects whose \built{Total} vanishes due to the so-called `$U(1)$-decoupling identity' (see e.g.~\cite{Mangano:1990by}):
\eq{\mathcal{A}(\b{1},2,3,\ldots,n\text{-}1,\r{n})+\mathcal{A}(\b{1},3,\ldots,n\text{-}1,\r{n},2)+\ldots+\mathcal{A}(\b{1},\r{n},2,3,\ldots,n\text{-}1){=}0.\vspace{-15pt}}
\mathematicaBox{
\mathematicaSequence{\funL[1]{u1DecouplingIdentity}\brace{6}}{\{\funL{amp}[1,2,3,4,5,6],\funL{amp}[1,3,4,5,6,2],\funL{amp}[1,4,5,6,2,3],\\\phantom{\{}\funL{amp}[1,5,6,2,3,4],\funL{amp}[1,6,2,3,4,5]\}}
}\vspace{-10pt}
}

\defnBox{egKKIdentity}{\var{$n$}\pattern}{gives a randomly-generated \emph{example} of a Kleiss-Kuijf (`KK') relation \cite{KK} among ordered amplitudes (see also~\cite{Arkani-Hamed:2014bca}). This is expressed as \{\var{ampSeed},\var{imageTerms}\} where \var{imageTerms} is a sequence of those terms in the sum generated by \builtL{kkAmpReductionRule} and \var{ampSeed} is chosen via \mbox{\built{RandomChoice}\brace{\fun{dihedralAmps}\brace{\var{$n$}}}\,}.
\mathematicaBox{
\mathematicaSequence{egId=\funL[1]{egKKIdentity}\brace{6}}{\{\funL{amp}\![\g1,\r{2},\r3,\g6,\b4,\b5],\\
\phantom{\{}{-}\,\funL{amp}\![\g1,\r2,\r3,\b5,\b4,\g6],{-}\,\funL{amp}\![\g1,\r2,\b5,\r3,\b4,\g6],{-}\funL{amp}\![\g1,\r2,\b5,\b4,\r3,\g6],\\
\phantom{\{}{-}\funL{amp}[\g1,\b5,\r2,\r3,\b4,\g6],{-}\funL{amp}[\g1,\b5,\r2,\b4,\r3,\g6],{-}\funL{amp}\![\g1,\b5,\b4,\r2,\r3,\g6]\}}
\mathematicaSequence{egId/\!\!.\{\funL{amp}\brace{\var{$x$}\patternTwo}\,$\mapsto$\funL[]{componentAmpN}\brace{\{\stateYM{m},\stateYM{p},\stateYM{m},\stateYM{p},\stateYM{m},\stateYM{p}\},\{\var{$x$}\}}\}}{\scalebox{0.875}{$\left\{\rule{0pt}{20pt}\right.\!\!\displaystyle\frac{1582692675}{140365546076},\text{-}\frac{975}{56518},\text{-}\frac{450160}{26196093},\frac{4174355}{179589564},\frac{24531260}{503169111},\text{-}\frac{17580320}{355097547},\frac{2281450}{3400432497}\!\!\left.\rule{0pt}{20pt}\right\}$}}
\mathematicaSequence{Total\brace{\textnormal{\%}}}{0}
}\vspace{-10pt}
}

\defnBox{kkIdentities}{\var{$n$}}{returns a \built{List} of all independent ($(\var{n}{-}3)(\var{n}{-}2)!/2$ in number) of Kleiss-Kuijf (`KK') relations among the amplitudes in \mbox{\fun{dihedralAmps}\brace{$\var{n}$}\,.}
\mathematicaBox{
\mathematicaSequence[2]{egIdentities=\funL[1]{kkIdentities}\brace{6};\\
egIdentities/\!\!.\{\funL{amp}\brace{\var{$x$}\patternTwo}$\,\mapsto$\funL[]{componentAmpN}\brace{\{\stateYM{m},\stateYM{p},\stateYM{m},\stateYM{p},\stateYM{m},\stateYM{p}\},\{\var{$x$}\}}\};\\
Total/@\textnormal{\%}}{\{0\!,\!0\!,\!0\!,\!0\!,\!0\!,\!0\!,\!0\!,\!0\!,\!0\!,\!0\!,\!0\!,\!0\!,\!0\!,\!0\!,\!0\!,\!0\!,\!0\!,\!0\!,\!0\!,\!0\!,\!0\!,\!0\!,\!0\!,\!0\!,\!0\!,\!0\!,\!0\!,\!0\!,\!0\!,\!0\!,\!0\!,\!0\!,\!0\!,\!0\!,\!0\!,\!0\}}\\[-20pt]
\mathematicaSequence[1]{egMat=Coefficient[Total[\#],\funL[]{dihedralAmps}\brace{6}]\&/@egIdentities;\\
MatrixRank[egMat]}{36}
}\vspace{-10pt}
}

\defnBox{kkAmpReductionRule}{}{is a \built{Rule} built using the ordered-amplitude relations of Kleiss-Kuijf (`KK') to replace any amplitude \funL{amp}\brace{\var{legOrdering}\patternTwo} as a linear combination of those of the basis given by \fun{kkBasisAmps}.\vspace{-40pt}
}

\newpage
\vspace{10pt}\subsubsectionAppendix{BCJ Relations, Reductions, and KLT}{appendix:bcj_and_klt}\vspace{-10pt}

\defnBox{bcjBasisAmps}{\var{$n$}}{returns a \built{List} of the (\var{$n$}{-}3)! partially-ordered amplitudes of the form
\vspace{-10pt}\eq{\fun{bcjBasisAmps}\brace{\var{n}}\,\,\Leftrightarrow\big\{\funL{amp}\brace{1,\r{\sigma}_{2},\ldots,\r{\sigma}_{\var{n}-2},\var{n}\text{-}1,\var{n}}\,\,\,\rule[-1.75pt]{0.65pt}{11pt}\,\,\r{\sigma}\in\mathfrak{S}\left([2,\var{n}{-}2]\right)\big\}\,.\vspace{-5pt}}
This list is identical to a re-ordering of \fun{kltBasisAmpsAlpha}\brace{\var{$n$}}\,.\vspace{-5pt}
}

\defnBox{bcjIdentitySeed}{\var{$n$}\pattern}{returns a \built{List} of terms whose vanishing when summed represent the primary `BCJ identity' \cite{Bern:2008qj}:
\eq{\sum_{\r{a}=1}^{\var{n\text{-}2}}\left[\big(\sum_{\b{b}=1}^{\r{a}}s_{\b{b}\,\,\var{n\text{-}1}}\big)\mathcal{A}_{\var{n}}\!\big(1,\ldots,\r{a},\var{n\text{-}1},\r{a}\text{+}1,\ldots,\var{n}\big)\right]{=}0\,.\label{bcj_identity_seed}}
Permutations of (\ref{bcj_identity_seed}) over all $(\var{n}\text{-}2)!$ relabelings from $\sigma\!\in\!\mathfrak{S}([2,\var{n}\text{-}1])$ generate $(\var{n}\text{-}3)\!\times\!(\var{n}\text{-}3)!$ independent functional relations among the partial amplitudes of the KK basis given by \fun{kkBasisAmps}, leaving an independent space of $(\var{n}\text{-}3)!$ partial amplitudes. 
}

\defnBox{egBCJIdentity}{\var{$n$}\pattern}{returns \built{RandomChoice}\brace{\fun{bcjIdentities}\brace{\var{$n$}}}\,.\vspace{-10pt}}\vspace{-7pt}

\defnBox{bcjIdentities}{\var{$n$}\pattern}{returns a \built{List} of all independent ($(\var{n}{-}3)(\var{n}{-}3)!$ in number) of BCJ identities satisfied by the amplitudes in \mbox{\fun{kkBasisAmps}\brace{$\var{n}$}\,.}
\mathematicaBox{
\mathematicaSequence[1]{egIdentities=\funL[]{evaluate}\brace{\funL[1]{bcjIdentities}\brace{6}};\\
egIdentities/\!\!.\{\funL{amp}\brace{\var{$x$}\patternTwo}$\,\mapsto$\funL[]{componentAmpN}\brace{\{\stateYM{m},\stateYM{p},\stateYM{m},\stateYM{p},\stateYM{m},\stateYM{p}\},\{\var{$x$}\}}\};\\
Total/@\textnormal{\%}}{\{0,0,0,0,0,0,0,0,0,0,0,0,0,0,0,0,0,0\}}\\[-20pt]
\mathematicaSequence[1]{egMat=Coefficient[Total[\#],\funL[]{kkBasisAmps}\brace{6}]\&/@egIdentities;\\
MatrixRank[egMat]}{18}
}\vspace{-14pt}
}

\defnBox{bcjAmpReductionRule}{}{is a \built{Rule} that employs instances of (\ref{bcj_identity_seed}) to \built{Replace} any partial amplitude into the basis given by \fun{bcjBasisAmps}\brace{}\, (which is the same---up to reordering---as the basis \fun{kltBasisAmpsAlpha}\brace{}\,).\
\mathematicaBox{
\mathematicaSequence{\funL[]{nice}\brace{\funL{amp}\brace{1,4,3,5,2,6}/\!\!.\builtL[1]{bcjAmpReductionRule}}}{$\rule{0pt}{22pt}\displaystyle\phantom{\text{-}}\frac{s_{1\,2}}{s_{2\,6}}\mathcal{A}_6\!\big(1,2,3,4,5,6\big){+}\frac{s_{1\,2}{+}s_{2\,3}}{s_{2\,6}}\mathcal{A}_6\!\big(1,3,2,4,5,6\big)$\\
$\displaystyle\text{-}\frac{(s_{2\,5}\text{+}s_{2\,6})(s_{1\,3}\text{+}s_{1\,4}\text{+}s_{3\,4})}{s_{2\,6}s_{1\,3\,4}}\mathcal{A}_6\!\big(1,3,4,2,5,6\big)$}
}\vspace{-14pt}
}

\defnBox{kltBasisAmpsAlpha}{\var{$n$}\pattern}{returns a \built{List} of the (\var{$n$}{-}3)! partially-ordered amplitudes in the `standard' $\alpha$ KLT basis, but \emph{reordered} so that the \funL{kltKernel} is \emph{block-diagonal} with respect to those of \fun{kltBasisAmpsBeta}\brace{\var{$n$}}\,.}

\defnBox{kltBasisAmpsBeta}{\var{$n$}\pattern}{returns a \built{List} of the (\var{$n$}{-}3)! partially-ordered amplitudes in the default KLT `$\beta$' basis. That is, all partial amplitudes whose arguments are ordered according to the form $\funL{amp}\brace{1,\r{x\rule[-1.05pt]{14.5pt}{.65pt}},\var{n},\r{y\rule[-1.05pt]{14.5pt}{.65pt}},\var{n\text{-}1}}$ 
where $|\{\r{x}\}|{=}\lceil(\var{n}\text{-}3)/2\rceil$ and $|\{\r{y}\}|{=}\lfloor(\var{n}\text{-}3)/2\rfloor$. \vspace{-40pt}
}

\defnBox{kltKernelInverse}{\var{$n$}\pattern}{returns the \built{Inverse} of the KLT kernel for the paring \{\fun{kltBasisAmpsAlpha}\brace{\var{$n$}}\,,\,\fun{kltBasisAmpsBeta}\brace{\var{$n$}}\,\}. As the entries of the \emph{inverse-kernel} can be computed combinatorially quite efficiently, it is often easier to define the KLT \emph{kernel} to be the inverse \fun{kltKernelInverse}.  See {e.g.}~\cite{Cachazo:2013gna} for more details.
}

\defnBox[]{kltKernelInverse}{\{\var{$\alpha$}\patternTwo\fun{amp}\},\{\var{$\beta$}\patternTwo\fun{amp}\}}{returns the entry of the KLT-kernel-inverse $m(\var{\alpha}|\var{\beta})$ in the notation of \cite{Mizera:2016jhj} (see also~\cite{Cachazo:2013gna,Mizera:2019blq}).\\[-10pt]

If a list of permuted amplitudes are given, then \fun{kltKernelInverse} returns the \built{Array} for their pairings.}

\defnBox{kltKernel}{\var{$n$}\pattern}{gives the KLT kernel for the paring of partial amplitude bases \{\fun{kltBasisAmpsAlpha}\brace{\var{$n$}}\,,\,\fun{kltBasisAmpsBeta}\brace{\var{$n$}}\,\}.}

\vspace{10pt}

\defnBox[]{kltKernel}{\{\var{$\alpha$}\patternTwo\fun{amp}\},\{\var{$\beta$}\patternTwo\fun{amp}\}}{gives the KLT kernel as an \built{Array} for the sets of basis amplitudes \{\{\var{$\alpha$}\},\{\var{$\beta$}\}\}.
}

\vspace{10pt}\subsectionAppendix{Analytic Expressions for Ordered superAmplitudes of sYM}{appendix:analytic_sym}\vspace{-14pt}

\defnBox{parkeTaylorAmp}{\var{$n$}\pattern}{gives the (single-element) \built{List} of terms contributing to the Parke-Taylor amplitude \cite{Parke:1986gb} for ordered (N${}^{k{=}0}$)MHV amplitude in sYM:
\mathematicaBox{\mathematicaSequence{\funL[1]{parkeTaylorAmp}[8]}{$\displaystyle\frac{\text{\texttt{1}}}{\text{\texttt{\funL{ab}[1,2]\funL{ab}[2,3]\funL{ab}[3,4]\funL{ab}[4,5]\funL{ab}[5,6]\funL{ab}[6,7]\funL{ab}[7,8]\funL{ab}[8,1]}}}$}}
\vspace{-10pt}}

\defnBox{parkeTaylorAmpBar}{\var{$n$}\pattern}{gives the (single-element) \built{List} of terms contributing to the parity-conjugate of the Parke-Taylor amplitude \cite{Parke:1986gb}---the N${}^{k=\var{n}\text{-}4}$MHV amplitude in sYM:
\mathematicaBox{\mathematicaSequence{\funL[1]{parkeTaylorAmpBar}[6]}{$\left.\displaystyle\funL{superFunction}\left[\frac{\text{\texttt{1}}}{\text{\texttt{\funL{sb}[1,2]\funL{sb}[2,3]\funL{sb}[3,4]\funL{sb}[4,5]\funL{sb}[5,6]\funL{sb}[6,1]}}},\right.\right.$
\{\{0,0,$\displaystyle\frac{\text{\texttt{\funL{sb}[4,5]}}}{\texttt{\funL{ab}[1,2]}},\frac{\text{\texttt{\funL{sb}[5,3]}}}{\texttt{\funL{ab}[1,2]}},\frac{\text{\texttt{\funL{sb}[3,4]}}}{\texttt{\funL{ab}[1,2]}}$,0\},\{0,0,0,$\displaystyle\frac{\text{\texttt{\funL{sb}[5,6]}}}{\texttt{\funL{sb}[4,5]}},\frac{\text{\texttt{\funL{sb}[6,4]}}}{\texttt{\funL{sb}[4,5]}}1$\}\}$\left.\left.\rule{0pt}{15pt}\right]\right.$}}
Generally, the \var{$n$}-point $\bar{\text{MHV}}$ amplitude in sYM is encoded by the package as:
\eq{\mathcal{A}_{\var{n}}^{(k{=}\var{n}\text{-}4)}\equivR\frac{1}{\sb{1\,2}\sb{2\,3}\cdots\sb{\var{n}\,1}}\delta^{(\var{n}\text{-}4)\times4}\!\big(C^{\bar{\text{MHV}}}_{\var{n}}\!\cdot\!\tilde\eta\big)\delta^{2\!\times\!4}\!\big(\lambda\!\cdot\!\tilde\eta\big)\delta^{2\!\times\!2}\!\big(\lambda\!\cdot\!\tilde\lambda\big)}
where the $(k\text{-}4)\!\times\!\var{n}$ matrix $C^{\bar{\text{MHV}}}_{\var{n}}$ is given by 
\eq{C_{\var{n}}^{\bar{\text{MHV}}}\equivR\left(\begin{array}{@{}ccccccccccc@{}}\deemph{0}&\deemph{0}&\frac{\sb{4\,5}}{\ab{1\,2}}&\frac{\sb{5\,3}}{\ab{1\,2}}&\frac{\sb{3\,4}}{\ab{1\,2}}&\deemph{0}&\deemph{\cdots}&\deemph{0}\\
\deemph{0}&\deemph{0}&\deemph{0}&\frac{\sb{5\,6}}{\sb{4\,5}}&\frac{\sb{6\,4}}{\sb{4\,5}}&1&\deemph{\ddots}&\deemph{\vdots}\\
\deemph{\vdots}&\deemph{\ddots}&\deemph{\ddots}&\deemph{\ddots}&\ddots&\ddots&\ddots&\deemph{0}~\\
\deemph{0}&\deemph{\cdots}&\deemph{\cdots}&\deemph{\cdots}&\deemph{0}&\frac{\sb{\var{n}{-}1\,\var{n}}}{\sb{\var{n}{-}2\,\var{n}{-}1}}&\frac{\sb{\var{n}\,\var{n}{-}2}}{\sb{\var{n}{-}2\,\var{n}{-}1}}&1\end{array}\right)\,.\vspace{-10pt}}
\vspace{-50pt}}

\defnBox{ymAmp}{\var{$n$}\pattern,\var{$k$}\pattern,\var{legOrderingList}\pattern\optArg{\!\{\}\,}}{returns a \built{List} of \funL{superFunction} contributions to the \var{$n$}-particle N${}^{\var{k}}$MHV partial tree-level amplitude in sYM \emph{expressed in terms of \textbf{spinor} variables} for the external leg ordering $\{1,\ldots,\var{n}\}$ or \{\var{legOrderingList}\} if specified.\\[-10pt]

For each instance of $(\var{n},\var{k})$ \fun{ymAmp} is computed internally first and then stored in memory for future evaluation or other use. These expressions are obtained by `uplifting' the formulae of \funL{ymSuperTwistorAmp} involving super momentum-twistor variables into spinor-dependent expressions.\\[-5pt]

The output of \fun{ymAmp} should be viewed as \emph{the} \textbf{standard} expression for each amplitude. It is obtained using a very specific, default scheme for iterated on-shell (`BCFW') recursion---one for which the first and last legs of all nested amplitudes are shifted using the same parity of the bridge. This particular recursion scheme enjoys some nice features, including the guarantee of potential (if not always realized) simplification for the resulting terms appearing in recursion. These simplifications are guaranteed to be realized for all amplitudes with multiplicity $\var{n}\!\leq\!10$ for any N${}^{\var{k}}$MHV degree through the use of certain reference formulae for the spinor-variable versions of some $R$-invariants.\\[-10pt]
}

\defnBox{ymAmpRandom}{\var{$n$}\pattern,\var{$k$}\pattern,\var{legOrderingList}\pattern\optArg{\!\{\}\,}}{returns a \built{List} of \funL{superFunction} contributions to the \var{$n$}-particle N${}^{\var{k}}$MHV partial tree-level amplitude in sYM \emph{expressed in terms of \textbf{spinor} variables} for the external leg ordering $\{1,\ldots,\var{n}\}$ or \{\var{legOrderingList}\}. In contrast to \fun{ymAmp}, however, \fun{ymAmpRandom} uses a \emph{random} choices of BCFW recursion (parity and leg pairings) at each successive stage of recursion. As such, the output of \fun{ymAmpRandom} are \emph{not} stored to memory, and can be viewed as a source of independent, internal verification of the correctness of the `standard' reference formulae provided by \fun{ymAmp}.\\[-10pt] 

It is worth noting that, because only a small subset of particular recursive schemes enjoy the same simplifications to spinor expressions as the default scheme, the output of \fun{ymAmpRandom} can be substantially more `complex'---as measured by \built{LeafCount}, say---relative to the output of \fun{ymAmp}. This additional complexity has consequences for time required for evaluation and other subsequent operations. 
\\[-4pt]}

\newpage
\vspace{0pt}\subsubsectionAppendix{Tree Amplitudes of superYang-Mills Theory in Twistor Variables}{appendix:twistor_amps_for_sym}\vspace{-10pt}

\defnBox{ymTwistorAmp}{\var{$n$}\pattern,\var{$k$}\pattern}{returns a \built{List} of (products of) \funL{R}-invariants sum gives the \var{$n$}-particle N${}^{\var{k}}$MHV tree amplitude in sYM in terms of (super)momentum-twistors, as obtained by the default recursion scheme.\\[-10pt]

These expressions are the primary input for \fun{ymAmp}, as recursion of analytic superAmplitudes is more efficiently implemented in momentum-twistor variables.\\[-18pt]
}

\defnBox{ymTwistorAmpRandom}{\var{$n$}\pattern,\var{$k$}\pattern}{returns a \built{List} of (products of) \funL{R}-invariants sum gives the \var{$n$}-particle N${}^{\var{k}}$MHV tree amplitude in sYM in terms of (super)momentum-twistors, as obtained by \emph{randomly} choosing which pair of adjacent legs to deform, and \emph{randomly} choosing the parity of the BCFW deformation.\\[-10pt]

These expressions are the primary input for \fun{ymAmpRandom}, as recursion of analytic superAmplitudes is more efficiently implemented in momentum-twistor variables.\\[-18pt]}

\defnBox{ymSuperTwistorAmp}{\var{$n$}\pattern,\var{$k$}\pattern}{returns a \built{List} of \funL{superFunction} objects whose sum gives the \var{$n$}-particle N${}^{\var{k}}$MHV tree amplitude in sYM in terms of (super)momentum-twistors according to the output of \fun{ymTwistorAmp} and the definition of \funL{R}-invariants in (\ref{r_invariant_defined}).\\[-10pt]

\textbf{\emph{Note}}: the \fun{superFunction} object appearing in these expressions are \emph{expressed in terms of {(super)momentum-twistor variables}}; as such, these expressions \textbf{should not be used in combination with} \funL{component}, or \funL{permuteLegs}, or many other functions that implicitly or explicitly \emph{require} that the \fun{superFunction} objects and the kinematic expressions on which they depend to be spinor expressions.\\[-10pt]

The primary role of \fun{ymSuperTwistorAmp} is as an intermediate expression (used, for example, in \funL{ymAmpN}).\\[-18pt] 
}

\defnBox{ymSuperTwistorAmpRandom}{\var{$n$}\pattern,\var{$k$}\pattern}{returns a \built{List} of \funL{superFunction} objects whose sum gives the \var{$n$}-particle N${}^{\var{k}}$MHV tree amplitude in sYM in terms of (super)momentum-twistors according to the output of \fun{ymTwistorAmpRandom} and the definition of \funL{R}-invariants in (\ref{r_invariant_defined}).\\[-10pt]

\textbf{\emph{Note}}: the \funL{superFunction} objects appearing in these expressions are \emph{expressed in terms of {(super)momentum-twistor variables}}; as such, these expressions \textbf{should not be used in combination with} \funL{component}, or \funL{permuteLegs}, or many other functions that implicitly or explicitly \emph{require} that \fun{superFunction} objects and the kinematic expressions on which they depend to be spinor expressions.\\[-10pt]

The primary role of \fun{ymSuperTwistorAmp} is as an intermediate expression (used, for example, in \funL{ymAmpRandomN}).\vspace{-40pt}}

\newpage
\vspace{-0pt}\subsectionAppendix{Numeric Expressions for Ordered Partial superAmplitudes of sYM}{appendix:numeric_superAmps}\vspace{-10pt}

\defnBox{ymAmpN}{\var{$n$}\pattern,\var{$k$}\pattern,\var{legOrdering}\pattern\optArg{}}{gives a list of numeric \fun{superFunction}s which contribute to the \var{$n$}-point N${}^{{\var{k}}}$MHV tree amplitude in sYM for the kinematic data encoded by the global variables representing spinors \builtL{Ls} and \builtL{Lbs}.\\[-10pt]

It is worth noting that \fun{ymAmpN}\brace{\var{$n$},\var{$k$}} is \emph{equivalent to} \textbf{but not identical with} the output of \funL{evaluate}\brace{\funL{ymAmp}\brace{\var{$n$},\var{$k$}}\,}\,. This is because, although \fun{ymAmpN} follows an \emph{identical recursion algorithm} to \funL{ymAmp}, \fun{ymAmpN} makes use of \emph{evaluated} twistor-variable expressions, without the time-expensive step of converting these to expressions defined in momentum space. Thus, \fun{ymAmpN} is generally faster and has a farther reach than \funL{ymAmp}:
\mathematicaBox{\mathematicaSequence[0]{\{\funL[]{timed}@\funL[]{evaluate}[\funL[]{ymAmp}[12,4]],\funL[]{timed}@\funL[1]{ymAmpN}[12,4]\};}{}
\\[-20pt]&\scalebox{0.6}{\hspace{-30pt}Evaluation of the function \texttt{\textbf{ymAmp[]}} for $\mathcal{A}^{(4)}_{12}$ required \textbf{6 seconds, 152 ms} to complete.}\\[-5pt]
&\scalebox{0.6}{\hspace{-30pt}Evaluation of the function \texttt{\textbf{ymAmpN[]}} for $\mathcal{A}^{(4)}_{12}$ required \textbf{1 seconds, 814 $\mathbf{\mu}$s} to complete.}\\[-5pt]
\mathematicaSequence{SameQ@@\textnormal{\%}}{False}
\mathematicaSequence{SameQ@@(\funL[]{standardizeSuperFunctions}@\textnormal{\%\%})}{True}
}
\textbf{\emph{Note}}: direct evaluation (via \funL{evaluate}\texttt{[]}) of an \emph{already computed analytic expression} is almost always faster than using \fun{ymAmpN}. 
\mathematicaBox{\mathematicaSequence[1]{\funL[]{useRandomKinematics}[12]\\\{\funL[]{timed}@\funL[]{evaluate}[\funL[]{ymAmp}[12,4]],\funL[]{timed}@\funL[1]{ymAmpN}[12,4]\};}{}
\\[-20pt]&\scalebox{0.6}{\hspace{-30pt}Evaluation of the function \texttt{\textbf{ymAmp[]}} for $\mathcal{A}^{(4)}_{12}$ required \textbf{6 seconds, 152 ms} to complete.}\\[-5pt]
&\scalebox{0.6}{\hspace{-30pt}Evaluation of the function \texttt{\textbf{ymAmpN[]}} for $\mathcal{A}^{(4)}_{12}$ required \textbf{1 seconds, 814 $\mathbf{\mu}$s} to complete.}\\[-5pt]
\mathematicaSequence[1]{\funL[]{useRandomKinematics}[12]\\\{\funL[]{timed}@\funL[]{evaluate}[\funL[]{ymAmp}[12,4]],\funL[]{timed}@\funL[1]{ymAmpN}[12,4]\};}{}
\\[-20pt]&\scalebox{0.6}{\hspace{-30pt}Evaluation of the function \texttt{\textbf{evaluate}} required \textbf{667 ms, 257 $\mathbf{\mu}$s} to complete.}\\[-5pt]
&\scalebox{0.6}{\hspace{-30pt}Evaluation of the function \texttt{\textbf{ymAmpN[]}} for $\mathcal{A}^{(4)}_{12}$ required \textbf{888 ms, 702 $\mathbf{\mu}$s} to complete.}\\[-5pt]
}
}

\defnBox{ymAmpRandomN}{\var{$n$}\pattern,\var{$k$}\pattern}{similar to the evaluation of \funL{ymAmpRandom}\brace{\var{$n$},\var{$k$}}\,, except that the evaluation takes place at the level of momentum-twistors and maps the corresponding \fun{superFunction}s to spinor-variables numerically by matrix multiplication. This saves considerable time relative to the time spent processing the analytic expressions given by \funL{ymAmpRandom}. 
}

\newpage
\vspace{-0pt}\subsectionAppendix{Analytic Expressions for \emph{Component} Amplitudes of sYM}{appendix:analytic_component_amps}\vspace{-10pt}
%

\vspace{-0pt}\defnBox{ymHelicityAmp}{\var{gluonHelicities}\patternTwo}{returns a \built{List} of terms resulting from BCFW recursion \emph{directly} for the gluon-component amplitude \emph{analytically} in terms of \emph{spinor} variables directly. This is an entirely independent implementation of recursion relative to that used by \funL{ymAmp}, which makes use of momentum-twistor variables. Moreover, the resulting formulae are almost always (dramatically) more compact and efficient than those obtained using, say,  \mbox{\funL{component}\brace{\var{gluonHelicities}}\,\brace{\funL{ymAmp}\brace{\var{$n$},\var{$k$}}}}\,. The reason for this is that the scheme employed guarantees that shifted expressions remain products of \emph{monomials} constructed out of kinematic invariants \funL{ab}, \funL{sb}, \funL{asb}. Even though this property is also true for \emph{super}amplitudes obtained in this scheme, the extraction of any particular component from a \funL{superFunction} requires the evaluation of determinants of the $\tilde\eta$-coefficients, which are difficult in general to realize as products of monomials of invariants.
As a simple example of this, consider
\mathematicaBox{
\mathematicaSequence{\funL[]{nice}\brace{\funL[]{component}\brace{\stateYM{p},\stateYM{m},\stateYM{p},\stateYM{m},\stateYM{p},\stateYM{m}}\,\brace{\funL[]{ymAmp}\brace{6,1}}}}{$\rule{0pt}{22pt}\displaystyle\Big\{\text{-}\frac{(\sb{3\,4}\ab{4\,6}\text{-}\ab{2\,6}\sb{2\,3})^4}{\ab{5\,6}\ab{6\,1}\sb{2\,3}\sb{3\,4}\langle1|(2\,3)|4]\langle5|(6\,1)|2]s_{5\,6\,1}},$\\
$\displaystyle\frac{(\ab{2\,4}\sb{4\,5}\text{-}\ab{2\,6}\sb{5\,6})^4}{\ab{1\,2}\ab{2\,3}\sb{4\,5}\sb{5\,6}\langle1|(2\,3)|4]\langle3|(1\,2)|6]s_{1\,2\,3}},$\\
$\displaystyle\frac{(\ab{2\,4}\sb{1\,2}\text{+}\ab{4\,6}\sb{6\,1})^4}{\ab{3\,4}\ab{4\,5}\sb{1\,2}\sb{6\,1}\langle3|(1\,2)|6]\langle5|(6\,1)|2]s_{6\,1\,2}}\Big\}$}
\mathematicaSequence{\funL[]{nice}\brace{\funL[1]{ymHelicityAmp}\brace{\stateYM{p},\stateYM{m},\stateYM{p},\stateYM{m},\stateYM{p},\stateYM{m}}}}{$\rule{0pt}{22pt}\displaystyle\Big\{\text{-}\frac{\sb{1\,5}^4\ab{2\,4}^4}{\ab{2\,3}\ab{3\,4}\sb{5\,6}\sb{6\,1}\langle2|(6\,1)|5]\langle4|(2\,3)|1]s_{5\,6\,1}},$\\
$\displaystyle\frac{\sb{1\,3}^4\ab{4\,6}^4}{\ab{4\,5}\ab{5\,6}\sb{1\,2}\sb{2\,3}\langle4|(2\,3)|1]\langle6|(1\,2)|3]s_{1\,2\,3}},$\\
$\displaystyle\frac{\sb{3\,5}^4\ab{2\,6}^4}{\ab{1\,2}\ab{6\,1}\sb{3\,4}\sb{4\,5}\langle6|(1\,2)|3]\langle2|(6\,1)|5]s_{6\,1\,2}}\Big\}$}
}
More generally, \fun{ymHelicityAmp} results in more compact analytic expressions more efficiently, with much reduced \built{LeafCount}. 
\mathematicaBox{
\mathematicaSequence{helExprn=\funL[1]{timed}@\funL[1]{ymHelicityAmp}\brace{\stateYM{p},\stateYM{p},\stateYM{p},\stateYM{m},\stateYM{m},\stateYM{p},\stateYM{p},\stateYM{p},\stateYM{m},\stateYM{m},\stateYM{m},\stateYM{p},\stateYM{p},\stateYM{m},\stateYM{m},\stateYM{m}};\\
cmptOfSuper=\funL[1]{timed}@\funL[1]{component}\brace{\stateYM{p},\stateYM{p},\stateYM{p},\stateYM{m},\stateYM{m},\stateYM{p},\stateYM{p},\stateYM{p},\stateYM{m},\stateYM{m},\stateYM{m},\stateYM{p},\stateYM{p},\stateYM{m},\stateYM{m},\stateYM{m}}[ \funL[]{timed}\brace{\funL[]{ymAmp}\brace{14,5}}];
}{}
&\scalebox{0.6}{\hspace{-30pt}Evaluation of \texttt{\textbf{ymHelicityAmp[]}} for $\mathcal{A}^{(5)}_{14}\big(+,+,+,-,-,+,+,+,-,-,-,+,+,-,-,-\big)$ required \textbf{3 seconds, 494 ms} to complete.}\\[-5pt]
&\scalebox{0.6}{\hspace{-30pt}Evaluation of \texttt{\textbf{ymAmp[]}} for $\mathcal{A}^{(5)}_{14}$ required \textbf{1 minutes, 41 seconds} to complete.}\\[-5pt]
&\scalebox{0.6}{\hspace{-30pt}Evaluation of the function \texttt{\textbf{component}} required \textbf{3 seconds, 504 ms} to complete.}\\[-5pt]
\mathematicaSequence{\{Length[\#],LeafCount[\#]\}\&/@\{helExprn,cmptOfSuper\}}{\{\{882,173\,751\},\{19404,2\,585\,289\}\}}
}\vspace{-20pt}\\[-55pt]}

\defnBox{helicityAmp}{\var{gluonHelicities}\patternTwo}{gives \built{Total} of \mbox{\fun{ymHelicityAmp}\brace{\var{gluonHelicities}}} when gluon helicities are specified $\{\stateYM{m},\stateYM{p}\}$. (This function will also work for graviton helicities $\{\stateGR{mm},\stateGR{pp}\}$, for which it gives the \built{Total} of \funL{grHelicityAmp}\brace{\var{helicities}}\,.)}

\defnBox{componentAmpTerms}{\var{componentStates}\patternTwo}{returns a \built{List} of terms contributing to the particular component amplitude involving the external \var{componentStates}.\\[-10pt]

Depending on the particular external states, \fun{componentAmpTerms} chooses \emph{different} underlying representations for the amplitude. If the \var{componentStates} include states that are \emph{not} gluons, then \fun{componentAmpTerms} simply calls \mbox{\funL{component}\brace{\var{componentStates}}\,\brace{\funL{ymAmp}\brace{\var{$n$},\var{$k$}}\,}}\,.\\[-10pt]

\emph{\textbf{However}}, if \var{componentStates} consists entirely of gluons (labelled by \{\stateYM{m},\stateYM{p}\}), then \fun{componentAmpTerms} calls \fun{ymHelicityAmp} \emph{but possibly for a rotated set of helicities (with the result appropriately rotated back to the case desired) so as to reduce the number of terms that result}. (To be clear, the resulting expression is not guaranteed to involve the \emph{fewest} terms in which this amplitude may be represented, but is almost always \emph{preferable} to the output of \fun{ymHelicitiyAmp}.)
}

\defnBox{componentAmp}{\var{componentStates}\patternTwo}{is \mbox{\built{Total}\brace{\fun{componentAmpTerms}\brace{\var{componentStates}}}\,.}}

\newpage
\vspace{0pt}\subsection[Local, Covariant, Helicity Partial Amplitudes in YM (via CHY)]{Local, Covariant, Helicity Partial Amplitudes in YM}\label{appendix:local_covariant_ym_amps}\vspace{-10pt}

\defnBox{ymLocalCovariantAmp}{\var{$n$}\pattern}{returns the sum of the terms encoded by \fun{ymLocalCovariantAmpData}\brace{\var{$n$}}.
}

\defnBox{ymLocalCovariantAmpData}{\var{$n$}\pattern}{returns a \built{List} of the form \{\texttt{expressionTerms},\{\texttt{numData}\}\} where \built{Total}\brace{\texttt{expressionTerms}} gives the $\var{n}$-particle pure-gluon tree amplitude for YM in terms of local poles arising from cubic, scalar Feynman graphs with coefficients \texttt{num}\brace{\var{i}\pattern} encoded by \texttt{numData}. For example,
\mathematicaBox{
\mathematicaSequence[1]{\{termList,numData\}=\funL[1]{ymLocalCovariantAmpData}\brace{5};\\
\funL[]{nice}\brace{\textnormal{\%}}}{$\rule{0pt}{22pt}\displaystyle\Big\{n_1\Big(\frac{1}{s_{1\,5}s_{2\,3}}\text{+}\frac{1}{s_{1\,2}s_{3\,4}}\text{+}\frac{1}{s_{1\,5}s_{3\,4}}\text{+}\frac{1}{s_{1\,2}s_{4\,5}}\text{+}\frac{1}{s_{2\,3}s_{4\,5}}\Big),\text{-}n_2\Big(\frac{1}{s_{1\,2}s_{3\,4}}\text{+}\frac{1}{s_{1\,5}s_{3\,4}}\Big),$\\
$\rule{0pt}{22pt}\displaystyle\text{-}n_3\Big(\frac{1}{s_{1\,5}s_{2\,3}}\text{+}\frac{1}{s_{2\,3}s_{4\,5}}\Big),\text{-}n_4\frac{1}{s_{1\,5}s_{3\,4}},\text{-}n_5\frac{1}{s_{1\,5}s_{2\,3}},n_6\Big(\frac{1}{s_{1\,5}s_{2\,3}}\text{+}\frac{1}{s_{1\,5}s_{3\,4}}\Big)\Big\}$}
\mathematicaSequence{numData[\![1]\!]}{\{\{num[1],-\funL{ss}[\funL{$\epsilon$}[1],\funL{$\epsilon$}[4]]\funL{ss}[\funL{$\epsilon$}[2],\funL{$\epsilon$}[1]]\funL{ss}[\funL{$\epsilon$}[3],\funL{$\epsilon$}[1]]-\ldots\}\}}
}
where the expressions for the \texttt{num}\brace{\var{i}\pattern} can be quite involved, depending on Lorentz-invariant squared-sums of momenta and polarization vectors.\\[-10pt]

To evaluate this expression for four-dimensional kinematics and specific polarizations, one can combine this data as follows:
\mathematicaBox{
\mathematicaSequence[3]{\{termList,numData\}=\funL[1]{ymLocalCovariantAmpData}\brace{5};\\
\mbox{\funL[]{evaluate}\brace{\{termList\!,numData\}/\!\!.\funL[]{helicityPolarizationRules}\brace{\stateYM{m},\stateYM{p},\stateYM{m},\stateYM{p},\stateYM{m}}};}\\
\{termSum,numRules\}=\{Total[termList],Rule@@@numData\}\!;\\
\funL[]{nice}\brace{\textnormal{\%}}}{$\displaystyle\Big\{\frac{138929}{26703}n_1\text{-}\frac{572}{621}n_2\text{-}\frac{982}{387}n_3\text{-}\frac{32}{621}n_4\text{-}\frac{4}{9}n_5\text{+}\frac{308}{621}n_6,\;\;\begin{array}{@{}c@{\rightarrow}l@{}}n_1&\frac{3575}{232}\\
n_2&\frac{460325}{9048}\\
n_3&\frac{1273325}{190008}\\
n_4&\text{-}\frac{29484625}{253344}\\
n_5&\text{-}\frac{14003125}{36192}\\
n_6&\text{-}\frac{107551175}{253344}\end{array}\Big\}$}
\mathematicaSequence{termSum/\!\!.numRules}{$\text{-}\frac{10440125}{640872}$}
\mathematicaSequence{\textnormal{\%}===\funL[]{componentAmpN}\brace{m,p,m,p,m}}{True}
}
}

\newpage
\vspace{-0pt}\subsectionAppendix{Numeric (or Evaluated) \emph{Component} Amplitudes of sYM}{appendix:numeric_component_amps}\vspace{-10pt}
%

\defnBox{helicityAmpN}{\var{gluonOrGravitonHelicities}\patternTwo}{gives \built{Total} of \fun{ymHelicityAmpN} or \fun{grHelicityAmpN} depending on whether the helicities specified are those of gluons or gravitons. 
}

\defnBox{ymHelicityAmpN}{\var{gluonHelicities}\patternTwo}{although similar to the evaluation of \funL{ymHelicityAmp}, \fun{ymHelicityAmpN} actually employs a \emph{numeric} recursion algorithm, where amplitudes appearing an successive stages of recursion are \emph{evaluated} for actively shifted spinors.\\[-10pt]

This strategy for recursion can appear to be the `fastest' one available---especially at large multiplicity. For example, we may compare 
\mathematicaBox{
\mathematicaSequence[1]{\funL[]{useRandomKinematics}\brace{14};\\
egCmpt=\funL[]{randomHelicityComponent}\brace{14,5}}{\{\stateYM{m},\stateYM{m},\stateYM{m},\stateYM{p},\stateYM{m},\stateYM{p},\stateYM{m},\stateYM{p},\stateYM{m},\stateYM{p},\stateYM{m},\stateYM{p},\stateYM{p},\stateYM{p}\}}
\mathematicaSequence{\funL[]{timed}@\funL[1]{ymHelicityAmpN}\brace{egCmpt};\\
\funL[]{timed}@\funL[1]{ymHelicityAmp}\brace{egCmpt};\\
\funL[]{timed}@\funL[]{evaluate}\brace{\textnormal{\%}};}{}
&\scalebox{0.58}{\hspace{-30pt}Evaluation of the function \texttt{\textbf{ymHelicityAmpN[]}} for $\mathcal{A}^{(5)}_{14}(-,-,-,+,-,+,-,+,-,+,-,+,+,+)$ required \textbf{5 seconds, 193 ms} to complete.}\\[-5pt]
&\scalebox{0.58}{\hspace{-30pt}Evaluation of the function \texttt{\textbf{ymHelicityAmp[]}} for $\mathcal{A}^{(5)}_{14}(-,-,-,+,-,+,-,+,-,+,-,+,+,+)$ required \textbf{1 minutes, 3 seconds} to complete.}\\[-5pt]
&\scalebox{0.58}{\hspace{-30pt}Evaluation of the function \texttt{\textbf{evaluate}} required \textbf{596 ms, 403 $\mu$s} to complete.}\\[-5pt]
}
Notice that although more than one minute was required to compute the helicity amplitude analytically, \emph{very} little time was required to \funL{evaluate} the expression. Thus, if one were interested in the amplitude evaluated at more than a single point in kinematic phase space (or in any analytic properties of the amplitude), the analytic expression is clearly superior, as \fun{ymHelictyAmpN} would require about 5 seconds every time.\\[-10pt]

A similar comment could be made for the costs/benefits associated with \funL{ymAmp} vs.\ \funL{ymHelicityAmp}: even if one were only interested in amplitudes involving gluons, the additional timed required to determine the superAmplitude---from which \emph{any} $\binom{n}{k\text{+}2}$ gluonic component amplitude may be extracted reasonably quickly---would be worth it. 
}

\defnBox{componentAmpTermsN}{\var{componentStates}\patternTwo}{returns a \built{List} of terms corresponding to \fun{ymHelicityAmpN} if \fun{componentStates} are all gluonic $\{\stateYM{m},\stateYM{p}\}$, or \mbox{\funL{component}\brace{\funL{ymAmpN}\brace{\var{$n$},\var{$k$}}}} otherwise. 
}

\defnBox{componentAmpN}{\var{componentStates}\patternTwo}{\mbox{\built{Total}\brace{\fun{componentAmpTermsN}\brace{\var{componentStates}}}.}
}

\newpage
\vspace{-0pt}\subsectionAppendix{Relations Among Component Amplitudes of sYM}{appendix:component_relations_sym}\vspace{-10pt}

\defnBox{inverseSoftFactorYM}{\var{$n$},\var{softLeg}\pattern}{gives the \emph{inverse} of the soft-factor \cite{Weinberg:1964ew}
\vspace{-5pt}\eq{\mathcal{A}_{\var{n}}\!(\ldots)=\big(\text{inverse-soft-factor}\big)\lim_{p_{\text{\var{softLeg}}}\!\to\!0}\mathcal{A}_{\var{n}+1}\!\big(\ldots,\text{\var{softLeg}},\ldots\big)\,,\vspace{-10pt}}
where |\var{softLeg}| labels the particle in $[\var{n}\text{+}1]$ whose momentum is becoming soft.\\[-10pt]

\textbf{\emph{Note}}: the \emph{sign} of \var{softLeg} determines whether this momentum is taken to be soft `holomorphically' or `anti-holomorphically'---whether its $\tilde\lambda$ or $\lambda$ is taken to zero, respectively. Below we give an example of each case:
\mathematicaBox{
\mathematicaSequence[3]{\funL[]{useReferenceKinematics}\brace{6};\\
egSoftSpinors=\funL[]{inverseSoftSpinors}[7];\\
spinorInverseSoftFactor=\funL[1]{inverseSoftFactorYM}[6,7];\\
\funL[]{nice}[\textnormal{\%}]}{$\rule{0pt}{22pt}\displaystyle\frac{\ab{6\,7}\,\ab{7\,1}}{\ab{6\,1}}$}
\mathematicaSequence{\funL[]{evaluateWithSpinors}\brace{egSoftSpinors}\,\brace{spinorInverseSoftFactor};\\
FullSimplify[\textnormal{\%}]}{$\rule{0pt}{22pt}\displaystyle\text{-}\frac{8085\,\epsilon^2}{77{+}2070\,\epsilon}$}
\mathematicaSequence[1]{polarizationSoftFactor=$\displaystyle\frac{\funL{ss}[\funL{$\epsilon$p}\texttt{[7],\funL{p}[1]]}}{\texttt{\funL{ss}[\funL{p}[7],\funL{p}[1]]}}\text{-}\frac{\funL{ss}[\funL{$\epsilon$p}\texttt{[7],\funL{p}[6]]}}{\texttt{\funL{ss}[\funL{p}[7],\funL{p}[6]]}}$;\\
FullSimplify[\funL[]{evaluateWithSpinors}\brace{egSoftSpinors}\,\brace{\textnormal{\%}}]}{$\displaystyle\rule{0pt}{22pt}\text{-}\frac{77{+}2070\,\epsilon}{8085\,\epsilon^2}$}
}
\mathematicaBox{
\mathematicaSequence[3]{\funL[]{useReferenceKinematics}\brace{6};\\
egSoftSpinors=\funL[]{inverseSoftSpinors}[-7];\\
spinorInverseSoftFactor=\funL[1]{inverseSoftFactorYM}[6,-7];\\
\funL[]{nice}[\textnormal{\%}]}{$\rule{0pt}{22pt}\displaystyle\frac{\sb{6\,7}\,\sb{7\,1}}{\sb{6\,1}}$}
\mathematicaSequence{\funL[]{evaluateWithSpinors}\brace{egSoftSpinors}\,\brace{spinorInverseSoftFactor};\\
FullSimplify[\textnormal{\%}]}{$\rule{0pt}{22pt}\displaystyle\frac{123000\,\epsilon^2}{154{+}1205\,\epsilon}$}
\mathematicaSequence[1]{polarizationSoftFactor=$\displaystyle\frac{\funL{ss}[\funL{$\epsilon$m}\texttt{[7],\funL{p}[1]]}}{\texttt{\funL{ss}[\funL{p}[7],\funL{p}[1]]}}\text{-}\frac{\funL{ss}[\funL{$\epsilon$m}\texttt{[7],\funL{p}[6]]}}{\texttt{\funL{ss}[\funL{p}[7],\funL{p}[6]]}}$;\\
FullSimplify[\funL[]{evaluateWithSpinors}\brace{egSoftSpinors}\,\brace{\textnormal{\%}}]}{$\displaystyle\rule{0pt}{22pt}\frac{125{+}1205\,\epsilon}{12300\,\epsilon^2}$}
}\\[-50pt]
}

\defnBox{egWardIdentity}{\var{seedComponentsStates}\patternTwo}{uses the \var{seedComponentStates} to find an example Ward identity often (but not always) involving those states.
\mathematicaBox{
\mathematicaSequence{egSeed=\funL[]{randomComponentStatesYM}\brace{1}\,\brace{6,1}}{\{\stateYM{psi}[1],\stateYM{psi}[1],\stateYM{psiBar}[1],\stateYM{psi}[1],\stateYM{psiBar}[1],\stateYM{psiBar}[1]\}}
\mathematicaSequence[1]{egTerms=\funL[1]{egWardIdentity}\brace{egSeed};\\
\funL[]{nice}\brace{egTerms}}{$\rule{0pt}{22pt}\displaystyle\big\{\mathcal{A}_{6}^{(1)}\!\big(|{\psi}\rangle_{({1})}^{\text{+}1/2},|{\psi}\rangle_{({1})}^{\text{+}1/2},|\bar{\psi}\rangle_{(\bar{1})}^{\text{-}1/2},|{\psi}\rangle_{({1})}^{\text{+}1/2},|\bar{\psi}\rangle_{(\bar{1})}^{\text{-}1/2},|\bar{\psi}\rangle_{(\bar{1})}^{\text{-}1/2}\big),$\\
$\displaystyle
\phantom{\big\{}\mathcal{A}_{6}^{(1)}\!\big(|{\psi}\rangle_{({2})}^{\text{+}1/2},|{\psi}\rangle_{({1})}^{\text{+}1/2},|\bar{\psi}\rangle_{(\bar{2})}^{\text{-}1/2},|{\psi}\rangle_{({1})}^{\text{+}1/2},|\bar{\psi}\rangle_{(\bar{1})}^{\text{-}1/2},|\bar{\psi}\rangle_{(\bar{1})}^{\text{-}1/2}\big),$\\
$\displaystyle\text{-}\mathcal{A}_{6}^{(1)}\!\big(|{\psi}\rangle_{({1})}^{\text{+}1/2},|{\psi}\rangle_{({2})}^{\text{+}1/2},|\bar{\psi}\rangle_{(\bar{2})}^{\text{-}1/2},|{\psi}\rangle_{({1})}^{\text{+}1/2},|\bar{\psi}\rangle_{(\bar{1})}^{\text{-}1/2},|\bar{\psi}\rangle_{(\bar{1})}^{\text{-}1/2}\big),$\\
$\displaystyle\text{-}\mathcal{A}_{6}^{(1)}\!\big(|{\psi}\rangle_{({1})}^{\text{+}1/2},|{\psi}\rangle_{({1})}^{\text{+}1/2},|\bar{\psi}\rangle_{(\bar{2})}^{\text{-}1/2},|{\psi}\rangle_{({2})}^{\text{+}1/2},|\bar{\psi}\rangle_{(\bar{1})}^{\text{-}1/2},|\bar{\psi}\rangle_{(\bar{1})}^{\text{-}1/2}\big)\big\}$}
\mathematicaSequence{egTerms/\!\!.\funL{amp}\brace{\var{$x$}\patternTwo}:>\funL[]{componentAmpN}\brace{\var{$x$}}}{$\rule{0pt}{22pt}\displaystyle\Big\{\text{-}\frac{41298509317}{126273931344},\frac{39365}{1026968},\frac{522665243}{1830056976},\frac{38785}{12418758}\Big\}$}
\mathematicaSequence{{Total}@\textnormal{\%}}{0}
}
All such Ward identities follow from Cramer's rule for minors---the fact that any particular column of $\tilde\eta$ coefficients $c_a\!\in\!\mathbb{C}^{k}$ can be decomposed into an arbitrary full-rank basis of $k$ other columns. Using the general representation of \funL{superFunction} objects and how {component} functions are extracted via \funL{component}, the generation of Ward identities is fairly trivial. 
}

\vspace{10pt}

\defnBoxTwo[]{egWardIdentity}{\var{$\mathcal{N}$}\pattern\optArg{4}\,}{\var{$n$}\pattern,\var{$k$}\pattern}{takes a randomly chosen initial state to seed a random example of a supersymmetric Ward identity (see e.g.~\cite{Grisaru:1977px,Grisaru:1976vm,Elvang:2009wd,Elvang:2010xn}): 
\mathematicaBox{
\mathematicaSequence[1]{egTerms=\funL[1]{egWardIdentity}\brace{4}\,\brace{6,1};\\
\funL[]{nice}\brace{egTerms}}{$\rule{0pt}{22pt}\displaystyle\big\{\mathcal{A}_{6}^{(1)}\!\big(+,|\bar{\psi}\rangle_{(\bar{2})}^{\text{-}1/2},|{\psi}\rangle_{(2)}^{\text{+}1/2},|\psi\rangle_{(2)}^{\text{+}1/2},-,|\bar{\psi}\rangle_{(\bar{2})}^{\text{-}1/2}\big),$\\
$\displaystyle
\,\text{-}\mathcal{A}_{6}^{(1)}\!\big(+,|\bar{\psi}\rangle_{(\bar{1})}^{\text{-}1/2},|{\psi}\rangle_{(1)}^{\text{+}1/2},|\psi\rangle_{(2)}^{\text{+}1/2},-,|\bar{\psi}\rangle_{(\bar{2})}^{\text{-}1/2}\big),$\\
$\displaystyle\phantom{\big\{}\mathcal{A}_{6}^{(1)}\!\big(+,|\bar{\psi}\rangle_{(\bar{1})}^{\text{-}1/2},|{\psi}\rangle_{(2)}^{\text{+}1/2},|\psi\rangle_{(1)}^{\text{+}1/2},-,|\bar{\psi}\rangle_{(\bar{2})}^{\text{-}1/2}\big)\big\}$}
\mathematicaSequence{egTerms/\!\!.\funL{amp}\brace{\var{$x$}\patternTwo}:>\funL[]{componentAmpN}\brace{\var{$x$}}}{$\rule{0pt}{22pt}\displaystyle\Big\{\frac{27708889}{67227003},\text{-}\frac{6220142222}{20369781909},\text{-}\frac{70182295}{657089739}\Big\}$}
\mathematicaSequence{{Total}@\textnormal{\%}}{0}
}\\[-30pt]
}

\newpage
\vspace{-0pt}\sectionAppendix{Color-Dressed Amplitudes of (pure) superYang-Mills}{appendix:colored_amps}\vspace{-0pt}
%
\vspace{-0pt}\subsectionAppendix{Color Factors: Bases and Reduction}{appendix:color_factors}\vspace{-10pt}

\defnBox{cyclicColorTraces}{\var{$n$}\pattern}{returns a \built{List} of color-index tensors of the form \mbox{\funL{tr}\brace{1,\var{perm}\patternTwo}} where \var{perm}$\in\!\mathfrak{S}([2,\var{n}])$.
}

\defnBox{basisColorFactors}{\var{$n$}\pattern}{returns a \built{List} of the (\var{$n$}{-}2)! color-factor tensors in the basis of (tree-level) color-tensors as arising in the DDM recursion scheme \cite{DelDuca:1999rs}. This basis consists of the \funL{colorFactor}\texttt{[}1,\var{$a{}_1$},\ldots,\var{$a{}_{\text{-}1}$},\var{$n$}\texttt{]} where $\{$\var{$a{}_1$},\ldots,\var{$a{}_{\text{-}1}$}$\}$$\!\in\!\mathfrak{S}[2,\ldots,\text{\var{$n$}}\text{-}1]$ is a permutation of the indices [\var{$n$}]$\backslash$$\{1,$\var{$n$}$\}$.
\mathematicaBox{\mathematicaSequence{\funL[1]{basisColorFactors}[5]}{\{\funL{colorFactor}\brace{1,2,3,4,5}\,,\funL{colorFactor}\brace{1,2,4,3,5}\,,\\
\phantom{\{}\funL{colorFactor}\brace{1,3,2,4,5}\,,\funL{colorFactor}\brace{1,3,4,2,5}\,,\\ 
\phantom{\{}\funL{colorFactor}\brace{1,4,2,3,5}\,,\funL{colorFactor}\brace{1,4,3,2,5}\,\}}}

Recall that these color tensors are defined by
\eq{\text{\funL{colorFactor}\texttt{[}\var{$a$},\var{$b_{1}$},\ldots,\var{$b_{\text{-}1}$},\var{$c$}\texttt{]}}\,\,\Leftrightarrow\,\sum_{\r{e_i}}f^{\text{\var{$a$}\,\var{$b{}_{1}$}}\,\r{e_1}}f^{\r{e_1}\,\text{\var{$b{}_{2}$}}\,\r{e_2}}\cdots f^{\r{e_{\text{-}3}}\,\text{\var{$b{}_{\text{-}2}$}}\,\r{e_{\text{-}2}}}f^{\r{e_{\text{-}1}}\,\text{\var{$b{}_{\text{-}1}$}\,\var{$c$}}},\vspace{-10pt}}
where $f^{c_1\,c_2\,c_3}$ are the structure constants of some Lie algebra $\mathfrak{g}$ and $c_a\!\in\![\dim(\mathfrak{g})]$ indexes the `color' of the $a$th gluon (in the adjoint representation of the algebra). 
}

\defnBox{colorFactorReductionRule}{}{a \built{Rule} that expands any color-tensor defined by (\ref{tree_like_color_tensors_defined}) into those of the basis given by \fun{basisColorFactors}\texttt{[}\var{$n$}\texttt{]}.
\mathematicaBox{\mathematicaSequence{\funL{colorFactor}@@\funL[]{randomPerm}[6]}{\funL{colorFactor}[3,5,2,6,1,4]}\mathematicaSequence{\textnormal{\%}/\!\!.\builtL[1]{colorFactorReductionRule}}{{-}\funL{colorFactor}\brace{1,4,2,3,5,6}{+}\funL{colorFactor}\brace{1,4,2,5,3,6} \hspace{80pt}~ {+}\funL{colorFactor}\brace{1,4,3,5,2,6}{-}\funL{colorFactor}\brace{1,4,5,3,2,6}}}
This rule makes use of only the Jacobi identity satisfied by the structure constants of any Lie algebra.\\[-10pt]

\textbf{\emph{Note}}: the rule \built{colorFactorReductionRule} \emph{also} implements the cyclic symmetry of color-\funL{tr}'s. 
}

\defnBox{colorFactorToTracesRule}{}{a \built{Rule} that expands any color-tensor represented by `\funL{colorFactor}' (as defined in (\ref{tree_like_color_tensors_defined})) into the `trace basis' color tensors consisting of tensors of the form%
\eq{\text{\fun{tr}\texttt{[}\var{a${}_{1}$},\,\ldots,\var{$a{}_{\text{-}1}$}\texttt{]}}\equivR\mathrm{tr}\big(T^{\text{\var{$a{}_{1}$}}}_{\mathbf{R}}\!\!\cdots T_{\mathbf{R}}^{\text{\var{$a{}_{\text{-}1}$}}}\big)}%
where $T_\mathbf{R}^{\text{\var{$c{}_{a}$}}}$ is taken as some particular $\dim(\mathbf{R})\!\times\!\dim(\mathbf{R})$ matrix encoding the \emph{generators} of the representation $\mathbf{R}$ of the Lie algebra $\mathfrak{g}$ and \var{$c{}_{a}$}$\,\in\![\dim(\mathfrak{g})]$ is an index encoding the `color' of the \var{$a$}th gluon; the normalization of these tensors are chosen so that they are related to the \emph{structure constants} defining the Lie algebra $\mathfrak{g}$ via
\eq{f^{c_1\,c_2\,c_3}\equivR\mathrm{tr}\big(T_\mathbf{R}^{c_1}T_\mathbf{R}^{c_2}T_\mathbf{R}^{c_3}\big){-}\mathrm{tr}\big(T_\mathbf{R}^{c_1}T_\mathbf{R}^{c_3}T_\mathbf{R}^{c_2}\big)\,.\vspace{-10pt}}
\mathematicaBox{\mathematicaSequence{\funL{colorFactor}[1,2,3]/\!\!.\builtL[1]{colorFactorToTracesRule}}{\funL{tr}\brace{1,2,3}-\funL{tr}\brace{1,3,2}}
\mathematicaSequence{\funL{colorFactor}@@\funL[]{randomPerm}[5]}{\funL{colorFactor}\brace{3,1,4,2,5}}
\mathematicaSequence{\textnormal{\%}/\!\!.\builtL[1]{colorFactorToTracesRule}}{{-}\funL{tr}\brace{1,2,5,4,3}\,{+}\funL{tr}\brace{1,3,2,5,4}\,{-}\funL{tr}\brace{1,3,4,2,5}\,{+}\mbox{\funL{tr}\brace{1,3,4,5,2}}\\
{-}\funL{tr}\brace{1,3,5,2,4}\,{+}\funL{tr}\brace{1,4,2,5,3}\,{-}\funL{tr}\brace{1,4,5,2,3}\,{+}\funL{tr}\brace{1,5,2,4,3}}}
For most of the physics literature, the representation $\mathbf{R}$ is taken as the $\mathbf{N}$-dimensional, fundamental representation of the gauge group $\mathfrak{g}\!=\!SU(N)$; in this case, the connection between \fun{colorFactor} and \fun{tr} tensors follows from the Fierz identity. However, the same relationship between them always exists when $\mathbf{R}$ is taken to be the adjoint representation of any Lie algebra---in which case the relation follows from the Jacobi identity alone. We choose to be agnostic about the representation to avoid any unnecessary reference to any particular $\mathfrak{g}$. 
}

\vspace{10pt}\subsectionAppendix{Graphical Representation of Color Factors}{appendix:draw_color_factors}\vspace{-10pt}

\defnBox{drawColorFactor}{\fun{colorFactor}\brace{\var{indexSequence}\patternTwo}}{graphically represents the abstract symbol \funL{colorFactor}\brace{} as a graph whose vertices represent structure constants with internal edges representing color indices being summed in the tensor definition of (\ref{tree_like_color_tensors_defined}):
\mathematicaBox{\mathematicaSequence{RandomChoice\brace{\funL[]{basisColorFactors}\brace{8}\,}}{\funL{colorFactor}\brace{1,7,5,3,4,6,2,8}}
\mathematicaSequence{
\funL[1]{drawColorFactor}\brace{\textnormal{\%}}}{\fig{-20pt}{0.8}{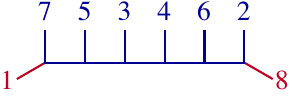}}}\vspace{-40pt}
}

\defnBox{drawColorTrace}{\fun{tr}\brace{\var{indexSequence}\patternTwo}}{graphically represents the abstract symbol \funL{tr}\brace{\var{indexList}} as a graph whose vertices represent the generators $T_{\mathbf{R}}$ of some Lie algebra in an arbitrary (matrix) representation $\mathbf{R}$, traced over its $\dim(\mathbf{R})$ internal indices.
\mathematicaBox{
\mathematicaSequence{\funL[1]{drawColorTrace}\brace{\funL{tr}\brace{3,6,2,4,5,1}}}{\fig{-28pt}{0.45}{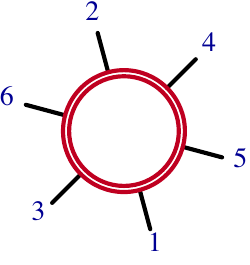}}
}\vspace{-0pt}
}

\vspace{-0pt}\subsectionAppendix{Analytic, Color-Dressed Amplitudes}{appendix:color_dressed_amplitudes}\vspace{-0pt}

\vspace{-0pt}\subsubsectionAppendix{Symbolic Color-Tensor Expansions of Amplitudes }{appendix:symbolic_reps}\vspace{-10pt}

\defnBox{colorDressedAmp}{\var{$n$}\pattern}{gives a \emph{symbolic} form for the \var{$n$}-particle color-dressed tree-amplitude amplitude of sYM in terms of color tensors and partial (ordered) amplitudes as obtained by on-shell recursion (equivalently, the DDM expansion \cite{DelDuca:1999rs}).
\eq{\fwbox{0pt}{\hspace{-45pt}\fun{colorDressedAmp}\brace{\var{n}}\,\,\,\Leftrightarrow\hspace{-18pt}\sum_{\r{\sigma}\in\mathfrak{S}([2,\var{n}{-}1])}\hspace{-10pt}\funL{colorFactor}\brace{1,\r{\sigma_1},\ldots,\r{\sigma_{\text{-}1}},\var{n}}\,\mathcal{A}_\var{n}\!\big(1,\r{\sigma_1},\ldots,\r{\sigma_{\text{-}1}},\var{n}\big)\,.}\label{color_factor_expansion}}
This representation involves $(\var{n}{-}2)!$ terms. 
}

\defnBox{colorTraceDressedAmp}{\var{$n$}\pattern}{gives a \emph{symbolic} form for the \var{$n$}-particle color-dressed tree-amplitude amplitude of sYM in terms of color `traces'
\eq{\fwbox{0pt}{\hspace{-45pt}\fun{colorTraceDressedAmp}\brace{\var{n}}\,\,\,\Leftrightarrow\hspace{-18pt}\sum_{\r{\sigma}\in\mathfrak{S}([2,\var{n}])}\hspace{-10pt}\funL{tr}\brace{1,\r{\sigma_1},\ldots,\r{\sigma_{\text{-}1}}}\,\mathcal{A}_\var{n}\!\big(1,\r{\sigma_1},\ldots,\r{\sigma_{\text{-}1}}\big)\,.}\label{color_trace_exapansion}}
where the sum over $(\var{n}{-}1)!$ permutations is more often expressed as `$\mathfrak{S}([\var{n}])/\mathbb{Z}_{\var{n}}$' (all permutations modulo cyclic).\\[-10pt]

Although the `trace' expansion is arguably more familiar (if quixotically so), it is worth noting that the equivalence between the trace expansion (\ref{color_factor_expansion}) and the \funL{colorFactor} expansion (\ref{color_factor_expansion}) relies upon (or can be viewed as implying) the so-called `KK relations':
\mathematicaBox{
\mathematicaSequence{colorTraceForm=\funL[1]{colorTraceDressedAmp}\brace{7};\\
Length[\textnormal{\%}]}{720}
\mathematicaSequence{colorFactorForm=\funL[1]{colorDressedAmp}\brace{7};\\
Length[\textnormal{\%}]}{120}
\mathematicaSequence{(colorFactorForm-colorTraceForm)/\!\!.\builtL[]{colorFactorToTracesRule};\\
Expand[\textnormal{\%}/\!\!.\builtL[]{kkAmpReductionRule}]}{0}
}
\vspace{-50pt}}

\newpage
\vspace{-0pt}\subsubsectionAppendix{Analytic Color-Dressed Amplitudes of sYM}{appendix:analytic_color_dresed_amps}\vspace{-15pt}

\defnBox{colorDressedYMampSeeds}{\var{$n$}\pattern,\var{$k$}\pattern}{returns a \built{List} of `seed' terms contributing to the color-dressed \var{$n$}-point N${}^{\text{\var{$k$}}}$MHV  YM amplitude as obtained via the `DDM' \cite{DelDuca:1999rs} recursion scheme---the basis of color tensors given by \mbox{\funL{basisColorFactors};} these `seeds' correspond to precisely those of the color-ordered partial amplitude \mbox{\funL{ymAmp}\brace{\var{$n$},\var{$k$}}}, combined with permuted leg images which leave invariant (or $GL(\var{k}{+}2)$-equivalent) the $\tilde\eta$-dependence of the terms appearing in the ordered amplitude. The complete color-dressed amplitude, given by \mbox{\fun{colorDressedYMamp}\brace{\var{$n$},\var{$k$}}}, is obtained by adding to the terms of \mbox{\fun{colorDressedYMampSeeds}\brace{\var{$n$},\var{$k$}}} those images with \emph{inequivalent} $\tilde\eta$-dependence. 
\mathematicaBox{
\mathematicaSequence{Length[\funL[1]{colorDressedYMampSeeds}[6,1]]}{3}\\[-22pt]
\mathematicaSequence{\funL[1]{colorDressedYMampSeeds}[6,1][\![2]\!]}{\funL{superFunction}$\left[\rule{0pt}{12pt}\right.$\\ \scalebox{0.95}{$\displaystyle{-}\frac{\text{\texttt{\funL{colorFactor}[1,2,3,5,4,6]}}}{\text{\funL{ab}[1,2]\funL{ab}[2,3]\funL{asb}[1,\funL{p}[2,3],5]\funL{asb}[3,\funL{p}[1,2],6]\funL{s}[1,2,3]\funL{sb}[4,5]\funL{sb}[4,6]}}$}\\ \scalebox{0.95}{$\displaystyle{+}\frac{\text{\texttt{\funL{colorFactor}[1,3,2,5,4,6]}}}{\text{\funL{ab}[1,3]\funL{ab}[2,3]\funL{asb}[1,\funL{p}[2,3],5]\funL{asb}[2,\funL{p}[1,3],6]\funL{s}[1,2,3]\funL{sb}[4,5]\funL{sb}[4,6]}}$} \scalebox{0.95}{$\displaystyle{-}\frac{\text{\texttt{\funL{colorFactor}[1,2,3,4,5,6]}}}{\text{\funL{ab}[1,2]\funL{ab}[2,3]\funL{asb}[1,\funL{p}[2,3],4]\funL{asb}[3,\funL{p}[1,2],6]\funL{s}[1,2,3]\funL{sb}[4,5]\funL{sb}[5,6]}}$} \scalebox{0.95}{$\displaystyle{-}\frac{\text{\texttt{\funL{colorFactor}[1,3,2,4,5,6]}}}{\text{\funL{ab}[1,3]\funL{ab}[2,3]\funL{asb}[1,\funL{p}[2,3],4]\funL{asb}[2,\funL{p}[1,3],6]\funL{s}[1,2,3]\funL{sb}[4,5]\funL{sb}[5,6]}}$\!\hspace{-1pt},} $\left.\rule{0pt}{12pt}\{\{0,0,0,\funL{sb}\t{[5,6]},\funL{sb}\t{[6,4]},\funL{sb}\t{[4,5]}\}\}\right]$}
}
\vspace{-10pt}}

\defnBox{colorDressedYMamp}{\var{$n$}\pattern,\var{$k$}\pattern}{gives a \built{List} of terms contributing to the color-dressed \var{$n$}-point N${}^{\var{k}}$MHV YM amplitude as obtained via the `DDM' \cite{DelDuca:1999rs} recursion scheme---the basis of color tensors given by \funL{basisColorFactors}\brace{\var{$n$}}. \textbf{Note}: these terms are grouped into \funL{superFunction} objects which share a common (or $\text{GL}(\var{k}{+}2)$-equivalent) $\tilde\eta$-dependence through $\delta^{\var{k}\!\times\!4}\big(C\!\cdot\!\tilde\eta\big)$.
\mathematicaBox{
\mathematicaSequence{\funL[1]{colorDressedYMamp}[5,1]}{\scalebox{0.875}{$\left\{\rule{0pt}{20pt}\right.$\funL{superFunction}$\displaystyle\left[\frac{\text{\texttt{\funL{colorFactor}[1,4,3,2,5]}}}{\text{\funL{sb}[1,4]\funL{sb}[2,3]\funL{sb}[2,5]\funL{sb}[3,4]\funL{sb}[5,1]}}\right.$} \scalebox{0.875}{$\displaystyle{-}\frac{\text{\texttt{\funL{colorFactor}[1,3,4,2,5]}}}{\text{\funL{sb}[1,3]\funL{sb}[2,4]\funL{sb}[2,5]\funL{sb}[3,4]\funL{sb}[5,1]}}$}\scalebox{0.875}{$\displaystyle{-}\frac{\text{\texttt{\funL{colorFactor}[1,4,2,3,5]}}}{\text{\funL{sb}[1,4]\funL{sb}[2,3]\funL{sb}[2,4]\funL{sb}[3,5]\funL{sb}[5,1]}}$}\\[5pt]\scalebox{0.875}{$\displaystyle{-}\frac{\text{\texttt{\funL{colorFactor}[1,2,4,3,5]}}}{\text{\funL{sb}[1,2]\funL{sb}[2,4]\funL{sb}[3,4]\funL{sb}[3,5]\funL{sb}[5,1]}}$}\scalebox{0.875}{$\displaystyle{-}\frac{\text{\texttt{\funL{colorFactor}[1,3,2,4,5]}}}{\text{\funL{sb}[1,3]\funL{sb}[2,3]\funL{sb}[2,4]\funL{sb}[4,5]\funL{sb}[5,1]}}$}\\[5pt]\scalebox{0.875}{$\displaystyle{+}\frac{\text{\texttt{\funL{colorFactor}[1,2,3,4,5]}}}{\text{\funL{sb}[1,2]\funL{sb}[2,3]\funL{sb}[3,4]\funL{sb}[4,5]\funL{sb}[5,1]}},$}\scalebox{0.875}{$\displaystyle\left.\{\{0,0,\frac{\text{\funL{sb}[4,5]}}{\text{\funL{ab}[1,2]}},\frac{\text{\funL{sb}[5,3]}}{\text{\funL{ab}[1,2]}},\frac{\text{\funL{sb}[3,4]}}{\text{\funL{ab}[1,2]}}\}\}\right]$$\left.\rule{0pt}{20pt}\right\}$}}
}
\vspace{-10pt}}

\vspace{10pt}\subsectionAppendix{Numeric Color-Dressed Amplitudes in superYang-Mills}{appendix:color_dressed_amplitudes}\vspace{-10pt}

\defnBox{colorDressedYMampN}{\var{$n$}\pattern,\var{$k$}\pattern}{similar to the correspondence between \funL{ymAmp} and \funL{ymAmpN}, in that some numerical routines are used to more efficiently obtain a form of the evaluated amplitude given the specified kinematic data.
}

\vspace{10pt}\subsectionAppendix{Local, Covariant, \emph{Color-Kinematic-Dual} Helicity Amplitudes in YM}{appendix:color_dressed_amplitudes}\vspace{-10pt}

\defnBox{ymLocalCovariantColorDressedAmp}{\var{$n$}\pattern}{returns the sum of the terms encoded by \fun{ymLocalCovariantColorDressedAmpData}\brace{\var{$n$}}.
}

\defnBox{ymLocalCovariantColorDressedAmpData}{\var{$n$}\pattern}{returns a \built{List} of terms \{\texttt{expressionTerms},\{\texttt{numData}\}\} where \built{Total}\brace{\texttt{expressionTerms}} gives a \emph{color-kinematic-dual}, \emph{\textbf{local, covariant}} expression for the color-dressed $\var{n}$-particle pure-gluon tree amplitude for YM. For example,
\mathematicaBox{\mathematicaSequence[1]{
\{termList,numData\}=\funL[1]{ymLocalCovariantColorDressedAmpData}[5];\\[-2pt]
\funL[]{nice}[termList/\!.Thread[Rule[\funL[]{basisColorFactors}\brace{5},c/@Range[6]]]]}{$\displaystyle\rule{0pt}{24pt}\left\{\frac{c_1\,n_{1}}{s_{1\,2}s_{4\,5}},\frac{c_2\,n_{2}}{s_{1\,2}s_{3\,5}},\frac{c_3\,n_{3}}{s_{1\,3}s_{4\,5}},\frac{c_4\,n_{4}}{s_{1\,3}s_{2\,5}},\frac{c_5\,n_{5}}{s_{1\,4}s_{3\,5}},\frac{c_6\,n_{6}}{s_{1\,4}s_{2\,5}},\frac{(c_1\text{-}c_2)(n_{1}\text{-}n_{2})}{s_{1\,2}s_{3\,4}},\frac{(c_3\text{-}c_4)(n_{3}\text{-}n_{4})}{s_{1\,3}s_{2\,4}},\right.$\\[4pt]
$\displaystyle\left.\frac{(c_5\text{-}c_6)(n_{5}\text{-}n_{6})}{s_{1\,4}s_{2\,3}},\frac{\!(c_1\text{-}c_3\text{-}c_5\text{+}c_6)(n_{1}\text{-}n_{3}\text{-}n_{5}\text{+}n_{6})\!}{s_{1\,5}s_{2\,3}},\frac{\!(c_2\text{-}c_3\text{+}n_4\text{-}n_5)(n_{2}\text{-}n_{3}\text{+}n_{4}\text{-}n_{5})\!}{s_{1\,5}s_{2\,4}},\right.$\\[4pt]
$\displaystyle\left.\frac{\!(c_1\text{-}c_2\text{-}c_4\text{+}c_6)(n_{1}\text{-}n_{2}\text{-}n_{4}\text{+}n_{6})\!}{s_{1\,5}s_{3\,4}},\frac{\!(c_1\text{-}c_3)(n_{1}\text{-}n_{3})\!}{s_{2\,3}s_{4\,5}},\frac{\!(c_2\text{-}c_5)(n_{2}\text{-}n_{5})\!}{s_{2\,4}s_{3\,5}},\frac{\!(c_4\text{-}c_6)(n_{4}\text{-}n_{6})\!}{s_{2\,5}s_{3\,4}}\right\}$\\[-10pt]}
\mathematicaSequence[2]{\funL[]{useReferenceKinematics}[5];\\[-3pt]
nNumData=\funL[]{evaluate}[numData/\!\!.\funL[]{helicityPolarizationRules}[\stateYM{m},\stateYM{p},\stateYM{m},\stateYM{p},\stateYM{m}]\!]\!]\!;\\[-3pt]
\funL[]{nice}[nNumData]}{$\rule{0pt}{55pt}\displaystyle\left(\rule{0pt}{50pt}\right.\begin{array}{lc}n_1&\frac{3575}{232}\\
n_2&\frac{460325}{9048}\\
n_3&\frac{1273325}{190008}\\
n_4&\text{-}\frac{29484625}{253344}\\
n_5&\text{-}\frac{14003125}{36192}\\
n_6&\text{-}\frac{107551175}{253344}\end{array}\left.\rule{0pt}{50pt}\right)$\\[-5pt]}
\mathematicaSequence{Expand[Total[\funL[]{evaluate}[termList]/\!\!.Rule@@@nNumData]]}{$\rule{0pt}{20pt}\displaystyle{-}\frac{10440125}{640872}\funL{colorFactor}\brace{1,2,3,4,5}\,{-}\frac{614125}{249642}\funL{colorFactor}\brace{1,2,4,3,5}$\\[5pt]
$\displaystyle{+}\frac{614125}{16254}\funL{colorFactor}\brace{1,3,2,4,5}\,{+}\frac{1228250}{247779}\funL{colorFactor}\brace{1,3,4,2,5}$\\[5pt]
$\displaystyle{+}\frac{1228250}{284013}\funL{colorFactor}\brace{1,4,2,3,5}\,{+}\frac{20880250}{5557329}\funL{colorFactor}\brace{1,4,3,2,5}$
}
\mathematicaSequence{\textnormal{\%}===Total[\funL[]{component}\brace{\stateYM{m},\stateYM{p},\stateYM{m},\stateYM{p},\stateYM{m}}\,\brace{\funL[]{colorDressedYMampN}\brace{5,1}}}{True}
}
Notice that this form is \emph{manifestly} color-kinematic dual, and directly generalizes to \funL{grLocalCovariantAmp}.\vspace{-40pt}
}

\newpage
\vspace{-0pt}\sectionAppendix{Tree Amplitudes of (pure) superGravity}{appendix:supergravity_trees}\vspace{-0pt}
%
\vspace{-0pt}\subsectionAppendix{Analytic Expressions for superAmplitudes in superGravity}{appendix:analytic_trees_in_sugra}\vspace{-0pt}

\vspace{0pt}\subsubsectionAppendix{Analytic Expressions for MHV Amplitudes in superGravity}{appendix:mhv_in_sugra}\vspace{-10pt}

\defnBox{grAmpBGK}{\var{$n$}\pattern}{returns the expression for MHV amplitudes in superGravity as derived by BGK in \cite{Berends:1988zp} (see also~\cite{Mason:2009afn,Spradlin:2008bu}); specifically,
\vspace{4pt}\eq{\fwbox{0pt}{\hspace{-30pt}\fun{grAmpBGK}\brace{\var{n}}\,\Leftrightarrow\hspace{-10pt}\sum_{\b\sigma\in\mathfrak{S}([\b{2},\b{n\text{-}2}])}\!\!\!\left(\frac{1}{\ab{\r1\,\r{n\text{-}1}}\ab{\r1\,\r{n}}^2\ab{\r{n}\,\r{n\text{-}1}}^2\ab{\r1\,\b{\sigma_2}}}\prod_{\g{a}=\b{2}}^{\b{n\text{-}2}}\frac{\langle\r{n}|(\b{\sigma_{\!\g{a}}}\cdots\b\sigma_{\!\g{n\text{-}2}}\,\r{n\text{-}1})|\b{\sigma}_{\!\g{a}}]}{\ab{\b\sigma_{\!\g{a}}\,\b{\sigma}_{\!\g{a{+}1}}}\ab{\b\sigma_{\!\g{a}}\,\r{n}}}\right)\,.}\label{bgk_formula}}

}

\defnBox{grAmpHodges}{\var{$n$}\pattern}{returns the expression for MHV amplitudes in superGravity as derived by Hodges in \cite{Hodges:2011wm}; specifically, 
\eq{\fun{grAmpHodges}\brace{\var{n}}\Leftrightarrow(\text{-}1)^{\sigma(\var{n},\r{\vec{\alpha}},\b{\vec{\beta}})}\frac{\left|\Phi\right|_{\r{\vec{\alpha}}}^{\b{\vec{\beta}}}}{\ab{\r{\alpha_1}\,\r{\alpha_2}}\ab{\r{\alpha_2\,\alpha_3}}\ab{\r{\alpha_3\,\alpha_1}}\ab{\b{\beta_1\,\beta_2}}\ab{\b{\beta_2\,\beta_3}}\ab{\b{\beta_3\,\beta_1}}}\,,\label{hodges_mhv_formula}}
where 
\eq{\sigma(\var{n},\r{\vec{\alpha}},\b{\vec{\beta}})\equivR1{+}\var{n}{+}\sum_{i=1}^3\!(\r{\alpha_i}\text{+}\b{\beta_i})\,,}
$\r{\vec{\alpha}},\b{\vec{\beta}}\!\in\!\binom{[\var{n}]}{3}$ are arbitrary three-tuples, and $\left|\Phi\right|_{\r{\vec{\alpha}}}^{\b{\vec{\beta}}}$ is the determinant of $\Phi\!\Leftrightarrow\,$\fun{hodgesPhiMat}\brace{\var{$n$}} after the rows $\{\r{\alpha_1},\r{\alpha_2},\r{\alpha_3}\}$ and columns $\{\b{\beta_1},\b{\beta_2},\b{\beta_3}\}$ have been removed. Hodges showed that this expression is permutation invariant and independent of the choices for $\r{\vec{\alpha}},\b{\vec{\beta}}$.\\[-10pt]

The function \fun{grAmpHodges} by default chooses the sets $\r{\vec{\alpha}}\!=\!\{\r{1},\r{2},\r{3}\}$ and $\b{\vec{\beta}}\!=\!\{\b{4},\b{5},\b{6}\}$ (with the later understood to be taken mod \var{$n$}). 
}

\defnBox{hodgesPhiMat}{\var{$n$}\pattern}{encodes the matrix $\Phi$ of (\ref{hodges_mhv_formula}) whose entries are given by 
\eq{\Phi_{a\,b}\equivR\left\{\begin{array}{@{}l@{$\;\;\;$}l}\frac{\ab{a\,b}}{\sb{a\,b}}&\text{if\;}a\neq b\\
\phi_a&\text{if\;}a=b\end{array}\right.\label{hodges_phi_mat_defined}}
where the diagonal entries are given by the `graviton (holomorphic) soft factors'
\eq{\phi_{\r{a}}\equivR\sum_{\b{j}\in[\var{n}]\backslash\{\r{a}\}}\frac{\ab{\r{a\text{-}1}\,\b{j}}\sb{\r{a}\,\b{j}}\ab{\b{j}\,\r{a\text{+}1}}}{\ab{\r{a}\,\b{j}}\ab{\r{a\text{-}1}\,\r{a}}\ab{\r{a\,a\text{+}1}}}\,.\label{graviton_holo_soft_factors}}
}

\vspace{10pt}\subsubsectionAppendix{Analytic Expressions for N${}^k$MHV superAmplitudes in superGravity}{appendix:general_case}\vspace{-10pt}

\defnBox{grAmp}{\var{$n$}\pattern,\var{$k$}\pattern}{returns a \built{List} of contributions to the \var{$n$}-particle N${}^{\var{k}}$MHV superGravity tree amplitude. For MHV and $\bar{\text{MHV}}$ amplitudes ($\var{k}\!\in\!\{0,\var{n}\text{-}4\}$), Hodges' formula is used (see e.g.~\fun{grAmpHodges} above), while for all other N$^{\var{k}}$MHV degrees \fun{grAmp}\brace{\var{$n$},\var{$k$}} is equivalent to \funL{grAmpBCFW}\brace{\var{$n$},\var{$k$}}\,.
}

\newpage
\vspace{10pt}\subsubsectionAppendix{On-Shell Recursion of Amplitudes in superGravity}{appendix:bcfw_for_sgr}\vspace{-10pt}

\defnBox{grAmpDenominators}{\var{$n$}\pattern,\var{$k$}\pattern}{returns a \built{List} of `double-copy \emph{denominators}' used in the construction of the BCFW representation of the \var{$n$}-particle N${}^{\var{k}}$MHV tree amplitude in superGravity as described in \cite{Bourjaily:2023ycy} (see also~\cite{structureOfOnShellGR}). Specifically, as every term in BCFW for sGR is \emph{proportional to} the square of the same term in sYM, we define this constant of proportionality it's \emph{denominator} $D$---for the reason that if we define the $i$th contribution to Yang-Mills to involve a \emph{numerator} $\mathfrak{n}_i\equivR D_i\,\mathfrak{f}_i$, then contribution in sYM is $\mathfrak{n}_i/D_i=\mathfrak{f}_i$, and the corresponding contribution to gravity will be $\mathfrak{n}_i^2/D_i$.\\[-10pt]

In the standard recursion scheme, these denominators are ensured to be products of monomials involving spinor invariants (angle/square brackets). 
\mathematicaBox{
\mathematicaSequence{egCmpt=\funL[]{randomComponentStatesGR}\brace{6,1}}{\{\stateGR{psi}[1,4,7],\stateGR{phiBar}[1,2,3,7],\stateGR{psiBar}[1,5,8],\\
\phantom{\{}\stateGR{psi}[1,3,5],\stateGR{psiBar}[3,4,6],\stateGR{phiBar}[1,4,5,7]\}}
\mathematicaSequence{\funL[]{nice}\brace{\funL{amp}@@egCmpt}}{$\mathcal{A}_{6}^{(1)}\!\big(|\psi\rangle_{(147)}^{\text{+}1/2},|\bar{\phi}\rangle_{(\bar{1237})}^{0},|\bar{\psi}\rangle_{(\bar{158})}^{\text{-}1/2},|\psi\rangle_{(135)}^{\text{+}1/2},|\bar{\psi}\rangle_{(\bar{346})}^{\text{-}1/2},|\bar{\phi}\rangle_{(\bar{1457})}^{0}\big)$}
\mathematicaSequence[1]{denList=\funL[1]{grAmpDenominators}\brace{6,1};\\
\funL[]{nice}\brace{denList}}{$\rule{0pt}{22pt}\displaystyle\Big\{{-}\frac{\sb{2\,3}\sb{5\,6}\ab{1\,2}\ab{3\,4}\ab{5\,6}\langle1|(2\,3)|4]\langle5|(6\,1)|2]}{\ab{1\,5}\langle1|(3\,4)|2]},$\\[3pt]
$\displaystyle\phantom{\Big\{{-}}\frac{\sb{2\,3}\sb{5\,6}\ab{1\,2}\ab{4\,5}\langle1|(2\,3)|4]\langle3|(1\,2)|6]s_{1\,2\,3}}{\langle1|(2\,3)|6]^2},$\\[3pt]
$\displaystyle\phantom{\Big\{}{-}\frac{\sb{1\,2}\sb{3\,4}\sb{5\,6}\ab{1\,2}\ab{4\,5}\langle3|(1\,2)|6]\langle5|(6\,1)|2]}{\sb{2\,5}\langle5|(1\,2)|6]}\Big\}$}
\mathematicaSequence[2]{permSeeds=denList.Power[\funL[]{ymAmp}[6,1],2]/\!.\builtL[]{superFunctionProductRule};\\
test=Total[\funL[]{permuteLegs}\brace{\#}\,\brace{permSeeds}\,\&/\!@Permutations[Range[2,5]]]\!;\\
\funL[]{evaluate}\brace{\funL[]{component}\brace{egCmpt}\,\brace{test}}}{$\rule{0pt}{20pt}\displaystyle{-}\frac{1137100091554}{375073120755}$}
\mathematicaSequence{\textnormal{\%}===\funL[]{componentAmpN}\brace{egCmpt}}{True}}
}

\defnBox{grAmpBCFW}{\var{$n$}\pattern,\var{$k$}\pattern}{returns a \built{List} of contributions to the \var{$n$}-particle N${}^{\var{k}}$MHV tree amplitude in superGravity as obtained using the BCFW recursion relations and as described in \cite{Bourjaily:2023ycy} (see also \cite{structureOfOnShellGR}). In particular, this represents the amplitude as a sum of \funL{superFunction} objects constructed as (non-local) `double-copies) of superYang-Mills amplitudes.\\[-10pt]

The only novelty relative to a na\"ive sum over the $(\var{n}{-}2)!$ permutations $\sigma\!\in\!\mathfrak{S}([2,\var{n}{-}1])$ which relabel the legs not deformed in the BCFW recursion is that \fun{grAmpBCFW} automatically groups the terms which share \mbox{($GL(\var{k}{+}2)$-)}equivalent fermionic parts. Thus, for example, there are 14 terms listed by \mbox{\fun{grAmpBCFW}\brace{6,1}\,,} which is considerably fewer than $3\!\times\!4!\!=\!72$. 
}

\defnBox{grAmpBCFWSeeds}{\var{$n$}\pattern,\var{$k$}\pattern}{returns the \built{List} of permutation `seeds' for the BCFW representation of the \var{$n$}-particle N${}^{\var{k}}$MHV tree amplitude in superGravity. These are defined essentially by %
\vspace{5pt}\eq{\fwbox{0pt}{\hspace{-40pt}\fun{grAmpBCFWSeeds}\brace{\var{n},\var{k}}\,\,\,\Leftrightarrow\,\fun{grAmpDenominators}\brace{\var{n},\var{k}}\,*\text{\built{Power}\brace{\fun{ymAmp}\brace{\var{$n$},\var{$k$}}\,,\,2}}\,.}}
Summing these `seeds' over all permutations $\sigma\!\in\!\mathfrak{S}([2,\var{n}{-}1])$ gives the tree amplitude. 
}

\defnBox{grAmpBCFWSeedGroups}{\var{$n$}\pattern,\var{$k$}\pattern}{for the $i$th `seed' given by \fun{grAmpBCFWSeeds}, there is a subgroup $\mathfrak{T}_i\!\subset\!\mathfrak{S}([2,\var{n}{-}1])$ of permutations which leaves the fermionic part of the \funL{superFunction} invariant (or $GL(\var{k}{+}2)$-equivalent). For example, 
\mathematicaBox{
\mathematicaSequence[1]{egFcn=First[\funL[1]{grAmpBCFWSeeds}\brace{6,1}];\\
\funL[]{nice}\brace{egFcn}}{$\displaystyle{-}\frac{\sb{5\,6}\ab{1\,2}\ab{2\,3}}{\sb{2\,3}\sb{3\,4}^2\ab{1\,5}\ab{5\,6}\ab{6\,1}^2\langle1|(2\,3)|4]\langle1|(3\,4)|2]\langle5|(6\,1)|2]s_{5\,6,1}^2},$\\
$\displaystyle\left(\begin{array}{@{}cccccc@{}}\fwbox{15pt}{0}&\fwbox{30pt}{\sb{3\,4}}&\fwbox{30pt}{\sb{4\,2}}&\fwbox{30pt}{\sb{2\,3}}&\fwbox{30pt}{0}&\fwbox{15pt}{0}\end{array}\right)\!\!\cdot\!\tilde\eta$}
}
This \funL{superFunction}'s fermionic part is invariant under all permutations of $\mathfrak{T}\!\equivR\mathfrak{S}(\{2,3,4\})$ of the external legs. Thus, we may combine them into a single \funL{superFunction}, whose bosonic part is a sum over these permutations $\mathfrak{T}$.\\[-10pt]

The output of \fun{grAmpBCFWSeedGroups} is therefore a \built{List} of \funL{superFunction} objects whose number is equal to the number of terms in \funL{ymAmp}\brace{\var{$n$},\var{$k$}}, but which have been combined over (term-specific) subgroups over the permutation group $\mathfrak{S}([2,\var{n}{-}1])$.
}

\newpage
\vspace{10pt}\subsubsectionAppendix{Representation of Amplitudes in superGravity via KLT}{appendix:klt_reps}\vspace{-10pt}

\defnBox{kltRepresentationSeeds}{\var{$n$}\pattern}{returns a \built{List} of `seed' contributions to the KLT representation of the \var{$n$}-particle tree amplitude in superGravity. This data is encoded (redundantly) similarly to how these expressions are typically expressed in the literature (see e.g.~\cite{Bern:1998sv}). Specifically, the output of \fun{kltRepresentationSeeds} is a \built{List} of \built{List}s of triples \{\t{leftAmp},\t{kinematicCoeff},\t{rightAmp}\}---where \t{leftAmp} is always \funL{amp}\brace{1,\ldots,\var{$n$}}\,. For example, 
\mathematicaBox{
\mathematicaSequence{\funL[]{nice}\brace{\funL[1]{kltRepresentationSeeds}\brace{6}}}{$\displaystyle\left(\begin{array}{@{}ccc@{}}\mathcal{A}_{6}\!\big(1,2,3,4,5,6\big)&{-}s_{1\,2}(s_{1\,3}{+}s_{2\,3})s_{4\,5}&\mathcal{A}_{6}\!\big(1,2,3,6,4,5\big)\\
\mathcal{A}_{6}\!\big(1,2,3,4,5,6\big)&{-}s_{1\,2}\,s_{1\,3}\,s_{4\,5}&\mathcal{A}_{6}\!\big(1,3,2,6,4,5\big)\\\end{array}\right)$}
}
which encodes the fact that (for any N${}^{k}$MHV degree), the gravity amplitude can be expressed as 
\eq{\begin{split}\mathcal{A}_6^{\text{GR}}=&\mathcal{A}_6^{\text{YM}}\!(123456)\big(\text{-}s_{1\,2}\,s_{4\,5}\big)\left[(s_{1\,3}{+}s_{2\,3})\mathcal{A}_{6}^{\text{YM}}\!(123645){+}s_{1\,3}\mathcal{A}_{6}^{\text{YM}}\!(132645)\right]\\
&\text{+`permutations of $\mathfrak{S}([2,4])$'}\end{split}}
This is data is essentially nothing but an organization of the output of 
\vspace{-10pt}$$\t{First[\funL{kltKernel}\brace{\var{$n$}}\,.\,\funL{kltBasisAmpsBeta}\brace{\var{$n$}}]}.\vspace{-24pt}$$
}

\defnBox{grAmpKLT}{\var{$n$}\pattern,\var{$k$}\pattern}{returns a \built{List} of terms contributing to the \var{$n$}-particle N${}^{\var{k}}$MHV tree amplitude in superGR, as obtained via the KLT representation. In essence, this is simply a manifestation of the KLT form  
\eq{\fwbox{0pt}{\hspace{-30pt}\mathcal{A}_n^{\text{GR}}=\big(\funL{kltBasisAmpsAlpha}\brace{\var{n}}\big).\funL{kltKernel}\brace{\var{n}}\,.\big(\funL{kltBasisAmpsAlpha}\brace{\var{n}}\big),}\label{klt_for_gr}}
where the (sYM) amplitudes appearing on either side are represented in terms of \funL{superFunction} terms by their BCFW recursed formulae as given by \mbox{\funL{ymAmp}\brace{\var{$n$},\var{$k$},\{\var{legOrdering}\}}\,.}\\[-10pt]

As the sYM amplitudes appearing in (\ref{klt_for_gr}) involve many terms, this formula involves (\funL{numberOfBCFWterms}\brace{\var{$n$},\var{$k$}}$\,)^2$ terms per non-vanishing entry of \mbox{\funL{kltKernel}\brace{\var{$n$}}\,;} the products of terms is expressed as a single \funL{superFunction} using \builtL{superFunctionProductRule}.
}

\defnBox{numberOfTermsInKLTforGR}{\var{$n$}\pattern,\var{$k$}\pattern\optArg{0}\,}{returns the number of terms non-trivial `terms' appearing in the KLT representation of an amplitude. If the optional argument \var{$k$} is 0 or not provided, then it essentially counts the number of non-vanishing entries in the matrix \funL{kltKernel}\brace{\var{$n$}}\,; that is, it counts each \mbox{`\funL{amp}\brace{\var{legOrdering}}\,'} object as one term, and each entry of the matrix \funL{kltKernel}\brace{\var{$n$}} as a single `term'. This is generous, obviously, but not far from a meaningful counting for MHV amplitudes:
\eq{\begin{split}\fun{numberOfTermsInKLTforGR}\brace{\var{n}}\,\,\,&\Leftrightarrow\,\,(\lfloor(\var{n}{-}3)/2\rfloor)!\,\times\!(\lceil(\var{n}{-}3)/2\rceil)!\!\times\!(\var{n}{-}3)!\\
&=\,\,\big((\var{n}{-}3)!\big)^2\bigg/\binom{\var{n}{-}3}{\lfloor(\var{n}{-}3)/2\rfloor}\end{split}\label{number_of_terms_in_klt_for_gr}}
However, a more practical number may be the number of \funL{superFunction} terms that appear in the representation for the N${}^{\var{k}}$MHV tree amplitude, which would be (in the best case scenario) given by \funL{numberOfBCFWterms}\brace{\var{$n$},\var{$k$}}\,. Thus, 
\eq{\fwbox{0pt}{\hspace{-50pt}\fun{numberOfTermsInKLTforGR}\brace{\var{n},\var{k}}\,\,\,\Leftrightarrow\,\,\left(\frac{\!(\var{n}-3)!\!}{\var{k}{+}1}\binom{\!\var{n}{-}3\!}{\var{k}}\binom{\!\var{n}{-}4\!}{\var{k}}\right)^2\!\!\!\!\bigg/\!\!\!\binom{\var{n}{-}3}{\!\lfloor(\var{n}{-}3)/2\rfloor\!}\!.}}
}

\newpage
\vspace{-0pt}\subsectionAppendix{Numeric Expressions for superAmplitudes in superGravity}{appendix:numeric_trees_in_sugra}\vspace{-10pt}

\defnBox{grAmpN}{\var{$n$}\pattern,\var{$k$}\pattern}{analogous to the relationship between \funL{ymAmpN} and \funL{ymAmp}, \fun{grAmpN} returns a \built{List} of numerically evaluated terms which contribute to the \var{$n$}-particle N${}^\var{k}$MHV tree amplitude in superGravity, in a form which is \emph{equivalent} (but \emph{not identical}) to the evaluation of the output of \fun{grAmp}. The differences in form of the output follow from \fun{grAmpN}'s exploitation of numeric efficiencies.\\[-10pt]

To illustrate the efficiencies gained by \fun{grAmpN}, consider any sufficiently high-multiplicity amplitude---e.g.,
\mathematicaBox{
\mathematicaSequence[1]{\funL[]{timed}\brace{grAmpN\brace{12,0}};\\
\funL[]{timed}\brace{\funL[]{evaluate}\brace{grAmp\brace{12,0}}};}{}\\[-18pt]
&\scalebox{0.75}{\hspace{-40pt}Evaluation of the function \texttt{\textbf{grAmpN}} for $\mathcal{A}_{12}^{(0)}$ required \textbf{7 ms, 841 $\mu$s} to complete.}\\[-5pt]
&\scalebox{0.75}{\hspace{-40pt}Evaluation of the function \texttt{\textbf{grAmp}} for $\mathcal{A}_{12}^{(0)}$ required \textbf{6 seconds, 145 ms} to complete.}\\[-5pt]
\mathematicaSequence{\textnormal{\%\%}===\textnormal{\%}}{True}
}
The near thousand-fold improvement of \fun{grAmpN} relative to the evaluation of \fun{grAmp} is easy to explain. While both algorithms make use of \funL{grAmpHodges} for the MHV amplitude \fun{grAmp} requires an \emph{analytic} expression for the determinant of $\Phi$ (\ref{hodges_phi_mat_defined})---an expression involving $(\var{n}{-}3)!$ terms; in contrast, \fun{grAmpN} first evaluates Hodges' $\Phi$ matrix before taking this determinant, which makes use of a built-in, polynomial-(cubic-)time complexity of computing determinants.\\[-10pt]

As with \fun{grAmp}, \fun{grAmpN} calls \funL{grAmpHodges} for MHV and $\bar{\text{MHV}}$ amplitudes, but \fun{grAmpBCFW} for all other N${}^{\var{k}}$MHV degrees. 
}

\defnBox{grAmpBCFWN}{\var{$n$}\pattern,\var{$k$}\pattern}{gives a \built{List} of numerically evaluated terms which contribute to the \var{$n$}-particle N${}^{\var{k}}$MHV tree amplitude in superGravity for the kinematics specified by the \built{Global} variables \builtL{Ls} and \builtL{Lbs}. Similar to the distinction between \funL{ymAmpN} and \funL{ymAmp}, \fun{grAmpBCFW} will return a \built{List} of \funL{superFunction}\hspace{-2pt}contributions \emph{equivalent} (but \emph{not identical}) to the evaluation of \mbox{\funL{grAmpBCFW}.} The distinction in form follows from the fact that \fun{grAmpBCFW} saves time by avoiding certain gauge-fixings and little-group rescaling that simplify analytic expressions but do not substantively benefit the extraction of components. 
}

\newpage
\vspace{10pt}\subsectionAppendix{Analytic Expressions for \emph{Component} Amplitudes in superGravity}{appendix:analytic_components_of_sGR}\vspace{-0pt}

\defnBox[]{helicityAmp}{\var{gravitonHelicities}\patternTwo}{is same as \mbox{\fun{grHelicityAmp}\brace{\var{gravitonHelicities}}\,.}
}

\vspace{-5pt}

\defnBox{grHelicityAmp}{\var{gravitonHelicities}\patternTwo}{returns a \built{List} of terms contributing to the \emph{graviton} helicity amplitude resulting from analytic (non-supersymmetric) BCFW recursion in terms of spinor variables directly. Like \funL{ymHelicityAmp}, this is an entirely distinct implementation of recursion relations relative to that used by \funL{grAmp}. As such, the resulting formulae are often quite different, and almost universally simpler.
\mathematicaBox{
\mathematicaSequence[2]{helicityRecursionTerms=\funL[1]{grHelicityAmp}\brace{\stateGR{mm},\stateGR{pp},\stateGR{mm},\stateGR{pp},\stateGR{mm},\stateGR{pp}}\,;\\
extractedFromSuperTerms=\funL[]{component}\brace{\stateGR{mm},\stateGR{pp},\stateGR{mm},\stateGR{pp},\stateGR{mm},\stateGR{pp}}\,\brace{\funL[]{grAmp}\brace{6,1}};\\
Length/@\{helicityRecursionTerms,extractedFromSuperTerms\}}{\{13,14\}}
\mathematicaSequence{LeafCount/@\{helicityRecursionTerms,extractedFromSuperTerms\}}{\{1263,4635\}}
\mathematicaSequence{\funL[]{evaluate}@\{helicityRecursionTerms,extractedFromSuperTerms\};\\
Length[Intersection@@\textnormal{\%}]}{0}
\mathematicaSequence{SameQ@@Total/@\textnormal{\%\%}}{True}
}
}

\defnBox[]{componentAmpTerms}{\var{componentStates}\patternTwo}{returns a \built{List} of terms contributing to the particular component amplitude involving the external \var{componentStates}.\\[-10pt]

Depending on the particular external states, \fun{componentAmpTerms} chooses \emph{different} underlying representations for the amplitude. If the \var{componentStates} include states that are \emph{not} gravitons, then \fun{componentAmpTerms} simply calls \mbox{\funL{component}\brace{\var{componentStates}}\,\brace{\funL{grAmp}\brace{\var{$n$},\var{$k$}}\,}}\,.
}

~

\defnBox[]{componentAmp}{\var{componentStates}\patternTwo}{is \mbox{\built{Total}\brace{\fun{componentAmpTerms}\brace{\var{componentStates}}}\,.}}

\newpage
\vspace{0pt}\subsectionAppendix{Local, Covariant, Color-Kinematic-Dual Graviton Amplitudes}{appendix:local_covariant_gr_amps}\vspace{-10pt}


\defnBox{doubleCopyNumerators}{\var{$n$}\pattern}{returns a \built{List} of the numerators of all distinct cubic graphs---as given by the output of \funL{cubicGraphDenominators}\brace{\var{$n$}}--which satisfy the Jacobi relations of color-factors naturally attached to each. Specifically, \fun{doubleCopyNumerators} gives a list of $(2\var{n}{-}5)!!$ (the number of cubic graphs) linear combinations of the $(\var{n}{-}2)!$ basis numerators \built{num}\brace{\var{$i$}\pattern}.\\[-10pt]

These are obtained by starting with a particular set of plane-embedded color-graph tensors (to fix signs), and using Jacobi relations to decompose these color tensors into the basis \funL{basisColorFactors}\brace{\var{$n$}}.
}

\defnBox{grLocalCovariantAmp}{\var{$n$}\pattern}{returns the sum of the terms encoded by \fun{grLocalCovariantAmpData}\brace{\var{$n$}}.
}

\defnBox{grLocalCovariantAmpData}{\var{$n$}\pattern}{returns a \built{List} of terms \{\texttt{termList},\{\texttt{numData}\}\} where \built{Total}\brace{\texttt{termList}} gives the $\var{n}$-particle pure-gluon tree amplitude for YM in terms of local poles arising from cubic, scalar Feynman graphs with coefficients \mbox{\texttt{num}\brace{\var{$i$}\pattern}} encoded by \texttt{numData}. For example,
\mathematicaBox{\mathematicaSequence[1]{\{amplitudeTermList,numData\}=\funL[1]{grLocalCovariantAmpData}[4];\\
\funL[]{nice}[amplitudeTermList]}{$\rule{0pt}{24pt}\displaystyle\left\{\frac{n_1^2}{s_{12}},\frac{n_2^2}{s_{13}},\frac{(n_1{-}n_2)^2}{s_{14}}\right\}$}\mathematicaSequence{numData[\![1]\!]}{\{\{num[1],-\funL{ss}\brace{\funL{$\epsilon$}\brace{1}\,,\funL{$\epsilon$}\brace{4}}\,\funL{ss}\brace{\funL{$\epsilon$}\brace{2}\,,\funL{$\epsilon$}\brace{1}}\,\funL{ss}\brace{\funL{$\epsilon$}\brace{3}\,,\funL{$\epsilon$}\brace{1}}\,-\ldots\}\}}}
These expressions can be \funL{evaluate}d for specific choices of helicities using \funL{helicityPolarizationRules}. For example, the \var{$5$}-graviton amplitude written in manifestly color-kinematic dual form would be given by
\mathematicaBox{\mathematicaSequence[1]{
\{amplitudeTermList,numData\}=\funL[1]{grLocalCovariantAmpData}[5];\\
\funL[]{nice}[amplitudeTermList]}{$\displaystyle\rule{0pt}{24pt}\left\{\frac{n_{1}^2}{s_{1\,2}s_{4\,5}},\frac{n_{2}^2}{s_{1\,2}s_{3\,5}},\frac{n_{3}^2}{s_{1\,3}s_{4\,5}},\frac{n_{4}^2}{s_{1\,3}s_{2\,5}},\frac{n_{5}^2}{s_{1\,4}s_{3\,5}},\frac{n_{6}^2}{s_{1\,4}s_{2\,5}},\frac{(n_{1}\text{-}n_{2})^2}{s_{1\,2}s_{3\,4}},\frac{(n_{3}\text{-}n_{4})^2}{s_{1\,3}s_{2\,4}},\frac{(n_{5}\text{-}n_{6})^2}{s_{1\,4}s_{2\,3}},\right.$\\
$\displaystyle\left.\frac{(n_{1}\text{-}n_{3}\text{-}n_{5}\text{+}n_{6})^2\!\!\!}{s_{1\,5}s_{2\,3}},\frac{(n_{2}\text{-}n_{3}\text{+}n_{4}\text{-}n_{5})^2\!\!\!}{s_{1\,5}s_{2\,4}},\frac{(n_{1}\text{-}n_{2}\text{-}n_{4}\text{+}n_{6})^2\!\!\!}{s_{1\,5}s_{3\,4}},\frac{(n_{1}\text{-}n_{3})^2\!\!\!}{s_{2\,3}s_{4\,5}},\frac{(n_{2}\text{-}n_{5})^2\!\!\!}{s_{2\,4}s_{3\,5}},\frac{(n_{4}\text{-}n_{6})^2\!\!}{s_{2\,5}s_{3\,4}}\right\}$\\[-10pt]}
\mathematicaSequence[2]{\funL[]{useReferenceKinematics}[5];\\[-3pt]
nNumData=\funL[]{evaluate}[numData/\!\!.\funL[]{helicityPolarizationRules}[\stateYM{m},\stateYM{p},\stateYM{m},\stateYM{p},\stateYM{m}]\!]\!]\!;\\[-3pt]
\funL[]{nice}[nNumData]}{$\rule{0pt}{55pt}\displaystyle\left(\rule{0pt}{50pt}\right.\begin{array}{lc}n_1&\frac{3575}{232}\\
n_2&\frac{460325}{9048}\\
n_3&\frac{1273325}{190008}\\
n_4&\text{-}\frac{29484625}{253344}\\
n_5&\text{-}\frac{14003125}{36192}\\
n_6&\text{-}\frac{107551175}{253344}\end{array}\left.\rule{0pt}{50pt}\right)$\\[-5pt]}
\mathematicaSequence{Total[\funL[]{evaluate}[amplitudeTermList]/\!\!.Rule@@@nNumData]}{$\rule{0pt}{20pt}\displaystyle\frac{8084954166453125}{21518333557056}$}
\mathematicaSequence{\funL[]{componentAmpN}[\stateGR{mm},\stateGR{pp},\stateGR{mm},\stateGR{pp},\stateGR{mm}]}{$\rule{0pt}{20pt}\displaystyle\frac{8084954166453125}{21518333557056}$}
}
This representation is found using the reduction algorithm described in \cite{Bjerrum-Bohr:2016axv} (see also \cite{Baadsgaard:2015hia,Baadsgaard:2015ifa,Baadsgaard:2015voa,Bjerrum-Bohr:2016juj}), and in principle works for arbitrary \var{$n$} (provided the user is sufficiently patient).
}

\vspace{-0pt}\subsectionAppendix{Numeric Expressions for \emph{Component} Amplitudes in superGravity}{appendix:numeric_components_of_sGR}\vspace{-0pt}

\defnBox[]{helicityAmpN}{\var{gravitonHelicities}\patternTwo}{gives \built{Total} of \fun{grHelicityAmpN}.}

\vspace{-10pt}~\vspace{-10pt}

\defnBox{grHelicityAmpN}{\var{gravitonHelicities}\patternTwo}{returns a \built{List} of \emph{numerically}-recursed terms contributing to the graviton helicity amplitude for \{\var{gravitonHelicities}\}.\\[-10pt]

Similar to \funL{ymHelicityAmpN}, \fun{grHelicityAmpN} employs an entirely distinct algorithm---one which begins with the globally defined spinor data \{\builtL{Ls},\builtL{Lbs}\} and iteratively \emph{shifting} that data for input into numeric amplitudes involving fewer particles. This has the advantage that absolutely no symbolic manipulations are required (and hence is relatively fast), but has a serious disadvantage that the entire recursion pathway must be run every time for any kinematics. 

It is instructive to compare this with the evaluation of \fun{grHelicityAmp}.
\mathematicaBox{
\mathematicaSequence[2]{v1Terms=\funL[]{timed}@\funL[1]{grHelicityAmpN}\brace{\stateGR{mm},\stateGR{pp},\stateGR{mm},\stateGR{pp},\stateGR{mm},\stateGR{pp},\stateGR{mm},\stateGR{pp},\stateGR{mm}};\\
\funL[]{timed}@\funL[]{grHelicityAmp}\brace{\stateGR{mm},\stateGR{pp},\stateGR{mm},\stateGR{pp},\stateGR{mm},\stateGR{pp},\stateGR{mm},\stateGR{pp},\stateGR{mm}};\\
v2Terms=\mbox{\funL[]{timed}@\funL[]{evaluate}\brace{\textnormal{\%}};}}{}
&\scalebox{0.58}{\hspace{-35pt}Evaluation of the function \texttt{\textbf{grHelicityAmpN[]}} for $\mathcal{A}^{(3)}_{9}(--,++,--,++,--,++,--,++,--)$ required \textbf{17 seconds, 903 ms} to complete.}\\[-5pt]
&\scalebox{0.58}{\hspace{-35pt}Evaluation of the function \texttt{\textbf{grHelicityAmp[]}} for $\mathcal{A}^{(3)}_{9}(--,++,--,++,--,++,--,++,--)$ required \textbf{51 seconds, 776 ms} to complete.}\\[-5pt]
&\scalebox{0.58}{\hspace{-35pt}Evaluation of the function \texttt{\textbf{evaluate}} required \textbf{1 seconds, 948 ms} to complete.}\\[-5pt]
\mathematicaSequence{SameQ@@Total/@\{v1Terms,v2Terms\}}{True}
}
Thus, even though numeric recursion is nearly three times faster than symbolic, subsequent evaluation is considerably cheaper. Thus, if more than one data-point were required, the start-up cost of the analytic expression would be worth it. 
}

~

\defnBox[]{componentAmpTermsN}{\var{componentStates}\patternTwo}{returns a \built{List} of terms corresponding to \fun{grHelicityAmpN} if \fun{componentStates} are all gravitons $\{\stateGR{mm},\stateGR{pp}\}$, or \mbox{\funL{component}\brace{\funL{grAmpN}\brace{\var{$n$},\var{$k$}}}} otherwise. 
}

~

\defnBox[]{componentAmpN}{\var{componentStates}\patternTwo}{\mbox{is \built{Total}\brace{\fun{componentAmpTermsN}\brace{\var{componentStates}}}.}
}

\newpage
\vspace{-0pt}\subsectionAppendix{Relations Among Component Amplitudes of sGR}{appendix:component_amplitude_relations_sGR}\vspace{-10pt}

\defnBox{inverseSoftFactorGR}{\var{$n$}\pattern,\var{softLeg}\pattern}{gives the \emph{inverse}-soft factor \cite{Weinberg:1964ew}
\vspace{-5pt}\eq{\mathcal{A}_{\var{n}}\!(\ldots)=\big(\text{inverse-soft-factor}\big)\lim_{p_{\text{\var{softLeg}}}\!\to\!0}\mathcal{A}_{\var{n}+1}\!\big(\ldots,\text{\var{softLeg}},\ldots\big)\,,\vspace{-5pt}}
where |\var{softLeg}| labels the particle in $[\var{n}\text{+}1]$ whose momentum is becoming soft.\\[-10pt]

\textbf{\emph{Note}}: the \emph{sign} of \var{softLeg} determines whether this momentum is taken to be soft `holomorphically' or `anti-holomorphically'---whether its $\tilde\lambda$ or $\lambda$ is taken to zero, respectively. In the former case, this is given by $1/\phi_a$ as given in (\ref{graviton_holo_soft_factors}):
\mathematicaBox{
\mathematicaSequence[3]{\funL[]{useReferenceKinematics}\brace{6};\\
egSoftSpinors=\funL[]{inverseSoftSpinors}[7];\\
spinorInverseSoftFactor=\funL[1]{inverseSoftFactorGR}[6,7];\\
\funL[]{nice}[\textnormal{\%}]}{$\rule{0pt}{22pt}\displaystyle1/\left(\frac{\sb{7\,2}\ab{2\,1}\ab{6\,2}}{\ab{6\,7}\ab{7\,1}\ab{7\,2}}{+}\frac{\sb{7\,3}\ab{3\,1}\ab{6\,3}}{\ab{6\,7}\ab{7\,1}\ab{7\,3}}{+}\frac{\sb{7\,4}\ab{4\,1}\ab{6\,4}}{\ab{6\,7}\ab{7\,1}\ab{7\,4}}{+}\frac{\sb{7\,5}\ab{5\,1}\ab{6\,5}}{\ab{6\,7}\ab{7\,1}\ab{7\,5}}\right)$}
\mathematicaSequence{\funL[]{evaluateWithSpinors}\brace{egSoftSpinors}\,\brace{spinorInverseSoftFactor};\\
FullSimplify[\textnormal{\%}]}{$\rule{0pt}{22pt}\displaystyle\text{-}\frac{1346538336\,\epsilon^3}{20940073{+}62337600\,\epsilon\,({-}22{+}653\,\epsilon)}$}
\mathematicaSequence[1]{polarizationSoftFactor=Total@$\displaystyle\left(\rule{0pt}{22pt}\right.\hspace{-5pt}\frac{\texttt{ss}[\funL[]{$\epsilon$p}\texttt{[7],\funL{p}[\#]]}^2}{\texttt{\funL{ss}[\funL{p}[7],\funL{p}[\#]]}}$\&/\!@Range[6]$\hspace{-5pt}\left.\rule{0pt}{22pt}\right)$;\\
FullSimplify[\funL[]{evaluateWithSpinors}\brace{egSoftSpinors}\,\brace{\textnormal{\%}}]}{$\displaystyle\rule{0pt}{22pt}\text{-}\frac{20940073{+}62337600\,\epsilon\,({-}22{+}653\,\epsilon)}{1346538336\,\epsilon^3}$}
}
\mathematicaBox{
\mathematicaSequence[3]{\funL[]{useReferenceKinematics}\brace{6};\\
egSoftSpinors=\funL[]{inverseSoftSpinors}[-7];\\
spinorInverseSoftFactor=\funL[1]{inverseSoftFactorYM}[6,-7];\\
\funL[]{nice}[\textnormal{\%}]}{$\rule{0pt}{22pt}\displaystyle1/\left(\frac{\ab{7\,2}\sb{2\,1}\sb{6\,2}}{\sb{6\,7}\sb{7\,1}\sb{7\,2}}{+}\frac{\ab{7\,3}\sb{3\,1}\sb{6\,3}}{\sb{6\,7}\sb{7\,1}\sb{7\,3}}{+}\frac{\ab{7\,4}\sb{4\,1}\sb{6\,4}}{\sb{6\,7}\sb{7\,1}\sb{7\,4}}{+}\frac{\ab{7\,5}\sb{5\,1}\sb{6\,5}}{\sb{6\,7}\sb{7\,1}\sb{7\,5}}\right)$}
\mathematicaSequence{\funL[]{evaluateWithSpinors}\brace{egSoftSpinors}\,\brace{spinorInverseSoftFactor};\\
FullSimplify[\textnormal{\%}]}{$\rule{0pt}{22pt}\displaystyle\text{-}\frac{447092100\,\epsilon^3}{771933089{+}4864860\,\epsilon({-}308{+}815\,\epsilon)}$}
\mathematicaSequence[1]{polarizationSoftFactor=Total@$\displaystyle\left(\rule{0pt}{22pt}\right.\hspace{-5pt}\frac{\texttt{ss}[\funL[]{$\epsilon$m}\texttt{[7],\funL{p}[\#]]}^2}{\texttt{\funL{ss}[\funL{p}[7],\funL{p}[\#]]}}$\&/\!@Range[6]$\hspace{-5pt}\left.\rule{0pt}{22pt}\right)$;\\
FullSimplify[\funL[]{evaluateWithSpinors}\brace{egSoftSpinors}\,\brace{\textnormal{\%}}]}{$\displaystyle\rule{0pt}{22pt}\text{-}\frac{771933089{+}4864860\,\epsilon({-}308{+}815\,\epsilon)}{447092100\,\epsilon^3}$}
}\\[-90pt]
}

\defnBox{egE77doubleSoftData}{\var{$n$}\pattern,\var{$k$}\pattern}{gives a randomly-chosen example of a double-soft identity following from the on-shell $E_{7(7)}$ symmetry of sGR. The data for the example is presented as a \built{List} of of the form
\eq{\text{\{\t{parentComponentAmp},\t{daughterCoeffs},\t{daughterComponentAmps}\}}}
which amounts to a explicit, component-amplitude-level translation of the discussion in ref.~\cite{ArkaniHamed:2008gz} (see also \cite{Cremmer:1978ds,Cremmer:1978km,Cremmer:1979up,deWit:1977fk,deWit:1982bul,Gaillard:1981rj,Adler:1964um,Weinberg:1966kf}). Indexing the list of \t{daughterCoeffs} and \t{daughterComponentAmps} by $i$, this corresponds to the claim that 
\eq{\lim_{\substack{p_{\var{n}{+}1}\to0\\p_{\var{n}{+}2}\to0}}\mathcal{A}_{\var{n}{+}2}\!\big(\text{\var{parentComponents}}\big)=\sum_{i}\text{\var{c}}_i\,\mathcal{A}_{\var{n}}\!\big(\text{\var{daughterComponents}}_i\big)\,.\label{e77_relation}}

These examples always involve the two soft particles being the particular scalars $\{|\phi\rangle_{(1\,2\,3\,4)}^{0},|\phi\rangle_{(1\,5\,6\,7)}^{0}\}$ of $\mathcal{N}\!=\!8$ sGR, with momenta indexed by $\{\var{n}{+}1,\var{n}{+}2\}$. (Any other non-trivial double-soft relation is related to this one by $R$-symmetry; other choices for scalars to be taken soft either result in vanishings or simple soft-graviton divergences in the limit.) The other states of \var{parentComponents} are chosen randomly by \fun{egE77doubleSoftData}.\\[-10pt]

Notice that this pair of scalars carries two units of $R$-charge for $I\!=\!1$, and has $R$-charge of 0 for $I\!=\!8$. Thus, the remaining \var{$n$} particles' of \var{parentComponents} $R$-charges must have an excess $I\!=\!8$ index and one fewer $I\!=\!1$ index relative to the $R$-charges of any non-vanishing component amplitude, as `legal' (or non-vanishing) sets of component states must have the same number of $R$-charges for each index $I\!\in\![8]$. Thus, the various \{\var{daughterComponents}\} on the RHS of (\ref{e77_relation}) must sum over all the ways of trading the excessive $I\!=\!1$ index for the absent $I\!=\!8$ index. 
\mathematicaBox{
\mathematicaSequence[1]{\funL[1]{egE77doubleSoftData}\brace{5,0};\\
\funL[]{nice}@\textnormal{\%}}{$\displaystyle\left\{\rule{0pt}{16pt}\right.\hspace{-3pt}\mathcal{A}_7^{(1)}\!\big(|\gamma\rangle_{(\bar{48})}^{\text{-}1},|\psi\rangle_{(278)}^{\text{+}1/2},|\bar{\psi}\rangle_{(\bar{127})}^{\text{-}1/2},|\gamma\rangle_{(48)}^{\text{+}1},++,|\phi\rangle_{(1\,2\,3\,4)}^{0},|\phi\rangle_{(1\,5\,6\,7)}^{0}\big),$\\
~\hspace{6pt}$\displaystyle\phantom{\{}\hspace{1pt}\left\{\frac{s_{6\,2}}{s_{6\,2}{+}s_{7\,2}},\text{-}\frac{s_{6\,3}}{s_{6\,3}{+}s_{7\,3}},\frac{s_{6\,4}}{s_{6\,4}{+}s_{7\,4}}\right\},$\\
$\displaystyle\phantom{\{}\hspace{0pt}\left\{\rule{0pt}{16pt}\right.\hspace{-4pt}\mathcal{A}_5^{(0)}\!\big(|\gamma\rangle_{(\bar{48})}^{\text{-}1},|\psi\rangle_{(178)}^{\text{+}1/2},|\bar{\psi}\rangle_{(\bar{127})}^{\text{-}1/2},|\gamma\rangle_{(48)}^{\text{+}1},++\big),$\\
$\displaystyle\phantom{\{}\hspace{7pt}\mathcal{A}_5^{(0)}\!\big(|\gamma\rangle_{(\bar{48})}^{\text{-}1},|\psi\rangle_{(278)}^{\text{+}1/2},|\bar{\psi}\rangle_{(\bar{228})}^{\text{-}1/2},|\gamma\rangle_{(48)}^{\text{+}1},++\big),$\\[-4pt]
$\displaystyle\phantom{\{}\hspace{7pt}\mathcal{A}_5^{(0)}\!\big(|\gamma\rangle_{(\bar{48})}^{\text{-}1},|\psi\rangle_{(278)}^{\text{+}1/2},|\bar{\psi}\rangle_{(\bar{127})}^{\text{-}1/2},|\gamma\rangle_{(14)}^{\text{+}1},++\big)\hspace{-5pt}\left.\rule{0pt}{18pt}\right\}\hspace{-5pt}\left.\rule{0pt}{18pt}\right\}$}
}
Examples of how these identities can be checked using the package are given in the demonstration notebook. 
}

\defnBox[]{egWardIdentity}{\var{seedComponentStates}\patternTwo}{uses the \var{seedComponentStates} to find an example Ward identity (see e.g.~\cite{Grisaru:1977px,Grisaru:1976vm,Elvang:2009wd,Elvang:2010xn}) often (but not always) involving those states. Analogous to \fun{egWardIdentity} above.
}

\defnBoxTwo{egWardIdentityGR}{\var{$\mathcal{N}$}\pattern\optArg{8}\,}{\var{$n$}\pattern,\var{$k$}\pattern}{takes a randomly chosen initial state to seed a random example of a supersymmetric Ward identity: 
\mathematicaBox{
\mathematicaSequence[1]{egTerms=\funL[1]{egWardIdentityGR}\brace{8}\,\brace{6,1};\\
\funL[]{nice}\brace{egTerms}}{$\rule{0pt}{22pt}\displaystyle\big\{\mathcal{A}_{6}^{(1)}\!\big(\fwbox{40pt}{|\bar{\psi}\rangle_{(\bar{134})}^{\text{-}1/2}},\fwbox{40pt}{|\bar{\psi}\rangle_{(\bar{126})}^{\text{-}1/2}},\fwbox{40pt}{|\gamma\rangle_{(68)}^{\text{+}1}},\fwbox{40pt}{|\phi\rangle_{(1237)}^{0}},\fwbox{40pt}{|\gamma\rangle_{(14)}^{\text{+}1}},\fwbox{40pt}{\bar{\gamma}\rangle_{(\bar{78})}^{\text{-}1}}\big),$\\
$\displaystyle
\,\text{-}\mathcal{A}_{6}^{(1)}\!\big(\fwbox{40pt}{|\bar{\psi}\rangle_{(\bar{348})}^{\text{-}1/2}},\fwbox{40pt}{|\bar{\psi}\rangle_{(\bar{126})}^{\text{-}1/2}},\fwbox{40pt}{|\gamma\rangle_{(68)}^{\text{+}1}},\fwbox{40pt}{|\bar{\phi}\rangle_{(\bar{1456})}^{0}},\fwbox{40pt}{|\gamma\rangle_{(14)}^{\text{+}1}},\fwbox{40pt}{\bar{\gamma}\rangle_{(\bar{78})}^{\text{-}1}}\big),$\\
$\displaystyle
\,\phantom{\text{-}}\mathcal{A}_{6}^{(1)}\!\big(\fwbox{40pt}{|\bar{\psi}\rangle_{(\bar{134})}^{\text{-}1/2}},\fwbox{40pt}{|\bar{\psi}\rangle_{(\bar{268})}^{\text{-}1/2}},\fwbox{40pt}{|\gamma\rangle_{(68)}^{\text{+}1}},\fwbox{40pt}{|\bar{\phi}\rangle_{(\bar{1456})}^{0}},\fwbox{40pt}{|\gamma\rangle_{(14)}^{\text{+}1}},\fwbox{40pt}{\bar{\gamma}\rangle_{(\bar{78})}^{\text{-}1}}\big),$\\
$\displaystyle\phantom{\big\{}\mathcal{A}_{6}^{(1)}\!\big(\fwbox{40pt}{|\bar{\psi}\rangle_{(\bar{134})}^{\text{-}1/2}},\fwbox{40pt}{|\bar{\psi}\rangle_{(\bar{126})}^{\text{-}1/2}},\fwbox{40pt}{|\gamma\rangle_{(16)}^{\text{+}1}},\fwbox{40pt}{|\bar{\phi}\rangle_{(\bar{1456})}^{0}},\fwbox{40pt}{|\gamma\rangle_{(14)}^{\text{+}1}},\fwbox{40pt}{\bar{\gamma}\rangle_{(\bar{78})}^{\text{-}1}}\big)\big\}$}
\mathematicaSequence{egTerms/\!\!.\funL{amp}\brace{\var{$x$}\patternTwo}:>\funL[]{componentAmpN}\brace{\var{$x$}}}{\scalebox{0.825}{$\rule{0pt}{22pt}\displaystyle\Big\{\frac{6407665443847}{483863965200,7939651451171},\frac{7939651451171}{529573865520},\text{-}\frac{19688985665651221}{1467978755221440},\text{-}\frac{46627994481930163}{3145668761188800}\Big\}$}}
\mathematicaSequence{{Total}@\textnormal{\%}}{0}
}\\[-30pt]
}

\newpage
\vspace{-0pt}\sectionAppendix{Tree Amplitudes of Other Theories}{appendix:other_theories}\vspace{-0pt}
%
\vspace{-0pt}\subsectionAppendix{Feynman Diagrams: Local Poles and Denominators}{appendix:local_structures}\vspace{-10pt}

\defnBox{physicalPolesPlanar}{\var{$n$}\pattern}{returns a \built{List} of physical (i.e.~local) inequivalent Mandelstam invariants \funL{s}\brace{\var{$a$},\ldots,\var{$b$}} that can appear in the denominators of planar Feynman diagrams. These number $\tfrac{1}{2}\var{n}(\var{n}{-}3)$. 
\mathematicaBox{
\mathematicaSequence{\funL[1]{physicalPolesPlanar}\brace{6}}{\{\funL{s}\brace{1,2}\,,\funL{s}\brace{2,3}\,,\funL{s}\brace{3,4}\,,\funL{s}\brace{4,5}\,,\funL{s}\brace{5,6}\,,\funL{s}\brace{6,1}\,,\\
\phantom{\{}\funL{s}\brace{1,2,3}\,,\funL{s}\brace{2,3,4}\,,\funL{s}\brace{3,4,5}\,\}}
}
}

\defnBox{cubicGraphDenominatorsPlanar}{\var{$n$}\pattern}{returns a \built{List} of all the denominators (products of Mandelstam invariants) that can appear in planar, cubic Feynman diagrams. The sum of the inverse of the output of \fun{cubicGraphDenominatorsPlanar}\brace{\var{$n$}} is equal to the partial/ordered tree amplitude in $\phi^3$ field theory. Note, however, it would differ from the output of \funL{phi3Amp}\brace{\var{$n$}} by conventions regarding the labeling of Mandelstam invariants:
\mathematicaBox{
\mathematicaSequence[1]{testExprn=Total[1/\funL[1]{cubicGraphDenominatorsPlanar}\brace{6}];
SameQ[testExprn,\funL[]{phi3Amp}\brace{6}]}{False}
\mathematicaSequence{SameQ@@\funL[]{canonicalizeSpinors}\brace{6}\,@\{testExprn,\funL[]{phi3Amp}\brace{6}\}}{True}
}
The number of cubic graph denominators is given by the Catalan number $C_{\var{n}{-}2}$, which is also given by \funL{numberOfPlaneGraphs}\brace{\var{$n$},3}. 
}

\defnBox{cubicGraphDenominators}{\var{$n$}\pattern}{returns a \built{List} of all the denominators (as products of Mandelstam invariants) that can arise in local (not necessarily planar) Feynman diagrams. The sum of the inverses of the output would give the complete amplitude of scalar $\phi^3$ field theory.\\[-10pt]

The output of \fun{cubicGraphDenominators} is somewhat peculiar---first listing the denominators associated with the Feynman graphs whose color-factors would correspond to \funL{basisColorFactors}\brace{$n$} (in that order). This is done in order to simplify expressions for color-kinematic-dual representations of tree amplitudes in sYM and sGR.\\[-10pt]

The number of \fun{cubicGraphDenominators}\brace{\var{$n$}} is given by $(2\var{n}{-}5)!!$. 
}

\defnBox{numberOfPlaneGraphs}{\var{$n$}\pattern,\var{$p$}\pattern\optArg{3}\,}{returns the number of \emph{plane} tree graphs with \var{$n$} external legs and involving \var{$p$}-valent vertices. (A \emph{plane} graph is a \emph{planar} graph endowed with a specific plane embedding---or, a fixed ordering to the external legs.) This returns the number of contributions to \fun{scalarAmp}\brace{\var{$p$}}\,\brace{\var{$n$}}\,, and is equal to 
\eq{\fun{numberOfPlaneGraphs}\brace{\var{n},\var{p}}\,\,\,\equivR\frac{1}{\var{n}{-}1}\binom{\r{v}(\var{p}{-}1)}{\var{n}{-}2}\,}
where $\r{v}\equivR\tfrac{\var{n}{-}2}{\var{p}{-}2}$ is the number of vertices, with $\r{v}\!\in\!\mathbb{Z}$.
}

\defnBox{numberOfYMfeynmanDiagrams}{\var{$n$}\pattern}{returns the number of cubic and quartic Feynman diagrams (non-planar) that contribute to the color-dressed $n$-particle tree amplitude in pure $\mathcal{N}\!=\!0$ YM theory:
\mathematicaBox{
\mathematicaSequence{\funL[1]{numberOfYMfeynmanDiagrams}/@Range[3,12]}{\{1,4,25,220,2485,34300,559405,10525900,224449225,5348843500\}}
}
}

\vspace{-0pt}\subsectionAppendix{Amplitudes of Scalar Field Theories}{appendix:scalar_field_theory}\vspace{-0pt}

\vspace{0pt}\subsubsectionAppendix{\texorpdfstring{(Ordered) Tree Amplitudes of Scalar $\phi^p$ Field Theory}{(Ordered) Tree Amplitudes of Scalar phi-p Field Theory}}{appendix:scalar_phi_p_thy}\vspace{-10pt}

\defnBoxTwo{scalarAmp}{\var{$p$}\pattern}{\var{$n$}\pattern}{returns the \var{$n$}-particle partial tree amplitude in scalar $\phi^{\var{p}}$ field theory with the external legs ordered $\{1,\ldots,\var{n}\}$.\\[-10pt]
}

\defnBoxTwo[]{scalarAmp}{\var{$p$}\pattern}{\var{legOrdering}\patternTwo}{returns the \var{$n$}-particle partial tree amplitude in scalar $\phi^{\var{p}}$ field theory with the external legs ordered $\{\var{\text{legOrdering}}\}$.
}

\defnBox{phi3Amp}{\var{$n$}\pattern}{returns the \var{$n$}-particle partial tree amplitude in scalar $\phi^{3}$ field theory with external legs ordered $\{1,\ldots,\var{n}\}$.  
}

\defnBox{phi3AmpCyclicSeeds}{\var{$n$}\pattern}{gives a \built{List} of cyclic \emph{seeds} whose cyclic sum gives the \var{$n$}-particle tree amplitude in scalar $\phi^3$ field theory.
\mathematicaBox{
\mathematicaSequence{\funL[]{nice}\brace{\funL[1]{phi3AmpCyclicSeeds}\brace{6}\,}}{$\left\{\rule{0pt}{16pt}\right.\displaystyle\!\!\! \frac{1}{s_{2\,3}\,s_{4\,5}\,s_{1\,2\,3}},\frac{1}{2\,s_{1\,2}\,s_{4\,5}\,s_{1\,2\,3}},\frac{1}{2\,s_{2\,3}\,s_{5\,6}\,s_{1\,2\,3}},\frac{1}{3\,s_{1\,2}\,s_{4\,5}\,s_{5\,6}}\left.\rule{0pt}{16pt}\right\}$}
}
\textbf{\emph{Note}}: \emph{all} cyclic images should be summed to obtain the amplitude from these seeds. In the example above, the factors of $\tfrac{1}{2}$ in the second and third terms and the factor of $\tfrac{1}{3}$ in the last are all symmetry factors to reflect their duplication within the cyclic sum.
}

\newpage
\vspace{0pt}\subsubsectionAppendix{Amplitudes in the Non-Linear Sigma Model}{appendix:nlsm_amps}\vspace{-5pt}

As described in \cite{hiddenZeroes} (see also \cite{hiddenZeroes2,hiddenZeroes3}), amplitudes in the non-linear sigma model can be computed directly from the partial (ordered) amplitudes of scalar $\phi^3$ field theory in the following way. Suppose that all arguments of Mandelstam invariants are consecutively ordered, \funL{s}\brace{\var{$a$},\var{$a{+}1$},\ldots,\var{$a{+}{q}$}}\,, then taking
\eq{\fun{s}\brace{\var{a},\ldots,\var{a{+}q}}\,\,\,\mapsto\fun{$\hat{\text{s}}$}\brace{\var{a},\ldots,\var{a{+}q}}\equivR\left\{\begin{array}{@{}l@{$\;\;\;$}l@{}}\fun{s}\brace{\var{a},\ldots,\var{a{+}q}}\,{+}(-1)^{\var{a}}\r{\eta}&\text{if\,\,}\var{q}\text{\,\,odd}\\\fun{s}\brace{\var{a},\ldots,\var{a{+}q}}&\text{if\,\,}\var{q}\text{\,\,even}\end{array}\right.}
the non-linear sigma model amplitude is \emph{minus} the coefficient of $\r\eta^{2{-}n}$ of $\mathcal{A}^{\phi^3}_{n}\!\big(\!\!\{\fun{$\hat{\text{s}}$}\brace{\var{a},\ldots,\var{a{+}q}}\,\}\!\!\big)$ in the limit $\r\eta\!\to\!\infty$.

\defnBox{nlsmAmp}{\var{$n$}\pattern}{returns \built{Total}\brace{\fun{nlsmAmpTerms}\brace{\var{$n$}}\,}.}

~

\defnBox[]{nlsmAmp}{\{\var{legOrdering}\patternTwo\}}{returns \funL{permuteLegs}\brace{\var{legOrdering}}\,\brace{\fun{nlsmAmp}\brace{\var{$n$}}}.}

\defnBox{nlsmAmpTerms}{\var{$n$}\pattern}{returns a \built{List} of terms contributing to the \var{$n$}-particle tree amplitude of the non-linear sigma model.
\mathematicaBox{
\mathematicaSequence{\funL[]{nice}\brace{\funL[1]{nlsmAmpTerms}\brace{6}\,}}{$\left\{\rule{0pt}{16pt}\right.\displaystyle\!\!\! s_{1\,2},s_{2\,3},s_{3\,4},s_{4\,5},s_{5\,6},s_{6\,1},{-}\frac{(s_{1\,2}{+}s_{2\,3})(s_{4\,5}{+}s_{5\,6})}{s_{1\,2\,3}},$\\\mbox{~\hspace{2pt}}$\hspace{10pt}\displaystyle\!\!\!{-}\frac{(s_{2\,3}{+}s_{3\,4})(s_{5\,6}{+}s_{6\,1})}{s_{5\,6\,1}},{-}\frac{(s_{3\,4}{+}s_{4\,5})(s_{1\,2}{+}s_{6\,1})}{s_{6\,1\,2}}\left.\rule{0pt}{16pt}\right\}$}
}
}

\defnBox{nlsmAmpCyclicSeeds}{\var{$n$}\pattern}{returns a \built{List} of cyclic \emph{seeds} whose cyclic sum gives the \var{$n$}-particle tree amplitude in the non-linear sigma model.
\mathematicaBox{
\mathematicaSequence{\funL[]{nice}\brace{\funL[1]{nlsmAmpCyclicSeeds}\brace{8}\,}}{$\left\{\rule{0pt}{16pt}\right.\displaystyle\!\!\! 2\,s_{1\,2},\frac{1}{2}s_{1\,2\,3\,4},\text{-}\frac{(s_{1\,2}{+}s_{2\,3})(s_{4\,5}{+}s_{5\,6}{+}s_{6\,7}{+}s_{7\,8}{+}s_{1\,2\,3\,4}{+}s_{8\,1\,2\,3})}{s_{1\,2\,3}},$\\\mbox{~\hspace{2pt}}$\hspace{10pt}\displaystyle\!\!\!{-}\frac{(s_{1\,2}{+}s_{2\,3})(s_{4\,5}{+}s_{5\,6})(s_{7\,8}{+}s_{8\,1\,2\,3})}{s_{1\,2\,3}\,s_{4\,5\,6}},\frac{(s_{1\,2}{+}s_{2\,3})(s_{5\,6}{+}s_{6\,7})(s_{1\,2\,3\,4}{+}s_{8\,1\,2\,3})}{2\,s_{1\,2\,3}\,s_{5\,6\,7}}\left.\rule{0pt}{16pt}\right\}$}
}
\textbf{\emph{Note}}: \emph{all} cyclic images should be summed to obtain the amplitude from these seeds. In the example above, the factors of $\tfrac{1}{2}$ in the second and last terms are symmetry factors to account for their duplication in the cyclic sum.
}

\vspace{10pt}\subsectionAppendix{Other Quantum Field Theories Obtained as Double Copies}{appendix:other_theories_as_double_copies}\vspace{-10pt}
\defnBox{specialGalileonAmp}{\var{$n$}\pattern}{returns the \var{$n$}-particle tree amplitude in the special Galileon theory, as constructed via the KLT double-copy:
\vspace{-5pt}\begin{align}&\hspace{-130pt}\fun{specialGalileonAmp}\brace{\var{n}}\,\,\,\Leftrightarrow\\
&\scalebox{0.9}{$\fwboxL{200pt}{\hspace{-150pt}\big(\funL{nlsmAmp}@@@\funL{kltBasisAmpsAlpha}\brace{\var{n}}\big)\,.\funL{kltKernel}\brace{\var{n}}\,.\big(\funL{nlsmAmp}@@@\funL{kltBasisAmpsBeta}\brace{\var{n}}\big)\,.}$}\nonumber
\end{align}\vspace{-30pt}
}

\defnBox{diracBornInfeldAmp}{\var{$n$}\pattern}{returns the \var{$n$}-particle tree amplitude of superDirac-Born-Infeld field theory, as constructed via the KLT double-copy:
\begin{align}&\hspace{-130pt}\fun{diracBornInfeldAmp}\brace{\var{n}}\,\,\,\Leftrightarrow\\
&\scalebox{0.9}{$\fwboxL{200pt}{\hspace{-150pt}\big(\t{Total[}\funL{ymAmp}\brace{\var{n},\var{n}/2{-}2,\{\t{\#\#}\}}\t{]\&@@@}\funL{kltBasisAmpsAlpha}\brace{\var{n}}\big)\,\,.\phantom{.\funL{kltKernel}\brace{\var{n}}\,.\big(\funL{nlsmAmp}@@@\funL{kltBasisAmpsBeta}\brace{\var{n}}\big)}}$}\nonumber\\
&\scalebox{0.9}{$\fwboxL{200pt}{\hspace{20pt}\funL{kltKernel}\brace{\var{n}}\,.\big(\funL{nlsmAmp}@@@\funL{kltBasisAmpsBeta}\brace{\var{n}}\big)\,.}$}\nonumber
\end{align}\vspace{-20pt}
}

\vspace{-0pt}\sectionAppendix{Twistor String Theory and the Scattering Equations}{appendix:chy_rsv_et_al}\vspace{-0pt}
%
\vspace{-0pt}\subsectionAppendix{General Aspects of the Scattering Equation Formalism}{appendix:general_chy_stuff}\vspace{-10pt}

\defnBox{scatteringEquations}{\var{$n$}\pattern}{gives the \emph{scattering equations} for \var{$n$} particles:
\eq{\fun{scatteringEquations}\brace{\var{n}}\,\,\,\Leftrightarrow\,\,\big\{S_1,\ldots,S_{\var{n}}\big\}}
where the $\r{a}$th scattering equation, $S_\r{a}$, is given by
\eq{S_{\r{a}}\equivR\sum_{\b{b}\neq \r{a}}\frac{s_{\r{a}\,\b{b}}}{\fun{$\sigma$}\brace{\r{a},\b{b}}}=0\,\label{scattering_equation_a}}
where \fun{$\sigma$}\brace{\var{$a$},\var{$b$}}$\equivR$\fun{$\sigma$}\brace{\var{$a$}}${-}$\fun{$\sigma$}\brace{\var{$b$}}\,, and \fun{$\sigma$}\brace{$a$}$\,\in\!\mathbb{P}^1$ is an auxiliary variable associated with the particle indexed by $a\!\in\![\var{n}]$. These variables are \emph{determined} (up to an $SL(2,\mathbb{C})$ redundancy) by the requirement that (\ref{scattering_equation_a}) is satisfied for all $a\!\in\![\var{n}]$.
}

\defnBox{scatteringEquationsJacobian}{\var{$n$}\pattern}{in the context of the scattering equation formalism of CHY, one is often interested in integrals of the form 
\eq{\int\!\!\Omega[\vec{\sigma}]\,\,\delta^{n{-}3}\!(\mathbf{S}_{rst})=\sum_{\vec{\sigma}^*}\frac{1}{\det(\mathbf{J}_{rst})}\left.\Omega[\vec{\sigma}]\rule{0pt}{18pt}\right|_{\vec{\sigma}=\vec{\sigma}^*}\quad\text{where}\quad\mathbf{J}\equivR\{\partial S_{a}/\partial \sigma_b\}}
where $\mathbf{J}'_{rst}$ is the matrix $\mathbf{J}$ with rows/columns $\{r,s,t\}\!\in\![n]$ removed. (Due to the $SL(2,\mathbb{C})$ redundancy in the coordinates $\sigma_a$, the full $(\var{n}\!\times\!\var{n})$ matrix $\mathbf{J}$ is only rank $(\var{n}{-}3)$.) As the determinant of $\mathbf{J}$ is antisymmetric in the ordering of the rows/columns, some care must be taken to ensure that no sign mistakes are made.  

Our conventions are to always use the $SL(2,\mathbb{C})$ redundancy to set $\{\sigma_1,\sigma_2,\sigma_3\}\!\to\!\{\infty,0,1\}$, and thus to define the effective Jacobian $\mathbf{J}'$ by eliminating the first three of its rows/columns. This introduces a minus sign depending on $\var{n}$, which we have chosen to absorb into the definition of the matrix $\mathbf{J}$. Thus, the output of \fun{scatteringEquationsJacobian} is the matrix
\eq{{-}\,\texttt{D[\fun{scatteringEquations}\brace{\var{$n$}}[\![4;;{-}1]\!],\{\funL{$\sigma$}/@Range[4,\var{$n$}]\}]}\,.\vspace{-20pt}}
}

\defnBox{solveScatteringEquations}{\var{$n$}\pattern\optArg{},\var{$k$}\pattern\optArg{}}{when the optional argument $\var{k}$ is not specified, returns a \built{List} of $(\var{n}{-}3)!$ pairs $\{\t{solnRules},\t{solutionJacobian}\}$ \emph{for the current, globally specified kinematics} encoded by \builtL{Ls} and \builtL{Lbs}, where \t{solnRules} provides a solution to the scattering equations and \t{solutionJacobian} is the corresponding solution's (determinant) of the Jacobian of the scattering equations evaluated at that particular solution. 
\mathematicaBox{
\mathematicaSequence{solnData=\funL[]{timed}@\funL[1]{solveScatteringEquations}\brace{7};\\
Length[solnData]}{24}
\mathematicaSequence{First[solnData]}{\big\{\big\{$\sigma$\brace{1}$\,\rightarrow\!\!\infty$,$\sigma$\brace{2}$\,\rightarrow\!\!0$,$\sigma$\brace{3}$\rightarrow\!\!1$,$\sigma$\brace{4}$\rightarrow\!\frac{5}{4}$,$\sigma$\brace{5}$\rightarrow\!\frac{27}{20}$,$\sigma$\brace{6}$\rightarrow\!\frac{7}{5}$,$\sigma$\brace{7}$\rightarrow\!\frac{10}{7}$\big\},\\
$\rule{0pt}{20pt}\displaystyle\frac{55071241349532800}{24057}$\big\}}
\mathematicaSequence{Det\brace{\funL[]{evaluate}\brace{\funL[1]{scatteringEquationsJacobian}\brace{7}\,}}/\!\!.First[solnData][\![1]\!]}{$\rule{0pt}{20pt}\displaystyle\frac{55071241349532800}{24057}$}
\mathematicaSequence[1]{checkSolns=\funL[]{evaluate}\brace{\funL[1]{scatteringEquations}\brace{7}}/\!\!.solnData[\![All,1]\!];\\
DeleteDuplicates\brace{Chop\brace{checkSolns,Power\brace{10,-900}\,}\,}}{\{\{0,0,0,0,0,0,0\}\}}
}
The call to `\built{Chop}' above is necessary because the most (all but the first and last) are only computed numerically to a \built{Precision} of around 1000 digits.\\[-10pt]

Whenever four-dimensional kinematics are used, the $(\var{n}{-}3)!$ solutions are arranged into sets of size \fun{numberOfScatteringEquationSolutions}\brace{\var{$n$},\var{$k$}} for \mbox{$\var{k}\!\in\![0,\var{n}{-}4]$}. The restriction of the scattering integral to sum over only these \var{$k$}-indexed subsets of solutions will always be rational, but may or may not be meaningful. However, this indexing is especially helpful for amplitudes of YM and GR, for which N${}^{\var{k}}$MHV component amplitudes will only have support on the corresponding subset of solutions.\\[-10pt]

When the optional argument $\var{k}\!\in\![0,\var{n}{-}4]$ is specified, it returns only those solutions in the relevant sector indexed by $\var{k}$.\\[-10pt]

The argument $\var{n}$ is also optional, as it \emph{must} be equal to \built{Length}\brace{\builtL{Ls}}. 
}

\defnBox{numberOfScatteringEquationSolutions}{\var{$n$}\pattern,\var{$k$}\pattern}{returns the number of solutions in the subset indexed by \var{$k$}:
\vspace{-5pt}\eq{\fun{numberOfScatteringEquationSolutions}\brace{\var{n},\var{k}}\,\,\,\Leftrightarrow\,\left\langle\begin{array}{@{}c@{}}\var{n}{-}3\\\var{k}\end{array}\right\rangle\vspace{-20pt}}
}

\defnBox{directCHYintegration}{\var{integrand}\pattern,\var{summedQ}\pattern\optArg{\built{True}}\,}{requires that \var{integrand} be some integrand relevant to $\var{n}=$\built{Length}\brace{\builtL{Ls}}, and returns the (rationalized) result of 
\eq{\int\!\!\!\Omega[\vec{\sigma}]\,\,(\text{\var{integrand}})\,\,\delta^{\var{n}{-}3}\!\big(\mathbf{S}_{rst}\big)\,.}
\mathematicaBox{
\mathematicaSequence[1]{egIntegrand=\funL[]{biadjointScalarCHYintegrand}\brace{7};\\
\funL[]{nice}\brace{egIntegrand}}{$\displaystyle\frac{1}{(1\,2)^2\,(2\,3)^2(3\,4)^2(4\,5)^2(5\,6)^2(6\,7)^2(7\,1)^2}$}
\mathematicaSequence{\funL[1]{directCHYintegration}\brace{egIntegrand}}{$\displaystyle\frac{310571971461476568023}{207118043370225025038000}$}
\mathematicaSequence{\textnormal{\%}===\funL[]{evaluate}\brace{\funL[]{phi3Amp}\brace{7}\,}}{True}
}

As the entire integral is localized, this amounts to a sum over particular solutions. If the optional argument \var{summedQ} is unspecified, it defaults to \built{True}; in this case, all $(\var{n}{-}3)!$ solutions are included.\\[-10pt]

If \var{summedQ} is set to \built{False}, then \fun{directCHYintegration} returns a list of length $\var{n}{-}3$ corresponding to the partial sum over solutions indexed by $\var{k}\!\in\![0,\var{n}{-}4]$. 
\mathematicaBox{
\mathematicaSequence{helList=\funL[]{randomHelicityComponent}\brace{7,\#}\,\&/@Range[0,3]}{\{\!\{\stateYM{p},\stateYM{p},\stateYM{p},\stateYM{m},\stateYM{p},\stateYM{m},\stateYM{p}\},\{\stateYM{m},\stateYM{p},\stateYM{m},\stateYM{p},\stateYM{p},\stateYM{p},\stateYM{m}\},\{\stateYM{p},\stateYM{m},\stateYM{m},\stateYM{p},\stateYM{p},\stateYM{m},\stateYM{m}\},\{\stateYM{m},\stateYM{p},\stateYM{m},\stateYM{m},\stateYM{p},\stateYM{m},\stateYM{m}\}\!\}}
\mathematicaSequence[1]{\funL[]{evaluate}\brace{\funL[]{ymCHYintegrand}\brace{7}/\!\!.\funL[]{helicityPolarizationRules}/@helList};\\
\funL[1]{directCHYintegration}\brace{\#,False}\,\&/@\textnormal{\%};\\
\funL[]{nice}[\textnormal{\%}]}{$\displaystyle\left(\begin{array}{@{}cccc@{}}\frac{243}{560}&0&0&0\\
0&\frac{27982196720540251}{3117464078414400}&0&0\\
0&0&\frac{10095465473426621568887}{50664305288822695470000}&0\\
0&0&0&\frac{25626566889}{4205327360000}\end{array}\right)$}
\mathematicaSequence{\funL[]{componentAmpN}@@@helList}{$\displaystyle\Big\{\frac{243}{560},\frac{27982196720540251}{3117464078414400},\frac{10095465473426621568887}{50664305288822695470000},\frac{25626566889}{4205327360000}\Big\}$}
}
}

\newpage
%
\vspace{-0pt}\subsectionAppendix{Specific Ingredients of the Scattering Equation Formalism of CHY}{appendix:specific_to_chy_suff}\vspace{-0pt}

\defnBox[]{chyCycle}{\var{$n$}\pattern}{is equivalent to \fun{chyCycle}\brace{\built{Range}[\var{$n$}]}\,.}

\defnBox{chyCycle}{\{\var{$a$},\var{$b$}\patternTwo,\var{$c$}\}}{returns $1/(\fun{$\sigma$}\brace{\var{a},\var{b}}\cdots\fun{$\sigma$}\brace{\var{c},\var{a}})$\,.}

~

\defnBox[]{chyCycle}{\{\var{permA}\patternTwo\},\{\var{permB}\patternTwo\}}{is \fun{chyCycle}\brace{\{\var{permA}\}}\,\fun{chyCycle}\brace{\{\var{permB}\}}\,.}

\defnBox{chyAmatrix}{\var{$n$}\pattern}{returns the matrix `$A$' for \var{$n$} particles defined by Cachazo, He and Yuan in \cite{Cachazo:2013gna,Cachazo:2013hca,Cachazo:2013iaa,Cachazo:2013iea,Dolan:2013isa,Dolan:2014ega}; specifically, this is a matrix
\eq{\fun{chyAmatrix}\brace{\var{n}}\,\,\,\Leftrightarrow\,\mathbf{A}\equivR\{A_{a\,b}\}\quad\text{where}\quad A_{a\,b}\equivR\left\{\begin{array}{@{}ll}\frac{s_{a\,b}}{\sigma_{a\,b}}&\text{if\;}a\neq b\\
0&\text{if\;}a=b\end{array}\right.\vspace{-10pt}\label{chy_a_matrix}}
}

\defnBox{chyPsiMatrix}{\var{$n$}\pattern,\var{abstractQ}\pattern\optArg{\built{True}}\,}{returns the matrix `$\Psi$' for \var{$n$} particles defined by Cachazo, He and Yuan in \cite{Cachazo:2013gna,Cachazo:2013hca,Cachazo:2013iaa,Cachazo:2013iea,Dolan:2013isa,Dolan:2014ega} to describe Yang-Mills amplitudes in the scattering equation formalism. 
This matrix is often described as the $(2\var{n})\!\times\!(2\var{n})$ matrix
\eq{\fun{chyPsiMatrix}\brace{\var{n}}\,\,\,\Leftrightarrow\,\,\mathbf{\Psi}\equivR\left(\begin{array}{@{}cc@{}}\mathbf{A}&\text{-}\mathbf{C}^{T}\\
\mathbf{C}&\mathbf{B}\end{array}\right)}
where $\mathbf{A}$ is given in (\ref{chy_a_matrix}), $\mathbf{B}$ is the same as $\mathbf{A}$ but with \mbox{\funL{s}\brace{\var{$a$},\var{$b$}}$\,\mapsto$\funL{ss}\brace{\funL{$\epsilon$}\brace{\var{a}}\,,\,\funL{$\epsilon$}\brace{\var{$b$}}}\,,} and where $\mathbf{C}$ is essentially the same again but with  \funL{s}\brace{\var{$a$},\var{$b$}}$\,\mapsto$\funL{ss}\brace{\funL{$\epsilon$}\brace{\var{a}}\,,\,\funL{p}\brace{\var{$b$}}}\,; but the diagonal entries of $\mathbf{C}$ are somewhat involved (and not especially unique).\\[-10pt]

As it is often useful to ignore the diagonal entries of $\mathbf{C}$, the default option of \fun{chyPsiMatrix} is to leave it \emph{abstract}---due to the default value of the optional second argument \var{abstractQ} being \built{True}.
\mathematicaBox{
\mathematicaSequence{\funL[]{nice}\brace{\funL[1]{chyPsiMatrix}\brace{4}}}{$\displaystyle\left(\begin{array}{@{}c@{}c@{}c@{}c@{}c@{}c@{}c@{}c@{}}0&\frac{s_{1\,2}}{(1\,2)}&\frac{s_{1\,3}}{(1\,3)}&\frac{s_{1\,4}}{(1\,4)}&c_{1\,1}&\frac{(\epsilon_2{+}p_1)^2}{(1\,2)}&\frac{(\epsilon_3{+}p_1)^2}{(1\,3)}&\frac{(\epsilon_4{+}p_1)^2}{(1\,4)}\\[-0pt] 
\text{-}\frac{s_{1\,2}}{(1\,2)}\phantom{\text{-}}&0&\frac{s_{2\,3}}{(2\,3)}&\frac{s_{2\,4}}{(2\,4)}&\text{-}\frac{(\epsilon_1{+}p_2)^2}{(1\,2)}\phantom{\text{-}}&c_{2\,2}&\frac{(\epsilon_3{+}p_2)^2}{(2\,3)}&\frac{(\epsilon_4{+}p_2)^2}{(2\,4)}\\[-0pt] 
\text{-}\frac{s_{1\,3}}{(1\,3)}\phantom{\text{-}}&\text{-}\frac{s_{2\,3}}{(2\,3)}\phantom{\text{-}}&0&\frac{s_{3\,4}}{(3\,4)}&\text{-}\frac{(\epsilon_1{+}p_3)^2}{(1\,3)}\phantom{\text{-}}&\text{-}\frac{(\epsilon_2{+}p_3)^2}{(2\,3)}\phantom{\text{-}}&c_{3\,3}&\frac{(\epsilon_4{+}p_3)^2}{(3\,4)}\\[-0pt] 
\text{-}\frac{s_{1\,4}}{(1\,4)}\phantom{\text{-}}&\text{-}\frac{s_{2\,4}}{(2\,4)}\phantom{\text{-}}&\text{-}\frac{s_{3\,4}}{(3\,4)}\phantom{\text{-}}&0&\text{-}\frac{(\epsilon_1{+}p_4)^2}{(1\,4)}\phantom{\text{-}}&\text{-}\frac{(\epsilon_2{+}p_4)^2}{(2\,4)}\phantom{\text{-}}&\text{-}\frac{(\epsilon_3{+}p_4)^2}{(3\,4)}\phantom{\text{-}}&c_{4\,4}\\[-0pt] 
\text{-}c_{1\,1}\phantom{\text{-}}&\frac{(\epsilon_1{+}p_2)^2}{(1\,2)}&\frac{(\epsilon_1{+}p_3)^2}{(1\,3)}&\frac{(\epsilon_1{+}p_4)^2}{(1\,4)}&0&\frac{(\epsilon_1{+}\epsilon_2)^2}{(1\,2)}&\frac{(\epsilon_1{+}\epsilon_3)^2}{(1\,3)}&\frac{(\epsilon_1{+}\epsilon_4)^2}{(1\,4)}\\[-0pt] 
\text{-}\frac{(\epsilon_2{+}p_1)^2}{(1\,2)}\phantom{\text{-}}&\text{-}c_{2\,2}\phantom{\text{-}}&\frac{(\epsilon_2{+}p_3)^2}{(2\,3)}&\frac{(\epsilon_2{+}p_4)^2}{(2\,4)}&\text{-}\frac{(\epsilon_1{+}\epsilon_2)^2}{(1\,2)}\phantom{\text{-}}&0&\frac{(\epsilon_2{+}\epsilon_3)^2}{(2\,3)}&\frac{(\epsilon_2{+}\epsilon_4)^2}{(2\,4)}\\[-0pt] 
\text{-}\frac{(\epsilon_3{+}p_1)^2}{(1\,3)}\phantom{\text{-}}&\text{-}\frac{(\epsilon_3{+}p_2)^2}{(2\,3)}\phantom{\text{-}}&\text{-}c_{3\,3}\phantom{\text{-}}&\frac{(\epsilon_3{+}p_4)^2}{(3\,4)}&\text{-}\frac{(\epsilon_1{+}\epsilon_3)^2}{(1\,3)}\phantom{\text{-}}&\text{-}\frac{(\epsilon_2{+}\epsilon_3)^2}{(2\,3)}\phantom{\text{-}}&0&\frac{(\epsilon_3{+}\epsilon_4)^2}{(3\,4)}\\[-0pt] 
\text{-}\frac{(\epsilon_4{+}p_1)^2}{(1\,4)}\phantom{\text{-}}&\text{-}\frac{(\epsilon_4{+}p_2)^2}{(2\,4)}\phantom{\text{-}}&\text{-}\frac{(\epsilon_4{+}p_3)^2}{(3\,4)}\phantom{\text{-}}&\text{-}c_{4\,4}\phantom{\text{-}}&\text{-}\frac{(\epsilon_1{+}\epsilon_4)^2}{(1\,4)}\phantom{\text{-}}&\text{-}\frac{(\epsilon_2{+}\epsilon_4)^2}{(2\,4)}\phantom{\text{-}}&\text{-}\frac{(\epsilon_3{+}\epsilon_4)^2}{(3\,4)}\phantom{\text{-}}&0
\end{array}\right)$}
}
}

\newpage
\vspace{0pt}\subsubsection*{Reduced Pfaffians in the CHY Formalism}\label{appendix:nlsm_amps}\vspace{-10pt}

\defnBox{chyPfaffianPrime}{\var{matrix}\pattern,\{\var{$i$}\pattern,\var{$j$}\pattern\}}{returns what is sometimes called the \emph{reduced Pfaffian} of the matrix \var{matrix}---that is, the Pfaffian where the matrix's rows and columns $\{\var{i},\var{j}\}$ have been removed. In the CHY formalism, this comes with an additional factor of $(\text{-}1)^{\var{i}\text{+}\var{j}}/$\funL{$\sigma$}\brace{\var{$i$},\var{$j$}}\,, which is included also in the output of \fun{chyPfaffianPrime}. 
}

\defnBox{chyAPfaffianPrime}{\var{$n$},\var{$ij$}\pattern\optArg{}}{returns `$\text{Pf'}\!(\mathbf{A})$' which appears, for example in the construction of the NLSM integrand in the representation of CHY; the deleted columns/rows are $\{\var{i},\var{j}\}\!\subset\![\var{n}]$. If the optional argument $\var{ij}$ is unspecified, then the first and last rows are removed by default. 
}

\defnBox{chyPsiPfaffianPrime}{\var{$n$}\pattern,\var{$ij$}\pattern\optArg{}}{returns `$\text{Pf'}\!(\Psi)$' which appears, for example in the construction of the YM integrand in the representation of CHY; where the deleted columns/rows are $\{\var{i},\var{j}\}\!\subset\![\var{n}]$. If the optional argument $\var{ij}$ is unspecified, then these are taken to be $\{\var{i},\var{j}\}\!\mapsto\!\{1,\var{n}\}$. 
}

\defnBox{chyAPfaffianPrime2InCycles}{\var{$n$}}{returns a \built{List} of terms contributing to $(\text{Pf'}\!(\mathbf{A}))^2$ (aka the determinant), where each term appearing has been decomposed according to the algorithm of \cite{Bjerrum-Bohr:2016axv} to involve complete \var{$n$}-cycles. These terms are listed according to $\{$\t{coeff},\{\var{$\alpha$}\patternTwo\}\} where \t{coeff} is the coefficient of the term $1/(\sigma_{\var{\alpha_1}\,\var{\alpha_2}}\sigma_{\var{\alpha_2}\,\var{\alpha_3}}\cdots\sigma_{\var{\alpha_{\text{-}1}}\,\var{\alpha_1}})$ in the decomposition of $(\text{Pf'}\!(\mathbf{A}))^2$.\\[-10pt]

This is useful because the CHY-integration over products of such simple, complete cycles can be done analytically: it results in \funL{kltKernelInverse}\brace{\var{$\alpha$},\var{$\beta$}}. 
}

\defnBox{chyPsiPfaffianPrimeInCycles}{\var{$n$}}{returns a \built{List} of terms contributing to $\text{Pf'}\!(\Psi)$ after decomposition according to the algorithm of cite soas to involve only complete $\var{n}$-cycles. These are listed according to  $\{$\t{coeff},\{\var{$\alpha$}\patternTwo\}\} where \t{coeff} is the coefficient of the term $1/(\sigma_{\var{\alpha_1}\,\var{\alpha_2}}\sigma_{\var{\alpha_2}\,\var{\alpha_3}}\cdots\sigma_{\var{\alpha_{\text{-}1}}\,\var{\alpha_1}})$ in the decomposition of $\text{Pf'}\!(\Psi)$.\\[-10pt]

This is useful because the CHY-integration over products of such simple, complete cycles can be done analytically: it results in \funL{kltKernelInverse}\brace{\var{$\alpha$},\var{$\beta$}}. For Yang-Mills, this gives rise to a manifestly color-kinematic dual representation. 
}

\newpage
\vspace{-0pt}\subsectionAppendix{Representing Amplitudes in the Scattering Equation Formalism}{appendix:chy_amps}\vspace{-10pt}

\defnBox{biadjointScalarCHYintegrand}{\var{$n$}\pattern}{returns the representation of the ordered, partial \var{$n$}-particle tree amplitude of $\phi^3$ field theory in the scattering equation formalism of CHY:
\eq{\fun{biadjointScalarCHYintegrand}\brace{\var{n}}\,\,\,\Leftrightarrow\,\left(\frac{1}{\sigma_{1\,2}\sigma_{2\,3}\cdots\sigma_{\var{n}\,1}}\right)^2\,.\vspace{-10pt}}
\mathematicaBox{
\mathematicaSequence[1]{
integrand=\funL[1]{biadjointScalarCHYintegrand}\brace{6};\\
\funL[]{nice}\brace{integrand}}{$\displaystyle\frac{1}{(1\,2)^2(2\,3)^2(3\,4)^2\,(4\,5)^2\,(5\,6)^2\,(6\,1)^2}$}
\mathematicaSequence[1]{\funL[]{useReferenceKinematics}\brace{6};\\
\funL[]{directCHYintegration}\brace{integrand}}{$\rule{0pt}{20pt}\displaystyle\frac{892012548475}{4580937620424}$}
\mathematicaSequence{\funL[]{evaluate}\brace{\funL[]{phi3Amp}\brace{6}}}{$\displaystyle\frac{892012548475}{4580937620424}$}
}
}

\defnBox{nlsmCHYintegrand}{\var{$n$}\pattern}{returns the representation of the ordered, partial \var{$n$}-particle tree amplitude of the non-linear sigma model in the scattering equation formalism of CHY:
\eq{\fun{nlsmCHYintegrand}\brace{\var{n}}\,\,\,\Leftrightarrow\,\frac{1}{\sigma_{1\,2}\sigma_{2\,3}\cdots\sigma_{\var{n}\,1}}(\text{Pf'}(\mathbf{A}))^2\,.\vspace{-10pt}}
\mathematicaBox{
\mathematicaSequence[1]{
integrand=\funL[1]{nlsmCHYintegrand}\brace{6};\\
\funL[]{nice}\brace{integrand}}{$\displaystyle\frac{1}{(1\,2)(2\,3)(3\,4)\,(4\,5)\,(5\,6)\,(6\,1)^2}\left(\frac{s_{2\,5}s_{3\,4}}{(2\,5)(3\,4)}{-}\frac{s_{2\,4}s_{3\,5}}{(2\,4)(3\,5)}{+}\frac{s_{2\,3}s_{4\,5}}{(2\,3)(4\,5)}\right)^2$}
\mathematicaSequence[1]{\funL[]{useReferenceKinematics}\brace{6};\\
\funL[]{directCHYintegration}\brace{integrand}}{$\rule{0pt}{20pt}\displaystyle\frac{740186713}{30433320}$}
\mathematicaSequence{\funL[]{evaluate}\brace{\funL[]{nlsmAmp}\brace{6}}}{$\displaystyle\frac{740186713}{30433320}$}
}
}

\defnBox{specialGalileonCHYintegrand}{\var{$n$}\pattern}{returns the representation of the \var{$n$}-particle tree amplitude of the Special Galileon field theory in the scattering equation formalism of CHY:
\eq{\fun{specialGalileonCHYintegrand}\brace{\var{n}}\,\,\,\Leftrightarrow\,(\text{Pf'}(\mathbf{A}))^4\,.\vspace{-10pt}}
\mathematicaBox{
\mathematicaSequence[1]{
integrand=\funL[1]{specialGalileonCHYintegrand}\brace{6};\\
\funL[]{nice}\brace{integrand}}{$\displaystyle\frac{1}{(1\,6)^4}\left(\frac{s_{2\,5}s_{3\,4}}{(2\,5)(3\,4)}{-}\frac{s_{2\,4}s_{3\,5}}{(2\,4)(3\,5)}{+}\frac{s_{2\,3}s_{4\,5}}{(2\,3)(4\,5)}\right)^4$}
\mathematicaSequence[1]{\funL[]{useReferenceKinematics}\brace{6};\\
\funL[]{directCHYintegration}\brace{integrand}}{$\rule{0pt}{20pt}\displaystyle\frac{4873781060219512306435390420649}{54997987753751793897262500}$}
\mathematicaSequence{\funL[]{evaluate}\brace{\funL[]{specialGalileonAmp}\brace{6}}}{$\displaystyle\frac{4873781060219512306435390420649}{54997987753751793897262500}$}
}
}

\defnBox{diracBornInfeldCHYintegrand}{\var{$n$}\pattern}{returns the representation of the \var{$n$}-particle tree amplitude of Dirac-Born-Infeld field theory in the scattering equation formalism of CHY:
\eq{\fun{specialGalileonCHYintegrand}\brace{\var{n}}\,\,\,\Leftrightarrow\,\text{Pf'}\!(\Psi)(\text{Pf'}(\mathbf{A}))^2\,.\vspace{-10pt}}
\mathematicaBox{
\mathematicaSequence[1]{
eg=\funL[1]{specialGalileonCHYintegrand}\brace{6};\\
\funL[]{directCHYintegration}\brace{eg/\!\!.\mbox{\funL[]{helicityPolarizationRules}\brace{\stateYM{m},\stateYM{p},\stateYM{m},\stateYM{p},\stateYM{m},\stateYM{p}}}}}{$\rule{0pt}{20pt}\displaystyle\text{-}\frac{359037941455903955653982723581}{154874333514565051614691200}$}
\mathematicaSequence{\funL[]{component}\brace{\stateYM{m},\stateYM{p},\stateYM{m},\stateYM{p},\stateYM{m},\stateYM{p}}\,\brace{\funL[]{evaluate}\brace{\funL[]{specialGalileonAmp}\brace{6}}}}{$\displaystyle\text{-}\frac{359037941455903955653982723581}{154874333514565051614691200}$}
}
}

\defnBox{ymCHYintegrand}{\var{$n$}\pattern}{returns the representation of the ordered, partial \var{$n$}-particle tree amplitude of non-supersymmetric Yang-Mills field theory in the scattering equation formalism of CHY:
\eq{\fun{ymCHYintegrand}\brace{\var{n}}\,\,\,\Leftrightarrow\,\frac{1}{\sigma_{1\,2}\sigma_{2\,3}\cdots\sigma_{\var{n}\,1}}\text{Pf'}\!(\Psi)\,.\vspace{-10pt}}
\mathematicaBox{
\mathematicaSequence[1]{
eg=\funL[1]{ymCHYintegrand}\brace{6};\\
\funL[]{directCHYintegration}\brace{eg/\!\!.\mbox{\funL[]{helicityPolarizationRules}\brace{\stateYM{m},\stateYM{p},\stateYM{m},\stateYM{p},\stateYM{m},\stateYM{p}}}}}{$\rule{0pt}{20pt}\displaystyle\text{-}\frac{464575984240415}{164913754335264}$}
\mathematicaSequence{\funL[]{componentAmpN}\brace{\stateYM{m},\stateYM{p},\stateYM{m},\stateYM{p},\stateYM{m},\stateYM{p}}}{$\displaystyle\text{-}\frac{464575984240415}{164913754335264}$}
}
}

\defnBox{grCHYintegrand}{\var{$n$}\pattern}{returns the representation of the \var{$n$}-particle tree amplitude of non-supersymmetric Gravity field theory in the scattering equation formalism of CHY:
\eq{\fun{grCHYintegrand}\brace{\var{n}}\,\,\,\Leftrightarrow\,(\text{Pf'}\!(\Psi))^2\,.\vspace{-10pt}}
\mathematicaBox{
\mathematicaSequence[1]{
eg=\funL[1]{grCHYintegrand}\brace{6};\\
\funL[]{directCHYintegration}\brace{eg/\!\!.\mbox{\funL[]{helicityPolarizationRules}\brace{\stateYM{m},\stateYM{p},\stateYM{m},\stateYM{p},\stateYM{m},\stateYM{p}}}}}{$\rule{0pt}{20pt}\displaystyle\frac{10021477263897884263649814417383981100505}{58223047135110819422758189970155094016}$}
\mathematicaSequence{\funL[]{componentAmpN}\brace{\stateGR{mm},\stateGR{pp},\stateGR{mm},\stateGR{pp},\stateGR{mm},\stateGR{pp}}}{$\displaystyle\frac{10021477263897884263649814417383981100505}{58223047135110819422758189970155094016}$}
}
}

\newpage
\vspace{-0pt}\subsectionAppendix{Grassmannian Representation of the Twistor String}{appendix:grassmannian_twistor_string}\vspace{-10pt}

\defnBox{gaugeFixedGrassmannianRepresentative}{\var{$n$}\pattern,\var{$k$}\pattern}{gives \fun{gaugeFixedGrassmannianRepresentative}\brace{\var{$n$},Range[\var{$k{+}2$}]}.
}
\\[-10pt]~

\defnBox[]{gaugeFixedGrassmannianRepresentative}{\var{$n$}\pattern,\{\var{gaugeFixedColumns}\patternTwo\}}{returns a $(\var{k}{+}2)\!\times\!\var{n}$ matrix parameterized by independent Pl\"ucker coordinates \funL{m}\brace{} where the columns labeled by \var{gaugeFixedColumns}$\,\in\!\binom{[\var{n}]}{\var{k}{+}2}$ are set to the identity.\\[-10pt] 
\mathematicaBox{
\mathematicaSequence{egMat=\funL[1]{gaugeFixedGrassmannianRepresentative}\brace{8,\{1,3,5,7\}};\\
\funL[]{nice}\brace{egMat}}{$\displaystyle\left(\begin{array}{@{}cccccccc@{}}
1&(2\,3\,5\,7)&0&(4\,3\,5\,7)&0&(6\,3\,5\,7)&0&(8\,3\,5\,7)\\ 
0&(1\,2\,5\,7)&1&(1\,4\,5\,7)&0&(1\,6\,5\,7)&0&(1\,8\,5\,7)\\ 
0&(1\,3\,2\,7)&0&(1\,3\,4\,7)&1&(1\,3\,6\,7)&0&(1\,3\,8\,7)\\ 
0&(1\,3\,5\,2)&0&(1\,3\,5\,4)&0&(1\,3\,5\,6)&1&(1\,3\,5\,8)
\end{array}\right)$}
}
Notice that each Pl\"ucker coordinate is the corresponding \emph{minor} of the matrix. 
}

\defnBox{grassmannianKinematicEquations}{\var{cMatrix}\pattern}{for a $(\var{k}{+}2)\!\times\!\var{n}$ matrix \var{cMatrix}, the kinematic equations are simply: 
\eq{\fwbox{0pt}{\hspace{-40pt}\fun{grassmannianKinematicEquations}\brace{\var{C}}\,\,\,\Leftrightarrow\,\text{\built{Join}\brace{\var{$C$}.\builtL{Lbs},\,\built{NullSpace}\brace{\var{$C$}}.\builtL{Ls}}}\,.}\vspace{-20pt}\label{gkn_kinematic_constratins}}
}

\defnBox{grassmannianTwistorStringMeasure}{\var{$n$}\pattern,\var{$k$}\pattern}{in the Grassmannian formulation of the connected prescription of the twistor string (see e.g.~\cite{ArkaniHamed:2009dg,Nandan:2009cc,Bourjaily:2010kw}), the \var{$n$}-particle N${}^{\var{k}}$MHV tree amplitude in sYM is given as an integral of the form 
\eq{\fwbox{420pt}{\hspace{-20pt}\mathcal{A}_n^{(k)}\equivL\int\!\!\!\frac{d^{(\var{k}+2)\times\var{n}}{C}}{\mathrm{vol}(GL(\var{k}{+}2))}\,\,M(C)\,\delta^{\var{k}(\var{n}{-}\var{k}{-}4)}\!\big(\vec{V}(C)\big)\delta^{(\var{k}{+}2)\!\times\!2}\!\big(C\!\cdot\!\tilde\lambda\big)\delta^{2\!\times\!(\var{n}{-}\var{k}{-}2)\!\times\!2}\!\big(\lambda\!\cdot\!C^{\perp}\big)\delta^{2\!\times\!4}\!\big(\lambda\!\cdot\!\tilde\eta\big)}\label{grassmannian_twistor_string_form}}
where the \emph{measure} is 
\eq{M(C)\,\,\Leftrightarrow\,\,\fun{grassmannianTwistorStringMeasure}\brace{\var{n},\var{k}}\,,}
and where the constraints $\vec{V}(C)$ are given by \mbox{\fun{veroneseContourConditions}\brace{\var{$n$},\var{$k$}}\,.} Notice that $M(C)$ is a \funL{superFunction} object.
\mathematicaBox{
\mathematicaSequence{\funL[]{nice}\brace{\funL[1]{grassmannianTwistorStringMeasure}\brace{6,1}}}{$\displaystyle\frac{(5\,1\,3)}{(1\,2\,3)(3\,4\,5)(5\,6\,1)}\,,\,\left(\fwbox{20pt}{0}\fwbox{20pt}{0}\fwbox{20pt}{1}\fwbox{40pt}{(1\,2\,4)}\fwbox{40pt}{(1\,2\,5)}\fwbox{40pt}{(1\,2\,6)}\right)\!\!\cdot\!\tilde\eta$}
}
}

\defnBox{veroneseContourConditions}{\var{$n$}\pattern,\var{$k$}\pattern}{returns a \built{List} of $\var{k}(\var{n}{-}\var{k}{-}4)$ polynomials of Pl\"ucker coordinates which encode the constraints $\vec{V}(C)$ in (\ref{grassmannian_twistor_string_form}). The vanishing of these polynomials ensures that the columns of the matrix $C$ lie on a degree $(\var{k}{+}1)$ curve in $\mathbb{P}^{\var{k}{+}1}$. 
}

\defnBox{solveTwistorStringEquations}{\var{$n$}\pattern,\var{$k$}\pattern}{returns a \built{List} of pairs \{\t{soln},\t{jacobian}\} to the $\delta$-function constraints in (\ref{grassmannian_twistor_string_form}).
\mathematicaBox{
\mathematicaSequence[5]{egMat=\funL[1]{gaugeFixedGrassmannianRepresentative}\brace{6,1};\\
measure=\funL[1]{grassmannianTwistorStringMerasure}\brace{6,1};
measure=measure/\!\!.\{\funL{m}\brace{\var{$x$}\patternTwo}$\,\mapsto$Det[egMat[\![All,\{\var{$x$}\}]\!]]\};\\
solnData=\funL[1]{solveTwistorStringEquations}\brace{6\,1};\\
intResult=Total[(measure/\!\!.\#1)/\#2\&@@@solnData];\\
\funL[]{rationalize}\brace{\funL[]{component}\brace{\stateYM{m},\stateYM{p},\stateYM{m},\stateYM{p},\stateYM{m},\stateYM{p}}\,\brace{intResult}}
}{$\rule{0pt}{20pt}\displaystyle\text{-}\frac{464575984240415}{164913754335264}$}
\mathematicaSequence{\funL[]{componentAmpN}\brace{\stateYM{m},\stateYM{p},\stateYM{m},\stateYM{p},\stateYM{m},\stateYM{p}}}{$\displaystyle\text{-}\frac{464575984240415}{164913754335264}$}
}
}

\defnBox{fromGrassmannianToScatteringEqnSolns}{\var{$n$}\pattern,\var{$k$}\pattern}{returns a \built{Rule} which translates the solution to the grassmannian twistor string solutions to the scattering equation formalism. 
\mathematicaBox{
\mathematicaSequence[1]{map=\funL[1]{fromGrassmannianToScatteringEqnSolns}\brace{8,2};\\
\funL[]{nice}\brace{map}}{$\displaystyle\left(\begin{array}{@{}llll@{}}\sigma_1\to&\infty&\sigma_5\to&\text{-}\frac{(1\,6\,3\,4)(6\,2\,5\,4)}{(1\,5\,6\,4)(6\,2\,3\,4)}\\
\sigma_2\to&0&\sigma_6\to&\text{-}\frac{(1\,5\,3\,4)(6\,2\,5\,4)}{(1\,5\,6\,4)(5\,2\,3\,4)}\\
\sigma_3\to&1&\sigma_7\to&\text{-}\frac{(1\,5\,3\,4)(1\,6\,3\,4)(6\,2\,5\,4)(7\,2\,3\,4)}{(1\,5\,6\,4)(1\,7\,3\,4)(5\,2\,3\,4)(6\,2\,3\,4)}\\
\sigma_4\to&\frac{(1\,5\,3\,6)(6\,2\,5\,4)}{(1\,5\,6\,4)(6\,2\,3\,5)}&\sigma_8\to&\text{-}\frac{(1\,5\,3\,4)(1\,6\,3\,4)(6\,2\,5\,4)(8\,2\,3\,4)}{(1\,5\,6\,4)(1\,8\,3\,4)(5\,2\,3\,4)(6\,2\,3\,4)}\end{array}\right)$}
\mathematicaSequence[4]{egMat=\funL[1]{gaugeFixedGrassmannianRepresentative}\brace{8,2};\\
solnData=\funL[1]{solveTwistorStringEquations}\brace{8,2};\\
seSolns=map/\!\!.\{\funL{m}\brace{\var{$x$}\patternTwo}\,\,$\mapsto$Det[egMat[\![All,\{\var{$x$}\}]\!]]\}/\!\!.solnData[\![All,1]\!];\\
\funL[]{evaluate}\brace{\funL[]{scatteringEquations}\brace{8}}/\!\!.seSolns;\\
DeleteDuplicates[Chop[\textnormal{\%},Power[10,-800]]]}{\{\{0,0,0,0,0,0,0,0\}\}}}
}

\newpage
\vspace{-0pt}\subsectionAppendix{The Connected Prescription for Twistor String Theory (`RSV')}{appendix:rsv}\vspace{-10pt}
\defnBox{rsvMatrix}{\var{$n$}\pattern,\var{$k$}\pattern}{returns the $(\var{k}{+}2)\!\times\!\var{n}$ matrix used in the formalism of RSV for the connected prescription of the twistor string
\eq{\hspace{-30pt}\fun{rsvMatrix}\brace{\var{n},\var{k}}\,\,\,\Leftrightarrow\left(\mathbbm{1}_{(\var{k}{+}2)\times(\var{k}{+}2)}|c_{\var{k}{+}3}\cdots\,c_{\var{n}}\right)\quad\text{where}\quad c_{a}\!\in\!\mathbb{C}^{\var{k}{+}2}\equivR\left(\begin{array}{@{}c@{}}\frac{1}{\xi{a}}\\\text{-}\frac{1}{\xi_a\sigma_a}\\
\frac{\xi_3}{\xi_a\sigma_{3\,a}}\\
\vdots\\
\frac{\xi_{\var{k}{+}2}}{\xi_a\sigma_{\var{k}{+}2\,a}}\end{array}\right)}
Notice that this matrix automatically satisfies the Veronese contour conditions:
\mathematicaBox{
\mathematicaSequence[2]{egMat=\funL[1]{rsvMatrix}\brace{8,2}/\!\!.\{\funL{$\sigma$}\brace{\var{$a$}\pattern,\var{$b$}\pattern}\,\,$\mapsto$\funL{$\sigma$}\brace{\var{$a$}}-\funL{$\sigma$}\brace{\var{$b$}}\};\\
\funL[]{veroneseContourConditions}\brace{8,2}/\!\!.\{\funL{m}\brace{\var{$x$}\patternTwo}\,\,$\mapsto$Det[egMat[\![All,\{\var{$x$}\}]\!]]\};\\
FullSimplify[\textnormal{\%}]
}{\{0,0,0,0\}}
}
}

\defnBox{rsvMeasure}{\var{$n$}\pattern,\var{$k$}\pattern}{in the RSV formalism for the connected prescription of the twistor string, the $\var{n}$-point N${}^{\var{k}}$MHV tree amplitude of sYM is given by 
\eq{\fwbox{420pt}{\hspace{-20pt}\mathcal{A}_n^{(k)}\equivL\int\!\!\!\frac{d^{(\var{k}+2)\times\var{n}}{C}}{\mathrm{vol}(GL(\var{k}{+}2))}\,\,M_{\text{RSV}}(C(\vec{\xi},\vec{\sigma}))\,\delta^{(\var{k}{+}2)\!\times\!2}\!\big(C(\vec{\xi},\vec{\sigma})\!\cdot\!\tilde\lambda\big)\delta^{2\!\times\!(\var{n}{-}\var{k}{-}2)\!\times\!2}\!\big(\lambda\!\cdot\!C^{\perp}(\vec{\xi},\vec{\sigma})\big)\delta^{2\!\times\!4}\!\big(\lambda\!\cdot\!\tilde\eta\big)}\label{rsv_twistor_string_form}}
where the \emph{measure} is 
\eq{M_{\text{RSV}}(C)\,\,\Leftrightarrow\,\,\fun{rsvMeasure}\brace{\var{n},\var{k}}\,.}

}

\defnBox{solveRSVequations}{\var{$n$}\pattern,\var{$k$}\pattern}{returns a \built{List} of pairs \{\t{soln},\t{jacobian}\} to the $\delta$-function constraints in (\ref{rsv_twistor_string_form}).
\mathematicaBox{
\mathematicaSequence[4]{egMat=\funL[1]{rsvMat}\brace{6,1};\\
measure=\funL[1]{rsvMeasure}\brace{6,1};\\
solnData=\funL[1]{solveRSVequations}\brace{6\,1};\\
intResult=Total[(measure/\!\!.\#1)/\#2\&@@@solnData];\\
\funL[]{rationalize}\brace{\funL[]{component}\brace{\stateYM{m},\stateYM{p},\stateYM{m},\stateYM{p},\stateYM{m},\stateYM{p}}\,\brace{intResult}}
}{$\rule{0pt}{20pt}\displaystyle\text{-}\frac{464575984240415}{164913754335264}$}
\mathematicaSequence{\funL[]{componentAmpN}\brace{\stateYM{m},\stateYM{p},\stateYM{m},\stateYM{p},\stateYM{m},\stateYM{p}}}{$\displaystyle\text{-}\frac{464575984240415}{164913754335264}$}
}
}

\defnBox{fromRSVtoScatteringEquationSolns}{\var{$n$}\pattern}{returns a \built{Rule} which translates the solutions of the RSV equations to the solutions to the scattering equations:
\mathematicaBox{
\mathematicaSequence[1]{map=\funL[1]{fromRSVtoScatteringEquationSolns}\brace{8};\\
\funL[]{nice}\brace{map}}{$\displaystyle\left(\begin{array}{@{}llll@{}}\sigma_1\to&\infty&\sigma_5\to&\frac{\sigma_4}{\sigma_3}\\
\sigma_2\to&0&\sigma_6\to&\frac{\sigma_6}{\sigma_3}\\
\sigma_3\to&1&\sigma_7\to&\frac{\sigma_7}{\sigma_3}\\
\sigma_4\to&\frac{\sigma_4}{\sigma_3}&\sigma_8\to&\frac{\sigma_8}{\sigma_3}\end{array}\right)$}
\mathematicaSequence[3]{solnData=\funL[1]{solveRSVequations}\brace{8,2};\\
seSolns=Thread[Rule[map[\![All,1]\!],(map[\![All,2]\!]/\!\!.solnData[\![All,1]\!]\!)\!]\!]\!;\\
\funL[]{evaluate}\brace{\funL[]{scatteringEquations}\brace{8}}/\!\!.seSolns;\\
DeleteDuplicates[Chop[\textnormal{\%},Power[10,-700]]]}{\{\{0,0,0,0,0,0,0,0\}\}}}
}

\newpage
\vspace{-0pt}\sectionAppendix{Positroids: Grassmannian Geometry \& On-Shell Functions}{appendix:positroids_appendix}\vspace{-0pt}
%
\vspace{-0pt}\subsectionAppendix{Positroid Labeling of superFunctions}{appendix:positroid_labeling}\vspace{-10pt}

\defnBox{positroidLabel}{\var{cMatrix}\pattern}{gives the positroid label for the positroid configuration encoded by the matrix \var{cMatrix}$\equivL(\var{c_1},\var{c_2},\ldots,\var{c_n})$. This is defined as follows: for each column $c_a$ there is a unique \emph{minimal} column $c_{\sigma(a)}$ with $\sigma(a)\!\geq\!a$ (with cyclic labeling understood) such that 
\eq{c_a\!\in\!\mathrm{span}\!\{c_{a{+}1},\ldots,c_{\sigma(a)}\}\,.}
Notice that if $c_a\!=\!\vec{0}\!\in\!\mathbb{C}^{k{+}2}$ then the smallest $\sigma(a)=a$, where the above range would correspond to the empty set $c_a\!\in\!\mathrm{span}\!\{\}$.\\[-10pt]

The \fun{positroidLabel} for this configuration would then be the list of images of $\sigma(a)$:
\eq{\fun{$\sigma$}\brace{\var{\sigma(1)},\ldots,\var{\sigma(n)}}\,.}
}

\defnBox[]{positroidLabel}{\fun{superFunction}\brace{\var{$f$}\pattern,\var{$C$}\pattern}}{is equivalent to \mbox{\fun{positroidLabel}\brace{\var{$\hat{C}$}}} where $\var{\hat{C}}\equivR$\built{Join}\brace{\built{Transpose}\brace{\funL{$\lambda$}}\,,\var{${C}$}}\,.
}

\defnBox{toTwistorPositroidLabels}{\fun{$\sigma$}\brace{\var{permutation}\patternTwo}}{the positroids associated with a \funL{superFunction} of spinors and those associated its image as a \funL{superFunction} of momentum twistors are related by map 
\eq{\fun{$\sigma$}\brace{\text{\var{spinorPerm}}}\,\,\mapsto\fun{$\sigma'$}\brace{\text{\var{twistorPerm}}}}
where 
\eq{\fun{$\sigma'$}\brace{\var{a}}\,\,\equivR\fun{$\sigma$}\brace{\var{a{-}1}}\,\,{-}1\,,}
which maps the labels \funL{$\sigma$} of positroids in $G_+(\var{k{+}2},\var{n})$ to those \fun{$\sigma'$} labeling positroids in $G_+(\var{k},\var{n})$.
\fun{toTwisorPositroidLabels}\brace{\fun{$\sigma$}\brace{\var{spinorPerm}}}\, returns \mbox{\fun{$\sigma$}\brace{{\var{twistorPerm}}}}\,.\\[-10pt]
\mathematicaBox{
\mathematicaSequence{egCells=\funL[]{ymPositroidAmp}\brace{7,1}}{\{\funL{$\sigma$}\brace{4,5,6,9,7,8,10}\,,\funL{$\sigma$}\brace{3,5,6,7,9,8,11}\,,\funL{$\sigma$}\brace{4,6,5,7,9,8,10}\,,\\
\phantom{\{}\mbox{\funL{$\sigma$}\brace{3,4,6,7,8,9,12}}\,,\funL{$\sigma$}\brace{3,5,7,6,8,9,11}\,,\funL{$\sigma$}\brace{4,7,5,6,8,9,10}\,\}}
\mathematicaSequence{\funL[1]{toTwistorPositroidLabels}/@egCells}{\{\funL{$\sigma$}\brace{2,3,4,5,8,6,7}\,,\funL{$\sigma$}\brace{3,2,4,5,6,8,7}\,,\funL{$\sigma$}\brace{2,3,5,4,6,8,7}\,,\\
\phantom{\{}\funL{$\sigma$}\brace{4,2,3,5,6,7,8}\,,\funL{$\sigma$}\brace{3,2,4,6,5,7,8}\,,\funL{$\sigma$}\brace{2,3,6,4,5,7,8}\,\}}
\mathematicaSequence{\textnormal{\%}===\funL[1]{positroidLabel}/@\funL[]{ymSuperTwistorAmp}\brace{7,1}}{True}
}
\emph{\textbf{Note}}: if \fun{$\sigma$}\brace{\var{$a$}}$\,\,<\var{a}{+}2$ for any $\var{a}\!\in\![n]$, then there is no image of the positroid in the momentum-twistor Grassmannian; in this case, \fun{toTwistorPositroidLabels} returns \built{Nothing}.\vspace{-40pt}
}

\defnBox{toSpinorPositroidLabels}{\fun{$\sigma$}\brace{\var{permutation}\patternTwo}}{acts similarly to the function  to \fun{toTwistorPositroidLabels}; specifically, the positroid labels of \funL{superFunction} objects that depend on momentum-twistors can be mapped to those that label their spinor-dependent \funL{superFunction} images via the map
\vspace{-5pt}\eq{\fun{$\sigma'$}\brace{\text{\var{twistorPerm}}}\,\,\mapsto\fun{$\sigma$}\brace{\text{\var{spinorPerm}}}\vspace{-7pt}}
where 
\vspace{-5pt}\eq{\fun{$\sigma$}\brace{\var{a}}\,\,\equivR\fun{$\sigma'$}\brace{\var{a{+}1}}\,\,{+}1\,.\vspace{-10pt}}
\mathematicaBox{
\mathematicaSequence{egCells=\funL[]{positroidLabel}/@\funL[]{ymSuperTwistorAmp}\brace{7,1}}{\{\funL{$\sigma$}\brace{2,3,4,5,8,6,7}\,,\funL{$\sigma$}\brace{3,2,4,5,6,8,7}\,,\funL{$\sigma$}\brace{2,3,5,4,6,8,7}\,,\\
\phantom{\{}\funL{$\sigma$}\brace{4,2,3,5,6,7,8}\,,\funL{$\sigma$}\brace{3,2,4,6,5,7,8}\,,\funL{$\sigma$}\brace{2,3,6,4,5,7,8}\,\}}
\mathematicaSequence{\funL[1]{toSpinorPositroidLabels}/@egCells}{\{\funL{$\sigma$}\brace{4,5,6,9,7,8,10}\,,\funL{$\sigma$}\brace{3,5,6,7,9,8,11}\,,\funL{$\sigma$}\brace{4,6,5,7,9,8,10}\,,\\
\phantom{\{}\mbox{\funL{$\sigma$}\brace{3,4,6,7,8,9,12}}\,,\funL{$\sigma$}\brace{3,5,7,6,8,9,11}\,,\funL{$\sigma$}\brace{4,7,5,6,8,9,10}\,\}}
\mathematicaSequence{\textnormal{\%}===\funL[]{ymPositroidAmp}\brace{7,1}}{True}
}
}

\vspace{10pt}\subsectionAppendix{General Manipulations of Positroid Labels}{appendix:manipulations_of_positroid_labeling}\vspace{-10pt}

\defnBox{decoratePermutation}{\fun{$\sigma$}\brace{\var{permutation}\patternTwo}}{an \emph{ordinary} \var{permutation} $\sigma_0\!:\!\mathbb{Z}_\var{n}\!\mapsto\!\mathbb{Z}_\var{n}$ can be labelled by the images $\{\sigma_0(1),\ldots,\sigma_0(\var{n})\}$. There are obviously $\var{n}!$ ordinary permutations. A \emph{decorated} permutation is a generalization of permutations for which fixed points are distinguished by two possible `decorations' of the identity. To distinguish between possible decorations of the identity, we conventionally label a \emph{decorated} permutation by a map $\sigma\!:\!\mathbb{Z}_\var{n}\!\mapsto\!\mathbb{Z}_{2\var{n}}$ with the requirement that $\sigma(a)\!\geq\!a$ for all $a\!\in\!\mathbb{Z}_\var{n}$. The cases $\sigma(a)\!\in\!\{a,a{+}\var{n}\}$ then represent the possible decorations of the identity.\\[-10pt]

Positroid varieties are labelled by \emph{decorated} permutations, and we conventionally require that $\sigma(a)\!\geq\!a$ for all $a\!\in\![\var{n}]$.\\[-10pt]

The number of decorated permutations is obviously larger than the number of ordinary permutations: they number $\var{n}!(1{+}\tfrac{1}{1!}{+}\tfrac{1}{2!}{+}\tfrac{1}{3!}{+}\ldots{+}\frac{1}{\var{n}!})$. Thus, there is no unique way to `decorate' an ordinary permutation. However, for permutations \emph{without} fixed points, we can enforce the convention that $\sigma(a)\!\geq\!a$ by simply adding $\var{n}$ to any image $\sigma_0(a)\!<\!a$---so that $\sigma(a)\!=\!\sigma_0(a){+}\var{n}$. This is done by the function \fun{decoratePermutation}.\\[-10pt]
}

\defnBox{parityConjugatePositroid}{\fun{$\sigma$}\brace{\var{permutation}\patternTwo}}{returns the positroid label \funL{$\sigma$}' for the \emph{parity-conjugate} positroid; that is, for any positroid $C_\sigma\!\in\!\mathrm{Gr}(k,n)$, there is a unique \emph{dual} positroid $C'_{\sigma'}\!\in\!\mathrm{Gr}(n\text{-}k,n)$ such that $C'_{\sigma'}\!\cdot\!C_\sigma{=}0$. 
\mathematicaBox{
\mathematicaSequence{defaultTreeCells=\funL[1]{ymPositroidAmp}[7,2]}{$\{\funL{$\sigma$}\![5,6,7,9,10,8,11],\funL{$\sigma$}\![4,6,7,9,8,10,12],\funL{$\sigma$}\![3,6,7,8,9,11,12],$\\ $\phantom{\{}\funL{$\sigma$}\![5,7,6,8,10,9,11],\funL{$\sigma$}\![4,7,6,8,9,10,12],\funL{$\sigma$}\![5,7,8,6,9,10,11]\}$}
\mathematicaSequence{conjugatedCells=\funL[1]{parityConjugatePositroid}/@\funL[1]{ymPositroidAmp}[7,1]}{$\{\funL{$\sigma$}\![6,4,7,8,9,10,12],\funL{$\sigma$}\![6,5,8,7,9,10,11],\funL{$\sigma$}\![6,5,7,8,10,9,11],$\\ $\phantom{\{}\funL{$\sigma$}\![5,6,8,9,7,10,11],\funL{$\sigma$}\![5,6,8,7,9,11,10],\funL{$\sigma$}\![5,6,7,8,10,11,9]\}$}
\mathematicaSequence{\funL[]{useReferenceKinematics}[7];\\
defaultFcns=\funL[]{fromPositroidsToSuperFunctionsN}/@defaultTreeCells;\\
conjugatedFcns=\funL[]{fromPositroidsToSuperFunctionsN}/conjugatedCells;}{}
\mathematicaSequence{egCmpt=\funL[]{randomComponentStates}[7,2]}{\{\stateYM{psiBar}[4],\stateYM{phi}[1,2],\stateYM{psiBar}[2],\stateYM{psi}[4],\stateYM{psiBar}[1],\stateYM{psiBar}[1],\stateYM{psi}[1]\}}
\mathematicaSequence{\funL[]{component}[egCmpt][defaultFcns]}{$\rule{0pt}{22pt}\displaystyle\left\{\text{-}\frac{31875}{402248},\text{-}\frac{10299756850}{30586805361},0,0,0,0\right\}$}
\mathematicaSequence{Total[\textnormal{\%}]}{$\rule{0pt}{20pt}\displaystyle\text{-}\frac{1061389475}{4121958456}$}
\mathematicaSequence{\funL[]{component}[egCmpt][conjugatedFcns]\\Total[\textnormal{\%}]}{$\rule{0pt}{22pt}\displaystyle\left\{\text{-}\frac{30202625}{1027265976},\text{-}\frac{771207925}{3114750672},\frac{55961575}{4020816261},\frac{2975}{532656},0,0\right\}$}
\mathematicaSequence{Total[\textnormal{\%}]}{$\rule{0pt}{20pt}\displaystyle\text{-}\frac{1061389475}{4121958456}$}
}
}

\defnBoxTwo{rotatePositroidLabel}{\var{$r$}\pattern}{\fun{$\sigma$}\brace{\var{permutation}\patternTwo}}{gives the positroid label for a geometry where the columns have been cyclically rotated \emph{to the right} by the transformation $c_{a}\!\mapsto\!c_{a{-}r}$. For \var{$r$}{=}1, for example, this corresponds to the rotation where $(c_1,\ldots,c_n)\!\mapsto(c_n,c_1,\ldots,c_{n{-}1})$.
\mathematicaBox{\mathematicaSequence{egCell=\funL[]{randomPositroid}[8,2]}{\funL{$\sigma$}\brace{4,6,8,9,7,11,10,13}}
\mathematicaSequence[1]{egMatrix=\funL[]{explicify}\brace{\funL[1]{positroidRepresentative}\brace{egCell}}\,;\\\funL[]{nice}\brace{egMatrix}}{$\rule{0pt}{39pt}\left(\begin{array}{@{}cccccccc@{}}1&196&15523&35340&0&0&0&0\\
0&1&209&11780&\text{-}26\phantom{\text{-}}&\text{-}10062\phantom{\text{-}}&\text{-}1326\phantom{\text{-}}&0\\
0&0&1&155&617&84366&0&0\\
0&0&0&0&1&387&51&272\end{array}\right)$}
\mathematicaSequence[1]{rotatedMatrix=RotateRight[egMatrix,\{0,0,3\}];\\\funL[]{nice}\brace{rotatedMatrix}}{$\rule{0pt}{39pt}\left(\begin{array}{@{}cccccccc@{}}
0&0&0&1&196&15523&35340&0\\
\text{-}10062\phantom{\text{-}}&\text{-}1326\phantom{\text{-}}&0&0&1&209&11780&\text{-}26\phantom{\text{-}}\\
84366&0&0&0&0&1&155&617\\
387&51&272&0&0&0&0&1\end{array}\right)$}
\mathematicaSequence{\funL[1]{positroidLabel}\brace{rotatedMatrix}}{\funL{$\sigma$}\brace{6,5,8,7,9,11,12,10}}
\mathematicaSequence{\funL[1]{rotatePositroidLabel}[3][egCell]}{\funL{$\sigma$}\brace{6,5,8,7,9,11,12,10}}
}
To better understand how rotations are encoded, consider the following:
\mathematicaBox{\mathematicaSequence[1]{egFcn=RandomChoice\brace{\funL[]{ymAmp}\brace{10,3}};\\
egFcnLabel=\funL[1]{positroidLabel}\brace{egFcn}}{\rule{0pt}{14pt}\funL{$\sigma$}\brace{4,6,8,12,10,13,9,11,15,17}}
\mathematicaSequence[1]{\funL[1]{positroidLabel}\brace{\funL[]{permuteLegs}\brace{Mod[Range[10]+1,10,1]}}\,\brace{egFcn}}{\funL{$\sigma$}\brace{8,5,7,9,13,11,14,10,12,16}}
\mathematicaSequence[1]{\funL[1]{rotatePositroidLabel}\brace{1}\,\brace{egFcnLabel}}{\funL{$\sigma$}\brace{8,5,7,9,13,11,14,10,12,16}}
}
}

\defnBox{positroidCyclicRep}{\fun{$\sigma$}\brace{\var{permutation}\patternTwo}}{returns a canonical representative positroid related to \funL{$\sigma$} by a cyclic relabeling of the external legs:
\eq{\begin{split}&\fun{positroidCyclicRep}\brace{\fun{$\sigma$}\brace{\var{perm}}}\equivR\\
&\hspace{20pt}\text{\built{Sort}\brace{\built{NestList}\brace{\funL[1]{rotatePositroidLabel}\brace{1}\,\brace{\t{\#}}\,\t{\&},\funL{$\sigma$}\brace{\var{perm}}\,,\var{$n$}}\,}\,\brace{\brace{1}}}\,.\end{split}}\vspace{-10pt}
}

\vspace{-0pt}\subsectionAppendix{Positroid Geometry and Combinatorial Data}{appendix:geometry_and_combinatorics}\vspace{-0pt}
%
\vspace{0pt}\subsubsectionAppendix{General Characteristics of Positroid Varieties}{appendix:general_characteristics}\vspace{-10pt}

\defnBox{positroidRank}{\fun{$\sigma$}\brace{\var{permutation}\patternTwo}}{every positroid variety is labelled by a \emph{decorated} permutation encoded by \{\var{permutation}\}$\equivR\{\sigma(1),\ldots,\sigma(n)\}$ with the property that $\sigma(a)\!\geq\!a$ for all $a\!\in\![\var{n}]$. These label subspaces of $Gr(\var{k}{+}2,\var{n})$ where $\var{k}$ is given by \fun{positroidRank}.\\[-10pt]

\textbf{\emph{Note}}: this function behaves similarly to the function `\built{permK}' of the package \texttt{positroids.m} \cite{Bourjaily:2012gy}; specifically,  \mbox{\fun{positroidRank}\brace{\funL{$\sigma$}\brace{\var{perm}}}$\equivR$\built{permK}\brace{\{\var{perm}\}}\,$-2$.}
}

\defnBox{positroidGeometry}{\fun{$\sigma$}\brace{\var{permutation}\patternTwo}}{similar to `\built{permToGeometry}' of the package \texttt{positroid.m} \cite{Bourjaily:2012gy}; specifically \fun{positroidGeometry} tabulates the geometric structure of the positroid configuration labelled by \funL{$\sigma$}\brace{\var{permutation}}, by listing all the consecutive ranges of columns which span spaces of a given rank. 
\mathematicaBox{
\mathematicaSequence{\funL[1]{positroidGeometry}\brace{\funL{$\sigma$}\brace{5,4,11,6,9,8,15,10}}}{$\rule{0pt}{30pt}\displaystyle\begin{array}{c}(\bb{1})\,(2)\,(\bb{3})\,(4)\,(\bb{5})\,(6)\,(\bb{7})\,(8)\\
(\bb{2},\bb{3},\bb{4})\,(4,5,6)\,(\bb{6},\bb{7},\bb{8})\,(8,1,2)\\
(\bb{4},\bb{5},\bb{6},\bb{7},\bb{8},\bb{1},\bb{2})(\bb{8},\bb{1},\bb{1},\bb{3},\bb{4},\bb{5},\bb{6})\end{array}$}
}
To understand the meaning of this output, the $k$th row lists consecutive sequences of column numbers which span a space of rank $k$. The sequences highlighted in \bb{blue} denote \emph{maximal} planes---meaning that the rank will increase if a column is added either to the beginning or end of the sequence. For more information, consult \cite{Bourjaily:2012gy}.
}

\defnBox{positroidNecklace}{\fun{$\sigma$}\brace{\var{permutation}\patternTwo}}{returns a \built{List} of $\var{n}$ of `minors' which correspond to the `necklace' of the positroid. This list of subsets is said to be \emph{maximally weakly separated}. The $a$th element of the necklace is the lexicographically minimal subset of $\var{k}{+}2$ columns, starting with the $a$th column and cyclic labeling understood, whose determinant is non-vanishing for the positroid configuration labelled by \mbox{\funL{$\sigma$}\brace{\var{permutation}}\,.}
\mathematicaBox{
\mathematicaSequence{egCell=\funL[]{randomPositroid}\brace{8,2}}{\funL{$\sigma$}\brace{4,6,7,10,8,11,9,13}}
\mathematicaSequence{\funL[1]{positroidNeckalce}\brace{egCell}}{\{\funL{m}\brace{1,2,3,5}\,,\funL{m}\brace{2,3,4,5}\,,\funL{m}\brace{3,4,5,6}\,,\funL{m}\brace{4,5,6,7}\,,\\
\phantom{\{}\funL{m}\brace{5,6,7,2}\,,\funL{m}\brace{6,7,8,2}\,,\funL{m}\brace{7,8,2,3}\,,\funL{m}\brace{8,1,2,3}\,\}}
}
The `necklace' of a positroid uniquely characterizes the cell. 
}

\defnBox{positroidNecklaceToPerm}{\{\var{necklaceMinors}\patternTwo\fun{m}\}}{returns the positroid label \mbox{\funL{$\sigma$}\brace{\var{permutation}}} of the cell whose necklace would be given by \{\var{necklaceMinors}\}. 
}

\defnBox{positroidTropicalFaceLabels}{\fun{$\sigma$}\brace{\var{permutation}\patternTwo}}{returns a the \built{List} of \emph{tropical labels} for the positroid labeled by \mbox{\funL{$\sigma$}\brace{\var{permutation}}}. These number one greater than the dimension of the positroid, and consist of pairs $\{\alpha,A\}$ with $\alpha\!\in\![\var{k}{+}2]$ and $A\!\in\![\var{n}]$.
\mathematicaBox{
\mathematicaSequence{egCell=\funL[]{randomPositroid}\brace{8,2}}{\funL{$\sigma$}\brace{3,5,7,8,9,12,10,14}}
\mathematicaSequence{\funL[1]{positroidTropicalLabels}\brace{egCell}}{\{\{2,3\},\{1,3\},\{3,5\},\{2,5\},\{4,7\},\{1,5\},\\
\phantom{\{}\{1,4\},\{1,2\},\{2,6\},\{3,7\},\{3,6\},\{4,8\},\{4,6\}\}}
}
These tropical labels defined in terms of the cluster $\mathcal{A}$ coordinates associated with the faces of some plabic graph representative for the positroid. They are \emph{useful} because while the \emph{set} of tropical labels is the same for \emph{all} plabic graph representatives, a square move results in a transposition of the ordering of these tropical labels. Thus, the number of square moves which can connect any pair of plabic graph representatives can be determined by the decomposition of their permuted relative ordering into sequences of transpositions. Even more importantly, if we associate a coordinate chart with the face variables of a graph, the \emph{signature} of the relative ordering of the faces' tropical label then determines the relative orientation of the two charts.
}

\newpage
%
\vspace{0pt}\subsubsectionAppendix{Canonical Coordinate Charts and Positroid Representatives}{appendix:canonical_charts_and_representatives}\vspace{-10pt}

\defnBox{positroidBridgeChain}{\fun{$\sigma$}\brace{\var{permutation}\patternTwo},\var{randomQ}\pattern\optArg{\built{False}}\,}{returns a \built{List} consisting of a pair \{\t{bridgeChain},\fun{$\sigma_0$}\} where \fun{$\sigma_0$} is a decoration of the identity (labeling a zero-dimensional cell) and \t{bridgeChain}$\equivL$\{\fun{$\tau_1$},\fun{$\tau_2$},\ldots,\fun{$\tau_{d}$}\} is a \built{List} of transpositions \funL{$\tau$}\brace{\var{$a$},\var{$b$}} such that 
\eq{\fun{$\sigma$}\brace{\var{permutation}}\,\,\,\Leftrightarrow \fun{$\tau_1$}\circ\fun{$\tau_2$}\circ\cdots\circ\fun{$\tau_d$}\circ\fun{$\sigma_0$}\,.\vspace{-20pt}}
\mathematicaBox{
\mathematicaSequence{egCell=\funL[]{randomPositroid}\brace{6,1}}{\funL{$\sigma$}\brace{5,4,6,7,8,9}}
\mathematicaSequence{\funL[1]{positroidBridgeChain}\brace{egCell}}{\{\{\funL{$\tau$}\brace{2,3}\,,\funL{$\tau$}\brace{1,2}\,,\funL{$\tau$}\brace{3,4}\,,\funL{$\tau$}\brace{2,3}\,,\funL{$\tau$}\brace{1,2}\,,\funL{$\tau$}\brace{3,5}\,,\funL{$\tau$}\brace{2,3}\,,\funL{$\tau$}\brace{3,6}\,\},\\
\phantom{\{}\funL{$\sigma$}\brace{7,8,9,4,5,6}\,\}}
}
This can be achieved in many ways. If the optional argument \var{randomQ} is set to \built{False}, its default value, then \fun{positroidBridgeChain} returns what is called the \emph{lexicographical} bridge chain. Choosing \built{True} for the optional argument \var{randomQ} will give a randomly-generated sequence of transpositions which generate \mbox{\funL{$\sigma$}\brace{\var{permutation}}\,.}
}

\defnBox{positroidBridgeChainToFaceLabels}{\{\var{transpositionChain}\patternTwo\fun{$\tau$}\}}{for any given `bridge chain' for a positroid, there is a representative plabic graph, with each bridge associated with a particular cluster $\mathcal{A}$-coordinate which (through a very slight abuse of notation) can be considered a Pl\"ucker coordinate. In fact, as there is always one face more than the dimension of the variety, and hence the number of bridges, there will be one excess `face label' associated with a given bridge chain. 
\mathematicaBox{
\mathematicaSequence{egChain=\funL[1]{positroidBridgeChain}\brace{\funL{$\sigma$}\brace{5,4,6,7,8,9}};}{\{\{\funL{$\tau$}\brace{2,3}\,,\funL{$\tau$}\brace{1,2}\,,\funL{$\tau$}\brace{3,4}\,,\funL{$\tau$}\brace{2,3}\,,\funL{$\tau$}\brace{1,2}\,,\funL{$\tau$}\brace{3,5}\,,\funL{$\tau$}\brace{2,3}\,,\funL{$\tau$}\brace{3,6}\,\},\\
\phantom{\{}\funL{$\sigma$}\brace{7,8,9,4,5,6}\,\}}
\mathematicaSequence{\funL[1]{positroidBridgeChainToFaceLabels}\brace{egChain}}{\{\funL{m}\brace{3,4,5}\,,\funL{m}\brace{2,3,5}\,,\funL{m}\brace{4,5,6}\,,\funL{m}\brace{3,5,6}\,,\funL{m}\brace{2,3,6}\,,\funL{m}\brace{1,5,6}\,,\\
\phantom{\{}\funL{m}\brace{1,3,6}\,,\funL{m}\brace{1,2,6}\,,\funL{m}\brace{1,2,3}\,\}}
}
}

\defnBox{positroidBridgeChainToTropicalLabels}{\{\var{transpositionChain}\patternTwo\fun{$\tau$}\}}{returns a list of \emph{tropical} labels $\{\alpha,A\}$---with $\alpha\!\in\![\var{k}{+}2]$ and $A\!\in\![\var{n}]$---associated with the canonical coordinates generated from the \var{transpositionChain}. This is useful because the \emph{set} of tropical labels is unique; and sign-changing cluster mutations lead to transpositions of the ordering of their sets of tropical labels. Hence, the relative orientation of two charts can be determined by the relative signature of their tropical labels. 
\mathematicaBox{
\mathematicaSequence{egCell=\funL[]{randomPositroid}\brace{8,2}}{\funL{$\sigma$}\brace{5,6,9,7,11,8,10,12}}
\mathematicaSequence[2]{egChains=\funL[]{positroidBridgeChain}\brace{egCell,True}\,\&/@Range\brace{50};\\
tropicalLabels=\funL[1]{positroidBridgeChainToTropicalLabels}/@egChains;\\
SameQ@@\{Sort/@tropicalLabels\}}{True};
\mathematicaSequence{tropicalSigns=Signature\brace{Ordering\brace{\#}}\,\&/@tropicalLabels}{\{1,1,1,-1,-1,1,1,-1,-1,-1,1,1,1,-1,-1,-1,1,-1,1,1,1,-1,1,-1,1,1,\\
\phantom{\{}-1,1,1,-1,-1,-1,1,-1,-1,-1,-1,-1,1,-1,-1,1,1,1,-1,-1,-1,1,-1,1\}}
\mathematicaSequence[2]{egCharts=\funL[]{bridgeChainToRepresentative}/@egChains;\\
\funL[]{directPositroidEvaluation}/@egCharts;\\
refSigns=\textnormal{\%}[\![All,1]\!]/\textnormal{\%}[\![1,1]\!]}{\{1,1,1,-1,-1,1,1,-1,-1,-1,1,1,1,-1,-1,-1,1,-1,1,1,1,-1,1,-1,1,1,\\
\phantom{\{}-1,1,1,-1,-1,-1,1,-1,-1,-1,-1,-1,1,-1,-1,1,1,1,-1,-1,-1,1,-1,1\}}
\mathematicaSequence{DeleteDuplicates\brace{\textnormal{\%}/tropicalSigns}}{\{1\}}
}
(To be clear, only the \emph{relative} orientation is meaningful; thus, had the last evaluation returned \t{\{-1\}}, it would have been just as meaningful a check.) 
}

\defnBox{positroidRepresentative}{\fun{$\sigma$}\brace{\var{permutation}\patternTwo},\var{randomQ}\pattern\optArg{\built{False}}\,}{provides a representative, parameterized matrix $C(\vec{\alpha})\!\in\!\mathrm{Gr}(k{+}2,n)$ for the positroid labelled by \funL{$\sigma$}\brace{\var{permutation}} where $\vec{\alpha}$ are \emph{canonical coordinates} (those for which the positroid volume form is $d\!\log$) with signs dictated by the requirement that when $\vec{\alpha}\!\in\!\mathbb{R}_+^d$, $C(\vec{\alpha})\!\in\!\mathrm{Gr}_+(k{+}2,n)$. This information is stored in a \funL{positroid} object:
\mathematicaBox{\mathematicaSequence{\funL[1]{positroidRepresentative}\brace{$\sigma$\brace{3,7,6,8,9,10,13,12}}}{\funL{positroid}\Big[\Big\{\{1,\funL{$\alpha$}\![8]\text{+}\hspace{0pt}\funL{$\alpha$}\![10]\text{+}\hspace{0pt}\funL{$\alpha$}\![12],\\
\!(\funL{$\alpha$}\![8]\text{+}\hspace{0pt}\funL{$\alpha$}\![10])\funL{$\alpha$}\![11],\funL{$\alpha$}\![8]$\!\alpha$[9],\funL{$\alpha$}\![2],\funL{$\alpha$}\![2]$\!\alpha$[5],\funL{$\alpha$}\![2]$\!\alpha$[3],0\},\\
\{0,1,\funL{$\alpha$}\![11],\funL{$\alpha$}\![6]\text{+}\funL{$\alpha$}\![9],\funL{$\alpha$}\![6]\funL{$\alpha$}\![7],0,0,0\},\\
\{0,0,0,1,\funL{$\alpha$}\![4]\text{+}\funL{$\alpha$}\![7],\funL{$\alpha$}\![4]\funL{$\alpha$}\![5],0,0\},\\
\{0,0,0,0,1,\funL{$\alpha$}\![5],\funL{$\alpha$}\![3],\funL{$\alpha$}\![1]\}\Big\}\Big]
}
\mathematicaSequence{\funL[]{nice}\brace{\textnormal{\%}}}{$\frac{1}{\alpha_1\alpha_2\alpha_3\alpha_4\alpha_5\alpha_6\alpha_7\alpha_8\alpha_9\alpha_{10}\alpha_{11}\alpha_{12}}$,\scalebox{0.85}{$\left(\begin{array}{@{}cccccccc@{}}1&\alpha_{8}\text{+}\alpha_{10}\text{+}\alpha_{12}&(\alpha_{8}\text{+}\alpha_{10})\alpha_{11}&\alpha_{8}\alpha_{9}&\alpha_{2}&\alpha_{2}\alpha_{5}&\alpha_{2}\alpha_{3}&0\\0&1&\alpha_{11}&\alpha_{6}\text{+}\alpha_{9}&\alpha_{6}\alpha_{7}&0&0&0\\0&0&0&1&\alpha_{4}\text{+}\alpha_{7}&\alpha_{4}\alpha_{5}&0&0\\0&0&0&0&1&\alpha_{5}&\alpha_{3}&\alpha_{1}\end{array}\right)\!\cdot\!\tilde\eta$}}}

If the optional argument \var{randomQ} is set to \built{True}, then the resulting, parameterized matrix representative will be obtained via a \emph{random} transposition-chain bridge construction. 
}

\defnBox{bridgeChainToRepresentative}{\{\var{transpositionChain}\patternTwo\fun{$\tau$}\}}{returns a representation of the subspace $C$ in the grassmannian parameterized by bridge coordinates associated with the \var{transpositionChain}; the output is given as a \mbox{\funL{positroid}\brace{$C(\vec{\alpha})$}} object. 
}

\defnBox{positiveChartQ}{\fun{positroid}\brace{\var{matrix}\pattern}}{returns \built{True} if all ordered minors of the matrix $C$ are \emph{positive} for $\alpha_i\!\in\!\mathbb{R}_{+}$ and \built{False} if otherwise.
\mathematicaBox{
\mathematicaSequence[2]{egCells=\funL[]{allPositroids}\brace{8,2};\\
egCharts=\funL[1]{positroidRepresentative}/@egCells;
And@@(\funL[1]{positiveChartQ}/@egCharts)}{True}
}
}

\newpage
\vspace{-0pt}\subsectionAppendix{Kinematic Support, Factorizability, and Tree-Like Positroids}{appendix:positroid_support_and_factorizability}\vspace{-10pt}

\defnBox{positroidKinematicSupport}{\fun{$\sigma$}\brace{\var{permutation}\patternTwo}}{returns the number of solutions to the \funL{grassmannianKinematicEquations} (\ref{gkn_kinematic_constratins}) for a representative $C_{\funL{$\sigma$}}$ of the positroid labelled by \funL{$\sigma$}. 
\mathematicaBox{\mathematicaSequence{egCells=\{\\
\funL{$\sigma$}\brace{6,5,8,7,10,9,12,11}\,,\\
\funL{$\sigma$}\brace{8,9,6,7,11,10,14,15,12,13}\,,\\
\funL{$\sigma$}\brace{9,7,6,11,8,13,10,16,12,15,14}\,,\\
\funL{$\sigma$}\brace{15,14,8,7,21,20,19,13,12,26,25,24,18,17,31,30,29,23,22,36}\,\};\\
\mbox{\{\funL[]{positroidRank}\brace{\#},Length[List@@\#],\funL[]{positroidDimension}\brace{\#}\}\&}/@egCells}{\{\{2,8,12\},\{3,10,16\},\{3,11,18\},\{8,20,36\}\}}
\mathematicaSequence{\funL[1]{positroidKinematicSupport}/@egCells}{\{2,0,3,34\}}
\mathematicaSequence[1]{eqnList=(\funL[]{useReferenceKinematics}\brace{Length@(List@@\#)};\\
\funL[]{grassmannianKinematicEquations}\brace{\\\mbox{~\hspace{43pt}}\funL[]{positroidRepresentative}\brace{\#}})\,\&/@egCells;\\
Length\brace{NSolve\brace{\#}}\,\&/@eqnList}{\{2,0,3,34\}}
}
This function is only defined for positroids of dimension $(2\var{n}{-}4)$. 
}

\defnBox{nonVanishingPositroidQ}{\fun{$\sigma$}\brace{\var{permutation}\patternTwo}}{returns \built{True} if there are any solutions to the \funL{grassmannianKinematicEquations} (\ref{gkn_kinematic_constratins}); that is, it is equivalent to the test \fun{positroidKinematicSupport}\brace{\funL{$\sigma$}\brace{\var{permutation}}}\,>0. 
}

\defnBox{positroidFactorizations}{\fun{$\sigma$}\brace{\var{permutation}\patternTwo}}{returns a \built{List} of all possible \emph{factorizations} of a given positroid, encoded as a the list of terms encoded as \{\fun{$\sigma_L$},\fun{$\sigma_R$},\{\var{$a$},\var{$b$}\},\var{conjugatedBridgeQ}\}. The meaning of this data is that it can be obtained by a BCFW bridge on legs $\{\var{a},\var{b}\}$ or $\{\var{b},\var{a}\}$ if \var{conjugatedBridgeQ} is \built{False} or \built{True}, respectively. 
\mathematicaBox{
\mathematicaSequence{egCell=\funL[]{randomPositroid}\brace{8,2}}{\funL{$\sigma$}\brace{3,6,7,9,8,12,10,13}}
\mathematicaSequence{\funL[1]{positroidFactorizations}\brace{egCell}}{\{\{\funL{$\sigma$}\brace{3,4,5}\,,\funL{$\sigma$}\brace{3,5,7,6,8,9,11}\,,\{4,5\},\built{False}\},\\
\phantom{\{}\{\funL{$\sigma$}\brace{5,7,6,9,8,10,11}\,,\funL{$\sigma$}\brace{2,3,4}\,,\{2,3\},\built{False}\},\\
\phantom{\{}\{\funL{$\sigma$}\brace{3,5,7,6,8,9,11}\,,\funL{$\sigma$}\brace{3,4,5}\,,\{5,6\},\built{True}\},\\
\phantom{\{}\{\funL{$\sigma$}\brace{4,5,7,6,8,9}\,,\funL{$\sigma$}\brace{3,4,5,6}\,,\{3,4\},\built{True}\},\\
\phantom{\{}\{\funL{$\sigma$}\brace{2,3,4}\,,\funL{$\sigma$}\brace{5,7,6,9,8,10,11}\,,\{1,2\},\built{True}\}\}
}
\mathematicaSequence{\mbox{\funL[]{rotatePositroidLabel}\brace{\!\#3[\![1]\!]\!}\,\brace{\built{positroids{\`{}}BCFWbridge}\brace{\#1,\#2,\#4}}\&}@@@\textnormal{\%}}{\{\funL{$\sigma$}\brace{3,6,7,9,8,12,10,13}\,,\mbox{\funL{$\sigma$}\brace{3,6,7,9,8,12,10,13}}\,,\\
\phantom{\{}\mbox{\funL{$\sigma$}\brace{3,6,7,9,8,12,10,13}\,,}\funL{$\sigma$}\brace{3,6,7,9,8,12,10,13}\,,\\
\phantom{\{}\funL{$\sigma$}\brace{3,6,7,9,8,12,10,13}\}}
\mathematicaSequence{\#===egCell\&/@\textnormal{\%}}{\{\built{True},\built{True},\built{True},\built{True},\built{True}\}}
}\vspace{-00pt}
}

\defnBox{treePositroidQ}{\fun{$\sigma$}\brace{\var{permutation}\patternTwo}}{returns \built{True} if there exists any sequence of factorizations in which connects the positroid to a sequence of BCFW bridges applied to three-particle positroids \funL{$\sigma$}\brace{2,3,4} or \funL{$\sigma$}\brace{3,4,5}\,.
\mathematicaBox{
\mathematicaSequence[1]{egCell=\funL{$\sigma$}\brace{4,5,9,7,8,12,10,11,15};\\
\funL[]{positroidKinematicSupport}\brace{egCell}}{1}
\mathematicaSequence{\funL[1]{treePositroidQ}\brace{egCell}}{False}
}
For a given cell, the existence of a `factorization' configuration in its boundary is easy to test for combinatorially: if a boundary configuration exists for which the permutation can be divided into a left/right ranges of legs for which exactly one image of the left range is within the right and vice versa, the boundary is said to be a factorization. The function \fun{treePositroidQ} simply tests the existence of such cells recursively.}

\newpage
\vspace{0pt}\subsectionAppendix{Drawing Representative Plabic Graphs of Positroids}{appendix:plabic_graph_drawing}\vspace{-10pt}

\defnBox{drawPlabicGraph}{\fun{$\sigma$}\brace{\var{permutation}\patternTwo},\var{detailedQ}\pattern\optArg{\built{False}}\,}{returns a graphical representation of a plabic graph whose left-right path permutation would be given by \funL{$\sigma$}\brace{\var{permutation}}\,. This is drawn in a way that should make its connection to bridge chains manifest:
\mathematicaBox{
\mathematicaSequence[1]{egCell=\funL{$\sigma$}\brace{6,5,7,10,8,11,9,12};\\
\funL[1]{positroidBridgeChain}\brace{egCell}}{\{\{\mbox{\funL{$\tau$}\brace{2,3}\,,}\mbox{\funL{$\tau$}\brace{1,2}\,,}\mbox{\funL{$\tau$}\brace{3,4}\,,}\mbox{\funL{$\tau$}\brace{2,3}\,,}\mbox{\funL{$\tau$}\brace{4,5}\,,}\mbox{\funL{$\tau$}\brace{3,4}\,,}\\
\phantom{\{\{}\mbox{\funL{$\tau$}\brace{1,3}\,,}\mbox{\funL{$\tau$}\brace{4,6}\,,}\mbox{\funL{$\tau$}\brace{3,4}\,,}\mbox{\funL{$\tau$}\brace{4,7}\,,}\mbox{\funL{$\tau$}\brace{1,4}\,,}\mbox{\funL{$\tau$}\brace{4,8}\,}\},\funL{$\sigma$}\brace{9,10,11,12,5,6,7,8}\,\}}
\mathematicaSequence{\funL[1]{drawPlabicGraph}\brace{egCell}}{\fig{-78pt}{0.5}{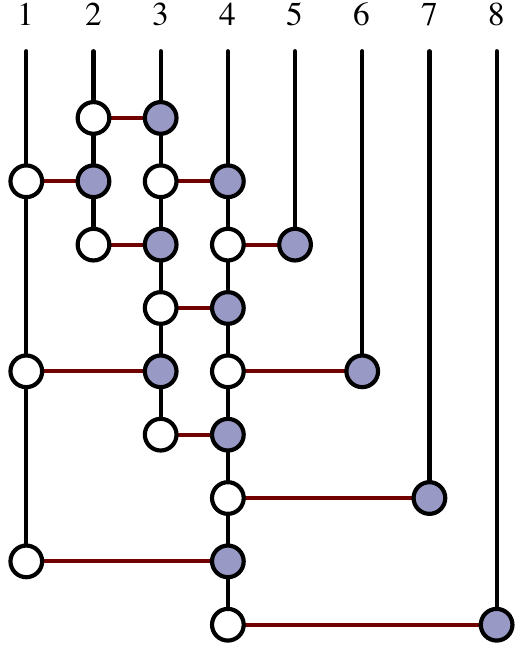}}
}
If the optional argument of \var{detailedQ} is specified as \built{True} then \fun{drawPlabicGraph} will return this plabic graph representative with a great deal of additional information:
\mathematicaBox{
\mathematicaSequence[1]{egCell=\funL{$\sigma$}\brace{6,5,7,10,8,11,9,12};\\
\funL[1]{drawPlabicGraph}\brace{egCell,\built{True}}}
{\fig{-78pt}{0.5}{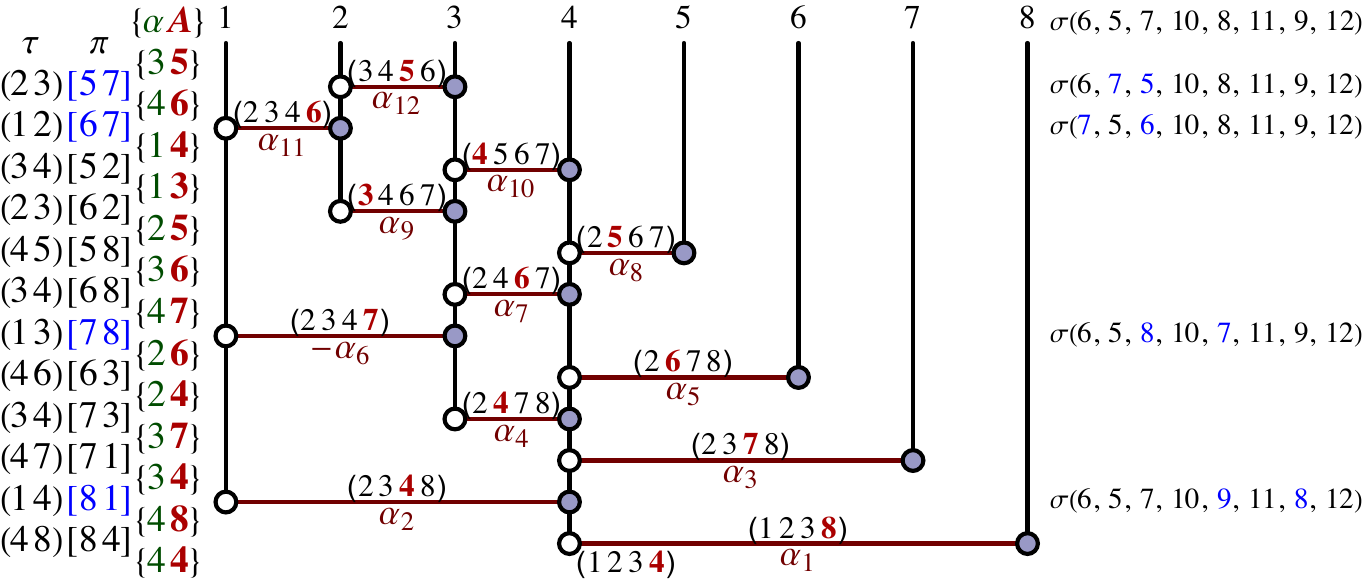}}
}
Here, the sequence of bridges \funL{$\tau$}, transpositions \funL{$\pi$} are listed along side with the each bridge-variables' corresponding tropical label $\{\alpha,A\}$; on the graph are labelled the bridges $\alpha_i$ and the face labels for each bridge; and on the right are the particular boundary elements \emph{accessible} from this chart---namely, those co-dimension one positroids accessible by setting the corresponding bridge variables to zero.\\[-50pt]~\vspace{-40pt}
}

\defnBox{drawPlabicGraphRandom}{\fun{$\sigma$}\brace{\var{permutation}\patternTwo}}{returns a plabic graph representative obtained via a randomly chosen bridge-chain sequence. 
\mathematicaBox{
\mathematicaSequence[1]{egCell=\funL{$\sigma$}\brace{5,6,9,7,11,8,10,12};\\
\funL[1]{drawPlabicGraphRandom}\brace{egCell}}{
\fig{-78pt}{0.5}{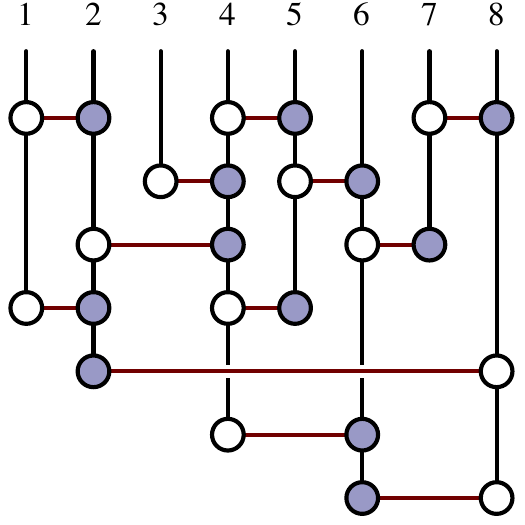}
}
}
}

\defnBox{drawPlabicGraphAtlas}{\fun{$\sigma$}\brace{\var{permutation}\patternTwo},\var{detailedQ}\pattern\optArg{\built{False}}\,}{returns a list of $\var{n}$ plabic graph representatives which correspond to the lexicographic \emph{atlas} of `canonical' coordinate charts.
\mathematicaBox{
\mathematicaSequence{egCell=\funL[]{randomPositroid}\brace{6,1}}{\funL{$\sigma$}\brace{3,5,6,7,8,10}}
\mathematicaSequence{\funL[1]{drawPlabicGraphAtlas}\brace{egCell}}{
$\left\{\rule{0pt}{50pt}\right.$\fig{-58pt}{0.425}{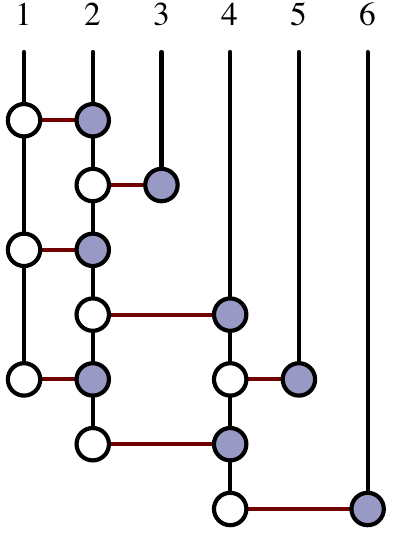}\;,\;\fig{-58pt}{0.425}{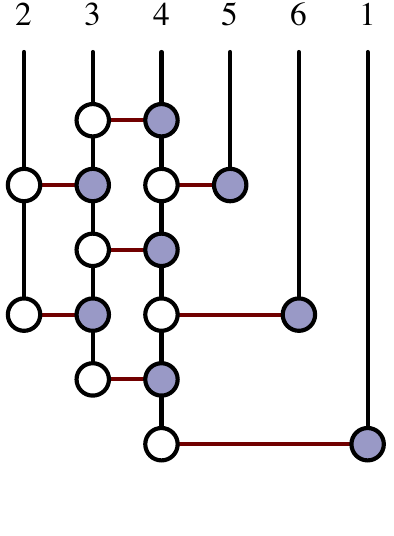}\;,\;\fig{-58pt}{0.425}{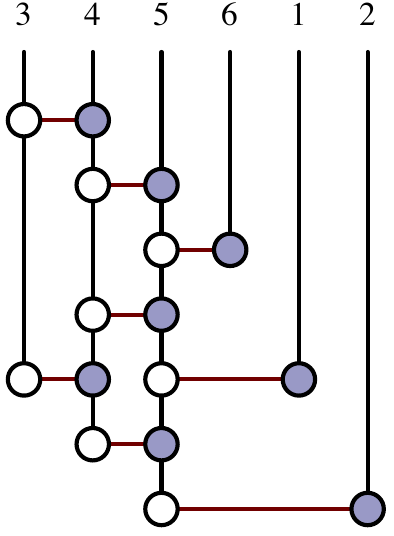}\;,\\
\phantom{\{\{}\fig{-58pt}{0.425}{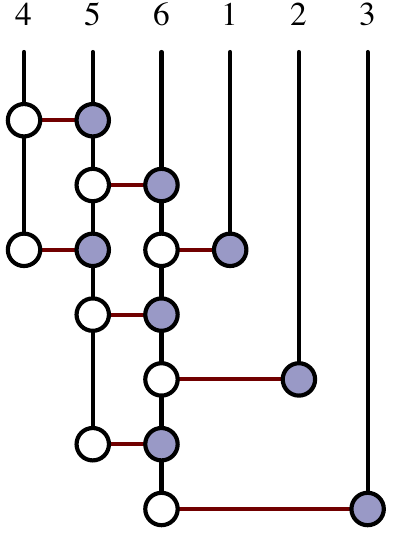}\;,\fig{-58pt}{0.425}{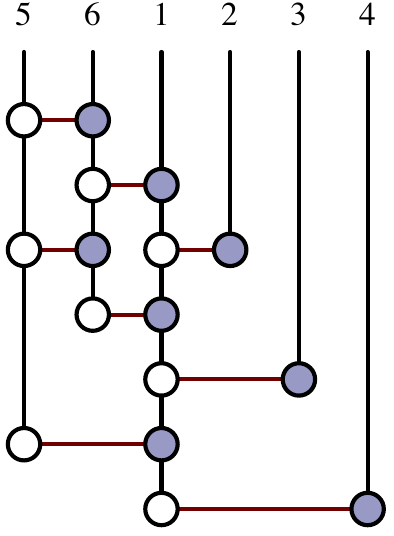}\;,\;\fig{-44pt}{0.425}{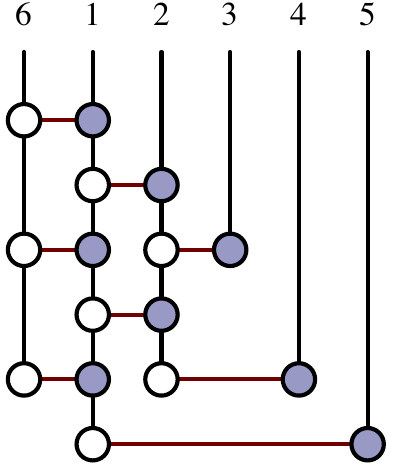}$\left.\rule{0pt}{40pt}\right\}$}
}
}

\newpage
\vspace{-0pt}\subsectionAppendix{Combinatorial Recursion Relations for Tree Amplitudes in sYM}{appendix:combinatorial_bcfw}\vspace{-10pt}

\defnBox{ymPositroidAmp}{\var{$n$}\pattern,\var{$k$}\pattern}{returns a \built{List} of positroid labels \funL{$\sigma$}\brace{\var{permutation}}\, which label contributions to the \var{$n$}-particle N${}^{\var{k}}$MHV tree amplitude in sYM using the same set of recursive choices as made by \fun{ymAmp}, but obtained \emph{entirely} and intrinsically using the combinatorial version of tree-level on-shell recursion (see e.g.~\cite{ArkaniHamed:book,Bourjaily:2012gy}).\\[-5pt]
}

\defnBox{ymPositroidAmpRandom}{\var{$n$}\pattern,\var{$k$}\pattern}{returns a \built{List} of positroid labels \funL{$\sigma$}\brace{\var{permutation}}\, which label contributions to the \var{$n$}-particle N${}^{\var{k}}$MHV tree amplitude in sYM using a \emph{random} choice of adjacent legs and a \emph{random} choice of bridge parity at each successive stage of on-shell recursion. As with \fun{ymPositroidAmp}, this manifestation of on-shell recursion is \emph{entirely} and intrinsically combinatorial and is by far the fastest-to-\emph{obtain} representation of any amplitude in sYM (although the conversion from positroid labels to explicit \funL{superFunction} objects requires non-trivial additional time).}

~

\defnBox{allRecursedYMpositroidTrees}{\var{$n$}\pattern,\var{$k$}\pattern}{returns a \built{List} of \emph{\textbf{all}} distinct sets of positroids \funL{$\sigma$}\brace{\var{perm}}\, \emph{obtainable via BCFW recursion} whose sum gives the  \var{$n$}-particle N${}^{\var{k}}$MHV tree amplitude in sYM. To be clear, a formula is `obtainable via on-shell recursion' if it can be represented as the sum over factorized trees involving shifted adjacent legs with arbitrary bridge parity.\\[-10pt]

To be clear, this does not represent \emph{all} `triangulations' of the amplituhedron (nor even the all `true' triangulations). Starting at 8-point N${}^2$MHV, for example there are merely 2624 20-term formulae resulting from BCFW, although there are an additional 638 distinct 20-term formulae (with all plus signs) resulting from homological identities within the Grassmannian. }

\newpage
\vspace{-0pt}\subsectionAppendix{Analytic \fun{superFunctions} for Tree-Like Positroids}{appendix:analytic_superfunctions_for_tree_like_positroids}\vspace{-10pt}

\defnBox{spinorFormOfPositroid}{\fun{$\sigma$}\brace{\var{permutation}\patternTwo}}{gives an analytic \funL{superFunction} of spinor variables whose corresponding \funL{positroidLabel} is  \mbox{\funL{$\sigma$}\brace{\var{permutation}}\,.}
\mathematicaBox{
\mathematicaSequence{egCell=\funL[]{randomPositroid}\brace{8,3}}{\funL{$\sigma$}\brace{6,8,7,9,11,12,10,13}}
\mathematicaSequence[1]{egFcn=\funL[1]{spinorFormOfPositroid}\brace{egCell};\\
\funL[]{nice}\brace{egFcn}}{$\displaystyle\frac{1}{\sb{1\,2}\sb{3\,4}\sb{4\,5}\sb{5\,6}\sb{8\,1}\sb{6|(3\,4\,5)|(1\,2)|8}\langle7|(8\,1)|2]\langle7|(8\,1\,2)|3]s_{812}s_{7812}},$\\
$\rule{0pt}{30pt}\displaystyle\left(\begin{array}{@{}cccccccc@{}}0&0&\text{-}s_{4\,5\,6}\phantom{\text{-}}&\text{-}\langle4|(56)3]\phantom{\text{-}}&\text{-}\langle5|46)|3]\phantom{\text{-}}&\text{-}\langle6|(45)3]\phantom{\text{-}}&0&0\\
\sb{28}&\sb{81}&0&0&0&0&0&\sb{12}\\
0&0&0&\frac{\sb{5\,6}}{s_{4\,5\,6}}&\text{-}\frac{\sb{4\,6}}{s_{4\,5\,6}}\phantom{\text{-}}&\frac{\sb{4\,5}}{s_{4\,5\,6}}&0&0\end{array}\right)\!\!\cdot\!\tilde\eta$}
\mathematicaSequence{\funL[]{positroidLabel}@egFcn}{\funL{$\sigma$}\brace{6,8,7,9,11,12,10,13}}
}
\textbf{\emph{Note}}: this function is \emph{\textbf{only} defined when} \mbox{\funL{$\sigma$}\brace{\var{permutation}}} labels a \emph{tree-like} positroid (as tested by \funL{treePositroidQ}). This works by recursively constructing the cell in terms of its factorizations. 
}

\defnBox{twistorFormOfPositroid}{\fun{$\sigma$}\brace{\var{permutation}\patternTwo}}{returns an \funL{R}-invariant expression for the cell labeled by the \emph{spinor-space positroid label} \mbox{\funL{$\sigma$}\brace{\var{permutation}}\,.}
\mathematicaBox{
\mathematicaSequence{egCell=\funL[]{randomPositroid}\brace{8,3}}{\funL{$\sigma$}\brace{6,8,7,9,11,12,10,13}}
\mathematicaSequence{egRs=\funL[1]{twistorFormOfPositroid}\brace{egCell}}{\funL{R}\brace{1,2,3,7,8}\,\funL{R}\brace{\funL{cap}\brace{\{2,3\},\{8,7,1\}\!},3,4,6,7}\,\\
\funL{R}\brace{\funL{cap}\brace{\!\{3,4\}\!,\!\{7,6,\funL{cap}\brace{\!\{2,3\}\!,\!\{8,7,1\}\!}\}\!},4,5,6,\\
\phantom{R\,\,}\funL{cap}\brace{\!\{7,6\}\!,\!\{4,3,\funL{cap}\brace{\!\{2,3\}\!,\!\{8,7,1\}\!}\}}}}
\mathematicaSequence{\funL[]{positroidLabel}\brace{egRs}}{\funL{$\sigma$}\brace{4,5,7,6,8,10,11,9}}
\mathematicaSequence{\funL[]{toSpinorPositroidLabels}\brace{\textnormal{\%}}\,===egCell}{True}
}
\textbf{\emph{Note}}: this function is \emph{\textbf{only} defined when} \mbox{\funL{$\sigma$}\brace{\var{permutation}}} labels a \emph{tree-like} positroid (as tested by \funL{treePositroidQ}). This works by recursively constructing the cell in terms of its factorizations.  
}

\defnBox{fromPositroidsToSuperFunctions}{\var{expression}\pattern}{applies \fun{spinorFormOfPositroid} to all tree-like positroids in \var{expression}.\vspace{-40pt}
}

\newpage
\vspace{-0pt}\subsectionAppendix{Numeric \fun{superFunctions} for non-Vanishing Positroids}{appendix:numeric_superFunctions_of_positroids}\vspace{-10pt}

\defnBox{directPositroidEvaluation}{\fun{$\sigma$}\brace{\var{permutation}\patternTwo}}{uses the canonical coordinate chart output by \mbox{\funL{positroidRepresentative}\brace{\funL{$\sigma$}\brace{\var{permutation}}}} and the \funL{grassmannianKinematicEquations} to \emph{directly}, numerically, evaluate the integral
\eq{\mathfrak{f}_{\sigma}\equivR\int\!\prod_{i=1}^{2\var{n}{-}4}\frac{d\,\alpha_i}{\alpha_i}\,\,\delta^{(\var{k}{+}2)\times\!4}\!\big(C(\vec{\alpha})\!\cdot\!\tilde\eta\big)\delta^{(\var{k}{+}2)\times\!2}\!\big(C(\vec{\alpha})\!\cdot\!\tilde\lambda\big)\delta^{2\!\times\!(\var{n}{-}\var{k}{-}2)}\!\big(\lambda\!\cdot\!C(\vec{\alpha})^{\perp}\!\big)}
When there are multiple solutions to the kinematic equations, a \emph{sum} of \funL{superFunction}s for each solution is given.
\mathematicaBox{
\mathematicaSequence{egFcns=FullSimplify\brace{\funL[1]{directPositroidEvaluation}\brace{\\
\mbox{\funL{$\sigma$}\brace{6,5,8,7,10,9,12,11}}}}\,;\\
\funL[]{nice}\brace{egFcns}}{\rule{0pt}{22pt}\scalebox{0.8}{$\displaystyle\frac{113825197759077793463312894604995{-}7991222802561389170729855359\sqrt{233521709}}{1304756847081997232705917282831531568000},$}\\[4pt]
\scalebox{0.7}{$\displaystyle\left(\begin{array}{@{}cccccccc@{}}0&1&\frac{917{-}3\sqrt{233521709}}{73678}&0&\frac{2({-}249033195{+}13423\sqrt{233521709})}{1581166719}&\frac{688057350{+}286301\sqrt{233521709}}{29949157854}&\frac{9233679849{-}1200973\sqrt{233521709})}{145467338148}&0\\[4pt]
0&0&0&1&\frac{26697{+}4\sqrt{233521709})}{42921}&\frac{{-}26743281{-}1657\sqrt{233521709})}{13820562}&\frac{5(153543{+}8\sqrt{233521709})}{987183}&0\end{array}\right)\!\!\cdot\!\tilde\eta$}\scalebox{0.8}{+}\\[10pt]
\scalebox{0.8}{$\displaystyle\frac{113825197759077793463312894604995{+}7991222802561389170729855359\sqrt{233521709}}{1304756847081997232705917282831531568000},$}\\[4pt]
\scalebox{0.7}{$\displaystyle\left(\begin{array}{@{}cccccccc@{}}0&1&\frac{917{+}3\sqrt{233521709}}{73678}&0&\frac{2({-}249033195{-}13423\sqrt{233521709})}{1581166719}&\frac{688057350{-}286301\sqrt{233521709}}{29949157854}&\frac{9233679849{+}1200973\sqrt{233521709})}{145467338148}&0\\[4pt]
0&0&0&1&\frac{26697{-}4\sqrt{233521709})}{42921}&\frac{{-}26743281{+}1657\sqrt{233521709})}{13820562}&\frac{5(153543{-}8\sqrt{233521709})}{987183}&0\end{array}\right)\!\!\cdot\!\tilde\eta$}
}
\mathematicaSequence{FullSimplify\brace{\funL[]{component}\brace{\stateYM{m},\stateYM{p},\stateYM{m},\stateYM{p},\stateYM{m},\stateYM{p},\stateYM{m},\stateYM{p}}\,\brace{egFcns}}}{$\displaystyle\frac{252613095859308442172145010588204506110131}{98009096213067566565994112102555079651456000}$}
}

\textbf{\emph{Note}}: this function \emph{requires} that the positroid labelled by \funL{$\sigma$}\brace{\var{permutation}} has a multiplicity equal to \built{Length}\brace{\builtL{Ls}}\,.
}

\defnBox{fromPositroidsToSuperFunctionsN}{\var{expression}\pattern}{uses numerically-optimized, semi-analytic pathways for expressions involving tree-like positroids, and applies \fun{directPositroidEvaluation} for any other positroids with non-vanishing kinematic support.\\[-10pt]

\textbf{\emph{Note}}: this function \emph{requires} that any positroids appearing in \var{expression} correspond to a multiplicity equal to \built{Length}\brace{\builtL{Ls}}\,.
}

\newpage
\vspace{-0pt}\subsectionAppendix{Covering Relations \& the Positroid Stratification of the Grassmannian}{appendix:covering_relations_and_stratifications_of_positroids}\vspace{-0pt}
%
\vspace{0pt}\subsubsectionAppendix{Positroid Dimensionality, Boundaries, and Inverse-Boundaries}{appendix:boundaries_and_dimensionality}\vspace{-10pt}

\defnBox{positroidDimension}{\fun{$\sigma$}\brace{\var{permutation}\patternTwo}}{returns the \emph{dimension} of the subspace of the positive Grassmannian labelled by \mbox{\funL{$\sigma$}\brace{\var{permutation}}\,.}
}

\defnBox{positroidBoundaries}{\fun{$\sigma$}\brace{\var{permutation}\patternTwo}}{returns a \built{List} of all the positroids in the image of the covering relation $\partial$. That is, it gives all the positroid labels of configurations accessible as \emph{boundaries} of the positroid labelled by \mbox{\funL{$\sigma$}\brace{\var{permutation}}\,.}
}

\defnBox{positroidBoundaryInversions}{\fun{$\sigma$}\brace{\var{permutation}\patternTwo}}{returns a \built{List} of all \emph{inversions} \funL{$\pi$} that connect the positroid labelled by \mbox{\funL{$\sigma$}\brace{\var{permutation}}} to its boundary elements. 
}

\defnBox{positroidInverseBoundaries}{\fun{$\sigma$}\brace{\var{permutation}\patternTwo}}{returns a \built{List} of all the positroids in the image of the inverse covering relation $\partial^{{-}1}$. That is, it gives all the positroid labels of configurations accessible as \emph{inverse-boundaries} of the positroid labelled by \mbox{\funL{$\sigma$}\brace{\var{permutation}}\,.}
}

\vspace{10pt}\subsubsectionAppendix{Enumerating Examples of Positroid Varieties}{appendix:enumeration_and_random_choices_of_positroids}\vspace{-10pt}

\defnBox{numberOfPositroids}{\var{$n$}\pattern,\var{$k$}\pattern,\var{dim}\pattern\optArg{}}{uses the closed-form combinatorial formula of \cite{williams2004enumeration} to give the number of positroid configurations with a given dimension.\\[-10pt]

This can be used to check that the Euler characteristic of the completely non-negative part of the Grassmannian is 1:
\mathematicaBox{
\mathematicaSequence{Power[-1,\#]\funL[1]{numberOfPositroids}\brace{8,2,\#}\,\&/@Range\brace{0,16}}{\{70,-560,1960,-4200,6426,-7672,7532,-6272,\\
\phantom{\{}4522,-2856,1588,-776,330,-120,36,-8,1\}}
\mathematicaSequence{Total[\textnormal{\%}]}{1}
}
If the optional argument \var{dim} is not specified, it is taken to be $(2\var{n}{-}4)$---the dimension relevant for on-shell functions with isolated kinematic support. 
}

\defnBox{allPositroids}{\var{$n$}\pattern,\var{$k$}\pattern,\var{dim}\pattern\optArg{}}{returns a \built{List} of all \emph{decorated} permutations \fun{$\sigma$}\texttt{[}\var{permutation}\texttt{]} labeling positroid varieties in $G_+$(\var{$k$}${+}2$,\var{$n$}) of dimension \var{dim}. If the optional argument \var{dim} is not specified, it is taken to be $(2\var{n}{-}4)$---the dimension of positroid varieties associated with isolated on-shell functions. 
}

\defnBox{randomPositroid}{\var{$n$}\pattern,\var{$k$}\pattern,\var{dim}\pattern\optArg{}}{returns a randomly chosen positroid label for a variety of dimension \var{dim}; if \var{dim} is not provided, it is taken to be $(2\var{n}{-}4)$---the dimension of positroid varieties associated with isolated on-shell functions.  
}

\newpage
\vspace{-0pt}\subsectionAppendix{\emph{Oriented} Coordinate Charts and \emph{Signed} Covering Relations}{appendix:oriented_charts_and_signed_coverings}\vspace{-0pt}
%
\vspace{0pt}\subsubsectionAppendix{Canonically Oriented Atlases of Coordinate Charts}{appendix:canonical_charts_for_positroids}\vspace{-10pt}

\defnBox{positroidReferenceChartOrientation}{\fun{$\sigma$}\brace{\var{permutation}\patternTwo}}{every positroid variety is endowed with an \emph{oriented} volume form which may be written as 
\eq{\Omega_{\fun{$\sigma$}}(\vec{\alpha})\equivR\pm\frac{d\,\alpha_1}{\alpha_1}\wedge\cdots\wedge\frac{d\,\alpha_d}{\alpha_d}}
in terms of canonical coordinates $\vec{\alpha}$. However, different choices of canonical coordinates can be differently oriented. Thus, it is important to choose some specific (if arbitrary) choice for the orientation for \emph{some} set of canonical coordinates for every positroid variety; in terms of these reference orientations, the orientation of all other charts will be fixed.\\[-10pt]

When a positroid is \emph{tree-like}---as detected by \funL{treePositroidQ}\brace{\fun{$\sigma$}\brace{\var{permutation}}\,}---it can be recursively represented by attaching bridges to three-particle vertices, allowing its orientation to be recursively defined in terms of the (still arbitrary) orientation of the three-particle positroids. Moreover, parity conjugation requires consistency between the orientation chosen for \fun{$\sigma$}\brace{2,3,4} and \fun{$\sigma$}\brace{3,4,5}, leaving only one conventional (and overall) sign ambiguity to remain.\\[-10pt]

Thus, for all \emph{tree-like} positroids (those potentially arising in BCFW), \fun{positroidReferenceChartOrientation} returns the orientation of the reference chart \emph{relative to that of the volume form recursively defined}; for all other cells, it assigns the orientation of +1 to the reference chart.\\[-10pt]
\mathematicaBox{
\mathematicaSequence{egTree=\funL[]{ymPositroidAmpRandom}[7,2]}{\{\funL{$\sigma$}\brace{5,6,7,8,10,11,9},\funL{$\sigma$}\brace{6,5,7,8,9,11,10},\funL{$\sigma$}\brace{5,6,8,7,9,11,10},\\
\phantom{\{}\funL{$\sigma$}\brace{6,7,5,8,9,10,11},\funL{$\sigma$}\brace{6,5,7,9,8,10,11},\funL{$\sigma$}\brace{5,6,8,9,7,10,11}\,\}}
\mathematicaSequence{\mbox{evals=\funL[]{component}\brace{\stateYM{p},\stateYM{p},\stateYM{p},\stateYM{m},\stateYM{m},\stateYM{m},\stateYM{m}}\hspace{1pt}\brace{\funL[]{directPositroidEvaluation}/@egTree\hspace{-1pt}};\hspace{-20pt}}}{$\rule{0pt}{20pt}\displaystyle\left\{\frac{70236160}{1906084957},\text{-}\frac{530012591925}{25057531744973},0,\text{-}\frac{300000}{1478528597},\frac{194400}{5431286861},0\right\}$}
\mathematicaSequence{recursedSigns=\funL[1]{positroidReferenceChartOrientation}/@egTree}{\{-1,1,-1,1,-1,1\}}
\mathematicaSequence{recursedSigns.evals}{$\rule{0pt}{20pt}\displaystyle\text{-}\frac{61944808028165716133213}{196261953594267811250}$}
\mathematicaSequence{\textnormal{\%}===\funL[]{componentAmpN}\brace{\stateYM{p},\stateYM{p},\stateYM{p},\stateYM{m},\stateYM{m},\stateYM{m},\stateYM{m}}}{True}
}
}

\defnBox{positroidReferenceChartAtlas}{\fun{$\sigma$}\brace{\var{permutation}\patternTwo}}{the \emph{lexicographic} atlas consists of those $n$ charts obtained by the lexicographic bridge chain algorithm, but starting from column $a\!\in\![n]$ (with columns labeled cyclically).
\mathematicaBox{
\mathematicaSequence{egPositroid=\funL[]{randomPositroid}\brace{6,2,5}}{\funL{$\sigma$}\brace{6,7,4,8,9,11}}
\mathematicaSequence[1]{egAtlas=\funL[1]{positroidReferenceChartAtlas}\brace{egPositroid}\,;\\
\funL[]{nice}\brace{egAtlas}}{\scalebox{0.65}{\!\!$\fwboxR{0pt}{\left\{\rule{0pt}{32pt}\right.}\displaystyle\frac{1}{\alpha_1\alpha_2\alpha_3\alpha_4\alpha_5}\!,\!\!\left(\begin{array}{@{}cccccc@{}}1&\alpha_5&0&0&0&0\\
0&1&\alpha_3&\alpha_3\,\alpha_4&0&0\\
0&0&1&\alpha_4&\alpha_2&0\\
0&0&0&0&1&\alpha_1\end{array}\right)\!\!\cdot\!\tilde\eta,\!\!$} 
\scalebox{0.65}{\!\!$\displaystyle\frac{1}{\alpha_1\alpha_2\alpha_3\alpha_4\alpha_5}\!,\!\!\left(\begin{array}{@{}cccccc@{}}
0&1&\alpha_4&\alpha_4\alpha_5&0&0\\
0&0&1&\alpha_5&\alpha_3&0\\
0&0&0&0&1&\alpha_2\\
\text{-}\alpha_1&0&0&0&0&1\end{array}\right)\!\!\cdot\!\tilde\eta,\!\!$}
\scalebox{0.65}{\!\!$\displaystyle\frac{1}{\alpha_1\alpha_2\alpha_3\alpha_4\alpha_5}\!,\!\!\left(\begin{array}{@{}cccccc@{}}
0&0&1&\alpha_5&\alpha_4&0\\
0&0&0&0&1&\alpha_3\\
\!\!\!\text{-}\alpha_2\phantom{\text{-}}\!\!\!&0&0&0&0&1\\
\!\!\!\text{-}1\phantom{\text{-}}\!\!\!&\text{-}\alpha_1\phantom{\text{-}}\!\!\!&0&0&0&0\end{array}\right)\!\!\cdot\!\tilde\eta,\!\!$}\\

\scalebox{0.65}{\!\!$\displaystyle\frac{1}{\alpha_1\alpha_2\alpha_3\alpha_4\alpha_5}\!,\!\!\left(\begin{array}{@{}cccccc@{}}
0&0&\alpha_1&1&\alpha_5&0\\
0&0&0&0&1&\alpha_4\\
\!\!\!\text{-}\alpha_3\phantom{\text{-}}\!\!\!&0&0&0&0&1\\
\!\!\!\text{-}1\phantom{\text{-}}\!\!\!&\text{-}\alpha_2\phantom{\text{-}}\!\!\!&0&0&0&0\end{array}\right)\!\!\cdot\!\tilde\eta,\!\!$}
\scalebox{0.65}{\!\!$\displaystyle\frac{1}{\alpha_1\alpha_2\alpha_3\alpha_4\alpha_5}\!,\!\!\left(\begin{array}{@{}cccccc@{}}
0&0&0&0&1&\alpha_5\\
\!\!\!\text{-}\alpha_4\phantom{\text{-}}\!\!\!&0&0&0&0&1\\
\!\!\!\text{-}1\phantom{\text{-}}\!\!\!&\!\!\!\text{-}\alpha_3\phantom{\text{-}}\!\!\!&0&0&0&0\\
0&\!\!\!\text{-}1\phantom{\text{-}}\!\!\!&\text{-}\alpha_2\phantom{\text{-}}\!\!\!&\!\!\!\text{-}\alpha_1\phantom{\text{-}}\!\!\!&0&0\end{array}\right)\!\!\cdot\!\tilde\eta,\!\!$}
\scalebox{0.65}{\!\!$\displaystyle\frac{1}{\alpha_1\alpha_2\alpha_3\alpha_4\alpha_5}\!,\!\!\left(\begin{array}{@{}cccccc@{}}
\!\!\!\text{-}\alpha_5\phantom{\text{-}}\!\!\!&0&0&0&0&1\\
\!\!\!\text{-}1\phantom{\text{-}}\!\!\!&\!\!\!\text{-}\alpha_4\phantom{\text{-}}\!\!\!&0&0&0&0\\
0&\!\!\!\text{-}1\phantom{\text{-}}\!\!\!&\!\!\!\text{-}\alpha_2\phantom{\text{-}}\!\!\!&\!\!\!\text{-}\alpha_2\alpha_3\phantom{\text{-}}\!\!\!&0&0\\
0&0&\!\!\!\text{-}1\phantom{\text{-}}\!\!\!&\text{-}\alpha_3\phantom{\text{-}}\!\!\!&\!\!\!\text{-}\alpha_1\phantom{\text{-}}\!\!\!&\end{array}\right)\!\!\cdot\!\tilde\eta\!\!\!\fwboxL{0pt}{\left.\rule{0pt}{32pt}\right\}}$}
}
\mathematicaSequence{\funL[]{positiveChartQ}/@egAtlas}{\{True,True,True,True,True,True,True\}}
}
Notice that these charts naturally include a `twisted' cyclicity (for even N${}^k$MHV degrees) in order to preserve the positivity of each chart.\\[-10pt]

The key purpose of the canonical chart atlas for a positroid is that \emph{every boundary element of any positroid} is the zero-locus of a coordinate in at least one of the charts of the atlas. For these `accessible' boundaries, the parent's chart provides a canonical coordinate chart for the boundary whose orientation relative to the boundary's reference chart can be easily determined. Thus, provided we have chosen arbitrary but fixed reference chart orientations for a cell and all of its boundary elements, we may deduce an \emph{oriented} boundary map between a parent and each of its boundaries.
}

\defnBox{accessibleChartBoundaries}{\fun{positroid}\brace{\var{matrix}\pattern}}{returns a \built{List} of pairs \{$\alpha_i$,\fun{$\sigma'$}\} where the locus $\alpha_i\!\to\!0$ corresponds to the particular co-dimension one \emph{boundary} configuration labeled by \fun{$\sigma'$}.
\mathematicaBox{
\mathematicaSequence{egCell=\funL[]{randomPositroid}\brace{6,1}}{\funL{$\sigma$}\brace{4,6,5,7,8,9}}
\mathematicaSequence[1]{chartAtlas=\funL[1]{positroidReferenceChartAtlas}\brace{egCell};\\
bdyLocii=\funL[1]{accessibleChartBoundaries}/@chartAtlas;\\
\funL[]{nice}@\textnormal{\%}}{$\displaystyle\left\{\rule{0pt}{22pt}\right.$\scalebox{0.8}{$\displaystyle\left(\begin{array}{@{}cc@{}}\alpha_8 & \funL{$\sigma$}\brace{6,4,5,7,8,9}\\
\alpha_7 & \funL{$\sigma$}\brace{5,6,4,7,8,9}\\
\alpha_5 & \funL{$\sigma$}\brace{4,6,7,5,8,9}\\
\alpha_4 & \funL{$\sigma$}\brace{4,7,5,6,8,9}\end{array}\right)$},\scalebox{0.8}{$\displaystyle\left(\begin{array}{@{}cc@{}}\alpha_8 & \funL{$\sigma$}\brace{4,6,7,5,8,9}\\
\alpha_7 & \funL{$\sigma$}\brace{4,7,5,6,8,9}\\
\alpha_4 & \funL{$\sigma$}\brace{4,6,5,8,7,9}\end{array}\right)$},\scalebox{0.8}{$\displaystyle\left(\begin{array}{@{}cc@{}}\alpha_8 & \funL{$\sigma$}\brace{4,6,7,5,8,9}\\
\alpha_6 & \funL{$\sigma$}\brace{4,6,5,8,7,9}\\
\alpha_4 & \funL{$\sigma$}\brace{4,6,5,7,9,8}\end{array}\right)$},\\
\phantom{\{\{\!}\scalebox{0.8}{$\displaystyle\left(\begin{array}{@{}cc@{}}\alpha_8 & \funL{$\sigma$}\brace{4,6,5,8,7,9}\\
\alpha_6 & \funL{$\sigma$}\brace{4,6,5,7,9,8}\\
\alpha_3 & \funL{$\sigma$}\brace{3,6,5,7,8,10}\end{array}\right)$},\scalebox{0.8}{$\displaystyle\left(\begin{array}{@{}cc@{}}\alpha_8 & \funL{$\sigma$}\brace{4,6,5,7,9,8}\\
\alpha_6 & \funL{$\sigma$}\brace{3,6,5,7,8,10}\\
\alpha_2 & \funL{$\sigma$}\brace{5,6,4,7,8,9}\end{array}\right)$},\scalebox{0.8}{$\displaystyle\left(\begin{array}{@{}cc@{}}\alpha_8 & \funL{$\sigma$}\brace{3,6,5,7,8,10}\\
\alpha_6 & \funL{$\sigma$}\brace{6,4,5,7,8,9}\\
\alpha_4 & \funL{$\sigma$}\brace{5,6,4,7,8,9}\\
\alpha_2 & \funL{$\sigma$}\brace{4,6,7,5,8,9}\end{array}\right)$}$\displaystyle\left.\rule{0pt}{22pt}\right\}$}
\mathematicaSequence[1]{bdyCells=\funL[]{positroidBoundaries}\brace{egCell};\\
\funL[]{complement}\brace{bdyCells,Join@@bdyLocii[\![All,All,2]\!]}}{\{\}}
}
}

\defnBox{positroidAtlasOrientations}{\fun{$\sigma$}\brace{\var{permutation}\patternTwo}}{gives the \emph{relative} orientation of the canonical coordinate charts of the lexicographic atlas of \fun{positroidReferenceChartAtlas}.
\mathematicaBox{
\mathematicaSequence[2]{explicitAtlas=\funL[1]{positroidReferenceChartAtlas}\brace{\fun{$\sigma$}\brace{6,7,5,8,9,10,11}\,};\\
\funL[]{directPositroidEvaluation}/@explicitAtlas;\\
\funL[]{standardizeSuperFunctions}\brace{\textnormal{\%}}[\![All,1]\!]}{\rule{0pt}{18pt}\scalebox{0.85}{$\displaystyle\Big\{\text{-}\frac{1715}{304027776},\frac{1715}{304027776},\text{-}\frac{1715}{304027776},\frac{1715}{304027776},\text{-}\frac{1715}{304027776},\text{-}\frac{1715}{304027776},\text{-}\frac{1715}{304027776}\Big\}$}}
\mathematicaSequence{\textnormal{\%}/(\textnormal{\%}[\![1]\!])}{\{1,-1,1,-1,1,1,1\}}
\mathematicaSequence{\funL[1]{positroidAtlasOrientations}\brace{\fun{$\sigma$}\brace{6,7,5,8,9,10,11}\,}}{\{1,-1,1,-1,1,1,1\}}
}
}

\newpage
\vspace{0pt}\subsubsectionAppendix{\emph{Signed} Covering Relations Induced for Oriented Positroids}{appendix:signed_boundaries}\vspace{-10pt}

\defnBox{positroidOrientedBoundaryRule}{}{a replacement \built{Rule} that, when applied to \mbox{\funL{$\sigma$}\brace{\var{permutation}\patternTwo}} returns the \emph{sum} of \emph{oriented} boundary elements of \fun{$\sigma$}.
\mathematicaBox{
\mathematicaSequence{egCell=\funL[]{randomPositroid}\brace{8,2}}{\funL{$\sigma$}\brace{5,6,7,10,11,8,9,12}}
\mathematicaSequence{egCell/\!\!.\builtL[1]{positroidOrientedBoundaryRule}}{\mbox{{-}\funL{$\sigma$}\brace{4,6,7,10,11,8,9,13}\,{+}\mbox{\funL{$\sigma$}\brace{5,6,7,10,11,8,12,9}}{-}\mbox{\funL{$\sigma$}\brace{5,6,7,10,11,9,8,12}}}\\
{+}\mbox{\funL{$\sigma$}\brace{5,6,7,10,12,8,9,11}}\,{+}\mbox{\funL{$\sigma$}\brace{5,6,7,11,10,8,9,12}}\\
{-}\funL{$\sigma$}\brace{5,6,8,10,11,7,9,12}\,{+}\funL{$\sigma$}\brace{5,6,10,7,11,8,9,12}\,{+}\mbox{\funL{$\sigma$}\brace{5,7,6,10,11,8,9,12}}\,{-}\mbox{\funL{$\sigma$}\brace{6,5,7,10,11,8,9,12}}}
\mathematicaSequence{\textnormal{\%}/\!\!.\builtL[1]{positroidOrientedBoundaryRule}}{0}
\mathematicaSequence[1]{allCells=\funL[]{allPositroids}\brace{8,2};\\
DeleteDuplicates\brace{allCells//\!\!.\builtL[1]{positroidOrientedBoundaryRule}}}{\{0\}}
}
}

\defnBox{positroidOrientedInverseBoundaryRule}{}{a replacement \built{Rule} that, when applied to \mbox{\funL{$\sigma$}\brace{\var{permutation}\patternTwo}} returns the \emph{sum} of \emph{oriented} inverse-boundary elements of \fun{$\sigma$}.
\mathematicaBox{
\mathematicaSequence{egCell=\funL[]{randomPositroid}\brace{9,3}}{\funL{$\sigma$}\brace{4,7,9,8,10,12,14,15,11}}
\mathematicaSequence{egCell/\!\!.\builtL[1]{positroidOrientedInverseBoundaryRule}}{\mbox{{-}\funL{$\sigma$}\brace{4,7,8,9,10,12,14,15,11}\,{-}\mbox{\funL{$\sigma$}\brace{4,7,9,8,10,11,14,15,12}}}\\
{-}\mbox{\funL{$\sigma$}\brace{4,7,9,8,10,12,11,15,14}}\,{+}\mbox{\funL{$\sigma$}\brace{4,7,9,8,10,12,14,11,15}}\\
{+}\mbox{\funL{$\sigma$}\brace{5,7,9,8,10,12,13,15,11}}\,{-}\funL{$\sigma$}\brace{6,7,9,8,10,12,14,13,11}}
\mathematicaSequence{\textnormal{\%}/\!\!.\builtL[1]{positroidOrientedInverseBoundaryRule}}{0}
\mathematicaSequence[1]{allCells=\funL[]{allPositroids}\brace{9,3};\\
DeleteDuplicates\brace{allCells//\!\!.\builtL[1]{positroidOrientedInverseBoundaryRule}}}{\{0\}}

}
}

\newpage
\vspace{-0pt}\subsectionAppendix{Functional Relations via Oriented Positroid Homology}{appendix:functional_identities_via_positroid_geometry}\vspace{-10pt}

\defnBox{positroidIdentities}{\fun{$\sigma$}\brace{\var{permutation}\patternTwo}}{uses \emph{oriented} boundary rules to construct \emph{signed} identities among positroid varieties. Specifically \fun{positroidIdentities} returns a \built{List} of `identities' \emph{involving} \mbox{\funL{$\sigma$}\brace{\var{permutation}}}, encoded as \{\t{leftTermList},\t{rightTermList}\} where \emph{given the canonical orientations of positroids} (including those that are \emph{not} tree-like(!)), the positive sum of \t{leftTermList} is equal to the positive sum of the \t{rightTermList}. 
\mathematicaBox{
\mathematicaSequence[1]{egCell=\funL{$\sigma$}\brace{5,9,8,7,11,12,10,13,15};\\
egIdData=RandomChoice\brace{\funL[1]{positroidIdentities}\brace{egCell}}}{\{\{\mbox{\funL{$\sigma$}\brace{5,9,8,7,11,12,10,13,15}},\mbox{\funL{$\sigma$}\brace{5,8,9,7,12,11,10,13,15}},\\
\phantom{\{\{}\mbox{\funL{$\sigma$}\brace{5,8,9,7,11,12,13,10,15}},\mbox{\funL{$\sigma$}\brace{5,8,9,6,11,12,10,13,16}}\,\},\\
\phantom{\{}\{\mbox{\funL{$\sigma$}\brace{8,5,9,7,11,12,10,13,15}},\mbox{\funL{$\sigma$}\brace{5,8,10,7,11,12,9,13,15}},\\
\phantom{\{\{}\mbox{\funL{$\sigma$}\brace{5,8,9,11,7,12,10,13,15}},\mbox{\funL{$\sigma$}\brace{5,8,9,7,11,13,10,12,15}},\\
\phantom{\{\{}\mbox{\funL{$\sigma$}\brace{5,8,9,7,11,12,10,15,13}},\mbox{\funL{$\sigma$}\brace{4,8,9,7,11,12,10,14,15}}\,\}\}
}
\mathematicaSequence[1]{nSuper=\funL[]{fromPositroidsToSuperFunctionsN}\brace{egIdData};\\
\funL[]{component}\brace{\stateYM{m},\stateYM{p},\stateYM{m},\stateYM{p},\stateYM{p},\stateYM{m},\stateYM{p},\stateYM{m},\stateYM{m}}\,\brace{nSuper}}{\rule{0pt}{18pt}\scalebox{0.675}{$\displaystyle\left\{\left\{\frac{517252646016379}{881914459751681107455},\frac{15026063618251415047513883375}{5975716213747750025276848850731008},{-}\frac{8589669794741}{2902573022254202880},\frac{22589493111}{13961639906138540}\right\}\!,\right.$}\\[5pt]
\scalebox{0.675}{$\phantom{\{}\displaystyle\left.\left\{\frac{166171947437}{31837992087179520},{-}\frac{21562012608885}{14930616769495508224},0,{-}\frac{3127686333707481}{2833137294297855193088},\right.\right.$}\\[5pt]
\scalebox{0.675}{$\phantom{\{}\displaystyle\left.\left.\frac{19560466324277036459}{3890289666308821595724960},{-}\frac{6882318213482544357634375}{1158729908896754640615570407424}\right\}\right\}$}}
\mathematicaSequence{Total/@\textnormal{\%}}{\scalebox{0.7}{$\displaystyle\left\{\frac{3663378985396816077535493629929302486789}{2081854369509564457988656981418939061568536576},\frac{3663378985396816077535493629929302486789}{2081854369509564457988656981418939061568536576}\right\}$}}
}
The above represents a (highly non-trivial) 10-term identity involving 1 non-tree-like positroid---namely, \mbox{\funL{$\sigma$}\brace{4,8,9,7,11,12,10,14,15}\,.} For the example above, a different \built{RandomChoice} of the \fun{positroidIdentities}\brace{\t{egCell}} could have been an identity involving one cell with \funL{positroidKinematicSupport} of 2. 
}

\defnBox{positroidTreeIdentities}{\fun{$\sigma$}\brace{\var{permutation}\patternTwo}}{returns a \built{List} of data encoding all non-trivial identities involving \mbox{\funL{$\sigma$}\brace{\var{permutation}}} \emph{and \textbf{only} \emph{other} tree-like positroids}.
}

\defnBox{numberOfIndependentSuperFunctions}{\var{$n$}\pattern,\var{$k$}\pattern}{returns the number of \emph{independent} \funL{superFunction}s for $\var{n}$-particle N${}^{\var{k}}$MHV scattering amplitudes that appear at arbitrary loop order. This number is empirical, but easily given as 
\eq{\fwbox{0pt}{\hspace{-40pt}\fun{numberOfIndependentSuperFunctions}\brace{\var{n},\var{k}}\,\,\,\Leftrightarrow\prod_{\r{j}=1}^{\var{k}}\binom{\var{n}{-}\r{j}}{4}\bigg/\binom{\var{k}{-}\r{j}{+}4}{4}\,.}}
Specifically, it can be computed as the \emph{total number} of `leading singularities' (of possible \funL{superFunction}s) minus the rank of all functional relations among them. To be clear the \emph{complete} set of on-shell functions is defined to be the total number of $G(\var{k}{+}2,\var{n})$, with each weighted by their \funL{positroidKinematicSupport}.
\mathematicaBox{
\mathematicaSequence[1]{nonVanishing=Select\brace{\funL[]{allPositroids}\brace{8,2},\funL[]{nonVanishingPositroidQ}};\\
Length\brace{\textnormal{\%}}}{178}
\mathematicaSequence{numOfSuperFcns=Total\brace{\funL[]{positroidKinematicSupport}/@nonVanishing}}{180}
\mathematicaSequence[2]{idList=\funL[]{allPositroids}\brace{8,2,13}\,/\!\!.\builtL[]{positroidOrientedBoundaryRule};\\
idMatrix=Coefficient\brace{\#,nonVanishing}\,\&/@idList\\
MatrixRank\brace{idMatrix}}{75}
\mathematicaSequence{numOfSuperFcns-\textnormal{\%}}{105}
\mathematicaSequence{\funL[1]{numberOfIndependentSuperFunctions}\brace{8,2}}{105}
}
}

\newpage
\vspace{-0pt}\sectionAppendix{Kinematic Data: Specification and \emph{Evaluation}}{appendix:evaluation}\vspace{-0pt}
%
\vspace{-0pt}\subsectionAppendix{Global Variables Required for Evaluation}{appendix:global_variables}\vspace{-10pt}
%

\defnBox{Ls}{}{a \built{Global} variable \emph{necessary} for evaluation that is \built{Set} by functions such as \funL{useReferenceKinematics}, \funL{useRandomKinematics}, \funL{useSpinors}, etc.\\[-12pt]

For $n$ particle kinematics, \built{Ls} is an $(n\!\times\!2)$ \built{Array} of numbers encoding \built{Ls}$\equivR\{\text{\fun{$\lambda$}}_\var{1},\ldots,\text{\fun{$\lambda$}}_\var{n}\}$ where each $\fun{$\lambda$}_{\var{a}}\equivR\{\fun{$\lambda$}_{\var{a}}^{\var{1}},\fun{$\lambda$}_{\var{a}}^{\var{2}}\}\equivR\{\fun{$\lambda$}\t{[}\var{a},\var{1}\t{]},\fun{$\lambda$}\t{[}\var{a},\var{2}\t{]}\}$ is a \built{List} of the two \emph{components} of the spinor $\fun{$\lambda$}_{\var{a}}$ for the $\var{a}$th particle. }

\defnBox{Lbs}{}{a \built{Global} variable \emph{necessary} for evaluation that is \built{Set} by functions such as \funL{useReferenceKinematics}, \funL{useRandomKinematics}, \funL{useSpinors}, etc.\\[-12pt]

For $n$ particle kinematics, \built{Lbs} is an $(n\!\times\!2)$ \built{Array} of numbers encoding \built{Lbs}$\equivR\{\text{\fun{$\tilde\lambda$}}_\var{1},\ldots,\text{\fun{$\tilde\lambda$}}_\var{n}\}$ where each $\fun{$\tilde\lambda$}_{\var{a}}\equivR\{\fun{$\tilde\lambda$}{}_{\var{a}}^{\var{\dot{1}}},\fun{$\tilde\lambda$}{}_{\var{a}}^{\var{\dot{2}}}\}\equivR\{\fun{$\lambda$b}\t{[}\var{a},\var{1}\t{]},\fun{$\lambda$b}\t{[}\var{a},\var{2}\t{]}\}$ is a \built{List} of the two \emph{components} of the spinor $\fun{$\tilde\lambda$}_{\var{a}}$ for the $\var{a}$th particle. }

\defnBox{Zs}{}{a \built{Global} variable \emph{necessary} for evaluation that is \built{Set} by functions such as  \funL{useReferenceKinematics}, \funL{useRandomKinematics}, \funL{useSpinors}, etc.\\[-12pt]

For $n$ particle kinematics, \built{Zs} is an $(n\!\times\!4)$ \built{Array} of numbers encoding \built{Zs}$\equivR\{\text{\fun{$z$}}_\var{1},\ldots,\text{\fun{$z$}}_\var{n}\}$ where each $\fun{$z$}_{\var{a}}\equivR\{\fun{$z$}{}_{\var{a}}^{\var{1}},\fun{$z$}{}_{\var{a}}^{\var{2}},\fun{$z$}{}_{\var{a}}^{\var{3}},\fun{$z$}{}_{\var{a}}^{\var{4}}\}\}$ is a \built{List} of the four (homogeneous) \emph{components} of the momentum-twistor $\fun{$z$}_{\var{a}}$.}

\defnBox{pList}{}{a \built{Global} variable \emph{not} required for evaluation that is \built{Set} by functions such as \funL{useReferenceKinematics}, \funL{useRandomKinematics}, \funL{useSpinors}, etc.\\[-12pt]

For $n$ particle kinematics, \built{pList} is an $(n\!\times\!2\!\times\!2)$ \built{Array} of numbers encoding \built{pList}$\equivR\{\text{\fun{p}}\t{[}\var{1}\t{]},\ldots,\text{\fun{p}}\t{[}\var{n}\t{]}\}$ where each $\fun{p}_{\var{a}}\equivR\{\{\fun{p}_\var{a}^{\var{1\,\dot{1}}},\fun{p}_{\var{a}}^{\var{1\,\dot{2}}}\},\{\fun{p}_{\var{a}}^{\var{2\,\dot{1}}},\fun{p}_{\var{a}}^{\var{2\,\dot{2}}}\}\}$ is a $(2\!\times\!2)$ \built{Array} encoding the momentum \mbox{\funL{p}\t{[}\var{$a$}\t{]}} in terms of its \emph{components} $\fun{p}^{\var{\alpha\,\dot\alpha}}_{\var{a}}$.\\
\textbf{\emph{Note}}: the variable \t{pList} is \emph{not} required for (or used by) evaluation.}

\newpage
\vspace{-0pt}\subsectionAppendix{Specification of Kinematic Data for Evaluation}{appendix:specifying_kinematic_data}\vspace{-10pt}
%

\defnBox{useReferenceKinematics}{\var{$n$}\pattern}{declares the \built{Global} variables \builtL{Ls}, \builtL{Lbs}, and \builtL{Zs} using the reference twistors defined by the homogeneous coordinates 
\eq{\builtL{Zs}\brace{\!\brace{\r{a}}}\,\,\,\Leftrightarrow\,\,\fun{z}\brace{\r{a}}\,\equivR\{\lambda_{\r{a}}^{\b{1}},\lambda_{\r{a}}^{\b{2}},\mu_{\r{a}}^{\b{1}},\mu_{\r{a}}^{\b{2}}\}\equivR\left\{1,\binom{\r{a}}{2},\binom{\r{a}{+}2}{5},\binom{\r{a}\text{+}3}{7}\right\}\,.}
with \builtL{Ls}\brace{\!\brace{\r{a}}}$\,\equivR$\builtL{Zs}\brace{\!\brace{\r{a},1;;2}}\, and 
\eq{\builtL{Lbs}\brace{\!\brace{\r{a}}}\,\,\,\Leftrightarrow\,\,\funL{$\lambda$b}_\r{a}\equivR\frac{\ab{\r{a}\,\r{a}{+}1}\mu_{\r{a}{-}1}{+}\ab{\r{a}{+}1\,\r{a}{-}1}\mu_{\r{a}}{+}\ab{\r{a}{-}1\,\r{a}}\mu_{\r{a}{+}1}}{\ab{\r{a}{-}1\,\r{a}}\ab{\r{a}\,\r{a}{+}1}}.}
These were chosen as reference values for several reasons: they are `positive' (rendering most \emph{ordered} kinematic invariants positive); they are non-singular for non-planar leg orderings; avoid many of the spurious poles that can arise in on-shell recursion of amplitudes; and they render many kinematic invariants relatively `small' (rational numbers involving modest integers);\\[-10pt]

To appreciate the final criterion mentioned above, consider the evaluation of the alternating 8-particle N${}^2$MHV amplitude component:
\mathematicaBox{
\mathematicaSequence[1]{\funL[1]{useReferenceKinematics}\brace{8};\\
\funL[]{componentAmpN}\brace{\stateYM{m},\stateYM{p},\stateYM{m},\stateYM{p},\stateYM{m},\stateYM{p},\stateYM{m},\stateYM{p}}}{\rule{0pt}{20pt}$\displaystyle\text{-}\frac{6787653230559005806372923435157}{28036956340009088805349336320000}$}
\mathematicaSequence[1]{\funL[1]{useRandomKinematics}\brace{8};\\
\funL[]{componentAmpN}\brace{\stateYM{m},\stateYM{p},\stateYM{m},\stateYM{p},\stateYM{m},\stateYM{p},\stateYM{m},\stateYM{p}}}{\rule{0pt}{20pt}\scalebox{0.875}{$\displaystyle\text{-}\frac{635241489369444615971247008586261250519814035794918169212045675660018676619}{1717683606643819011881807904377981102739537571278984375000}$}}
}
As you can easily observe, the use of \emph{random} kinematics results in amplitudes that are generically ratios of larger-magnitude integers. 
}

\defnBox{useRandomKinematics}{\var{$n$}\pattern,\var{realQ}\pattern\optArg{\built{False}}\,}{declares the \built{Global} variables \builtL{Ls}, \builtL{Lbs}, and \builtL{Zs} using \emph{randomly} chosen values consistent with momentum-conserving, massless \var{$n$}-particle kinematics. By default, the optional second argument \var{realQ} is set to \built{False}, which ensures (perhaps ironically) that \builtL{Ls}, \builtL{Lbs}, and \builtL{Zs} will all have \emph{real-valued}, \emph{rational} components; as such, all amplitude components will also be \emph{real}-valued. The source of the irony is that \emph{real} spinors translate to \emph{complex}-valued \emph{four-momenta}.\\[-10pt]

If the optional second argument \var{realQ} is set to \built{True}, then \emph{rational} momentum-conserving, on-shell four-momenta will be randomly chosen and used to define \emph{complex} (and algebraic) spinors \builtL{Ls} and \builtL{Lbs} which satisfy the relation \mbox{\builtL{Ls}\brace{\!\brace{\r{a}}}$\,=\!\pm$\built{Conjugate}\brace{\builtL{Lbs}\brace{\!\brace{\r{a}}}}\,,} where the sign of the relation is positive for $\r{a}\!\in\![3,\ldots,\var{n}]$ and negative for $\r{a}\!\in\!\{1,2\}$ (corresponding to the first two particles having negative energy---or being `incoming').\\[-10pt]

One advantage to using `real' kinematics is that component amplitudes for conjugated states are complex conjugates of one another:
\mathematicaBox{
\mathematicaSequence[1]{\funL[1]{useRandomKinematics}\brace{6,True};\\
\funL[]{evaluate}\brace{\{\funL{ab}\brace{2,3}\,,\funL{sb}\brace{2,3}\}}}{\rule{0pt}{22pt}$\displaystyle\left\{\frac{{-}3596{-}5652\,\mathbbm{i}}{\sqrt{473}},\frac{3596{-}5652\,\mathbbm{i}}{\sqrt{473}}\right\}$}
\mathematicaSequence{\funL[]{componentAmpN}\brace{\stateYM{m},\stateYM{p},\stateYM{m},\stateYM{p},\stateYM{m},\stateYM{p}}}{\rule{0pt}{18pt}\scalebox{0.9}{$\displaystyle\frac{352553811438167719166209993321331}{199591728828605016353758137877132800}{+}\frac{9607224877379577533035596761123}{12096468413854849475985341689523200}\mathbbm{i}$}}
\mathematicaSequence{\funL[]{componentAmpN}\brace{\stateYM{p},\stateYM{m},\stateYM{p},\stateYM{m},\stateYM{p},\stateYM{m}}}{\rule{0pt}{18pt}\scalebox{0.9}{$\displaystyle\frac{352553811438167719166209993321331}{199591728828605016353758137877132800}{-}\frac{9607224877379577533035596761123}{12096468413854849475985341689523200}\mathbbm{i}$}}
}
One moderate (if not relatively severe) disadvantage is that the non-rational spinor components (confined to particles labeled $\r{a}\!\in\!\{1,2\}$) can dramatically slow \textsc{Mathematica}'s internal algorithms and complicate the infinite-precision verification of identities.
\mathematicaBox{
\mathematicaSequence[2]{egIdentity=\funL[]{egWardIdentity}\brace{}\,\brace{10,3};\\
\funL[1]{useRandomKinematics}\brace{10,True};\\
\funL[]{timed}\brace{Total\brace{egIdentity/\!\!.\{\funL{amp}\brace{\var{$x$}\patternTwo}$\,\mapsto$\funL[]{componentAmpN}\brace{\var{$x$}}\}}}}{\rule{0pt}{24pt}0}
\\[-48pt]
&\scalebox{0.6}{\hspace{-30pt}Evaluation of the function \texttt{\textbf{Total}} required \textbf{11 seconds, 708 ms} to complete.}\\[12pt]
\mathematicaSequence[1]{\funL[1]{useRandomKinematics}\brace{10};\\
\funL[]{timed}\brace{Total\brace{egIdentity/\!\!.\{\funL{amp}\brace{\var{$x$}\patternTwo}$\,\mapsto$\funL[]{componentAmpN}\brace{\var{$x$}}\}}}}{\rule{0pt}{24pt}0}
\\[-48pt]
&\scalebox{0.6}{\hspace{-30pt}Evaluation of the function \texttt{\textbf{Total}} required \textbf{235 ms, 250 $\mu$s} to complete.}\\[12pt]
}
}

\defnBox{useSpinors}{\var{$\lambda$}\pattern,\var{$\lambda$b}\pattern}{uses the specified variables $\var{\lambda},\var{\lambda\text{b}}$ to \built{Set} the \built{Global} variables \builtL{Ls}, \builtL{Lbs}, and \builtL{Zs} for use in evaluation.\\[-10pt]

\textbf{\emph{Note}}: the function \fun{useSpinors} does \emph{\textbf{not}} verify that the kinematic data is momentum-conserving! If the user declares improper kinematics, then many computations may go awry. 
}

\defnBox{useTwistors}{\var{$z$s}\pattern}{uses the specified \built{List} of twistors \var{$z$s}---which must be a $\var{n}\!\times\!4$ of homogeneous coordinates for momentum twistors---to \built{Set} the \built{Global} variables \builtL{Ls}, \built{Lbs}, and \builtL{Zs} for use in evaluation. Unlike \fun{useSpinors}, for which momentum conservation is a non-trivial constraint, arbitrary momentum-twistors will define on-shell, momentum conserving kinematics. 
}

\vspace{10pt}\subsectionAppendix{Display of Kinematic Data Used for Evaluation}{appendix:display_of_kinematic_data}\vspace{-10pt}
%

\defnBox{showSpinors}{}{gives a \emph{formatted} display of the current spinors encoded by the \built{Global} variables \builtL{Ls} and  \builtL{Lbs}:
\mathematicaBox{\mathematicaSequence[1]{\funL[]{useReferenceKinematics}[8];\\\builtL[1]{showSpinors}}{$\begin{array}{@{}rcccccccc}\rule[-10pt]{0pt}{112pt}
\end{array}$}}
} 
\raisebox{50pt}[000pt]{\fwbox{250pt}{\hspace{-20pt}\begin{array}{@{}r|cccccccc|}\multicolumn{1}{c}{~}
&\lambda_1&\lambda_2&\lambda_3&\lambda_4&\lambda_5&\lambda_6&\lambda_7&\multicolumn{1}{c}{\lambda_8}\\\cline{2-9}\lambda^1&1&1&1&1&1&1&1&1\\
\lambda^2&0&1&3&6&10&15&21&28\\\cline{2-9}\multicolumn{1}{c}{~}\\\multicolumn{1}{c}{~}\\[-20pt]\multicolumn{1}{c}{~}
&\tilde{\lambda}_1&\tilde{\lambda}_2&\tilde{\lambda}_3&\tilde{\lambda}_4&\tilde{\lambda}_5&\tilde{\lambda}_6&\tilde{\lambda}_7&\multicolumn{1}{c}{\tilde{\lambda}_8}\\\cline{2-9}\tilde{\lambda}^{\dot{1}}&\text{-}9&\frac{1}{2}&\frac{7}{6}&\frac{25}{12}&\frac{13}{4}&\frac{14}{3}&\frac{19}{3}&\text{-}9\\
\tilde{\lambda}^{\dot{2}}&\text{-}\frac{165}{14}&0&\frac{1}{3}&\frac{17}{12}&\frac{77}{20}&\frac{42}{5}&16&\text{-}\frac{255}{14}\\\cline{2-9}
\end{array}}}\vspace{-10pt}

\defnBox{showTwistors}{}{gives a \emph{formatted} display of the current momentum-twistors encoded by the \built{Global} variable \builtL{Zs}:
\mathematicaBox{\mathematicaSequence[1]{\funL[]{useReferenceKinematics}[8];\\\builtL[1]{showTwistors}}{$\begin{array}{@{}rcccccccc}\rule[-10pt]{0pt}{80pt}
\end{array}$}}
} 
\raisebox{34pt}[000pt]{\fwbox{250pt}{\hspace{-20pt}\begin{array}{@{}r|cccccccc}
&z_1&z_2&z_3&z_4&z_5&z_6&z_7&\multicolumn{1}{c}{z_8}\\\cline{1-9}\lambda^1&1&1&1&1&1&1&1&1\\
\lambda^2&0&1&3&6&10&15&21&28\\
\mu^1&0&0&1&6&21&56&126&252\\
\mu^2&0&0&0&1&8&36&120&330\end{array}}}\vspace{-10pt}

\defnBox{showMomenta}{}{gives a \emph{formatted} display of the current momenta encoded by the \built{Global} variable \builtL{pList}:
\mathematicaBox{\mathematicaSequence[1]{\funL[]{useReferenceKinematics}[8];\\\builtL[1]{showMomenta}}{$\begin{array}{@{}rcccccccc}\rule[-10pt]{0pt}{45pt}
\end{array}$}}
} 
\raisebox{18pt}[000pt]{\scalebox{0.9}{\fwboxL{250pt}{\hspace{40pt}\begin{array}{@{}r|cccccccc}
&p_1&p_2&p_3&p_4&p_5&p_6&p_7&\multicolumn{1}{c}{p_8}\\\cline{1-9}~&\left(\begin{array}{cc}\text{-}9&\text{-}\frac{165}{14}\\0&0\end{array}\right)&\left(\begin{array}{cc}\frac{1}{2}&0\\\frac{1}{2}&0\end{array}\right)&\left(\begin{array}{cc}\frac{7}{6}&\frac{1}{3}\\\frac{7}{2}&1\end{array}\right)&\left(\begin{array}{cc}\frac{25}{12}&\frac{17}{12}\\\frac{25}{2}&\frac{17}{2}\end{array}\right)&\left(\begin{array}{cc}\frac{13}{4}&\frac{77}{20}\\\frac{65}{2}&\frac{77}{2}\end{array}\right)&
\left(\begin{array}{cc}\frac{14}{3}&\frac{42}{5}\\70&126\end{array}\right)&
\left(\begin{array}{cc}\frac{19}{3}&16\\133&336\end{array}\right)&
\left(\begin{array}{cc}\text{-}9&\text{-}\frac{255}{14}\\\text{-}252&\text{-}510\end{array}\right)\\
\end{array}}}}\vspace{-5pt}

\defnBox{showMomentaComponents}{}{gives a \emph{formatted} display of the \emph{four-component} momenta $p^\mu$ associated with the \built{Global} variable \builtL{pList}:
\mathematicaBox{\mathematicaSequence[1]{\funL[]{useReferenceKinematics}[8];\\\builtL[1]{showMomentaComponents}}{$\begin{array}{@{}rcccccccc}\rule[-10pt]{0pt}{80pt}
\end{array}$}}
\vspace{-10pt}
\raisebox{44pt}[000pt]{\scalebox{1}{\fwboxL{250pt}{\hspace{14pt}\begin{array}{@{}r|cccccccc}
&p_1^\mu&p_2^\mu&p_3^\mu&p_4^\mu&p_5^\mu&p_6^\mu&p_7^\mu&\multicolumn{1}{c}{p_8^\mu}\\\cline{1-9}
p^0&\text{-}\frac{9}{2}&\frac{1}{4}&\frac{13}{12}&\frac{127}{24}&\frac{167}{8}&\frac{196}{3}&\frac{1027}{6}&\text{-}\frac{519}{2}\\
p^1&\text{-}\frac{165}{28}&\frac{1}{4}&\frac{23}{12}&\frac{167}{24}&\frac{727}{40}&\frac{196}{5}&\frac{149}{2}&\text{-}\frac{3783}{28}\\
p^2&\text{-}\frac{165}{28} i&\text{-}\frac{1}{4}i&\text{-}\frac{19}{12}i&\text{-}\frac{133}{24} i&\text{-}\frac{573}{40}i&\text{-}\frac{154}{5}i&\text{-}\frac{117}{2} i&\frac{3273}{28}i\\
p^3&\text{-}\frac{9}{2}&\frac{1}{4}&\frac{1}{12}&\text{-}\frac{77}{24}&\text{-}\frac{141}{8}&\text{-}\frac{182}{3}&\text{-}\frac{989}{6}&\text{-}\frac{501}{2}
\end{array}}}\vspace{-0pt}}\vspace{0pt}}

\vspace{10pt}\subsectionAppendix{Randomly Generating On-Shell Kinematic Data}{appendix:randomly_generating_kinematic_data}\vspace{-10pt}
%
\defnBox{generateRandomSpinors}{\var{$n$}\pattern,\var{realQ}\pattern\optArg{\built{False}}\,}{returns a randomly-generated \built{List} \{\t{lambdas},\t{lambdaBars}\} of two $\var{n}\!\times\!2$ matrices which would give momentum-conserving four-momenta. If the optional second argument of \var{realQ} is unspecified, it defaults to \built{False} and these randomly-generated spinors will have rational, real-valued components; if \var{realQ} is set to \built{True}, then the \emph{four-component} momenta associated with the spinors will be real, but the spinors themselves will be complex and generally algebraic.\\[-10pt]

\textbf{\emph{Note}}: the function \fun{generateRandomSpinors} does \textbf{\emph{not}} \built{Set} the \built{Global} variables required for or used by evaluation!---it merely provides the user with a randomly-generated set of kinematic data consistent with \var{$n$}-particle massless scattering. 
}

\defnBox{generateRandomMomenta}{\var{$n$}\pattern}{returns a randomly-generated \built{List} \{\t{ps}\} of $2\!\times\!2$ Hermitian matrices with rational components, each of which has vanishing determinant and the sum of which is vanishes.\\[-10pt]

\textbf{\emph{Note}}: the function \fun{generateRandomMomenta} does \textbf{\emph{not}} \built{Set} the \built{Global} variables required for or used by evaluation!---it merely provides the user with a randomly-generated set of kinematic data consistent with \var{$n$}-particle massless scattering. 
}

\defnBox{generateRandomTwistors}{\var{$n$}\pattern}{returns a randomly-generated \built{List} of \{\t{zList}\} of 4-tuples of components of twistors.\\[-10pt]

\textbf{\emph{Note}}: the function \fun{generateRandomTwistors} does \textbf{\emph{not}} \built{Set} the \built{Global} variables required for ore used by evaluation!---it merely provides the user with a randomly-generated set of kinematic data consistent with \var{$n$}-particle massless scattering. 
}

\vspace{10pt}\subsectionAppendix{Mapping Between Kinematic Variables}{appendix:mapping_kinematic_data}\vspace{-10pt}
%

\defnBox{fromMomentumToSpinors}{\var{pMatrix}\pattern}{returns a pair of (complex little-group) representative spinors \{\fun{$\lambda$},\fun{$\lambda$b}\} whose \built{TensorProduct} gives \var{pMatrix}. 
\mathematicaBox{
\mathematicaSequence[2]{\funL[]{useRandomKinematics}\brace{6,True};\\
egP=RandomChoice\brace{\builtL[]{pList}};\\
\funL[]{nice}\brace{egP}}{\rule{0pt}{20pt}$\displaystyle\left(\begin{array}{@{}cc@{}}9&\text{-}24{+}45\mathbbm{i}\\
\text{-}24\text{-}45\mathbbm{i}&289\end{array}\right)$}
\mathematicaSequence{\funL[1]{fromMomentumToSpinors}\brace{egP}}{$\displaystyle\{\{1,\text{-$\tfrac{8}{3}$}{-}5\,\mathbbm{i}\},\{9,\text{-}24\text{+}45\mathbbm{i}\}\}$}
\mathematicaSequence{TensorProduct@@\textnormal{\%}===egP}{True}
}
\textbf{\emph{Note}}: the little-group redundancy is used to set $\lambda^1\mapsto1$; in the example above, this spoils the complex-conjugacy between $\lambda$ and $\tilde\lambda$; but this can always be restored (if desired) by a different choice of little group rescaling. 
}

\defnBox{toFourComponentMomenta}{\var{pMatrix}\pattern}{returns a \built{List} of the four components $p^\mu$ associated with the $2\!\times\!2$ matrix \var{pMatrix}.
\mathematicaBox{
\mathematicaSequence[2]{\funL[]{useRandomKinematics}\brace{6};\\
egP=RandomChoice\brace{\builtL{pList}};\\
\funL[]{nice}@egP}{\rule{0pt}{20pt}$\displaystyle\left(\begin{array}{@{}cc@{}}\frac{1257445}{36732513}&\frac{68635}{36732513}\\
\frac{1058390}{36732513}&\frac{57770}{36732513}\end{array}\right)$}
\mathematicaSequence{Det\brace{egP}}{0}
\mathematicaSequence{\funL[1]{toFourComponentMomenta}\brace{egP}}{\rule{0pt}{20pt}$\displaystyle\left\{\frac{438405}{24488342},\frac{375675}{24488342},\text{-}\frac{989755}{73465026}\mathbbm{i},\frac{1199675}{73465026}\right\}$}
\mathematicaSequence{\{1,-1,-1,-1\}.Power\brace{\textnormal{\%},2}}{0}
}
\vspace{-0pt}}

\defnBox{fromMomentumXToBiTwistor}{\var{pMatrix}\pattern}{takes a point in momentum-$x$ space and returns a canonically `gauge-fixed' pair of twistors \{\t{zA},\t{zB}\} that represents that points as a bi-twistor.
\mathematicaBox{
\mathematicaSequence[1]{\{xA,xB\}=RandomInteger[\{1,50\},\{2,2,2\}];\\
\funL[]{nice}/@\textnormal{\%}}{$\rule{0pt}{20pt}\displaystyle\left\{\left(\begin{array}{@{}cc@{}}2&42\\20&39\end{array}\right),\left(\begin{array}{@{}cc@{}}29&34\\50&13\end{array}\right)\right\}$}
\mathematicaSequence{Det\brace{xA-xB}}{-462}
\mathematicaSequence[1]{\{lineA,lineB\}=\funL[1]{fromMomentumXToBiTwistor}/@\{xA,xB\};\\
\funL[]{nice}/@\textnormal{\%}}{$\rule{0pt}{20pt}\displaystyle\left\{\left(\begin{array}{@{}cccc@{}}1&0&-20&-39\\0&1&2&42\end{array}\right),\left(\begin{array}{@{}cccc@{}}1&0&-50&-13\\0&1&29&34\end{array}\right)\right\}$}
\mathematicaSequence{Det\brace{Join\brace{lineA,lineB}}}{-462}
} 
}

\defnBox{fromBiTwistorToMomentumX}{\{\var{zA}\pattern,\var{zB}\pattern\}}{returns a $2\!\times\!2$ matrix representing the point $x$ in momentum space corresponding to the bi-twistor \{\var{zA},\var{zB}\}.
\mathematicaBox{
\mathematicaSequence[1]{xList=\funL[1]{fromBiTwistorToMomentumX}/@RotateRight\brace{Partition\brace{\builtL[]{Zs},2,1,1}};\\
pList===(\#2-\#1\&@@@Partition\brace{xList,2,1,1})}{True}
}
\vspace{-40pt}}

\newpage
\vspace{-0pt}\subsectionAppendix{Evaluation of Analytic Expressions}{appendix:evaluation_of_expressions}\vspace{-10pt}
%

\defnBox{evaluate}{\var{expression}\pattern}{uses the user-specified kinematic data (encoded by the \built{Global} variables \builtL{Ls}, \builtL{Lbs}, and \builtL{Zs} which are set by functions such as \funL{useReferenceKinematics} \funL{useRandomKinematics}, \funL{useSpinors}, etc.) to \emph{evaluate} all the otherwise-\emph{abstract} and \emph{undefined} symbols representing kinematic invariants.\\[-10pt]

\textbf{\emph{Note}}: if \emph{no} kinematic data has been specified, the appropriate number $\var{n}$ of external particles is assessed (perhaps incorrectly) by the package and \funL{useReferenceKinematics}\brace{\var{$n$}} is called---which \built{Set}s the global variables \builtL{Ls}, \builtL{Lbs}, and \builtL{Zs}. If an expression refers to particle indices \emph{beyond those} appropriate for the kinematics stored in \built{Ls}, \built{Lbs}, and \built{Zs}, then reference kinematics are \emph{temporarily used} for evaluation---without altering the globally stored kinematic data of \builtL{Ls}, \builtL{Lbs}, and \builtL{Zs}.\\[-10pt]

\textbf{\emph{Note}}: the function \fun{evaluate} cannot detect when an expression requires \emph{fewer} particles' kinematics than those stored by the \built{Global} variables \built{Ls}, etc. 
\mathematicaBox{\mathematicaSequence[1]{\funL[]{useReferenceKinematics}[7];\\Total/@\funL[1]{evaluate}[\funL[]{component}[\stateYM{m},\stateYM{p},\stateYM{m},\stateYM{p},\stateYM{m},\stateYM{p}][\\\funL[]{ymAmpRandom}[6,1]\&/@Range[2]]]\vspace{8pt}}{$\displaystyle\left\{\text{-}\frac{4563699803857304057942282807}{658184537737880177986080000},\text{-}\frac{96283330215367867044431}{32642303041160101344000}\right\}$}}
\textbf{\emph{Note}}: the function \fun{evaluate} does not detect whether the global kinematic data is well-formed or appropriate (e.g.~momentum-conserving). 
}

\defnBoxTwo{evaluatePermuted}{\var{permutedLegLabels}\patternTwo}{\var{expression}\pattern}{is a slightly more efficient implementation of what would otherwise be achieved by calling \mbox{\fun{evaluate}\brace{\funL{permuteLegs}\brace{\var{permutedLegLabels}}\,\brace{\var{expression}}}\,.} That is, \fun{evaluatePermuted} effectively redirects the analytic expressions appearing in \var{expression} to refer to different particle labeling (without losing track of stored evaluations).  
}

\defnBoxTwo{evaluateWithSpinors}{\{\var{$\lambda$s}\pattern,\var{$\tilde{\lambda}s$}\pattern\}}{\var{expression}\pattern}{uses the input spinors \{\var{$\lambda$s},\var{$\tilde\lambda$s}\} to define and evaluate all necessary kinematic invariants appearing in \var{expression} \emph{without altering the globally-defined kinematic data} encoded by \builtL{Ls}, \builtL{Lbs}, \builtL{Zs} \emph{or} losing the temporarily stored values of the evaluations of various invariants for those \built{Global} variables.\\[-10pt]

One example where this may be useful would be when it is not desirable to alter an analytic expression.
\mathematicaBox{
\mathematicaSequence[2]{egFcn=\funL[]{componentAmp}\brace{\stateYM{m},\stateYM{p},\stateYM{m},\stateYM{p},\stateYM{m},\stateYM{p}};\\
shiftedExprn=\funL[]{bcfwShift}\brace{\{1,6\}}\,\brace{egFcn};\\
FullSimplify\brace{\funL[]{evaluate}\brace{shiftedExprn}}}{\rule{0pt}{20pt}$\displaystyle\frac{2088025}{961308({-}1{+}2\,\alpha)}{+}\frac{34300}{63327({-}3{+}4\,\alpha)}{+}\frac{90705633605}{2099859936({-}93{+}143\,\alpha)}$}
\mathematicaSequence{FullSimplify\brace{\funL[1]{evaluateWithSpinors}\brace{\funL[]{bcfwShiftSpinors}\brace{\{1,6\}}}\,\brace{egFcn}}}{\rule{0pt}{20pt}$\displaystyle\frac{2088025}{961308({-}1{+}2\,\alpha)}{+}\frac{34300}{63327({-}3{+}4\,\alpha)}{+}\frac{90705633605}{2099859936({-}93{+}143\,\alpha)}$}
}
}

\defnBoxTwo{evaluateWithTwistors}{\var{$z$List}\pattern}{\var{expression}\pattern}{uses the specified twistors \var{$z$List} to specify kinematic data at which the \var{expression} should be evaluated \emph{without} making any changes to the otherwise-defined \built{Global} kinematic parameters \builtL{Zs}, \builtL{Ls}, or \builtL{Lbs}. This function is especially useful for evaluating expressions over parametric families of on-shell, momentum-conserving kinematic data.
}

\defnBox{storeKinematicEvaluationsQ}{}{to speed up the function \fun{evaluate}, for example, intermediate values of kinematic invariants are by default stored in memory until reset by specifying new kinematic data. This improvement in speed, however, comes at a (reasonably small) cost of additional overhead in memory. This can be turned off by setting the \built{Global} variable \built{storeKinematicEvaluationsQ} to \built{False}. 
}

\newpage
\vspace{-0pt}\subsectionAppendix{On-Shell Modifications to Specified Kinematic Data}{appendix:modifications_to_kinematic_data}\vspace{-10pt}
%
\defnBox{bcfwShiftSpinors}{\{\var{$a$},\var{$b$}\}}{starts from the \built{Global} kinematic variables \{\builtL{Ls},\,\builtL{Lbs}\}, and returns the list \{\fun{$\hat{\lambda}$},\fun{$\hat{\lambda\text{b}}$}\} where \fun{$\hat{\lambda\text{b}}$}\brace{\var{$a$}}$\equivR\!$\funL{$\lambda$b}\brace{\var{$a$}}\,+\,$\r{\alpha}$\funL{$\lambda$b}\brace{\var{$b$}}, and  \fun{$\hat{\lambda}$}\brace{\var{$b$}}$\equivR\!$\funL{$\lambda$}\brace{\var{$b$}}\,$-$$\r{\alpha}$\funL{$\lambda$}\brace{\var{$a$}}, with all other spinors unchanged. 
\mathematicaBox{
\mathematicaSequence[1]{egAmp=\funL[]{componentAmp}\brace{\stateYM{p},\stateYM{m},\stateYM{m},\stateYM{m},\stateYM{p},\stateYM{p},\stateYM{p}}\,;\\
FullSimplify\brace{\funL[]{evaluate}\brace{\funL[]{bcfwShift}\brace{\{7,1\}}\,\brace{egAmp}}}}{$\rule{0pt}{24pt}\displaystyle\frac{10322533950389}{205058559972000(20\,\r\alpha{-}1)}{+}\frac{8732691}{2021600(156\,\r\alpha{-}19)}$}
\mathematicaSequence[1]{shiftedSpinors=\funL[1]{bcfwShiftSpinors}\brace{\{7,1\}};\\
FullSimplify\brace{\funL[]{evaluateWithSpinors}\brace{shiftedSpinors}\,\brace{egAmp}}}{$\rule{0pt}{24pt}\displaystyle\frac{10322533950389}{205058559972000(20\,\r\alpha{-}1)}{+}\frac{8732691}{2021600(156\,\r\alpha{-}19)}$}
}
}

\defnBox{inverseSoftSpinors}{\var{insertionSlot}\patternTwo}{starts from the \built{Global} kinematic variables \{\builtL{Ls},\,\builtL{Lbs}\} for $n$ particles, and returns a new set \{\fun{$\hat{\lambda}$},\fun{$\hat{\lambda\text{b}}$}\} for $(n{+}1)$-particle kinematics where the additional particle is indexed by |\var{insertionSlot}|$\in\![n{+}1]$ with a randomly-chosen momentum proportional to a new parameter $\epsilon$---for which the on-shell condition and momentum conservation are satisfied independent of $\epsilon$, allowing the $\epsilon\!\to\!0$ limit to be probed while remaining momentum-conserving.\\[-10pt]

In four dimensions, a massless particle can be made soft in two ways, depending on whether it's spinor \funL{$\lambda$} or its conjugate spinor \funL{$\lambda$b} vanishing in the limit $\epsilon\!\to\!0$. These two cases are distinguished by the sign of \var{insertionSlot} with positive corresponding to the former case, and if negative the later. 
\mathematicaBox{
\mathematicaSequence[2]{\funL[]{useReferenceKinematics}\brace{5}\,;\\
\funL[]{useSpinors}\brace{\funL[1]{inverseSoftSpinors}\brace{6}}\,;\\
\builtL[]{showSpinors}}{\rule[-15pt]{0pt}{45pt}~}
\mathematicaSequence[2]{\funL[]{useReferenceKinematics}\brace{5}\,;\\
\funL[]{useSpinors}\brace{\funL[1]{inverseSoftSpinors}\brace{-6}}\,;\\
\builtL[]{showSpinors}}{\rule[-15pt]{0pt}{45pt}~}
}
\raisebox{120pt}[000pt]{\fwboxL{250pt}{\hspace{20pt}\scalebox{0.85}{$\begin{array}{@{}r|cccccc|}\multicolumn{1}{c}{~}
&\fwbox{20pt}{\lambda_1}&\fwbox{20pt}{\lambda_2}&\fwbox{20pt}{\lambda_3}&\fwbox{20pt}{\lambda_4}&\fwbox{20pt}{\phantom{\tilde{\lambda}_5}\lambda_5\phantom{\tilde{\lambda}_5}}&\multicolumn{1}{c}{\fwbox{20pt}{\lambda_6}}\\\cline{2-7}\lambda^1&\big(1\text{-}\frac{6232}{27}\r\epsilon\big)&1&1&1&\big(1{+}\frac{10108}{27}\r\epsilon\big)&19\,\r\epsilon\\
\lambda^2&\text{-}\frac{3280}{27}\r\epsilon&1&3&6&\big(10{+}\frac{5320}{27}\,\r\epsilon\big)&10\,\r\epsilon\\\cline{2-7}
\end{array}$}\hspace{20pt}
\scalebox{0.85}{$\begin{array}{@{}r|cccccc|}\multicolumn{1}{c}{~}
&\fwbox{20pt}{\tilde{\lambda}_1}&\fwbox{20pt}{\tilde{\lambda}_2}&\fwbox{20pt}{\tilde{\lambda}_3}&\fwbox{20pt}{\tilde{\lambda}_4}&\fwbox{20pt}{\tilde{\lambda}_5}&\multicolumn{1}{c}{\tilde{\lambda}_6}\\\cline{2-7}
\tilde{\lambda}^{\dot{1}}&\text{-}\frac{21}{10}&\phantom{\big)}\frac{1}{2}\phantom{\big)}&\frac{7}{6}&\frac{25}{12}&\text{-}\frac{33}{20}&7\\
\tilde{\lambda}^{\dot{2}}&\text{-}\frac{4}{5}&0&\phantom{\big)}\frac{1}{3}\phantom{\big)}&\frac{17}{12}&\text{-}\frac{19}{20}&9\\\cline{2-7}
\end{array}$}
}}\vspace{-10pt}\\[-12pt]
\raisebox{20pt}[000pt]{\fwboxL{250pt}{\hspace{20pt}\scalebox{0.85}{$\begin{array}{@{}r|cccccc|}\multicolumn{1}{c}{~}
&\fwbox{20pt}{\lambda_1}&\fwbox{20pt}{\lambda_2}&\fwbox{20pt}{\lambda_3}&\fwbox{20pt}{\lambda_4}&\fwbox{20pt}{\phantom{\tilde{\lambda}_5}\lambda_5\phantom{\tilde{\lambda}_5}}&\multicolumn{1}{c}{\fwbox{20pt}{\lambda_6}}\\\cline{2-7}\lambda^1&\phantom{\big(}1\phantom{\big)}&1&1&1&1&12\\
\lambda^2&\phantom{\big(}0\phantom{\big(}&1&3&6&10&19\\\cline{2-7}
\end{array}$}\hspace{20pt}
\scalebox{0.85}{$\begin{array}{@{}r|cccccc|}\multicolumn{1}{c}{~}
&\fwbox{20pt}{\tilde{\lambda}_1}&\fwbox{20pt}{\tilde{\lambda}_2}&\fwbox{20pt}{\tilde{\lambda}_3}&\fwbox{20pt}{\tilde{\lambda}_4}&\fwbox{20pt}{\tilde{\lambda}_5}&\multicolumn{1}{c}{\tilde{\lambda}_6}\\\cline{2-7}
\tilde{\lambda}^{\dot{1}}&\text{-}\big(\frac{21}{10}\text{+}\frac{101}{2}\,\r\epsilon\big)&\phantom{\big)}\frac{1}{2}\phantom{\big)}&\frac{7}{6}&\frac{25}{12}&\text{-}\big(\frac{33}{20}\text{+}\frac{19}{2}\,\r\epsilon\big)&5\,\r\epsilon\\
\tilde{\lambda}^{\dot{2}}&\text{-}\big(\frac{4}{5}\text{+}\frac{606}{5}\,\r\epsilon\big)&0&\phantom{\big)}\frac{1}{3}\phantom{\big)}&\frac{17}{12}&\text{-}\big(\frac{19}{20}\text{+}\frac{114}{5}\,\r\epsilon\big)&12\,\r\epsilon\\\cline{2-7}
\end{array}$}
}}\vspace{-0pt}\\
\textbf{\emph{Note}}: the new spinors added are \emph{randomly} chosen each time. 
}

\newpage
\vspace{-0pt}\sectionAppendix{Documentation and Functions of the Package}{appendix:documentation_and_information}\vspace{-0pt}
%
\vspace{-0pt}\subsectionAppendix{Listing Defined Function Categories}{appendix:listing_defined_functions}\vspace{-10pt}
%

\defnBox{protectedSymbolList}{}{a \built{List} of symbols protected by the package. }

\defnBox{symbolsAndFunctionHeads}{}{an alphabetical \built{List} of all function \built{Head}s defined by the package---similar to the alphabetical listing in the \hyperlink{index}{index}.\\[-5pt] }

\defnBox[]{definedFunctionCategories}{\var{section}\pattern\optArg{},\var{subsection}\pattern\optArg{}}{As demonstrated in earlier in the \hyperlink{context_organization_of_appendix}{appendix}, the function \fun{definedFunctionCategories} summarizes the list of functions defined by the package similarly to the organization of this appendix. }

\vspace{10pt}\subsectionAppendix{Installing the \package\, Package}{appendix:installation}\vspace{-10pt}
%

\defnBox{installTreeAmplitudesPackage}{}{makes a copy of the file `\package\!\textbf{\t{.m}}' and places it into one of the directories of \built{\$Path} so that it can be called in future sessions without 
\mathematicaBox{
\mathematicaSequence{\builtL[1]{installTreeAmplitudesPackage}}{}
\\[-22pt]&\scalebox{0.6}{\hspace{-30pt}The \package\, package has been successfully installed to}\\[-7pt]
&\scalebox{0.6}{\hspace{-20pt} /Users/bourjaily/Library/Mathematica/Applications}\\[-7pt]
&\scalebox{0.6}{\hspace{-30pt}To use, simply evaluate: \t{<<\,\package\!\textbf{.m}}\,\,in any new notebook.}\\[-2pt]
}
\textbf{\emph{Note}}: the package cannot be `installed' \emph{from an existing install}; to \emph{re}-install the package, simply load \package\!\textbf{\t{.m}} from a \emph{different} location than the install-directory and call again \built{installTreeAmplitudesPacakge}\,. (This is equivalent to simply copying the file to one of the directories in \built{\$Path}.)
}

\addtocontents{toc}{\protect~\\[-20pt]\mbox{\vspace{0pt}}\protect\hrulefill\par}
\renewcommand{\indexname}{Alphabetic Glossary of Functions Defined by the Package}

\newpage
\hypertarget{index}{}\input{tools_for_tree_amplitudes.ind}

\newpage
\input{tools_for_tree_amplitudes.bbl}\addcontentsline{toc}{section}{References}\addtocontents{toc}{\protect~\\[-40pt]\mbox{\vspace{0pt}}\protect\par}
\end{document}

%% file: header_data.tex
\let\olditemize\itemize\renewcommand{\itemize}{\vspace{-2pt}\olditemize\setlength{\itemsep}{1pt}\setlength{\parskip}{0pt}\setlength{\parsep}{-0pt}}
\let\oldenumerate\enumerate\renewcommand{\enumerate}{\vspace{-4pt}\oldenumerate\setlength{\itemsep}{1pt}\setlength{\parskip}{0pt}\setlength{\parsep}{0pt}}
\makeatletter\renewcommand\section{\addtocontents{toc}{\protect\addvspace{-15.5\p@}}\@startsection {section}{1}{\z@}{-0.0ex \@plus .2ex \@minus 0.2ex}{1ex \@plus.1ex\@minus 0.5ex}{\normalfont\large\bfseries}}
\renewcommand\subsection{\addtocontents{toc}{\protect\addvspace{-2.25\p@}}\@startsection {subsection}{2}{\z@}{0.5ex \@plus .2ex \@minus 0.2ex}{0.75ex \@plus.1ex\@minus 0.5ex}{\normalfont\bfseries}}
\renewcommand\subsubsection{\addtocontents{toc}{\protect\addvspace{-2.25\p@}}\@startsection {subsubsection}{3}{\z@}{0.5ex \@plus .2ex \@minus 0.2ex}{0.75ex \@plus.1ex\@minus 0.5ex}{\normalfont\bfseries}}
\renewcommand\paragraph{\addtocontents{toc}{\protect\addvspace{-2.25\p@}}\@startsection {paragraph}{4}{\z@}{0.5ex \@plus .2ex \@minus 0.2ex}{0.75ex \@plus.1ex\@minus 0.5ex}{\normalfont\bfseries}}
\newcommand{\eq}[1]{\vspace{-4.5pt}\begin{equation}#1\vspace{-4.5pt}\end{equation}}

\newcommand{\fwbox}[2]{\text{\makebox[#1][c]{$\hspace{-150pt}\displaystyle#2\hspace{-150pt}$}}}
\newcommand{\fwboxL}[2]{\text{\makebox[#1][l]{$#2$}}}
\newcommand{\fwboxR}[2]{\text{\makebox[#1][r]{$#2$}}}
\newcommand{\equivR}{\fwbox{14.5pt}{\hspace{-0pt}\fwboxR{0pt}{\raisebox{0.47pt}{\hspace{1.25pt}:\hspace{-4pt}}}=\fwboxL{0pt}{}}}
\newcommand{\equivL}{\fwbox{14.5pt}{\fwboxR{0pt}{}=\fwboxL{0pt}{\raisebox{0.47pt}{\hspace{-4pt}:\hspace{1.25pt}}}}}
\newcommand{\fig}[3]{\raisebox{#1}{\includegraphics[scale=#2]{#3}}}

\newcommand{\bigger}[1]{\raisebox{-0.95pt}{\scalebox{1.25}{$#1$}}}
\newcommand{\mi}{\raisebox{0.75pt}{\scalebox{0.75}{$\hspace{-0.5pt}\,-\,\hspace{-0.5pt}$}}}

\renewcommand{\phi}{\varphi}
\renewcommand{\bar}{\overline}
\renewcommand{\hat}{\widehat}
\renewcommand{\tilde}{\widetilde}

\newcommand{\ab}[1]{\langle #1\rangle}
\renewcommand{\sb}[1]{[ #1]}

\newcommand{\newcap}{\mathrm{\raisebox{0.75pt}{{$\,\bigcap\,$}}}}
\newcommand{\tcap}{\scalebox{1}{$\!\newcap\!$}}
\newcommand{\tncap}{\scalebox{0.8}{$\!\newcap\!$}}

\newcommand{\package}{\textbf{\t{tree}\rule[-1.05pt]{7.5pt}{.75pt}\t{amplitudes}}}

\newcommand{\sectionAppendix}[2]{\section{\texorpdfstring{#1}{\Alph{section}.\, #1}\label{#2}}}
\newcommand{\subsectionAppendix}[2]{\subsection{\texorpdfstring{#1}{\Alph{section}.\arabic{subsection}\, #1}\label{#2}}}
\newcommand{\subsubsectionAppendix}[2]{\subsubsection{\texorpdfstring{#1}{\Alph{section}.\arabic{subsection}.\arabic{subsubsection}\, #1}\label{#2}}}

\definecolor{hred}{rgb}{0.575,0.0,0.225}
\definecolor{mhvBlue}{rgb}{0.3,0.2,0.75}
\definecolor{varcolor}{rgb}{0.1,0.55,0.25}
\definecolor{functioncolor}{rgb}{0.1,0.35,0.75}
\definecolor{paper_blue}{rgb}{0.3,0.2,0.75}
\definecolor{paper_red}{rgb}{0.65,0.1,0.15}
\definecolor{paper_green}{rgb}{0.05,0.35,0.125}
\definecolor{paper_grey}{gray}{0.375}
\definecolor{hblue}{rgb}{0,0,0.725}
\definecolor{hred}{rgb}{0.575,0.0,0.225}
\definecolor{dim}{rgb}{0.55,0.55,0.55}
\definecolor{deemph}{rgb}{0.25,0.25,0.25}
\definecolor{hteal}{rgb}{0.0,0.545,0.7451}
\definecolor{rindou1}{rgb}{0.4431,0.2862,0.7960}
\definecolor{rindou2}{rgb}{0.0078,0.1215,0.4392}
\definecolor{lapis}{rgb}{0.0.0470,0.2941,0.5568}
\definecolor{emerald}{rgb}{0.31, 0.78, 0.47}
\definecolor{pinegreen}{rgb}{0.0, 0.47, 0.44}
\definecolor{jade}{rgb}{0.0, 0.66, 0.42}
\definecolor{teal}{rgb}{0.0, 0.5, 0.5}
\definecolor{hblue}{rgb}{0,0,0.725}
\definecolor{hred}{rgb}{0.575,0.0,0.225}
\definecolor{hgreen}{rgb}{0.0,0.4,0.2}
\definecolor{hteal}{rgb}{0.0,0.545,0.7451}

\newcommand{\deemph}[1]{{\color{dim}#1}}
\renewcommand{\r}[1]{{\color{hred}#1}}
\renewcommand{\b}[1]{{\color{hblue}#1}}
\newcommand{\bb}[1]{{\color[rgb]{0,0,1}#1}}
\newcommand{\g}[1]{{\color{paper_green}#1}}

\newcounter{lineNo}
\def\inItemQ{0}
\renewcommand{\brace}[1]{\hspace{-2pt}{\texttt{[}}#1{\texttt{]}}\hspace{-3pt}}
\newcommand{\optArg}[1]{\textnormal{\texttt{:}}#1\!}
\newcommand{\head}[1]{{\color{black}\textnormal{{\texttt{#1}}}}}

\newcommand{\boxedList}[1]{\def\inItemQ{1}~\\[-12pt]\noindent\fwbox{24pt}{}\fwboxL{425pt}{\begin{minipage}[t]{0.932\textwidth}\hspace{-22pt}{$\bullet$\,\,#1}\,\,~\\[-12pt]\end{minipage}\hspace{\fill}~\def\inItemQ{0}\vspace{-0pt}\hspace{\fill}}}

\newcommand{\var}[1]{{{\color{varcolor}{\sl#1}}}}
\newcommand{\fun}[1]{\text{\texttt{\textbf{\texttt{\color{functioncolor}#1}}}}}
\newcommand{\built}[1]{{\color{black}\textbf{\texttt{#1}}}}
\renewcommand{\t}[1]{\texttt{#1}}
\newcommand{\rstar}{\raisebox{1.2pt}{$\star$}}

\newcommand{\pattern}{{\color{varcolor}\rule[-1.05pt]{7.5pt}{.65pt}}}
\newcommand{\patternTwo}{{\color{varcolor}\rule[-1.05pt]{12pt}{.65pt}}}
\newcommand{\patternThree}{{\color{varcolor}\rule[-1.05pt]{14.5pt}{.65pt}}}

\newcommand{\defnBox}[4][2]{\def\inItemQ{1}~\\[-12pt]\noindent\fwbox{24pt}{}\fwboxL{425pt}{\begin{minipage}[t]{0.932\textwidth}%
\ifthenelse{\equal{#1}{5}}{\hypertarget{fourBr:ab}{\hspace{-22pt}{$\bullet$\,\,\ifthenelse{\equal{#3}{}}{\built{#2}}{\fun{#2}}}}}{\ifthenelse{\equal{#1}{2}}{\hypertarget{#2}{\hspace{-22pt}{$\bullet$\,\,\ifthenelse{\equal{#3}{}}{\built{#2}}{\fun{#2}}}}}{\hspace{-22pt}{$\bullet$\,\,\ifthenelse{\equal{#3}{}}{\built{#2}}{\fun{#2}}}}}%
\ifthenelse{\equal{#1}{2}}{\ifthenelse{\equal{#3}{}}{\index{#2@\built{#2}|textbf}}{\index{#2@\fun{#2}|textbf}}}{\ifthenelse{\equal{#1}{5}}{\index{abFour@\fun{#2}\,(four-bracket)|textbf}}{\ifthenelse{\equal{#3}{}}{\index{#2@\built{#2}|textnormal}}{\index{#2@\fun{#2}|textnormal}}}}%
\ifthenelse{\equal{#3}{}}{\hspace{-1.75pt}}{\brace{#3}}%
\hspace{0pt}\textnormal{\texttt{:\!}}\,\,#4~\\[-12pt]\end{minipage}\hspace{\fill}~\def\inItemQ{0}\vspace{5pt}\hspace{\fill}}\\[-12pt]}

\newcommand{\defnBoxTwo}[5][2]{\def\inItemQ{1}~\\[-12pt]\noindent\fwbox{24pt}{}\fwboxL{425pt}{\begin{minipage}[t]{0.932\textwidth}\ifthenelse{\equal{#1}{2}}{\hypertarget{#2}{\hspace{-22pt}{$\bullet$\,\,\ifthenelse{\equal{#3}{}}{\built{#2}}{\fun{#2}}}}}{\hspace{-22pt}{$\bullet$\,\,\ifthenelse{\equal{#3}{}}{\built{#2}}{\fun{#2}}}}\ifthenelse{\equal{#1}{2}}{\ifthenelse{\equal{#3}{}}{\index{#2@\built{#2}|textbf}}{\index{#2@\fun{#2}|textbf}}}{\ifthenelse{\equal{#3}{}}{\index{#2@\built{#2}}}{\index{#2@\fun{#2}}}}\ifthenelse{\equal{#3}{}}{\hspace{-1.75pt}}{\brace{#3}}\hspace{1.5pt}\ifthenelse{\equal{#4}{}}{\hspace{-1.75pt}}{\brace{#4}}\hspace{0pt}\textnormal{\texttt{:\!}}\,\,#5~\\[-12pt]\end{minipage}\hspace{\fill}~\def\inItemQ{0}\vspace{5pt}\hspace{\fill}}\\[-12pt]}

\newcommand{\mathematicaBox}[1]{~\\[0pt]~\\[-12pt]\noindent\fwboxL{450pt}{\setcounter{lineNo}{0}\ifthenelse{\inItemQ=1}{\phantom{}\hspace{-24.55pt}}{\phantom{}\hspace{-0.5pt}}\boxed{\ifthenelse{\inItemQ=1}{\begin{minipage}[t]{1.0585\textwidth}}{\begin{minipage}{0.9865\textwidth}}\begin{tabular}{@{}l@{$\;\;\;$}p{11cm}@{}}#1\end{tabular}\end{minipage}\hspace{\fill}}}\hspace{\fill}\\[2pt]}

\newcommand{\mathematicaSequence}[3][0]{\stepcounter{lineNo}\ifthenelse{\value{lineNo}=1}{}{~\\[-12pt]}
{\color{paper_blue}{\scriptsize{\tt In[}\arabic{lineNo}{\tt]}}\raisebox{-0.65pt}{{\scriptsize{\tt\raisebox{0.75pt}{:}=}}}}&\begin{minipage}[t]{14cm}{\tt #2}\end{minipage}\\[2pt]
\addtocounter{lineNo}{#1}\ifthenelse{\equal{#3}{}}{}{{\color{paper_blue}{\scriptsize{\tt Out[\arabic{lineNo}]}}\raisebox{-0.65pt}{{\scriptsize{\tt=}}}}&\begin{minipage}[t]{14cm}{\tt #3}\end{minipage}\\[1pt]}}

\newcommand{\fourBr}[1][0]{\hyperlink{fourBr:ab}{\fun{ab}}\ifthenelse{\equal{#1}{0}}{\index{abFour@\fun{ab}\,(four-bracket)|textnormal}}{[#1]\index{abFour@\fun{ab}\,(four-bracket)|textnormal}}}
\newcommand{\funL}[2][2]{\ifthenelse{\equal{#1}{2}}{\hyperlink{#2}{\fun{#2}\hspace{0pt}}\index{#2@\fun{#2}}}{\ifthenelse{\equal{#1}{1}}{\hyperlink{#2}{\textbf{\texttt{{\color{black}#2}}}}}{\hyperlink{#2}{\textbf{\texttt{{\color{black}#2}}}}\index{#2@\fun{#2}}}}}

\newcommand{\builtL}[2][2]{\ifthenelse{\equal{#1}{2}}{\hyperlink{#2}{\built{#2}}\index{#2@\built{#2}}}{\ifthenelse{\equal{#1}{1}}{\hyperlink{#2}{\textbf{\texttt{{\color{black}#2}}}}}{\hyperlink{#2}{\textbf{\texttt{{\color{black}#2}}}\index{#2@\built{#2}}}}}}

\newcommand{\stateYM}[2][0]{\ifthenelse{\equal{#1}{0}}{\hyperlink{sYM:#2}{\fun{#2}}\index{$\a$a(sYM)#2@\fun{#2} (sYM state)}}{\hypertarget{sYM:#2}{\fun{#2}}\index{$\a$a(sYM)#2@\fun{#2} (sYM state)|textbf}}}

\newcommand{\stateGR}[2][0]{\ifthenelse{\equal{#1}{0}}{\hyperlink{sGR:#2}{\fun{#2}}\index{$\a$a(superGR)#2@\fun{#2} (sGR state)}}{\hypertarget{sGR:#2}{\fun{#2}}\index{$\a$a(superGR)#2@\fun{#2} (sGR state)|textbf}}}

%% file: tools_for_tree_amplitudes.bbl
\providecommand{\href}[2]{#2}\begingroup\raggedright\endgroup

%% file: tools_for_tree_amplitudes_penult.bbl
\begin{thebibliography}{100}

\bibitem{Parke:1986gb}
S.~J. Parke and T.~R. Taylor, ``{An Amplitude for $n$-Gluon Scattering},''
\href{http://dx.doi.org/10.1103/PhysRevLett.56.2459}{{\em Phys. Rev. Lett.}
  {\bf 56} (1986)  2459}.

\bibitem{Parke:1985pn}
S.~J. Parke and T.~R. Taylor, ``{Perturbative QCD Utilizing Extended
  Supersymmetry},'' \href{http://dx.doi.org/10.1016/0370-2693(85)91216-X}{{\em
  Phys. Lett. B} {\bf 157} (1985)  81}. [Erratum: Phys.Lett.B 174, 465 (1986)].

\bibitem{Parke:1985ax}
S.~J. Parke and T.~R. Taylor, ``{Gluonic Two Goes to Four},''
  \href{http://dx.doi.org/10.1016/0550-3213(86)90230-0}{{\em Nucl. Phys. B}
  {\bf 269} (1986)  410--420}.

\bibitem{BCF}
R.~Britto, F.~Cachazo, and B.~Feng, ``{New Recursion Relations for Tree
  Amplitudes of Gluons},''
  \href{http://dx.doi.org/10.1016/j.nuclphysb.2005.02.030}{{\em Nucl.Phys.}
  {\bf B715} (2005)  499--522},
\href{http://arxiv.org/abs/hep-th/0412308}{{ arXiv:hep-th/0412308 [hep-th]}}.

\bibitem{BCFW}
R.~Britto, F.~Cachazo, B.~Feng, and E.~Witten, ``{Direct Proof of Tree-Level
  Recursion Relation in Yang- Mills Theory},''
  \href{http://dx.doi.org/10.1103/PhysRevLett.94.181602}{{\em Phys. Rev. Lett.}
  {\bf 94} (2005)  181602},
\href{http://arxiv.org/abs/hep-th/0501052}{{ arXiv:hep-th/0501052}}.

\bibitem{Drummond:2006rz}
J.~M. Drummond, J.~Henn, V.~A. Smirnov, and E.~Sokatchev, ``{Magic identities
  for Conformal Four-Point Integrals},''
  \href{http://dx.doi.org/10.1088/1126-6708/2007/01/064}{{\em JHEP} {\bf 01}
  (2007)  064}, \href{http://arxiv.org/abs/hep-th/0607160}{{
  arXiv:hep-th/0607160}}.

\bibitem{Alday:2007hr}
L.~F. Alday and J.~M. Maldacena, ``{Gluon Scattering Amplitudes at Strong
  Coupling},'' \href{http://dx.doi.org/10.1088/1126-6708/2007/06/064}{{\em
  JHEP} {\bf 06} (2007)  064},
\href{http://arxiv.org/abs/0705.0303}{{ arXiv:0705.0303 [hep-th]}}.

\bibitem{Drummond:2008vq}
J.~Drummond, J.~Henn, G.~Korchemsky, and E.~Sokatchev, ``{Dual Superconformal
  Symmetry of Scattering Amplitudes in $\mathcal{N}\!=\!4$ super Yang-Mills
  Theory},'' \href{http://dx.doi.org/10.1016/j.nuclphysb.2009.11.022}{{\em
  Nucl. Phys.} {\bf B828} (2010)  317--374},
\href{http://arxiv.org/abs/0807.1095}{{ arXiv:0807.1095 [hep-th]}}.

\bibitem{Drummond:2009fd}
J.~M. Drummond, J.~M. Henn, and J.~Plefka, ``{Yangian Symmetry of Scattering
  Amplitudes in $\mathcal{N}\!=\!4$ Super Yang-Mills Theory},''
  \href{http://dx.doi.org/10.1088/1126-6708/2009/05/046}{{\em JHEP} {\bf 05}
  (2009)  046},
\href{http://arxiv.org/abs/0902.2987}{{ arXiv:0902.2987 [hep-th]}}.

\bibitem{ArkaniHamed:2009dn}
N.~Arkani-Hamed, F.~Cachazo, C.~Cheung, and J.~Kaplan, ``{A Duality For The
  $S$-Matrix},'' \href{http://dx.doi.org/10.1007/JHEP03(2010)020}{{\em JHEP}
  {\bf 1003} (2010)  020},
\href{http://arxiv.org/abs/0907.5418}{{ arXiv:0907.5418 [hep-th]}}.

\bibitem{Kaplan:2009mh}
J.~Kaplan, ``{Unraveling $\mathcal{L}_{n,k}$: Grassmannian Kinematics},''
  \href{http://dx.doi.org/10.1007/JHEP03(2010)025}{{\em JHEP} {\bf 1003} (2010)
   025},
\href{http://arxiv.org/abs/0912.0957}{{ arXiv:0912.0957 [hep-th]}}.

\bibitem{ArkaniHamed:2012nw}
N.~Arkani-Hamed, J.~L. Bourjaily, F.~Cachazo, A.~B. Goncharov, A.~Postnikov,
  and J.~Trnka, ``{Scattering Amplitudes and the Positive Grassmannian},''
\href{http://arxiv.org/abs/1212.5605}{{ arXiv:1212.5605 [hep-th]}}.

\bibitem{Bourjaily:2012gy}
J.~L. Bourjaily, ``{Positroids, Plabic Graphs, and Scattering Amplitudes in
  {\sc Mathematica}},''
\href{http://arxiv.org/abs/1212.6974}{{ arXiv:1212.6974 [hep-th]}}.

\bibitem{Arkani-Hamed:2014bca}
N.~Arkani-Hamed, J.~L. Bourjaily, F.~Cachazo, A.~Postnikov, and J.~Trnka,
  ``{On-Shell Structures of MHV Amplitudes Beyond the Planar Limit},''
  \href{http://dx.doi.org/10.1007/JHEP06(2015)179}{{\em JHEP} {\bf 06} (2015)
  179},
\href{http://arxiv.org/abs/1412.8475}{{ arXiv:1412.8475 [hep-th]}}.

\bibitem{ArkaniHamed:2010gg}
N.~Arkani-Hamed, J.~L. Bourjaily, F.~Cachazo, A.~Hodges, and J.~Trnka, ``{A
  {N}ote on {P}olytopes for {S}cattering {A}mplitudes},''
  \href{http://dx.doi.org/10.1007/JHEP04(2012)081}{{\em JHEP} {\bf 1204} (2012)
   081},
\href{http://arxiv.org/abs/1012.6030}{{ arXiv:1012.6030 [hep-th]}}.

\bibitem{Arkani-Hamed:2013jha}
N.~Arkani-Hamed and J.~Trnka, ``{The Amplituhedron},''
  \href{http://dx.doi.org/10.1007/JHEP10(2014)030}{{\em JHEP} {\bf 1410} (2014)
   30},
\href{http://arxiv.org/abs/1312.2007}{{ arXiv:1312.2007 [hep-th]}}.

\bibitem{Arkani-Hamed:2013kca}
N.~Arkani-Hamed and J.~Trnka, ``{Into the Amplituhedron},''
  \href{http://dx.doi.org/10.1007/JHEP12(2014)182}{{\em JHEP} {\bf 1412} (2014)
   182},
\href{http://arxiv.org/abs/1312.7878}{{ arXiv:1312.7878 [hep-th]}}.

\bibitem{Bourjaily:2018bbb}
J.~Bourjaily and H.~Thomas, ``{What is the Amplituhedron?},''
  \href{http://dx.doi.org/10.1090/noti1630}{{\em Not. Amer. Math. Soc.} {\bf
  65} (2018) no. 2, 167--169}.

\bibitem{ArkaniHamed:2010gh}
N.~Arkani-Hamed, J.~L. Bourjaily, F.~Cachazo, and J.~Trnka, ``{Local Integrals
  for Planar Scattering Amplitudes},''
  \href{http://dx.doi.org/10.1007/JHEP06(2012)125}{{\em JHEP} {\bf 1206} (2012)
   125},
\href{http://arxiv.org/abs/1012.6032}{{ arXiv:1012.6032 [hep-th]}}.

\bibitem{Bourjaily:2013mma}
J.~L. Bourjaily, S.~Caron-Huot, and J.~Trnka, ``{Dual-Conformal Regularization
  of Infrared Loop Divergences and the {\it Chiral} Box Expansion},''
  \href{http://dx.doi.org/10.1007/JHEP01(2015)001}{{\em JHEP} {\bf 1501} (2015)
   001},
\href{http://arxiv.org/abs/1303.4734}{{ arXiv:1303.4734 [hep-th]}}.

\bibitem{Bourjaily:2015jna}
J.~L. Bourjaily and J.~Trnka, ``{Local Integrand Representations of All
  Two-Loop Amplitudes in Planar SYM},''
  \href{http://dx.doi.org/10.1007/JHEP08(2015)119}{{\em JHEP} {\bf 08} (2015)
  119},
\href{http://arxiv.org/abs/1505.05886}{{ arXiv:1505.05886 [hep-th]}}.

\bibitem{Bourjaily:2023apy}
J.~L. Bourjaily and S.~Caron-Huot, ``{Loop Amplitude Integrands from Unitarity
  Cuts},'' \href{http://dx.doi.org/10.1103/PhysRevD.108.025008}{{\em Phys. Rev.
  D} {\bf 108} (2023) no. 2, 025008}, \href{http://arxiv.org/abs/2305.01673}{{
  arXiv:2305.01673 [hep-th]}}.

\bibitem{Bourjaily:2016mnp}
J.~L. Bourjaily, S.~Franco, D.~Galloni, and C.~Wen, ``{Stratifying On-Shell
  Cluster Varieties: the Geometry of Non-Planar On-Shell Diagrams},''
  \href{http://dx.doi.org/10.1007/JHEP10(2016)003}{{\em JHEP} {\bf 10} (2016)
  003},
\href{http://arxiv.org/abs/1607.01781}{{ arXiv:1607.01781 [hep-th]}}.

\bibitem{Bourjaily:2018omh}
J.~L. Bourjaily, E.~Herrmann, and J.~Trnka, ``{Amplitudes at Infinity},''
  \href{http://dx.doi.org/10.1103/PhysRevD.99.066006}{{\em Phys. Rev. D} {\bf
  99} (2019) no. 6, 066006}, \href{http://arxiv.org/abs/1812.11185}{{
  arXiv:1812.11185 [hep-th]}}.

\bibitem{Bourjaily:2019gqu}
J.~L. Bourjaily, E.~Herrmann, C.~Langer, A.~J. McLeod, and J.~Trnka,
  ``{All-Multiplicity Nonplanar Amplitude Integrands in Maximally
  Supersymmetric Yang-Mills Theory at Two Loops},''
  \href{http://dx.doi.org/10.1103/PhysRevLett.124.111603}{{\em Phys. Rev.
  Lett.} {\bf 124} (2020) no. 11, 111603},
  \href{http://arxiv.org/abs/1911.09106}{{ arXiv:1911.09106 [hep-th]}}.

\bibitem{Bourjaily:2019iqr}
J.~L. Bourjaily, E.~Herrmann, C.~Langer, A.~J. McLeod, and J.~Trnka,
  ``{Prescriptive Unitarity for Non-Planar Six-Particle Amplitudes at Two
  Loops},'' \href{http://dx.doi.org/10.1007/JHEP12(2019)073}{{\em JHEP} {\bf
  12} (2019)  073}, \href{http://arxiv.org/abs/1909.09131}{{ arXiv:1909.09131
  [hep-th]}}.

\bibitem{Bourjaily:2021hcp}
J.~L. Bourjaily, C.~Langer, and Y.~Zhang, ``{Illustrations of Integrand-Basis
  Building at Two Loops},''
  \href{http://dx.doi.org/10.1007/JHEP08(2022)176}{{\em JHEP} {\bf 08} (2022)
  176}, \href{http://arxiv.org/abs/2112.05157}{{ arXiv:2112.05157 [hep-th]}}.

\bibitem{Bourjaily:2021iyq}
J.~L. Bourjaily, C.~Langer, and Y.~Zhang, ``{All Two-Loop, Color-Dressed,
  Six-Point Amplitude Integrands in sYM},''
  \href{http://dx.doi.org/10.1103/PhysRevD.105.105015}{{\em Phys. Rev. D} {\bf
  105} (2022) no. 10, 105015}, \href{http://arxiv.org/abs/2112.06934}{{
  arXiv:2112.06934 [hep-th]}}.

\bibitem{Schwartz:2014sze}
M.~D. Schwartz, \href{http://dx.doi.org/doi.org/10.1017/9781139540940}{{\em
  {Quantum Field Theory and the Standard Model}}}.
\newblock Cambridge University Press, 3, 2014.

\bibitem{Zee:2010qce}
A.~Zee, {\em {Quantum Field Theory in a Nutshell: Second Edition}}.
\newblock Princeton University Press, 2, 2010.

\bibitem{Dixon:1996wi}
L.~J. Dixon, ``{Calculating Scattering Amplitudes Efficiently},''
\href{http://arxiv.org/abs/hep-ph/9601359}{{ arXiv:hep-ph/9601359}}.

\bibitem{Bedford:2005yy}
J.~Bedford, A.~Brandhuber, B.~J. Spence, and G.~Travaglini, ``{A Recursion
  Relation for Gravity Amplitudes},''
  \href{http://dx.doi.org/10.1016/j.nuclphysb.2005.016}{{\em Nucl. Phys. B}
  {\bf 721} (2005)  98--110}, \href{http://arxiv.org/abs/hep-th/0502146}{{
  arXiv:hep-th/0502146}}.

\bibitem{Alday:2008zza}
L.~F. Alday and J.~Maldacena, ``{Lectures on Scattering Amplitudes via
  AdS/CFT},'' \href{http://dx.doi.org/10.1063/1.2972015}{{\em AIP Conf. Proc.}
  {\bf 1031} (2008) no. 1, 43--60}.

\bibitem{Dixon:2013uaa}
L.~J. Dixon, \href{http://dx.doi.org/10.5170/CERN-2014-008.31}{``{A Brief
  Introduction to Modern Amplitude Methods},''} in {\em {Proceedings, 2012
  European School of High-Energy Physics (ESHEP 2012): La Pommeraye, Anjou,
  France, June 06-19, 2012}}, pp.~31--67.
\newblock 2014.
\newblock \href{http://arxiv.org/abs/1310.5353}{{ arXiv:1310.5353 [hep-ph]}}.
\newblock
\url{http://inspirehep.net/record/1261436/files/arXiv:1310.5353.pdf}.
\newblock

\bibitem{Cheung:2017pzi}
C.~Cheung, {\em {TASI Lectures on Scattering Amplitudes}},
  \href{http://dx.doi.org/10.1142/9789813233348_0008}{pp.~571--623}.
\newblock 2018.
\newblock \href{http://arxiv.org/abs/1708.03872}{{ arXiv:1708.03872 [hep-ph]}}.

\bibitem{HennPlefka}
J.~M. Henn and J.~C. Plefka,
  \href{http://dx.doi.org/https://doi.org/10.1007/978-3-642-54022-6}{{\em
  {Scattering Amplitudes in Gauge Theories}}}.
\newblock Lecture Notes in Physics. Springer Berlin, Heidelberg, 2014.

\bibitem{Strominger:2017zoo}
A.~Strominger, {\em {Lectures on the Infrared Structure of Gravity and Gauge
  Theory}}.
\newblock Princeton University Press, 3, 2017.
\newblock \href{http://arxiv.org/abs/1703.05448}{{ arXiv:1703.05448 [hep-th]}}.

\bibitem{Travaglini:2022uwo}
G.~Travaglini {\em et al.}, ``{The SAGEX Review on Scattering Amplitudes},''
  \href{http://dx.doi.org/10.1088/1751-8121/ac8380}{{\em J. Phys. A} {\bf 55}
  (2022) no. 44, 443001}, \href{http://arxiv.org/abs/2203.13011}{{
  arXiv:2203.13011 [hep-th]}}.

\bibitem{Bern:2022wqg}
Z.~Bern, J.~J. Carrasco, M.~Chiodaroli, H.~Johansson, and R.~Roiban, ``{The
  SAGEX Review on Scattering Amplitudes Chapter 2: An Invitation to
  Color-Kinematics Duality and the Double Copy},''
  \href{http://dx.doi.org/10.1088/1751-8121/ac93cf}{{\em J. Phys. A} {\bf 55}
  (2022) no. 44, 443003}, \href{http://arxiv.org/abs/2203.13013}{{
  arXiv:2203.13013 [hep-th]}}.

\bibitem{Badger:2023eqz}
S.~Badger, J.~Henn, J.~Plefka, and S.~Zoia, ``{Scattering Amplitudes in Quantum
  Field Theory},'' \href{http://arxiv.org/abs/2306.05976}{{ arXiv:2306.05976
  [hep-th]}}.

\bibitem{Elvang:2013cua}
H.~Elvang and Y.-t. Huang,
  \href{http://dx.doi.org/https://doi.org/10.1017/CBO9781107706620}{{\em
  {Scattering Amplitudes in Gauge Theory and Gravity}}}.
\newblock Cambridge University Press, 8, 2015.
\newblock \href{http://arxiv.org/abs/1308.1697}{{ arXiv:1308.1697 [hep-th]}}.

\bibitem{Carrasco:2021otn}
J.~J.~M. Carrasco, A.~Edison, and H.~Johansson, ``{Maximal Super-Yang-Mills at
  Six Loops via Novel Integrand Bootstrap},''
  \href{http://arxiv.org/abs/2112.05178}{{ arXiv:2112.05178 [hep-th]}}.

\bibitem{Bourjaily:2016evz}
J.~L. Bourjaily, P.~Heslop, and V.-V. Tran, ``{Amplitudes and Correlators to
  Ten Loops Using Simple, Graphical Bootstraps},''
  \href{http://dx.doi.org/10.1007/JHEP11(2016)125}{{\em JHEP} {\bf 11} (2016)
  125},
\href{http://arxiv.org/abs/1609.00007}{{ arXiv:1609.00007 [hep-th]}}.

\bibitem{Maitre:2007jq}
D.~Maitre and P.~Mastrolia, ``{S@M, a \textsc{Mathematica} Implementation of
  the Spinor-Helicity Formalism},''
  \href{http://dx.doi.org/10.1016/j.cpc.2008.05.002}{{\em Comput. Phys.
  Commun.} {\bf 179} (2008)  501--574}, \href{http://arxiv.org/abs/0710.5559}{{
  arXiv:0710.5559 [hep-ph]}}.

\bibitem{Dixon:2010ik}
L.~J. Dixon, J.~M. Henn, J.~Plefka, and T.~Schuster, ``{All Tree-Level
  Amplitudes in Massless QCD},''
  \href{http://dx.doi.org/10.1007/JHEP01(2011)035}{{\em JHEP} {\bf 1101} (2011)
   035},
\href{http://arxiv.org/abs/1010.3991}{{ arXiv:1010.3991 [hep-ph]}}.

\bibitem{Bourjaily:2010wh}
J.~L. Bourjaily, ``{Efficient Tree-Amplitudes in $\mathcal{N}\!=\!4$: Automatic
  BCFW Recursion in {\sc Mathematica}},''
\href{http://arxiv.org/abs/1011.2447}{{ arXiv:1011.2447 [hep-ph]}}.

\bibitem{Bern:2013pya}
Z.~Bern, L.~J. Dixon, F.~Febres~Cordero, S.~H\"oche, H.~Ita, D.~A. Kosower,
  D.~Ma\^\i{}tre, and K.~J. Ozeren, ``{The BlackHat Library for One-Loop
  Amplitudes},'' \href{http://dx.doi.org/10.1088/1742-6596/523/1/012051}{{\em
  J. Phys. Conf. Ser.} {\bf 523} (2014)  012051},
  \href{http://arxiv.org/abs/1310.2808}{{ arXiv:1310.2808 [hep-ph]}}.

\bibitem{Cullen:2013cka}
G.~Cullen {\em et al.}, ``{GoSam Applications for Automated NLO
  Calculations},'' \href{http://dx.doi.org/10.1088/1742-6596/523/1/012056}{{\em
  J. Phys. Conf. Ser.} {\bf 523} (2014)  012056},
  \href{http://arxiv.org/abs/1309.3741}{{ arXiv:1309.3741 [hep-ph]}}.

\bibitem{Cullen:2013mza}
G.~Cullen {\em et al.}, ``{GoSam @ LHC: Algorithms and Applications to Higgs
  Production},'' \href{http://dx.doi.org/10.22323/1.197.0029}{{\em PoS} {\bf
  RADCOR2013} (2013)  029}, \href{http://arxiv.org/abs/1312.1761}{{
  arXiv:1312.1761 [hep-ph]}}.

\bibitem{vanDeurzen:2014uaa}
H.~van Deurzen {\em et al.}, ``{Automated One-Loop Calculations with GoSam
  2.0},'' \href{http://dx.doi.org/10.22323/1.211.0021}{{\em PoS} {\bf LL2014}
  (2014)  021}, \href{http://arxiv.org/abs/1407.0922}{{ arXiv:1407.0922
  [hep-ph]}}.

\bibitem{Kuczmarski:2014ara}
J.~Kuczmarski, ``{SpinorsExtras - \textsc{Mathematica} Implementation of
  Massive Spinor-Helicity Formalism},''
\newblock 6, 2014.
\newblock \href{http://arxiv.org/abs/1406.5612}{{ arXiv:1406.5612 [hep-ph]}}.

\bibitem{Panzer:2014caa}
E.~Panzer, ``{Algorithms for the Symbolic Integration of Hyperlogarithms with
  Applications to Feynman Integrals},''
  \href{http://dx.doi.org/10.1016/j.cpc.2014.10.019}{{\em Comput. Phys.
  Commun.} {\bf 188} (2015)  148--166}, \href{http://arxiv.org/abs/1403.3385}{{
  arXiv:1403.3385 [hep-th]}}.
{\tt HyperInt} is obtainable at
  \href{https://bitbucket.org/PanzerErik/hyperint/wiki/Home}{this URL}.

\bibitem{AccettulliHuber:2023ldr}
M.~Accettulli~Huber, ``{SpinorHelicity4D: a \textsc{Mathematica} Toolbox for
  the Four-Dimensional Spinor-Helicity Formalism},''
  \href{http://arxiv.org/abs/2304.01589}{{ arXiv:2304.01589 [hep-th]}}.

\bibitem{Mertig:1990an}
R.~Mertig, M.~Bohm, and A.~Denner, ``{FEYN CALC: Computer Algebraic Calculation
  of Feynman Amplitudes},''
  \href{http://dx.doi.org/10.1016/0010-4655(91)90130-D}{{\em Comput. Phys.
  Commun.} {\bf 64} (1991)  345--359}.

\bibitem{Shtabovenko:2020gxv}
V.~Shtabovenko, R.~Mertig, and F.~Orellana, ``{FeynCalc 9.3: New features and
  improvements},'' \href{http://dx.doi.org/10.1016/j.cpc.2020.107478}{{\em
  Comput. Phys. Commun.} {\bf 256} (2020)  107478},
  \href{http://arxiv.org/abs/2001.04407}{{ arXiv:2001.04407 [hep-ph]}}.

\bibitem{Gleisberg:2003xi}
T.~Gleisberg, S.~Hoeche, F.~Krauss, A.~Schalicke, S.~Schumann, and J.-C.
  Winter, ``{SHERPA 1. $\alpha$: A Proof of Concept Version},''
  \href{http://dx.doi.org/10.1088/1126-6708/2004/02/056}{{\em JHEP} {\bf 02}
  (2004)  056}, \href{http://arxiv.org/abs/hep-ph/0311263}{{
  arXiv:hep-ph/0311263}}.

\bibitem{Alwall:2014hca}
J.~Alwall, R.~Frederix, S.~Frixione, V.~Hirschi, F.~Maltoni, O.~Mattelaer,
  H.~S. Shao, T.~Stelzer, P.~Torrielli, and M.~Zaro, ``{The Automated
  Computation of Tree-Level and Next-to-Leading Order Differential Cross
  Sections, and their Matching to Parton Shower Simulations},''
  \href{http://dx.doi.org/10.1007/JHEP07(2014)079}{{\em JHEP} {\bf 07} (2014)
  079}, \href{http://arxiv.org/abs/1405.0301}{{ arXiv:1405.0301 [hep-ph]}}.

\bibitem{Bogner:2018bvz}
C.~Bogner, S.~Borowka, T.~Hahn, G.~Heinrich, S.~P. Jones, M.~Kerner, A.~von
  Manteuffel, M.~Michel, E.~Panzer, and V.~Papara, ``{Loopedia, a Database for
  Loop Integrals},''
  \href{http://dx.doi.org/10.1088/1742-6596/1085/5/052003}{{\em J. Phys. Conf.
  Ser.} {\bf 1085} (2018) no. 5, 052003}.

\bibitem{Cachazo:2013gna}
F.~Cachazo, S.~He, and E.~Y. Yuan, ``{Scattering Equations and
  Kawai-Lewellen-Tye Orthogonality},''
  \href{http://dx.doi.org/10.1103/PhysRevD.90.065001}{{\em Phys. Rev.} {\bf
  D90} (2014) no. 6, 065001},
\href{http://arxiv.org/abs/1306.6575}{{ arXiv:1306.6575 [hep-th]}}.

\bibitem{Cachazo:2013hca}
F.~Cachazo, S.~He, and E.~Y. Yuan, ``{Scattering of Massless Particles in
  Arbitrary Dimensions},''
  \href{http://dx.doi.org/10.1103/PhysRevLett.113.171601}{{\em Phys. Rev.
  Lett.} {\bf 113} (2014) no. 17, 171601},
\href{http://arxiv.org/abs/1307.2199}{{ arXiv:1307.2199 [hep-th]}}.

\bibitem{Cachazo:2013iaa}
F.~Cachazo, S.~He, and E.~Y. Yuan, ``{Scattering in Three Dimensions from
  Rational Maps},'' \href{http://dx.doi.org/10.1007/JHEP10(2013)141}{{\em JHEP}
  {\bf 1310} (2013)  141},
\href{http://arxiv.org/abs/1306.2962}{{ arXiv:1306.2962 [hep-th]}}.

\bibitem{Cachazo:2013iea}
F.~Cachazo, S.~He, and E.~Y. Yuan, ``{Scattering of Massless Particles:
  Scalars, Gluons and Gravitons},''
  \href{http://dx.doi.org/10.1007/JHEP07(2014)033}{{\em JHEP} {\bf 1407} (2014)
   033},
\href{http://arxiv.org/abs/1309.0885}{{ arXiv:1309.0885 [hep-th]}}.

\bibitem{Dolan:2013isa}
L.~Dolan and P.~Goddard, ``{Proof of the Formula of Cachazo, He and Yuan for
  Yang-Mills Tree Amplitudes in Arbitrary Dimension},''
  \href{http://dx.doi.org/10.1007/JHEP05(2014)010}{{\em JHEP} {\bf 1405} (2014)
   010},
\href{http://arxiv.org/abs/1311.5200}{{ arXiv:1311.5200 [hep-th]}}.

\bibitem{Dolan:2014ega}
L.~Dolan and P.~Goddard, ``{The Polynomial Form of the Scattering Equations},''
  \href{http://dx.doi.org/10.1007/JHEP07(2014)029}{{\em JHEP} {\bf 1407} (2014)
   029},
\href{http://arxiv.org/abs/1402.7374}{{ arXiv:1402.7374 [hep-th]}}.

\bibitem{Cachazo:2014xea}
F.~Cachazo, S.~He, and E.~Y. Yuan, ``{Scattering Equations and Matrices: From
  Einstein To Yang-Mills, DBI and NLSM},''
  \href{http://dx.doi.org/10.1007/JHEP07(2015)149}{{\em JHEP} {\bf 07} (2015)
  149},
\href{http://arxiv.org/abs/1412.3479}{{ arXiv:1412.3479 [hep-th]}}.

\bibitem{Bjerrum-Bohr:2016axv}
N.~E.~J. Bjerrum-Bohr, J.~L. Bourjaily, P.~H. Damgaard, and B.~Feng,
  ``{Manifesting Color-Kinematics Duality in the Scattering Equation
  Formalism},'' \href{http://dx.doi.org/10.1007/JHEP09(2016)094}{{\em JHEP}
  {\bf 09} (2016)  094},
\href{http://arxiv.org/abs/1608.00006}{{ arXiv:1608.00006 [hep-th]}}.

\bibitem{Baadsgaard:2015hia}
C.~Baadsgaard, N.~E.~J. Bjerrum-Bohr, J.~L. Bourjaily, P.~H. Damgaard, and
  B.~Feng, ``{Integration Rules for Loop Scattering Equations},''
  \href{http://dx.doi.org/10.1007/JHEP11(2015)080}{{\em JHEP} {\bf 11} (2015)
  080},
\href{http://arxiv.org/abs/1508.03627}{{ arXiv:1508.03627 [hep-th]}}.

\bibitem{Baadsgaard:2015ifa}
C.~Baadsgaard, N.~E.~J. Bjerrum-Bohr, J.~L. Bourjaily, and P.~H. Damgaard,
  ``{Scattering Equations and Feynman Diagrams},''
  \href{http://dx.doi.org/10.1007/JHEP09(2015)136}{{\em JHEP} {\bf 09} (2015)
  136},
\href{http://arxiv.org/abs/1507.00997}{{ arXiv:1507.00997 [hep-th]}}.

\bibitem{Baadsgaard:2015voa}
C.~Baadsgaard, N.~E.~J. Bjerrum-Bohr, J.~L. Bourjaily, and P.~H. Damgaard,
  ``{Integration Rules for Scattering Equations},''
  \href{http://dx.doi.org/10.1007/JHEP09(2015)129}{{\em JHEP} {\bf 09} (2015)
  129},
\href{http://arxiv.org/abs/1506.06137}{{ arXiv:1506.06137 [hep-th]}}.

\bibitem{Bjerrum-Bohr:2016juj}
N.~E.~J. Bjerrum-Bohr, J.~L. Bourjaily, P.~H. Damgaard, and B.~Feng,
  ``{Analytic Representations of Yang-Mills Amplitudes},''
  \href{http://dx.doi.org/10.1016/j.nuclphysb.2016.10.012}{{\em Nucl. Phys.}
  {\bf B913} (2016)  964--986},
\href{http://arxiv.org/abs/1605.06501}{{ arXiv:1605.06501 [hep-th]}}.

\bibitem{Hodges:2011wm}
A.~Hodges, ``{New Expressions for Gravitational Scattering Amplitudes},''
  \href{http://dx.doi.org/10.1007/JHEP07(2013)075}{{\em JHEP} {\bf 07} (2013)
  075},
\href{http://arxiv.org/abs/1108.2227}{{ arXiv:1108.2227 [hep-th]}}.

\bibitem{Berends:1988zp}
F.~A. Berends, W.~T. Giele, and H.~Kuijf, ``{On Relations Between multi-Gluon
  and multi-Graviton Scattering},''
  \href{http://dx.doi.org/10.1016/0370-2693(88)90813-1}{{\em Phys. Lett. B}
  {\bf 211} (1988)  91--94}.

\bibitem{DelDuca:1999rs}
V.~Del~Duca, L.~J. Dixon, and F.~Maltoni, ``{New Color Decompositions for Gauge
  Amplitudes at Tree and Loop Level},''
  \href{http://dx.doi.org/10.1016/S0550-3213(99)00809-3}{{\em Nucl. Phys.} {\bf
  B571} (2000)  51--70},
\href{http://arxiv.org/abs/hep-ph/9910563}{{ arXiv:hep-ph/9910563 [hep-ph]}}.

\bibitem{Cvitanovic:1980bu}
P.~Cvitanovic, P.~G. Lauwers, and P.~N. Scharbach, ``{Gauge Invariance
  Structure of Quantum Chromodynamics},''
  \href{http://dx.doi.org/10.1016/0550-3213(81)90098-5}{{\em Nucl. Phys. B}
  {\bf 186} (1981)  165--186}.

\bibitem{Berends:1987cv}
F.~A. Berends and W.~Giele, ``{The Six Gluon Process as an Example of Weyl-Van
  Der Waerden Spinor Calculus},''
  \href{http://dx.doi.org/10.1016/0550-3213(87)90604-3}{{\em Nucl. Phys. B}
  {\bf 294} (1987)  700--732}.

\bibitem{Mangano:1987xk}
M.~L. Mangano, S.~J. Parke, and Z.~Xu, ``{Duality and Multi-Gluon
  Scattering},''
\href{http://dx.doi.org/10.1016/0550-3213(88)90001-6}{{\em Nucl. Phys.} {\bf
  B298} (1988)  653--672}.

\bibitem{Kosower:1987ic}
D.~Kosower, B.-H. Lee, and V.~P. Nair, ``{Multigluon Scattering: a String-Based
  Calculation},'' \href{http://dx.doi.org/10.1016/0370-2693(88)90085-8}{{\em
  Phys. Lett. B} {\bf 201} (1988)  85--89}.

\bibitem{Zeppenfeld:1988bz}
D.~Zeppenfeld, ``{Diagonalization of Color Factors},''
  \href{http://dx.doi.org/10.1142/S0217751X88000916}{{\em Int. J. Mod. Phys. A}
  {\bf 3} (1988)  2175--2179}.

\bibitem{Mangano:1988kk}
M.~L. Mangano, ``{The Color Structure of Gluon Emission},''
  \href{http://dx.doi.org/10.1016/0550-3213(88)90453-1}{{\em Nucl. Phys. B}
  {\bf 309} (1988)  461--475}.

\bibitem{Mangano:1990by}
M.~L. Mangano and S.~J. Parke, ``{Multiparton Amplitudes in Gauge Theories},''
  \href{http://dx.doi.org/10.1016/0370-1573(91)90091-Y}{{\em Phys. Rept.} {\bf
  200} (1991)  301--367},
\href{http://arxiv.org/abs/hep-th/0509223}{{ arXiv:hep-th/0509223}}.

\bibitem{KK}
R.~Kleiss and H.~Kuijf, ``{Multi-Gluon Cross-Sections and Five Jet Production
  at Hadron Colliders},''
\href{http://dx.doi.org/10.1016/0550-3213(89)90574-9}{{\em Nucl. Phys.} {\bf
  B312} (1989)  616}.

\bibitem{Bern:2008qj}
Z.~Bern, J.~Carrasco, and H.~Johansson, ``{New Relations for Gauge-Theory
  Amplitudes},'' \href{http://dx.doi.org/10.1103/PhysRevD.78.085011}{{\em Phys.
  Rev.} {\bf D78} (2008)  085011},
\href{http://arxiv.org/abs/0805.3993}{{ arXiv:0805.3993 [hep-ph]}}.

\bibitem{Cachazo:2005ca}
F.~Cachazo and P.~Svrcek, ``{Tree Level Recursion Relations in General
  Relativity},'' \href{http://arxiv.org/abs/hep-th/0502160}{{
  arXiv:hep-th/0502160}}.

\bibitem{Benincasa:2007qj}
P.~Benincasa, C.~Boucher-Veronneau, and F.~Cachazo, ``{Taming Tree Amplitudes
  In General Relativity},''
  \href{http://dx.doi.org/10.1088/1126-6708/2007/11/057}{{\em JHEP} {\bf 11}
  (2007)  057}, \href{http://arxiv.org/abs/hep-th/0702032}{{
  arXiv:hep-th/0702032}}.

\bibitem{Bjerrum-Bohr:2005xoa}
N.~E.~J. Bjerrum-Bohr, D.~C. Dunbar, H.~Ita, W.~B. Perkins, and K.~Risager,
  ``{MHV-Vertices for Gravity Amplitudes},''
  \href{http://dx.doi.org/10.1088/1126-6708/2006/01/009}{{\em JHEP} {\bf 01}
  (2006)  009}, \href{http://arxiv.org/abs/hep-th/0509016}{{
  arXiv:hep-th/0509016}}.

\bibitem{Spradlin:2008bu}
M.~Spradlin, A.~Volovich, and C.~Wen, ``{Three Applications of a Bonus Relation
  for Gravity Amplitudes},''
  \href{http://dx.doi.org/10.1016/j.physletb.2009.02.059}{{\em Phys. Lett. B}
  {\bf 674} (2009)  69--72}, \href{http://arxiv.org/abs/0812.4767}{{
  arXiv:0812.4767 [hep-th]}}.

\bibitem{Bianchi:2008pu}
M.~Bianchi, H.~Elvang, and D.~Z. Freedman, ``{Generating Tree Amplitudes in
  $\mathcal{N}\!=\!4$ SYM and $\mathcal{N}\!=\!8$ SG},''
  \href{http://dx.doi.org/10.1088/1126-6708/2008/09/063}{{\em JHEP} {\bf 0809}
  (2008)  063},
\href{http://arxiv.org/abs/0805.0757}{{ arXiv:0805.0757 [hep-th]}}.

\bibitem{Elvang:2007sg}
H.~Elvang and D.~Z. Freedman, ``{Note on Graviton MHV amplitudes},''
  \href{http://dx.doi.org/10.1088/1126-6708/2008/05/096}{{\em JHEP} {\bf 05}
  (2008)  096}, \href{http://arxiv.org/abs/0710.1270}{{ arXiv:0710.1270
  [hep-th]}}.

\bibitem{Drummond:2009ge}
J.~M. Drummond, M.~Spradlin, A.~Volovich, and C.~Wen, ``{Tree-Level Amplitudes
  in $\mathcal{N}\!=\!8$ Supergravity},''
  \href{http://dx.doi.org/10.1103/PhysRevD.79.105018}{{\em Phys. Rev.} {\bf
  D79} (2009)  105018},
\href{http://arxiv.org/abs/0901.2363}{{ arXiv:0901.2363 [hep-th]}}.

\bibitem{Mason:2009afn}
L.~J. Mason and D.~Skinner, ``{Gravity, Twistors and the MHV Formalism},''
  \href{http://dx.doi.org/10.1007/s00220-009-0972-4}{{\em Commun. Math. Phys.}
  {\bf 294} (2010)  827--862}, \href{http://arxiv.org/abs/0808.3907}{{
  arXiv:0808.3907 [hep-th]}}.

\bibitem{Drummond:2008cr}
J.~M. Drummond and J.~M. Henn, ``{All Tree-Level Amplitudes in
  $\mathcal{N}\!=\!4$ SYM},''
  \href{http://dx.doi.org/10.1088/1126-6708/2009/04/018}{{\em JHEP} {\bf 04}
  (2009)  018},
\href{http://arxiv.org/abs/0808.2475}{{ arXiv:0808.2475 [hep-th]}}.

\bibitem{Hodges:2009hk}
A.~Hodges, ``{Eliminating Spurious Poles from Gauge-Theoretic Amplitudes},''
  \href{http://dx.doi.org/10.1007/JHEP05(2013)135}{{\em JHEP} {\bf 1305} (2013)
   135},
\href{http://arxiv.org/abs/0905.1473}{{ arXiv:0905.1473 [hep-th]}}.

\bibitem{ArkaniHamed:2008yf}
N.~Arkani-Hamed and J.~Kaplan, ``{On Tree Amplitudes in Gauge Theory and
  Gravity},'' \href{http://dx.doi.org/10.1088/1126-6708/2008/04/076}{{\em JHEP}
  {\bf 0804} (2008)  076},
\href{http://arxiv.org/abs/0801.2385}{{ arXiv:0801.2385 [hep-th]}}.

\bibitem{Cohen:2010mi}
T.~Cohen, H.~Elvang, and M.~Kiermaier, ``{On-Shell Constructibility of Tree
  Amplitudes in General Field Theories},''
  \href{http://dx.doi.org/10.1007/JHEP04(2011)053}{{\em JHEP} {\bf 1104} (2011)
   053},
\href{http://arxiv.org/abs/1010.0257}{{ arXiv:1010.0257 [hep-th]}}.

\bibitem{Heslop:2016plj}
P.~Heslop and A.~E. Lipstein, ``{On-Shell Diagrams for $\mathcal{N}\!=\!8$
  superGravity Amplitudes},''
  \href{http://dx.doi.org/10.1007/JHEP06(2016)069}{{\em JHEP} {\bf 06} (2016)
  069}, \href{http://arxiv.org/abs/1604.03046}{{ arXiv:1604.03046 [hep-th]}}.

\bibitem{Herrmann:2016qea}
E.~Herrmann and J.~Trnka, ``{Gravity On-Shell Diagrams},''
  \href{http://dx.doi.org/10.1007/JHEP11(2016)136}{{\em JHEP} {\bf 11} (2016)
  136}, \href{http://arxiv.org/abs/1604.03479}{{ arXiv:1604.03479 [hep-th]}}.

\bibitem{Paranjape:2022ymg}
S.~Paranjape, J.~Trnka, and M.~Zheng, ``{Non-Planar BCFW Grassmannian
  Geometries},'' \href{http://dx.doi.org/10.1007/JHEP12(2022)084}{{\em JHEP}
  {\bf 12} (2022)  084}, \href{http://arxiv.org/abs/2208.02262}{{
  arXiv:2208.02262 [hep-th]}}.

\bibitem{Trnka:2020dxl}
J.~Trnka, ``{Towards the Gravituhedron: New Expressions for NMHV Gravity
  Amplitudes},'' \href{http://dx.doi.org/10.1007/JHEP04(2021)253}{{\em JHEP}
  {\bf 04} (2021)  253}, \href{http://arxiv.org/abs/2012.15780}{{
  arXiv:2012.15780 [hep-th]}}.

\bibitem{Brown:2022wqr}
T.~V. Brown, U.~Oktem, and J.~Trnka, ``{Poles at Infinity in On-Shell
  Diagrams},'' \href{http://dx.doi.org/10.1007/JHEP02(2023)003}{{\em JHEP} {\bf
  02} (2023)  003}, \href{http://arxiv.org/abs/2212.06840}{{ arXiv:2212.06840
  [hep-th]}}.

\bibitem{Paranjape:2023qsq}
S.~Paranjape and J.~Trnka, ``{Gravity Amplitudes from Double Bonus
  Relations},'' \href{http://dx.doi.org/10.1103/PhysRevLett.131.251601}{{\em
  Phys. Rev. Lett.} {\bf 131} (2023) no. 25, 251601},
  \href{http://arxiv.org/abs/2309.05710}{{ arXiv:2309.05710 [hep-th]}}.

\bibitem{Bourjaily:2023ycy}
J.~L. Bourjaily, N.~Kalyanapuram, K.~Patatoukos, M.~Plesser, and Y.~Zhang,
  ``{Gauge-Invariant Double Copies via Recursion},''
  \href{http://dx.doi.org/10.1103/PhysRevLett.131.191601}{{\em Phys. Rev.
  Lett.} {\bf 131} (2023) no. 19, 191601},
  \href{http://arxiv.org/abs/2307.02542}{{ arXiv:2307.02542 [hep-th]}}.

\bibitem{structureOfOnShellGR}
J.~L. Bourjaily, N.~Kalyanapuram, K.~Patatoukos, M.~Plesser, and Y.~Zhang,
  ``{On-Shell Structure of Gravitational Tree Amplitudes}.'' In preparation.

\bibitem{Bern:2010ue}
Z.~Bern, J.~J.~M. Carrasco, and H.~Johansson, ``{Perturbative Quantum Gravity
  as a Double Copy of Gauge Theory},''
  \href{http://dx.doi.org/10.1103/PhysRevLett.105.061602}{{\em Phys. Rev.
  Lett.} {\bf 105} (2010)  061602},
\href{http://arxiv.org/abs/1004.0476}{{ arXiv:1004.0476 [hep-th]}}.

\bibitem{Bern:2019prr}
Z.~Bern, J.~J.~M. Carrasco, M.~Chiodaroli, H.~Johansson, and R.~Roiban, ``{The
  Duality Between Color and Kinematics and its Applications},''
  \href{http://arxiv.org/abs/1909.01358}{{ arXiv:1909.01358 [hep-th]}}.

\bibitem{Adamo:2022dcm}
T.~Adamo, J.~J.~M. Carrasco, M.~Carrillo-Gonz\'alez, M.~Chiodaroli, H.~Elvang,
  H.~Johansson, D.~O'Connell, R.~Roiban, and O.~Schlotterer, ``{Snowmass White
  Paper: the Double Copy and its Applications},'' in {\em {Snowmass 2021}}.
\newblock 4, 2022.
\newblock \href{http://arxiv.org/abs/2204.06547}{{ arXiv:2204.06547 [hep-th]}}.

\bibitem{Bern:2023zkg}
Z.~Bern, J.~J.~M. Carrasco, M.~Chiodaroli, H.~Johansson, and R.~Roiban,
  ``{Supergravity Amplitudes, the Double Copy and Ultraviolet Behavior},''
  \href{http://arxiv.org/abs/2304.07392}{{ arXiv:2304.07392 [hep-th]}}.

\bibitem{KLT}
H.~Kawai, D.~Lewellen, and S.~Tye, ``{A Relation Between Tree Amplitudes of
  Closed and Open Strings},''
\href{http://dx.doi.org/10.1016/0550-3213(86)90362-7}{{\em Nucl. Phys.} {\bf
  B269} (1986)  1}.

\bibitem{Mizera:2016jhj}
S.~Mizera, ``{Inverse of the String Theory KLT Kernel},''
  \href{http://dx.doi.org/10.1007/JHEP06(2017)084}{{\em JHEP} {\bf 06} (2017)
  084}, \href{http://arxiv.org/abs/1610.04230}{{ arXiv:1610.04230 [hep-th]}}.

\bibitem{Mizera:2019blq}
S.~Mizera, ``{Kinematic Jacobi Identity is a Residue Theorem: Geometry of
  Color-Kinematics Duality for Gauge and Gravity Amplitudes},''
  \href{http://dx.doi.org/10.1103/PhysRevLett.124.141601}{{\em Phys. Rev.
  Lett.} {\bf 124} (2020) no. 14, 141601},
  \href{http://arxiv.org/abs/1912.03397}{{ arXiv:1912.03397 [hep-th]}}.

\bibitem{Chi:2021mio}
H.-H. Chi, H.~Elvang, A.~Herderschee, C.~R.~T. Jones, and S.~Paranjape,
  ``{Generalizations of the Double-Copy: the KLT Bootstrap},''
  \href{http://dx.doi.org/10.1007/JHEP03(2022)077}{{\em JHEP} {\bf 03} (2022)
  077}, \href{http://arxiv.org/abs/2106.12600}{{ arXiv:2106.12600 [hep-th]}}.

\bibitem{hiddenZeroes}
N.~Arkani-Hamed, Q.~Cao, J.~Dong, C.~Figueiredo, and S.~He, ``{Hidden Zeros of
  Particle/String Amplitudes and the Unity of Colored Scalar, Pions and
  Gluons},'' \href{http://arxiv.org/abs/2312.16282}{{ arXiv:2312.16282
  [hep-th]}}.

\bibitem{hiddenZeroes2}
N.~Arkani-Hamed, Q.~Cao, J.~Don, C.~Figueiredo, and S.~He, ``{Scalar-Scaffolded
  Gluons and the Combinatorial Origins of Yang-Mills Theory},''. To appear.

\bibitem{hiddenZeroes3}
N.~Arkani-Hamed, Q.~Cao, J.~Don, C.~Figueiredo, and S.~He,
  ``{NLSM$\subset\mathrm{Tr}(\varphi^3)$},''. To appear.

\bibitem{Witten:2003nn}
E.~Witten, ``{Perturbative Gauge Theory as a String Theory in Twistor Space},''
  \href{http://dx.doi.org/10.1007/s00220-004-1187-3}{{\em Commun. Math. Phys.}
  {\bf 252} (2004)  189--258},
\href{http://arxiv.org/abs/hep-th/0312171}{{ arXiv:hep-th/0312171}}.

\bibitem{RSV}
R.~Roiban, M.~Spradlin, and A.~Volovich, ``{A Googly Amplitude from the B-Model
  in Twistor Space},'' {\em JHEP} {\bf 04} (2004)  012,
\href{http://arxiv.org/abs/hep-th/0402016}{{ arXiv:hep-th/0402016}}.

\bibitem{Roiban:2004yf}
R.~Roiban, M.~Spradlin, and A.~Volovich, ``{On the Tree-Level $S$-Matrix of
  Yang-Mills Theory},''
  \href{http://dx.doi.org/10.1103/PhysRevD.70.026009}{{\em Phys. Rev.} {\bf
  D70} (2004)  026009},
\href{http://arxiv.org/abs/hep-th/0403190}{{ arXiv:hep-th/0403190}}.

\bibitem{Dixon:2005cf}
L.~J. Dixon, ``{Twistor String Theory and QCD},'' {\em PoS} {\bf HEP2005}
  (2006)  405,
\href{http://arxiv.org/abs/hep-ph/0512111}{{ arXiv:hep-ph/0512111}}.

\bibitem{Dolan:2007vv}
L.~Dolan and P.~Goddard, ``{Tree and Loop Amplitudes in Open Twistor String
  Theory},'' {\em JHEP} {\bf 06} (2007)  005,
\href{http://arxiv.org/abs/hep-th/0703054}{{ arXiv:hep-th/0703054}}.

\bibitem{Dolan:2009wf}
L.~Dolan and P.~Goddard, ``{Gluon Tree Amplitudes in Open Twistor String
  Theory},'' \href{http://dx.doi.org/10.1088/1126-6708/2009/12/032}{{\em JHEP}
  {\bf 0912} (2009)  032},
\href{http://arxiv.org/abs/0909.0499}{{ arXiv:0909.0499 [hep-th]}}.

\bibitem{Dolan:2010xv}
L.~Dolan and P.~Goddard, ``{General Split Helicity Gluon Tree Amplitudes in
  Open Twistor String Theory},''
  \href{http://dx.doi.org/10.1007/JHEP05(2010)044}{{\em JHEP} {\bf 1005} (2010)
   044},
\href{http://arxiv.org/abs/1002.4852}{{ arXiv:1002.4852 [hep-th]}}.

\bibitem{ArkaniHamed:2009dg}
N.~Arkani-Hamed, J.~Bourjaily, F.~Cachazo, and J.~Trnka, ``{Unification of
  Residues and Grassmannian Dualities},''
  \href{http://dx.doi.org/10.1007/JHEP01(2011)049}{{\em JHEP} {\bf 1101} (2011)
   049},
\href{http://arxiv.org/abs/0912.4912}{{ arXiv:0912.4912 [hep-th]}}.

\bibitem{Nandan:2009cc}
D.~Nandan, A.~Volovich, and C.~Wen, ``{A Grassmannian \'{E}tude in NMHV
  Minors},'' \href{http://dx.doi.org/10.1007/JHEP07(2010)061}{{\em JHEP} {\bf
  1007} (2010)  061},
\href{http://arxiv.org/abs/0912.3705}{{ arXiv:0912.3705 [hep-th]}}.

\bibitem{Bourjaily:2010kw}
J.~L. Bourjaily, J.~Trnka, A.~Volovich, and C.~Wen, ``{The Grassmannian and the
  Twistor String: Connecting All Trees in $\mathcal{N}\!=\!4$ SYM},''
  \href{http://dx.doi.org/10.1007/JHEP01(2011)038}{{\em JHEP} {\bf 1101} (2011)
   038},
\href{http://arxiv.org/abs/1006.1899}{{ arXiv:1006.1899 [hep-th]}}.

\bibitem{ArkaniHamed:book}
N.~Arkani-Hamed, J.~L. Bourjaily, F.~Cachazo, A.~B. Goncharov, A.~Postnikov,
  and J.~Trnka, \href{http://dx.doi.org/10.1017/CBO9781316091548}{{\em
  {Grassmannian Geometry of Scattering Amplitudes}}}.
\newblock Cambridge University Press, 2016.

\bibitem{Olson:2014pfa}
T.~M. Olson, ``{Orientations of BCFW Charts on the Grassmannian},''
  \href{http://dx.doi.org/10.1007/JHEP08(2015)120}{{\em JHEP} {\bf 08} (2015)
  120}, \href{http://arxiv.org/abs/1411.6363}{{ arXiv:1411.6363 [hep-th]}}.

\bibitem{Weinberg:2010fx}
S.~Weinberg, ``{Six-Dimensional Methods for Four-Dimensional Conformal Field
  Theories},'' \href{http://dx.doi.org/10.1103/PhysRevD.82.045031}{{\em Phys.
  Rev.} {\bf D82} (2010)  045031},
\href{http://arxiv.org/abs/1006.3480}{{ arXiv:1006.3480 [hep-th]}}.

\bibitem{Weinberg:2020nsn}
S.~Weinberg, ``{Massless Particles in Higher Dimensions},''
  \href{http://dx.doi.org/10.1103/PhysRevD.102.095022}{{\em Phys. Rev. D} {\bf
  102} (2020) no. 9, 095022}, \href{http://arxiv.org/abs/2010.05823}{{
  arXiv:2010.05823 [hep-th]}}.

\bibitem{Mason:2009qx}
L.~Mason and D.~Skinner, ``{Dual Superconformal Invariance, Momentum Twistors
  and Grassmannians},''
  \href{http://dx.doi.org/10.1088/1126-6708/2009/11/045}{{\em JHEP} {\bf 0911}
  (2009)  045},
\href{http://arxiv.org/abs/0909.0250}{{ arXiv:0909.0250 [hep-th]}}.

\bibitem{ArkaniHamed:2009vw}
N.~Arkani-Hamed, F.~Cachazo, and C.~Cheung, ``{The Grassmannian Origin Of Dual
  Superconformal Invariance},''
  \href{http://dx.doi.org/10.1007/JHEP03(2010)036}{{\em JHEP} {\bf 1003} (2010)
   036},
\href{http://arxiv.org/abs/0909.0483}{{ arXiv:0909.0483 [hep-th]}}.

\bibitem{Penrose:1962ij}
R.~Penrose, ``{Asymptotic Properties of Fields and Space-Times},''
\href{http://dx.doi.org/10.1103/PhysRevLett.10.66}{{\em Phys. Rev. Lett.} {\bf
  10} (1963)  66--68}.

\bibitem{Penrose:1967wn}
R.~Penrose, ``{Twistor Algebra},''
\href{http://dx.doi.org/10.1063/1.1705200}{{\em J. Math. Phys.} {\bf 8} (1967)
  345}.

\bibitem{Penrose:1968me}
R.~Penrose, ``{Twistor Quantization and Curved Space-Time},''
\href{http://dx.doi.org/10.1007/BF00668831}{{\em Int. J. Theor. Phys.} {\bf 1}
  (1968)  61--99}.

\bibitem{Penrose:1972ia}
R.~Penrose and M.~A.~H. MacCallum, ``{Twistor Theory: An Approach to the
  Quantization of Fields and Space-Time},''
\href{http://dx.doi.org/10.1016/0370-1573(73)90008-2}{{\em Phys. Rept.} {\bf 6}
  (1972)  241--316}.

\bibitem{ArkaniHamed:2008gz}
N.~Arkani-Hamed, F.~Cachazo, and J.~Kaplan, ``{What is the Simplest Quantum
  Field Theory?},'' \href{http://dx.doi.org/10.1007/JHEP09(2010)016}{{\em JHEP}
  {\bf 1009} (2010)  016},
\href{http://arxiv.org/abs/0808.1446}{{ arXiv:0808.1446 [hep-th]}}.

\bibitem{Berends:1987me}
F.~A. Berends and W.~T. Giele, ``{Recursive Calculations for Processes with $n$
  Gluons},''
\href{http://dx.doi.org/10.1016/0550-3213(88)90442-7}{{\em Nucl. Phys.} {\bf
  B306} (1988)  759}.

\bibitem{Cachazo:2004kj}
F.~Cachazo, P.~Svrcek, and E.~Witten, ``{MHV Vertices and Tree Amplitudes in
  Gauge Theory},'' \href{http://dx.doi.org/10.1088/1126-6708/2004/09/006}{{\em
  JHEP} {\bf 09} (2004)  006},
\href{http://arxiv.org/abs/hep-th/0403047}{{ arXiv:hep-th/0403047}}.

\bibitem{Risager:2005vk}
K.~Risager, ``{A Direct Proof of the CSW Rules},'' {\em JHEP} {\bf 12} (2005)
  003,
\href{http://arxiv.org/abs/hep-th/0508206}{{ arXiv:hep-th/0508206}}.

\bibitem{Elvang:2008vz}
H.~Elvang, D.~Z. Freedman, and M.~Kiermaier, ``{Proof of the MHV Vertex
  Expansion for All Tree Amplitudes in $\mathcal{N}\!=\!4$ SYM Theory},''
  \href{http://dx.doi.org/10.1088/1126-6708/2009/06/068}{{\em JHEP} {\bf 06}
  (2009)  068}, \href{http://arxiv.org/abs/0811.3624}{{ arXiv:0811.3624
  [hep-th]}}.

\bibitem{Kiermaier:2009yu}
M.~Kiermaier and S.~G. Naculich, ``{A Super MHV Vertex Expansion for
  $\mathcal{N}\!=\!4$ sYM Theory},''
  \href{http://dx.doi.org/10.1088/1126-6708/2009/05/072}{{\em JHEP} {\bf 05}
  (2009)  072}, \href{http://arxiv.org/abs/0903.0377}{{ arXiv:0903.0377
  [hep-th]}}.

\bibitem{ArkaniHamed:2009sx}
N.~Arkani-Hamed, J.~Bourjaily, F.~Cachazo, and J.~Trnka, ``{Local Spacetime
  Physics from the Grassmannian},''
  \href{http://dx.doi.org/10.1007/JHEP01(2011)108}{{\em JHEP} {\bf 1101} (2011)
   108},
\href{http://arxiv.org/abs/0912.3249}{{ arXiv:0912.3249 [hep-th]}}.

\bibitem{Risager:2008yz}
K.~Risager, ``{Unitarity and On-Shell Recursion Methods for Scattering
  Amplitudes},''
\href{http://arxiv.org/abs/0804.3310}{{ arXiv:0804.3310 [hep-th]}}.

\bibitem{sage}
W.~Stein {\em et al.}, {\em {S}age {M}athematics {S}oftware ({V}ersion 9.5)}.
\newblock The Sage Development Team, YYYY.
\newblock {\tt http://www.sagemath.org}.

\bibitem{Melia:2013bta}
T.~Melia, ``{Dyck Words and Multiquark Primitive Amplitudes},''
  \href{http://dx.doi.org/10.1103/PhysRevD.88.014020}{{\em Phys. Rev. D} {\bf
  88} (2013) no. 1, 014020}, \href{http://arxiv.org/abs/1304.7809}{{
  arXiv:1304.7809 [hep-ph]}}.

\bibitem{Melia:2013epa}
T.~Melia, ``{Getting More Flavor Out of One-Flavor QCD},''
  \href{http://dx.doi.org/10.1103/PhysRevD.89.074012}{{\em Phys. Rev. D} {\bf
  89} (2014) no. 7, 074012}, \href{http://arxiv.org/abs/1312.0599}{{
  arXiv:1312.0599 [hep-ph]}}.

\bibitem{Melia:2015ika}
T.~Melia, ``{Proof of a New Colour Decomposition for QCD Amplitudes},''
  \href{http://dx.doi.org/10.1007/JHEP12(2015)107}{{\em JHEP} {\bf 12} (2015)
  107}, \href{http://arxiv.org/abs/1509.03297}{{ arXiv:1509.03297 [hep-ph]}}.

\bibitem{Johansson:2015oia}
H.~Johansson and A.~Ochirov, ``{Color-Kinematics Duality for QCD Amplitudes},''
  \href{http://dx.doi.org/10.1007/JHEP01(2016)170}{{\em JHEP} {\bf 01} (2016)
  170},
\href{http://arxiv.org/abs/1507.00332}{{ arXiv:1507.00332 [hep-ph]}}.

\bibitem{Ochirov:2019mtf}
A.~Ochirov and B.~Page, ``{Multi-Quark Colour Decompositions from Unitarity},''
  \href{http://dx.doi.org/10.1007/JHEP10(2019)058}{{\em JHEP} {\bf 10} (2019)
  058}, \href{http://arxiv.org/abs/1908.02695}{{ arXiv:1908.02695 [hep-ph]}}.

\bibitem{Arkani-Hamed:2017jhn}
N.~Arkani-Hamed, T.-C. Huang, and Y.-t. Huang, ``{Scattering Amplitudes for All
  Masses and Spins},'' \href{http://dx.doi.org/10.1007/JHEP11(2021)070}{{\em
  JHEP} {\bf 11} (2021)  070}, \href{http://arxiv.org/abs/1709.04891}{{
  arXiv:1709.04891 [hep-th]}}.

\bibitem{ArkaniHamed:2010kv}
N.~Arkani-Hamed, J.~L. Bourjaily, F.~Cachazo, S.~Caron-Huot, and J.~Trnka,
  ``{The All-Loop Integrand For Scattering Amplitudes in Planar
  $\mathcal{N}\!=\!4$ SYM},''
  \href{http://dx.doi.org/10.1007/JHEP01(2011)041}{{\em JHEP} {\bf 1101} (2011)
   041},
\href{http://arxiv.org/abs/1008.2958}{{ arXiv:1008.2958 [hep-th]}}.

\bibitem{CaronHuot:2010zt}
S.~Caron-Huot, ``{Loops and Trees},''
  \href{http://dx.doi.org/10.1007/JHEP05(2011)080}{{\em JHEP} {\bf 1105} (2011)
   080},
\href{http://arxiv.org/abs/1007.3224}{{ arXiv:1007.3224 [hep-ph]}}.

\bibitem{Boels:2011tp}
R.~H. Boels and R.~S. Isermann, ``{New Relations for Scattering Amplitudes in
  Yang-Mills Theory at Loop Level},''
  \href{http://dx.doi.org/10.1103/PhysRevD.85.021701}{{\em Phys. Rev.} {\bf
  D85} (2012)  021701},
\href{http://arxiv.org/abs/1109.5888}{{ arXiv:1109.5888 [hep-th]}}.

\bibitem{Bourjaily:2017wjl}
J.~L. Bourjaily, E.~Herrmann, and J.~Trnka, ``{Prescriptive Unitarity},''
  \href{http://dx.doi.org/10.1007/JHEP06(2017)059}{{\em JHEP} {\bf 06} (2017)
  059},
\href{http://arxiv.org/abs/1704.05460}{{ arXiv:1704.05460 [hep-th]}}.

\bibitem{Bourjaily:2021ewt}
J.~L. Bourjaily, C.~Langer, and K.~Patatoukos, ``{Locally-Finite Quantities in
  sYM},'' \href{http://dx.doi.org/10.1007/JHEP04(2021)298}{{\em JHEP} {\bf 04}
  (2021)  298}, \href{http://arxiv.org/abs/2102.02821}{{ arXiv:2102.02821
  [hep-th]}}.

\bibitem{Benincasa:2015zna}
P.~Benincasa, ``{On-Shell Diagrammatics and the Perturbative Structure of
  Planar Gauge Theories},''
\href{http://arxiv.org/abs/1510.03642}{{ arXiv:1510.03642 [hep-th]}}.

\bibitem{Benincasa:2016awv}
P.~Benincasa and D.~Gordo, ``{On-Shell Diagrams and the Geometry of Planar
  $\mathcal{N}\!<\!4$ SYM Theories},''
\href{http://arxiv.org/abs/1609.01923}{{ arXiv:1609.01923 [hep-th]}}.

\bibitem{Bourjaily:2021ujs}
J.~L. Bourjaily, E.~Herrmann, C.~Langer, K.~Patatoukos, J.~Trnka, and M.~Zheng,
  ``{Integrands of Less-Supersymmetric Yang-Mills at One Loop},''
  \href{http://dx.doi.org/10.1007/JHEP03(2022)126}{{\em JHEP} {\bf 03} (2022)
  126}, \href{http://arxiv.org/abs/2112.06901}{{ arXiv:2112.06901 [hep-th]}}.

\bibitem{Bourjaily:2020qca}
J.~L. Bourjaily, E.~Herrmann, C.~Langer, and J.~Trnka, ``{Building Bases of
  Loop Integrands},'' \href{http://dx.doi.org/10.1007/JHEP11(2020)116}{{\em
  JHEP} {\bf 11} (2020)  116}, \href{http://arxiv.org/abs/2007.13905}{{
  arXiv:2007.13905 [hep-th]}}.

\bibitem{Franco:2012mm}
S.~Franco, ``{Bipartite Field Theories: from D-Brane Probes to Scattering
  Amplitudes},'' \href{http://dx.doi.org/10.1007/JHEP11(2012)141}{{\em JHEP}
  {\bf 1211} (2012)  141},
\href{http://arxiv.org/abs/1207.0807}{{ arXiv:1207.0807 [hep-th]}}.

\bibitem{Franco:2012wv}
S.~Franco, D.~Galloni, and R.-K. Seong, ``{New Directions in Bipartite Field
  Theories},'' \href{http://dx.doi.org/10.1007/JHEP06(2013)032}{{\em JHEP} {\bf
  1306} (2013)  032},
\href{http://arxiv.org/abs/1211.5139}{{ arXiv:1211.5139 [hep-th]}}.

\bibitem{Weinberg:1964ew}
S.~Weinberg, ``{Photons and Gravitons in $S$-Matrix Theory: Derivation of
  Charge Conservation and Equality of Gravitational and Inertial Mass},''
  \href{http://dx.doi.org/10.1103/PhysRev.135.B1049}{{\em Phys. Rev.} {\bf 135}
  (1964)  B1049--B1056}.

\bibitem{Grisaru:1977px}
M.~T. Grisaru and H.~N. Pendleton, ``{Some Properties of Scattering Amplitudes
  in Supersymmetric Theories},''
  \href{http://dx.doi.org/10.1016/0550-3213(77)90277-2}{{\em Nucl. Phys. B}
  {\bf 124} (1977)  81--92}.

\bibitem{Grisaru:1976vm}
M.~T. Grisaru, H.~N. Pendleton, and P.~van Nieuwenhuizen, ``{Supergravity and
  the S Matrix},'' \href{http://dx.doi.org/10.1103/PhysRevD.15.996}{{\em Phys.
  Rev. D} {\bf 15} (1977)  996}.

\bibitem{Elvang:2009wd}
H.~Elvang, D.~Z. Freedman, and M.~Kiermaier, ``{Solution to the Ward Identities
  for Superamplitudes},'' \href{http://dx.doi.org/10.1007/JHEP10(2010)103}{{\em
  JHEP} {\bf 1010} (2010)  103},
\href{http://arxiv.org/abs/0911.3169}{{ arXiv:0911.3169 [hep-th]}}.

\bibitem{Elvang:2010xn}
H.~Elvang, D.~Z. Freedman, and M.~Kiermaier, ``{SUSY Ward Identities,
  Superamplitudes, and Counterterms},''
  \href{http://dx.doi.org/10.1088/1751-8113/44/45/454009}{{\em J. Phys. A} {\bf
  44} (2011)  454009}, \href{http://arxiv.org/abs/1012.3401}{{ arXiv:1012.3401
  [hep-th]}}.

\bibitem{Bern:1998sv}
Z.~Bern, L.~J. Dixon, M.~Perelstein, and J.~Rozowsky, ``{Multileg One Loop
  Gravity Amplitudes from Gauge Theory},''
  \href{http://dx.doi.org/10.1016/S0550-3213(99)00029-2}{{\em Nucl. Phys.} {\bf
  B546} (1999)  423--479},
\href{http://arxiv.org/abs/hep-th/9811140}{{ arXiv:hep-th/9811140}}.

\bibitem{Cremmer:1978ds}
E.~Cremmer and B.~Julia, ``{The $\mathcal{N}\!=\!8$ Supergravity Theory. 1. The
  Lagrangian},'' \href{http://dx.doi.org/10.1016/0370-2693(78)90303-9}{{\em
  Phys. Lett. B} {\bf 80} (1978)  48}.

\bibitem{Cremmer:1978km}
E.~Cremmer, B.~Julia, and J.~Scherk, ``{Supergravity Theory in
  Eleven-Dimensions},''
  \href{http://dx.doi.org/10.1016/0370-2693(78)90894-8}{{\em Phys. Lett. B}
  {\bf 76} (1978)  409--412}.

\bibitem{Cremmer:1979up}
E.~Cremmer and B.~Julia, ``{The $SO(8)$ Supergravity},''
  \href{http://dx.doi.org/10.1016/0550-3213(79)90331-6}{{\em Nucl. Phys. B}
  {\bf 159} (1979)  141--212}.

\bibitem{deWit:1977fk}
B.~de~Wit and D.~Z. Freedman, ``{On $SO(8)$ Extended Supergravity},''
  \href{http://dx.doi.org/10.1016/0550-3213(77)90395-9}{{\em Nucl. Phys. B}
  {\bf 130} (1977)  105--113}.

\bibitem{deWit:1982bul}
B.~de~Wit and H.~Nicolai, ``{$\mathcal{N}\!=\!8$ Supergravity},''
  \href{http://dx.doi.org/10.1016/0550-3213(82)90120-1}{{\em Nucl. Phys. B}
  {\bf 208} (1982)  323}.

\bibitem{Gaillard:1981rj}
M.~K. Gaillard and B.~Zumino, ``{Duality Rotations for Interacting Fields},''
  \href{http://dx.doi.org/10.1016/0550-3213(81)90527-7}{{\em Nucl. Phys. B}
  {\bf 193} (1981)  221--244}.

\bibitem{Adler:1964um}
S.~L. Adler, ``{Consistency Conditions on the Strong Interactions Implied by a
  Partially Conserved Axial Vector Current},''
  \href{http://dx.doi.org/10.1103/PhysRev.137.B1022}{{\em Phys. Rev.} {\bf 137}
  (1965)  B1022--B1033}.

\bibitem{Weinberg:1966kf}
S.~Weinberg, ``{Pion Scattering Lengths},''
  \href{http://dx.doi.org/10.1103/PhysRevLett.17.616}{{\em Phys. Rev. Lett.}
  {\bf 17} (1966)  616--621}.

\bibitem{williams2004enumeration}
L.~K. Williams, ``{Enumeration of Totally Positive Grassmann Cells},'' 2004.

\end{thebibliography}
